\documentclass[10pt]{amsart}
\usepackage{amsmath,bm,natbib,amssymb,graphicx,epsfig,color}
\usepackage[left=3cm,right=3cm,top=3cm,bottom=3cm]{geometry}

\def \cal{\mathcal}

\newcommand{\real}{\ensuremath{\mathbb{R}}}

\newcommand{\ltwo}{\ensuremath{\mathbb{L}^2}}

\newcommand{\inner}[2]{\left\langle #1,#2 \right\rangle}

\numberwithin{equation}{section}
\theoremstyle{plain}
\newtheorem{proposition}{\small{Proposition}}
\theoremstyle{definition}
\newtheorem{example}{\small{Example}}
\newtheorem{definition}{\small{Definition}}
\theoremstyle{remark}

\usepackage{multirow}

\begin{document}

\title{Bayesian Sensitivity Analysis with Fisher-Rao Metric}
\author{Sebastian Kurtek}\address{Department of Statistics; 404 Cockins Hall; 1958 Neil Avenue; The Ohio State University; Columbus, OH 43210.}\email{kurtek.1@stat.osu.edu}
\author{Karthik Bharath} \address{Department of Statistics; 404 Cockins Hall; 1958 Neil Avenue; The Ohio State University; Columbus, OH 43210.} \email{bharath.4@osu.edu}
%
\date{}
\maketitle
\begin{abstract}
We propose a geometric framework to assess sensitivity of Bayesian procedures to modeling assumptions based on the nonparametric Fisher-Rao metric. While the framework is general in spirit, the focus of this article is restricted to metric-based diagnosis under two settings: assessing local and global robustness in Bayesian procedures to perturbations of the likelihood and prior, and identification of influential observations. The approach is based on the square-root representation of densities which enables one to compute geodesics and geodesic distances in analytical form, facilitating the definition of naturally calibrated local and global discrepancy measures. An important feature of our approach is the definition of a geometric $\epsilon$-contamination class of sampling distributions and priors via intrinsic analysis on the space of probability density functions. We showcase the applicability of our framework on several simulated toy datasets as well as in real data settings for generalized mixed effects models, directional data and shape data. 
\end{abstract}

\section{Introduction}
The main ingredients in a Bayesian model involve a likelihood $f(x|\theta)$ and a prior distribution $\pi(\theta)$, where $x$ is the given data and $\theta$ is a set of unknown parameters. Interest then is in performing inference on this set of parameters using the posterior distribution $p(\theta|x)\propto f(x|\theta)\pi(\theta)$ or some functional thereof. It is then necessarily important to develop diagnostic procedures to assess the influence of the data, prior and likelihood on posterior inference, of which the three main include: (1) detection of outlying or influential observations (robustness to perturbation of the data); (2) global sensitivity to the perturbation of the likelihood or the prior over a suitable class; and (3), local sensitivity to perturbations of the likelihood or the prior. Such assessments in the Bayesian setting have received considerable attention over the years; we refer the interested reader to \cite{RBA} for a detailed account of the foundational, methodological and implementation issues.

Global Bayesian sensitivity analysis is characterized by derivation of measures from variational properties of posterior functionals, such as their ranges, over a class of prior or likelihood perturbations (\cite{berger1994, berger2, BB, FS}). On the other hand, local Bayes robustness methods are based on the derivatives of posterior functionals with respect to a small perturbation of the likelihood or prior; see \cite{RM, GW, PGust} in this regard. Outlier detection in the Bayesian setting using divergence or other discrepancy measures to ascertain ``distances" between posteriors have been employed by several authors (\cite{PengDey,CP, PG, DeyBirmiwal,GD}). Since the posterior distribution contains all information about the unknown $\theta$, comparing the full model posterior with the posterior resulting following case-deletion seems reasonable. In the case of local perturbations of the likelihood and prior, and case-deletion measures using divergences, considerations of the geometry of the nonlinear manifold of densities would appear to be significant. This line of work, more from a frequentist perspective, has a rich history starting with seminal work of \cite{cook}; also see, for example, \cite{HZ1}, \cite{HZ3} and \cite{HZ2}. In a Bayesian setting, \cite{HZ} elegantly constructed a Riemannian-geometric Bayesian perturbation model, which provided a geometric background upon which different perturbations to a Bayesian model, currently used in practice, could be embedded onto.

In this paper, in similar spirit to the work by \cite{HZ}, we propose a comprehensive framework for Bayes sensitivity analysis based on the manifold of probability densities using the square-root representation; the framework is comprehensive in the sense that global, local and data perturbations to the Bayesian model are developed under the same geometric setup. Without striving for utmost generality we ensure that \emph{the entirety of our sensitivity analysis, the global, local and data perturbation, and subsequent inference, is performed intrinsically on the space of densities under a unified Riemannian metric}. The key difference to the approach by \cite{HZ} is that we utilize the nonparametric version of the Fisher-Rao Riemannian metric in contrast to the parametric version employed in their work. The advantage of working with the nonparametric Fisher-Rao metric is that, under the square-root transformation of the densities, the geometry of the space of probability density functions becomes the positive orthant of the Hilbert unit sphere, and the Riemannian metric reduces to the standard $\ltwo$ metric. This allows one to develop analytic tools for generating perturbations of density functions as well as computing geodesic distances, which are actual distances bounded above by $\pi/2$. Additionally, unlike the square-root representation used in this paper, the log transformation of densities used by \cite{HZ} requires the densities to be strictly positive. Our approach allows one to circumvent some issues associated with divergences like $\phi$-divergences (\cite{csiszar}), the Kullback-Leibler divergence (\cite{KL}) or the functional Bregman divergence (\cite{FSG}), and divergence-based measures: lack of symmetry (comparing the full model to the model under case-deletion is different from comparing the model under case-deletion to the full model); violation of the triangle inequality; and, the unboundedness and absence of natural scale.

We note that there exist several approaches in literature using the Hellinger distance to quantify differences between distributions; see for example, \cite{WH,BL,RB}. In fact, this distance can be viewed as the extrinsic version of the distance we utilize in this paper. The Hellinger distance also satisfies symmetry and triangle inequality and has an upper bound of $\sqrt{2}$. But, in order to utilize the intrinsic structure of the manifold of densities to define perturbation classes as well as local sensitivity measures, we choose to work with the intrinsic metric. Furthermore, the intrinsic metric respects the geometry of the space we are working on while the Hellinger distance provides distances in the ambient space.

As an alternative to global sensitivity measures for prior and likelihood perturbations currently available in literature, we propose a novel geometric $\epsilon$-contamination class, and propose measures based on geodesic distances. In the local setup, we propose local sensitivity measures for the Bayes factor, posterior mean (which can be easily extended to other posterior functionals) and the geodesic distance. Importantly, these sensitivity measures are derived using directional derivatives on the space of posterior distributions giving them a natural geometric calibration. For perturbations of the data, we propose to use the geodesic distance (under the Fisher-Rao metric) to measure differences between posterior distributions.

An important advantage of the proposed framework is that the geodesic distances are available in closed form and can hence be computed quickly and exactly. This stands in contrast to the geodesic distance used in \cite{HZ} which requires approximation via Dijkstra's algorithm (\cite{EWD:NumerMath59}). This severely affects situations wherein many such distances need to be computed, and the accuracy of the approximation can affect the inference. In higher dimension, this issue is exacerbated since the estimated distance heavily depends on the nature of the discretization of the space. Also, the approximations suffer from what is called metrication error: roughly, the distance computed using Dijkstra's algorithm does not generally converge to the true geodesic distance with increasing resolution of the grid (\cite{CohenK97}). We now summarize the main contributions of this paper as:
\begin{enumerate}
\item Definition of a novel $\epsilon$-contamination class for likelihood and prior sensitivity analysis based on the geometry of the space of probability density functions;
\item Definition of geometrically calibrated local and global sensitivity measures to the perturbation of the likelihood and prior;
\item Identification of influential observations using geodesic distances between posterior distributions.
\end{enumerate}

\section{Fisher-Rao Metric and Representation Space of Probability Density Functions}\label{geometric framework}

For simplicity, we restrict our attention to the case of univariate densities on $\mathbb{R}$. We note, however, that the framework is equally valid for all finite dimensional distributions. Denote by $\mathcal{P}$, the Banach manifold of probability density functions on $\mathbb{R}$, defined as ${\cal P}=\{p:\real\to\real_{\geq 0}|\int_\real p(x) dx = 1\}.$ The space $\mathcal{P}$ is not a vector space but a manifold with a boundary because any density function whose value is zero for any $x \in \real$ is a boundary element. For a point $p$ in the interior of ${\cal P}$, define the tangent space as $T_p({\cal P})=\{\delta p:\real\to \real | \int_\real \delta p(x) p(x) dx = 0\}.$ Intuitively, the tangent space at any point $p$ on the manifold ${\cal P}$ contains all possible perturbations of the density function $p$. For any two tangent vectors $\delta p_1, \delta p_2 \in T_p({\cal P})$, the nonparametric version of the Fisher-Rao Riemannian metric (simply referred to as Fisher-Rao metric hereafter) is given by (\cite{c-r-rao:45,amari85,kass-vos-geometry-book,srivastava-etal-Fisher-Rao-CVPR:2007}):
\begin{equation}\label{FRmetric}
\langle\langle\delta p_1 , \delta p_2\rangle\rangle_p = \int_\real
\delta p_1(x) \delta p_2(x) \frac{1}{p(x)} dx.
\end{equation}
An important property of this metric is that it is invariant to re-parameterization (\cite{Cencov82}). This metric has already proven to be very useful for various tasks in computer vision, shape analysis and functional data analysis (\cite{SJJ2007,SKJJ2011,KurtekJASA,EFA:10}).
One drawback in using the Fisher-Rao metric in practice is the difficulty associated with computing geodesic paths and distances. This difficulty stems from the fact that the Riemannian metric changes from point to point on the manifold. It is hence important to choose a suitable representation of the space $\mathcal{P}$ which simplifies these computations. Depending on the choice of the representation, the resulting Riemannian structure can have varying degrees of complexity requiring numerical techniques to approximate geodesics. Choices of representation include the CDF, the log density, etc. Unfortunately, none of these representations alleviate the problem of computing geodesics (\cite{SJJ2007}).

The square-root representation proposed by \cite{bhattacharya-43} provides an elegant solution to this problem. In particular, under this representation, the Fisher-Rao metric becomes the standard $\ltwo$ metric and the space of probability density functions becomes the positive orthant of the unit hypersphere in $\ltwo$ (see Appendix for more details). This leads to the following definition.
\begin{definition}\label{def:SRT}
Define a continuous mapping $\phi:{\cal P}\mapsto\Psi$ where the space $\Psi$ is the space containing the positive square-root of all possible density functions. Using this mapping, define the square-root transform (SRT) of probability density functions as $\phi(p)=\psi=+\sqrt{p}$. Note, that the inverse mapping is simply $\phi^{-1}(\psi)=p=\psi^2$.
\end{definition}
\noindent We omit the $+$ sign from the representation for notational convenience. The space of all square-root transform (SRT) representations of probability density functions is $\Psi=\{ \psi:\real \to \real_{\geq 0} | \int_\real |\psi(x)|^2 dx = 1\}$ and represents the positive orthant of the Hilbert sphere (\cite{lang-geometry}). Since the differential geometry of the sphere is well known, one can compute geodesic paths and distances between probability density functions analytically. Our general approach in the remainder of the paper will be to represent probability density functions using their SRT representation, compute quantities of interest on $\Psi$, and then map them back to ${\cal P}$ using the inverse mapping provided in Definition \ref{def:SRT}.

\subsection{Geometry of Unit Hilbert Hypersphere} \label{sec:geometry-background}

In this section, we describe the tools relevant to our analysis based on the geometry of $\Psi$. To begin, the $\ltwo$ Riemmanian metric on $\Psi$ is defined as $\langle \delta \psi_1 , \delta \psi_2 \rangle= \int_\real \delta \psi_1(x)\delta \psi_2(x) dx$, where $\delta \psi_1, \delta \psi_2 \in T_\psi(\Psi)$ and $T_{\psi}(\Psi) = \big\{ \delta\psi| \inner{\delta \psi}{\psi} = 0\big\}$. Next, we are interested in the geodesic path and distance between two points in $\Psi$. Observe that since we are on the unit infinite dimensional sphere, the geodesic distance between any two points is given by the angle between them. In other words, the {\it geodesic distance} between $\psi_1,\ \psi_2\in\Psi$, is given by $ d(\psi_1,\psi_2)=\theta=\cos^{-1}(\langle \psi_{1} , \psi_{2}\rangle)$. The {\it geodesic path} between $\psi_1$ and $\psi_2$ (indexed by $\tau \in [0,1]$) is given by $\eta(\tau) =\frac{1}{\sin(\theta)}[\sin(\theta-\tau\theta)\psi_{1}+\sin( \tau \theta)\psi_{2}]$. The restriction to the positive orthant of the unit sphere does not pose any additional difficulties: for two points $\psi_1,\ \psi_2\in\Psi$ the shortest geodesic between them is entirely contained in $\Psi$. It is easy to see that $\theta$ is bounded above by $\frac{\pi}{2}$, which imposes an upper bound on the geodesic distance between probability densities.

\begin{figure}[!h]
\begin{center}
\includegraphics[width=3in]{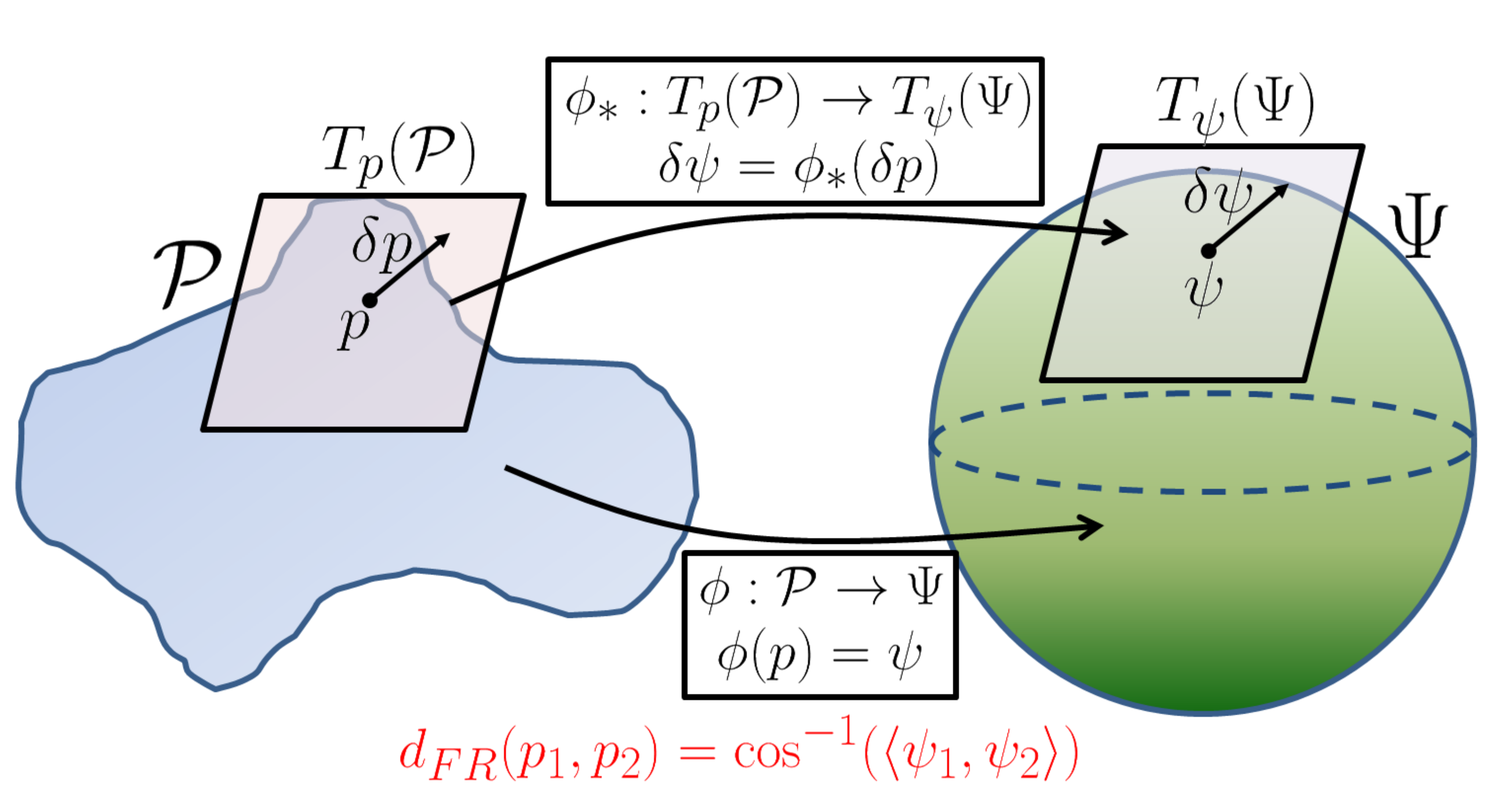}
\end{center}
\caption{Description of the square-root transformation from $\mathcal{P}$ to the positive orthant of the unit Hilbert sphere $\Psi$. On $\mathcal{P}$, at a point (density) $p$, its tangent space $T_p(\mathcal{P})$ is shown along with the corresponding tangent vector $\delta p$. These quantities are mapped to the tangent space of $\psi$ on $\Psi$ and the counterparts are displayed in a similar manner. Note the isometric property: $d_{FR}(p_1,p_2)=\cos^{-1}(\langle\psi_1,\psi_2 \rangle)$.}\label{geo_fig}
\end{figure}

In the proposed Bayes sensitivity analysis framework we will utilize various geometric tools including the exponential and inverse exponential maps. The exponential map at a point $\psi_1 \in \Psi$, denoted by $\exp: T_{\psi_1}(\Psi) \mapsto \Psi$, is defined as $\exp_{\psi_1}(\delta \psi) = \cos(\|\delta \psi\|)\psi_1+ \sin(\|\delta \psi\|)\frac{\delta \psi}{\|\delta \psi\|}.$ The purpose of this map is to map points from the tangent space to the representation space. The inverse exponential map, denoted by $\exp^{-1}_{\psi_1}: \Psi \mapsto T_{\psi_1}(\Psi)$, is given by $\exp^{-1}_{\psi_1}(\psi_2) = [\frac{\theta}{\sin(\theta)}\left(\psi_2 - \cos(\theta) \psi_1\right)],$ and can be used to map points from the representation space to the tangent space. Figure \ref{geo_fig} presents a pictorial description of the relationship between ${\cal P}$ and $\Psi$.

\begin{figure}[!ht]
\centering
\begin{tabular}{c c c}
Fisher-Rao Geodesic & Straight Line & Midpoint \\
\includegraphics[width=1.5in]{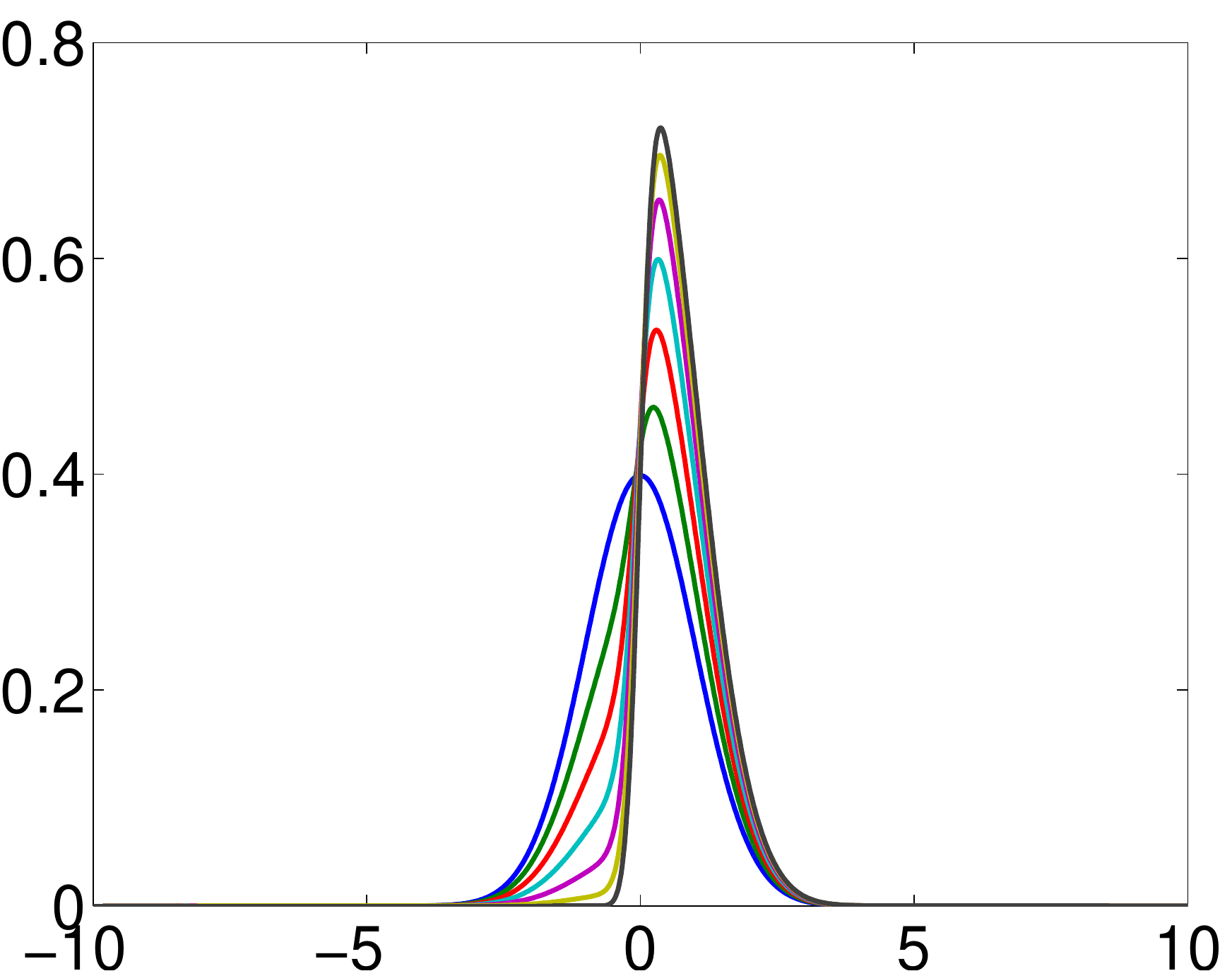}&\includegraphics[width=1.5in]{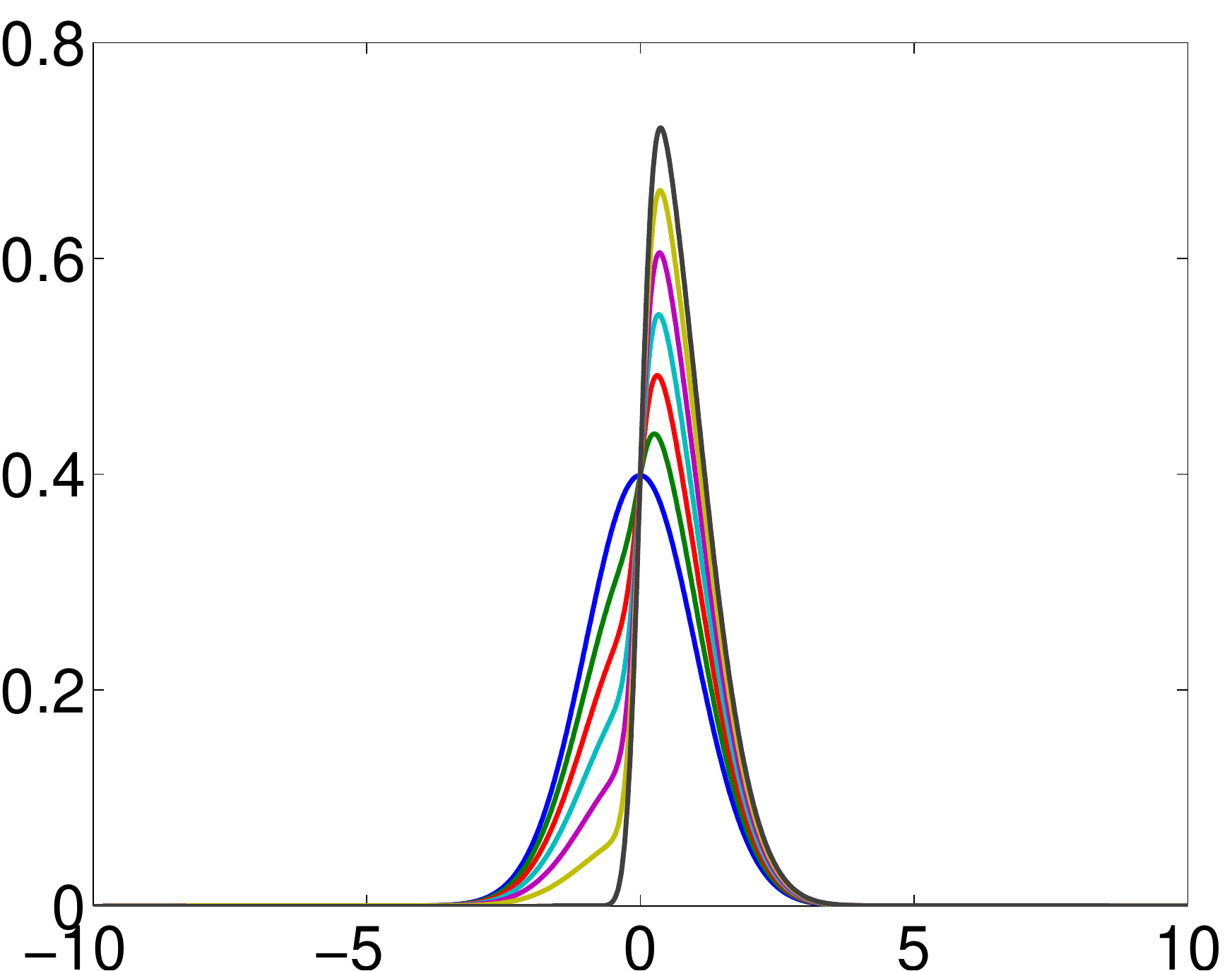}&\includegraphics[width=1.5in]{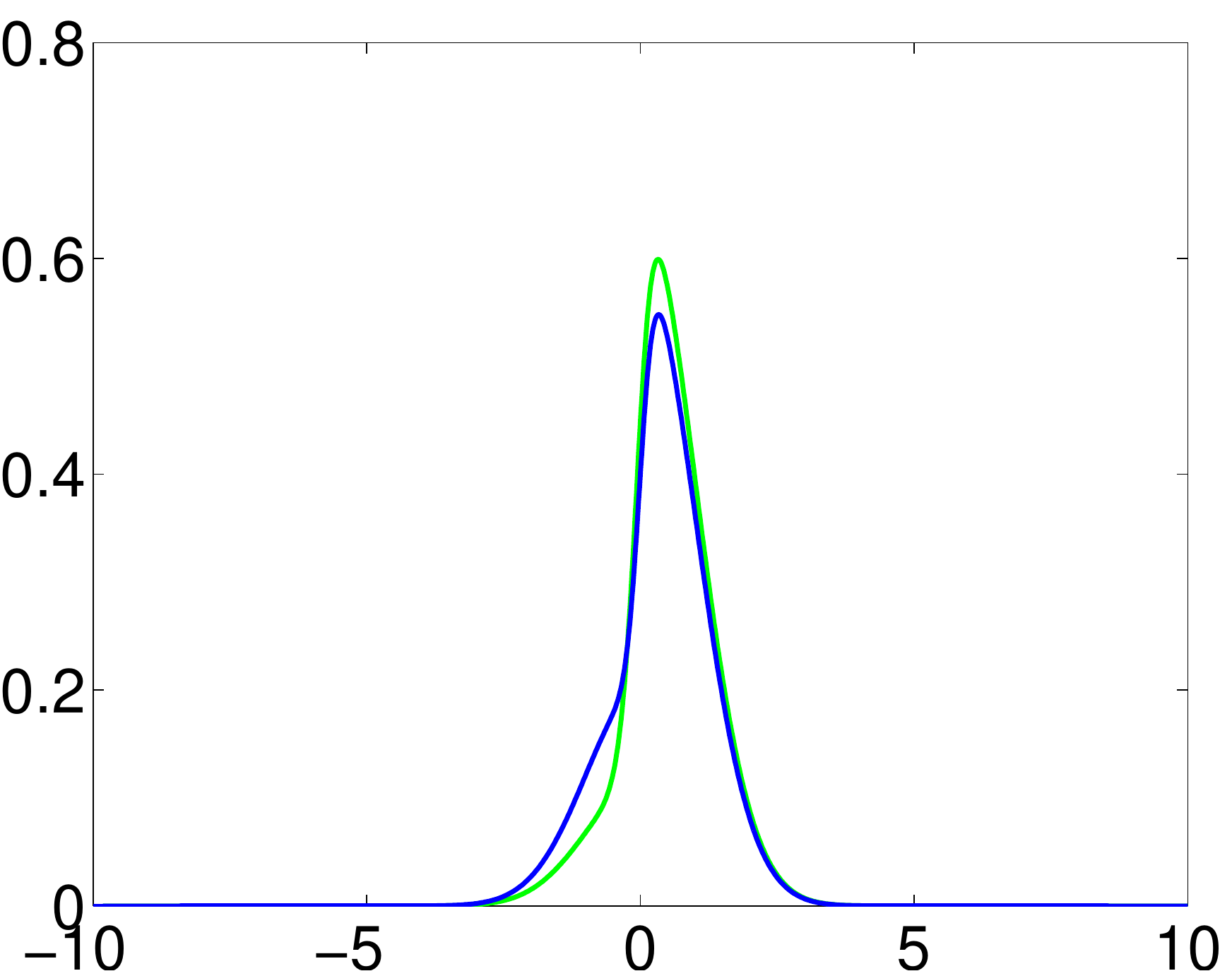}\\
\multicolumn{3}{c}{$d_{FR}(p_1,p_2)=0.6700$; $KL(p_1,p_2)=6.6692$; $KL(p_2,p_1)=0.5520$}\\
\vspace*{3mm}
\end{tabular}

\begin{tabular}{c c c c c c c c}
\includegraphics[width=.5in]{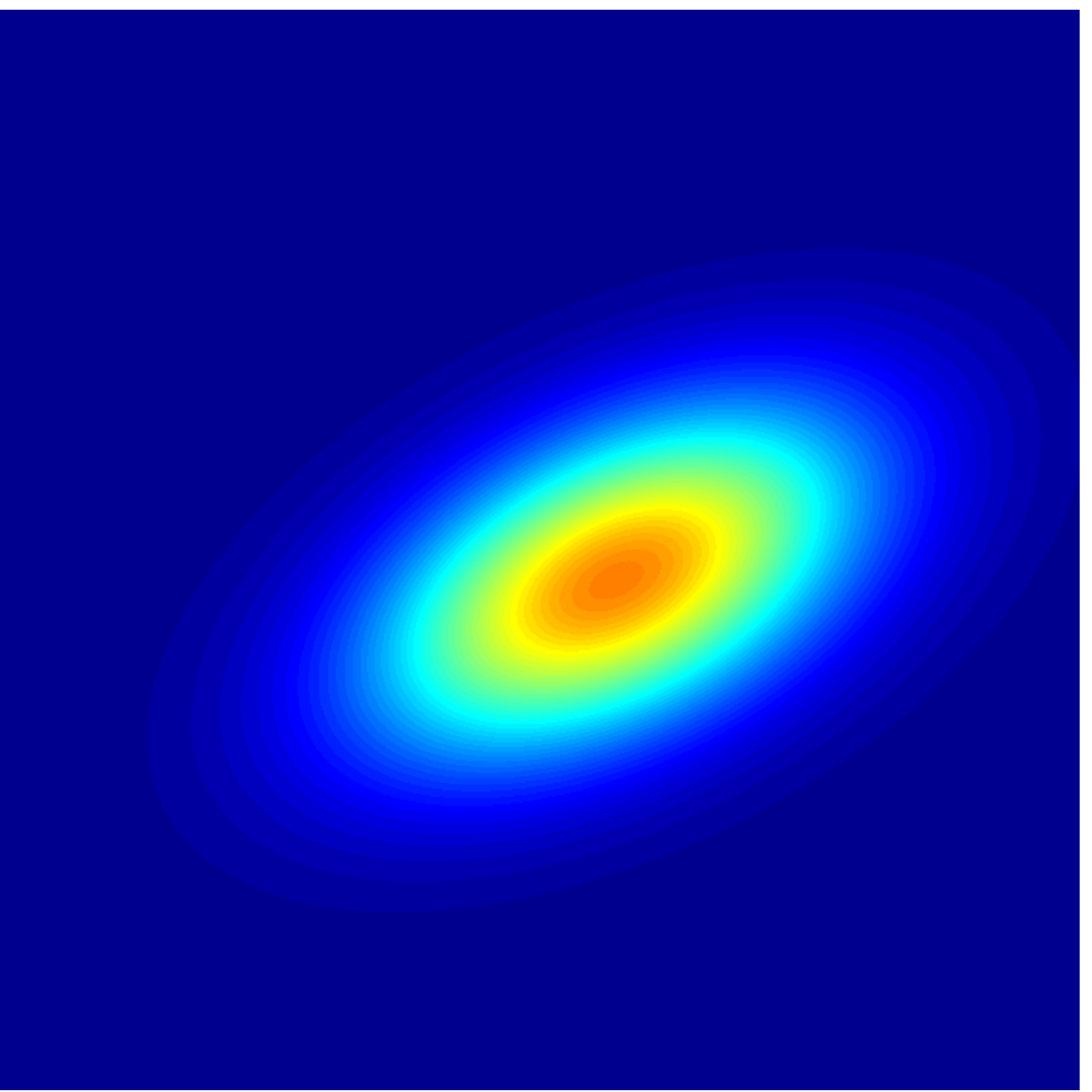}&\includegraphics[width=.5in]{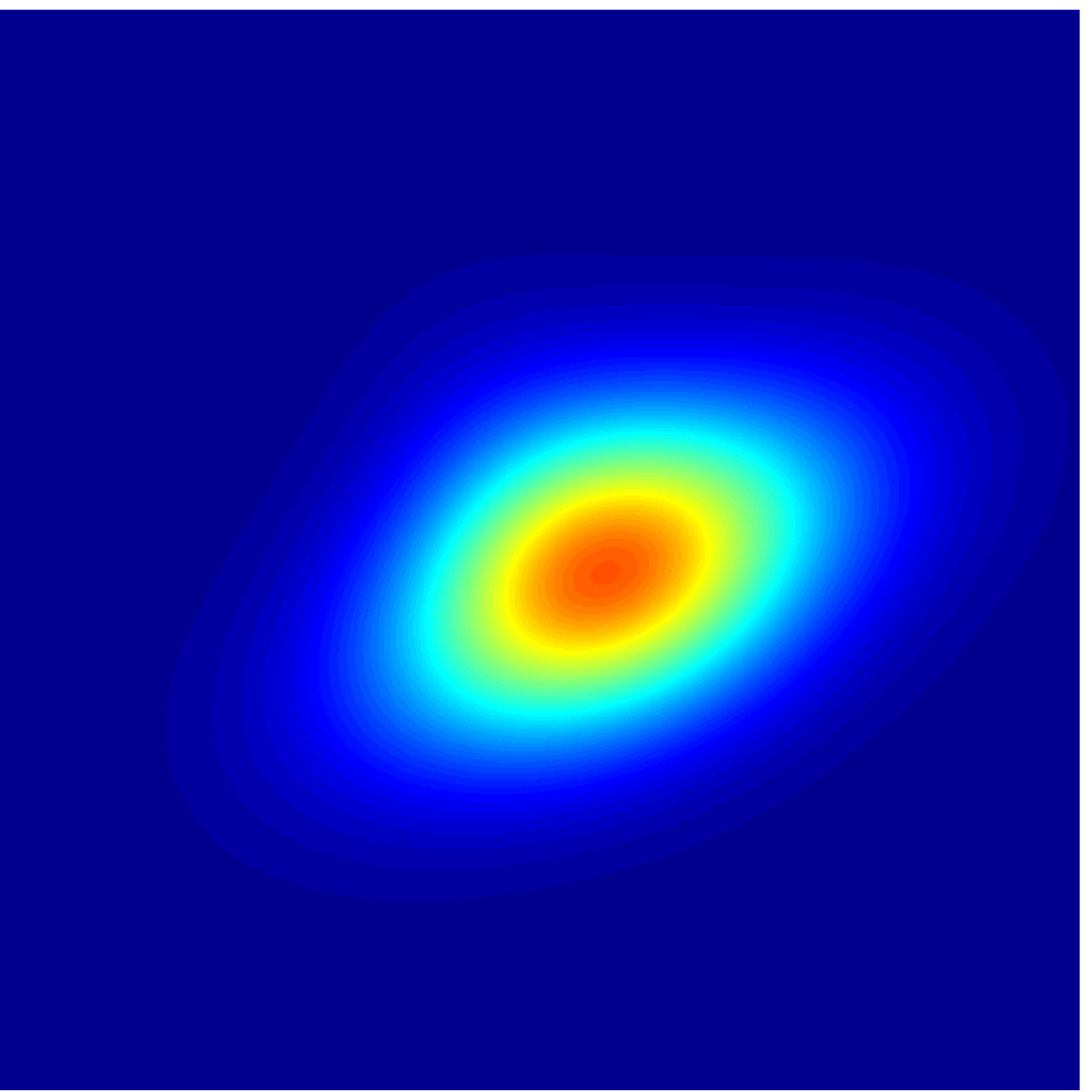}&\includegraphics[width=.5in]{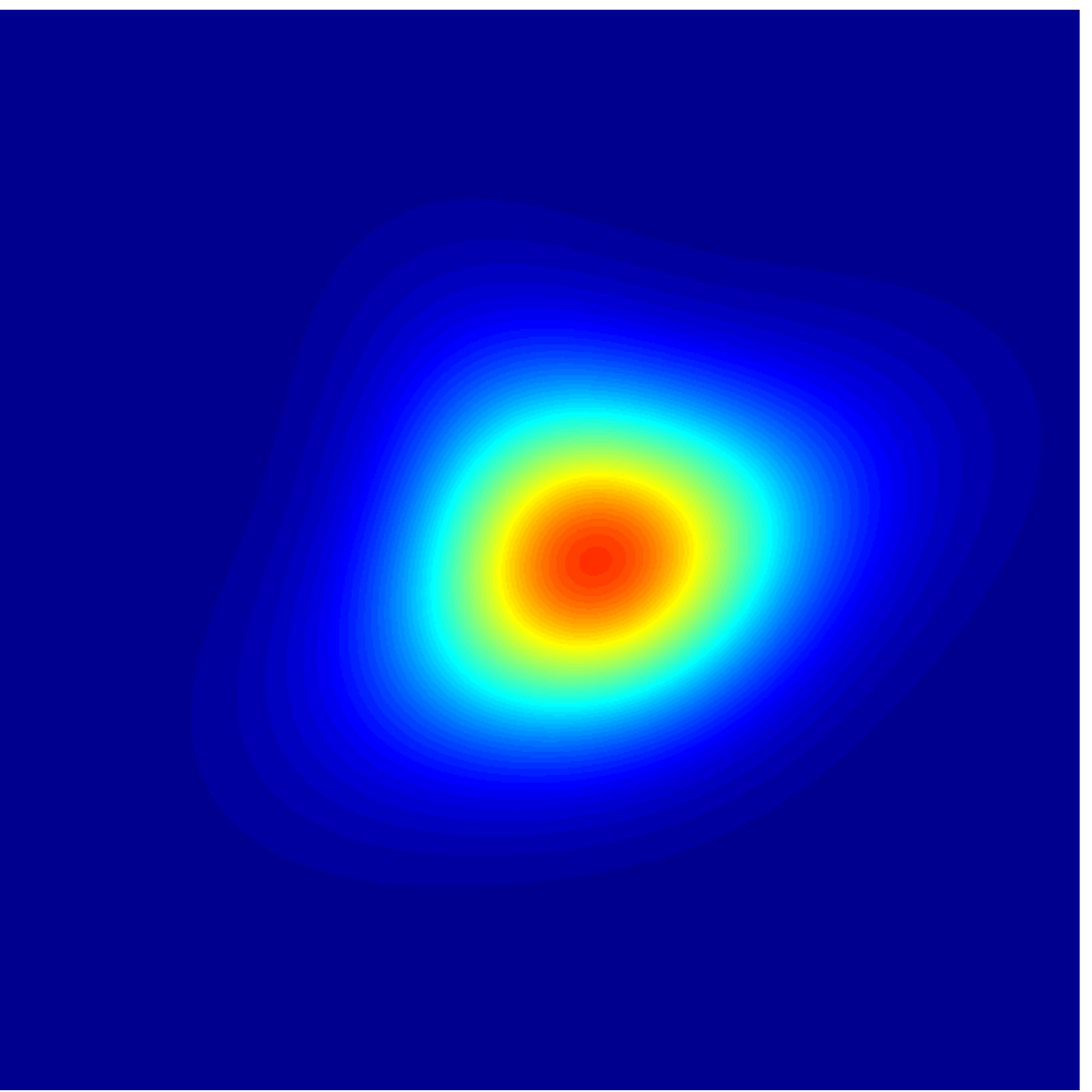}&\includegraphics[width=.5in]{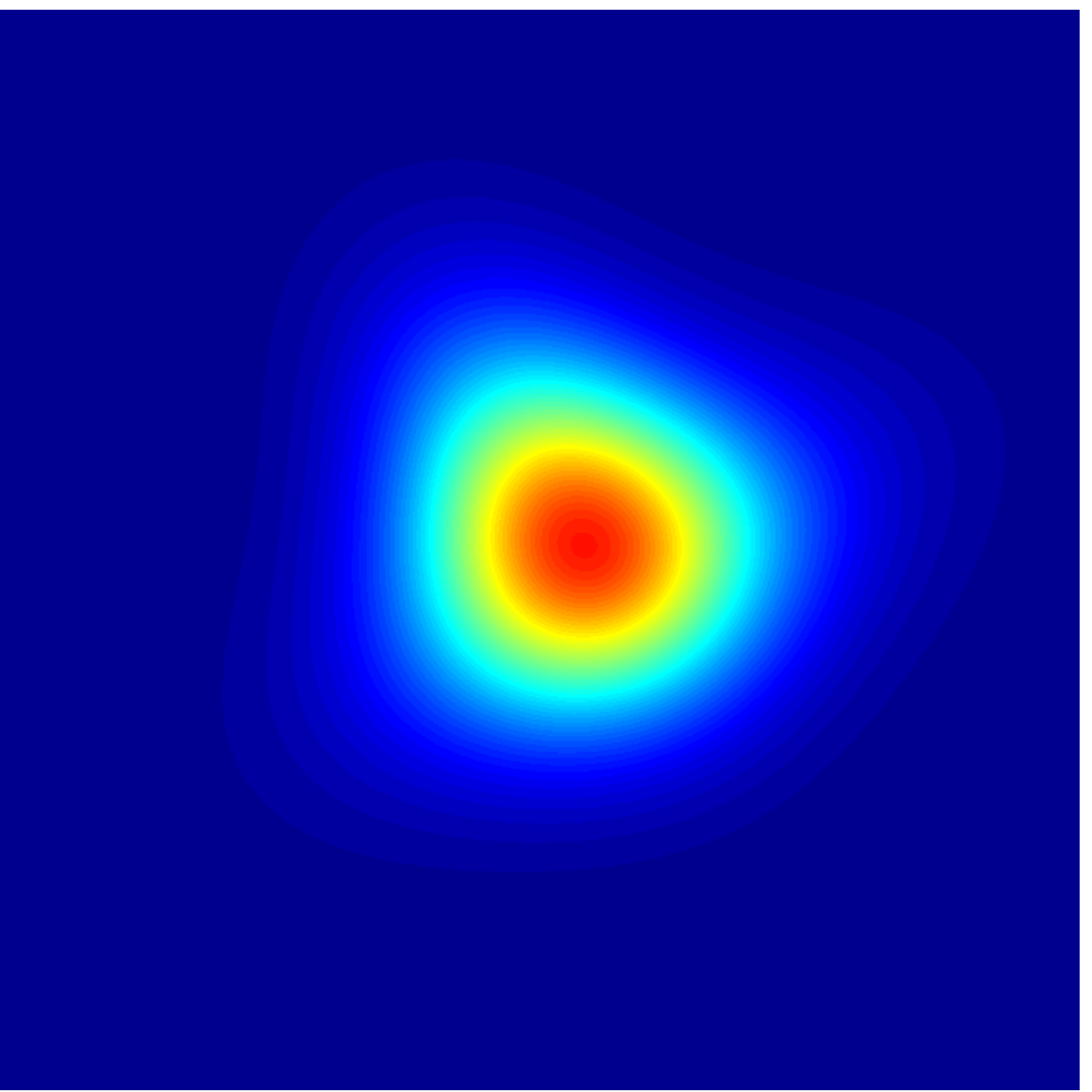}&\includegraphics[width=.5in]{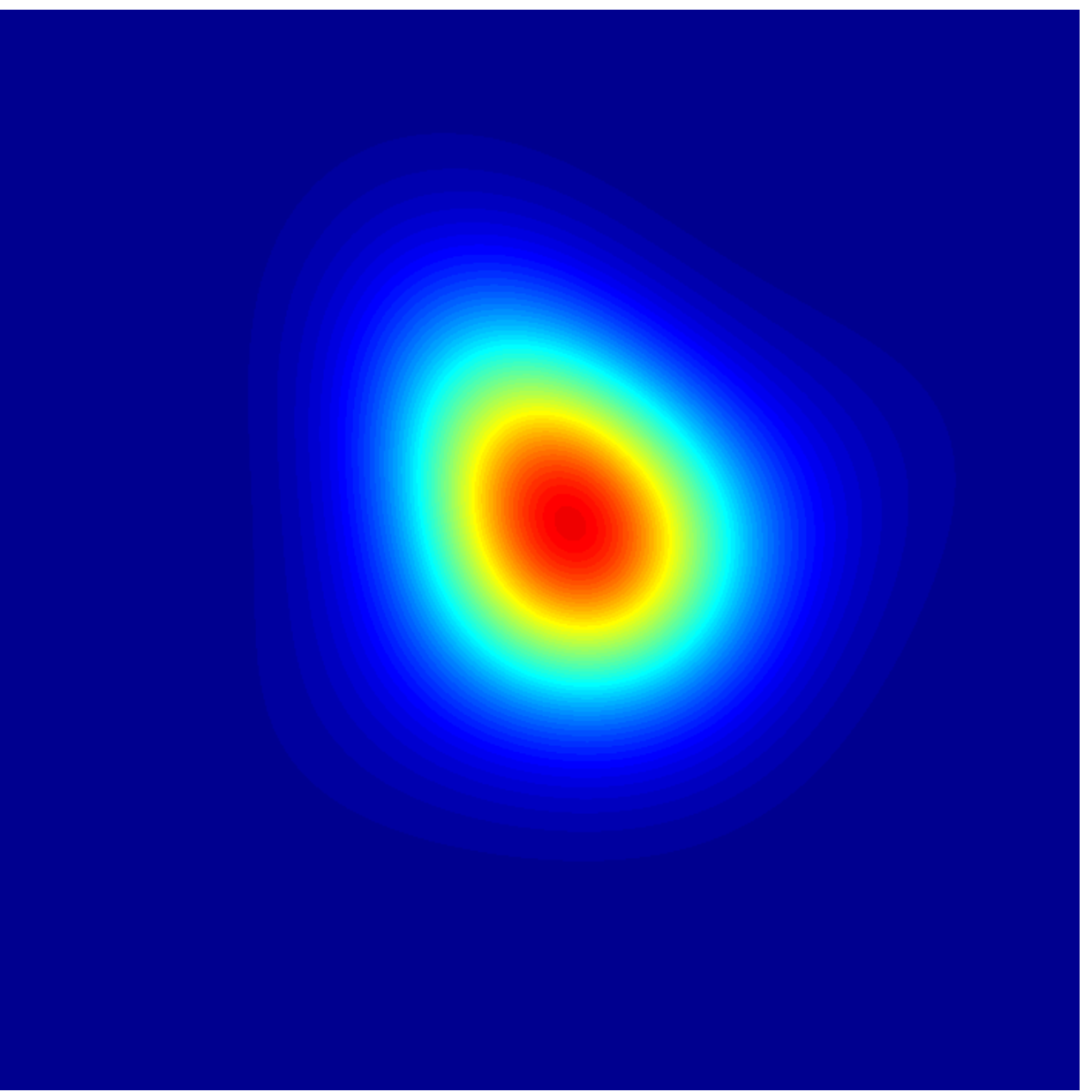}&\includegraphics[width=.5in]{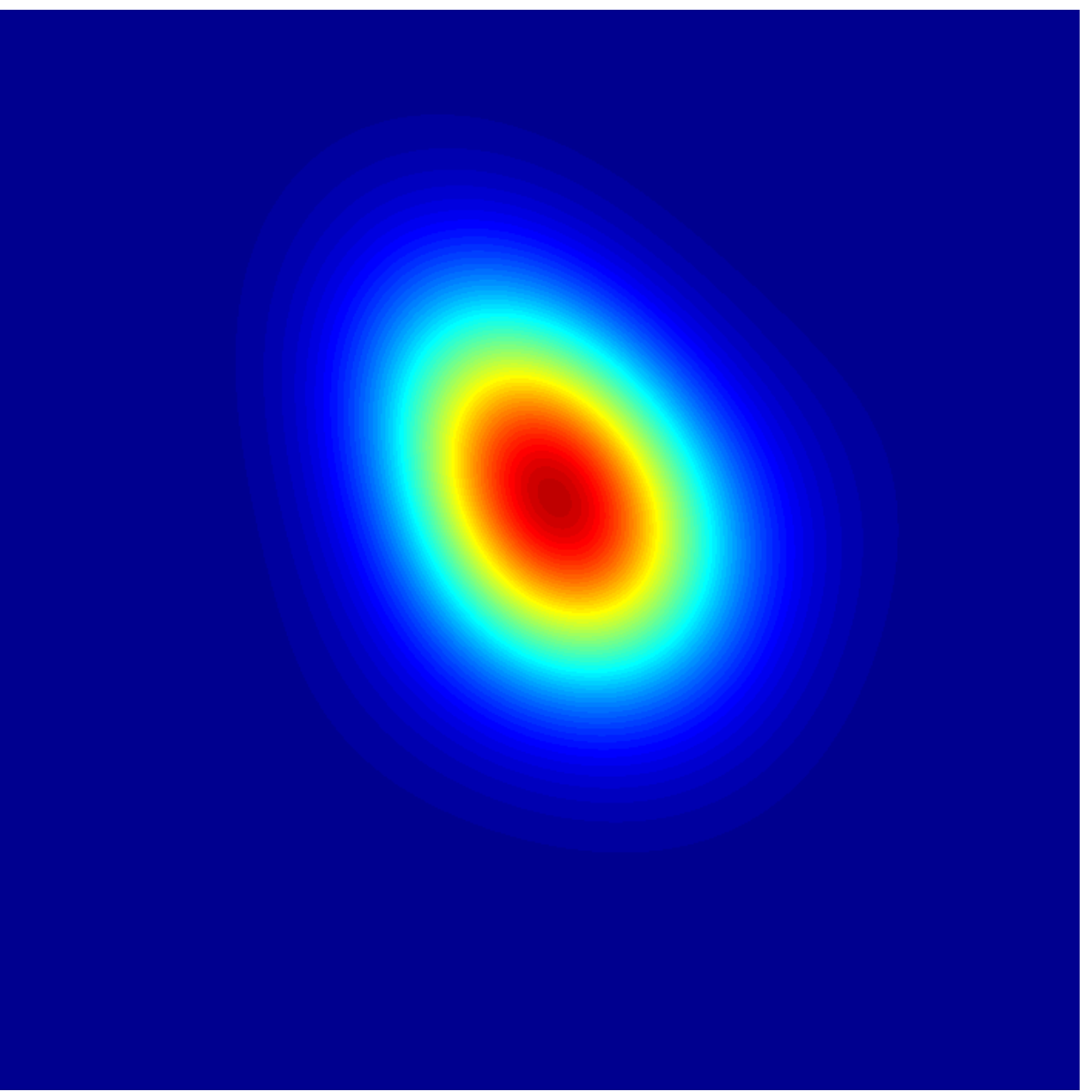}&\multirow{2}{*}[2.5em]{\includegraphics[width=1.2in]{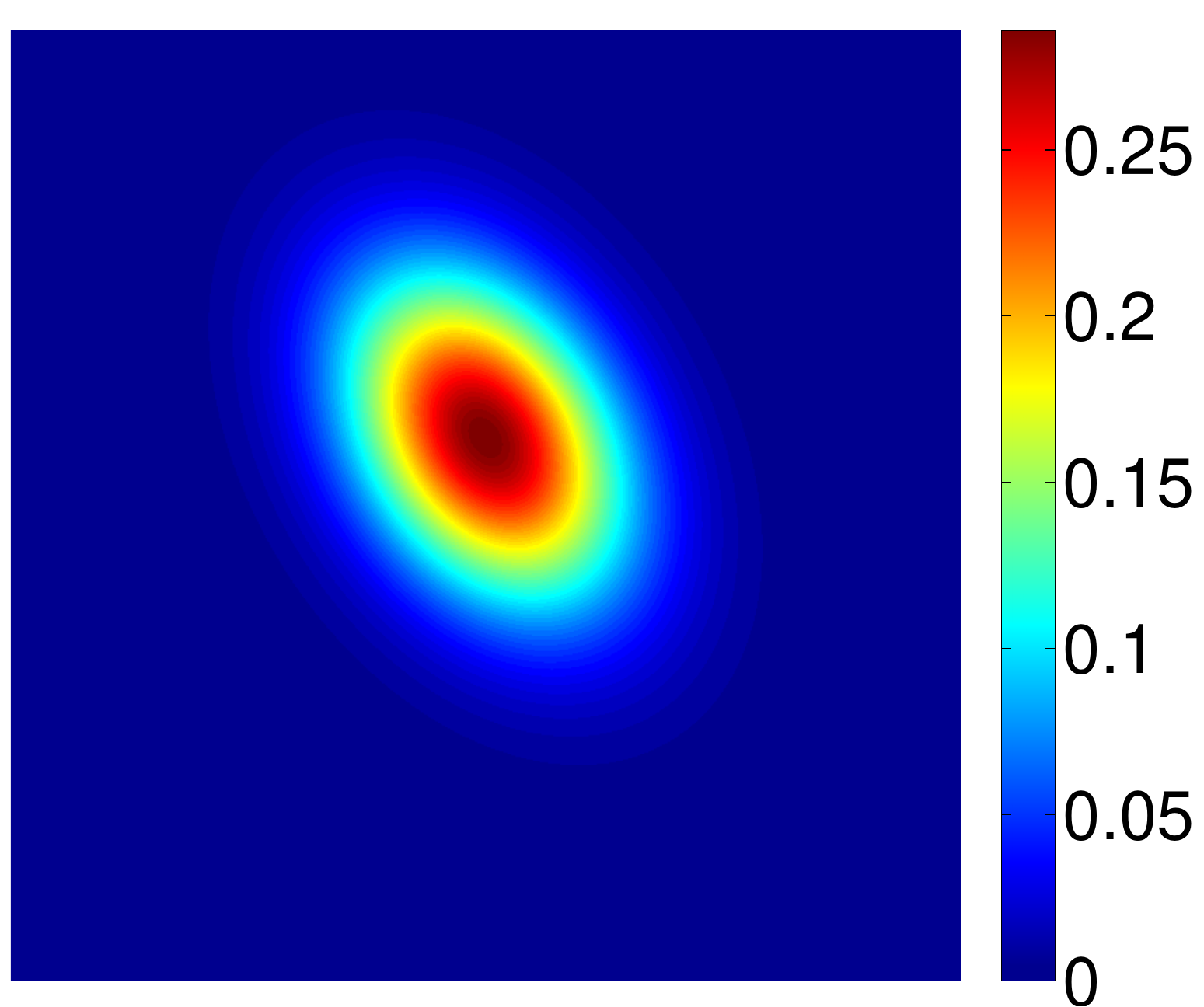}}\\
\includegraphics[width=.5in]{figures/I1}&\includegraphics[width=.5in]{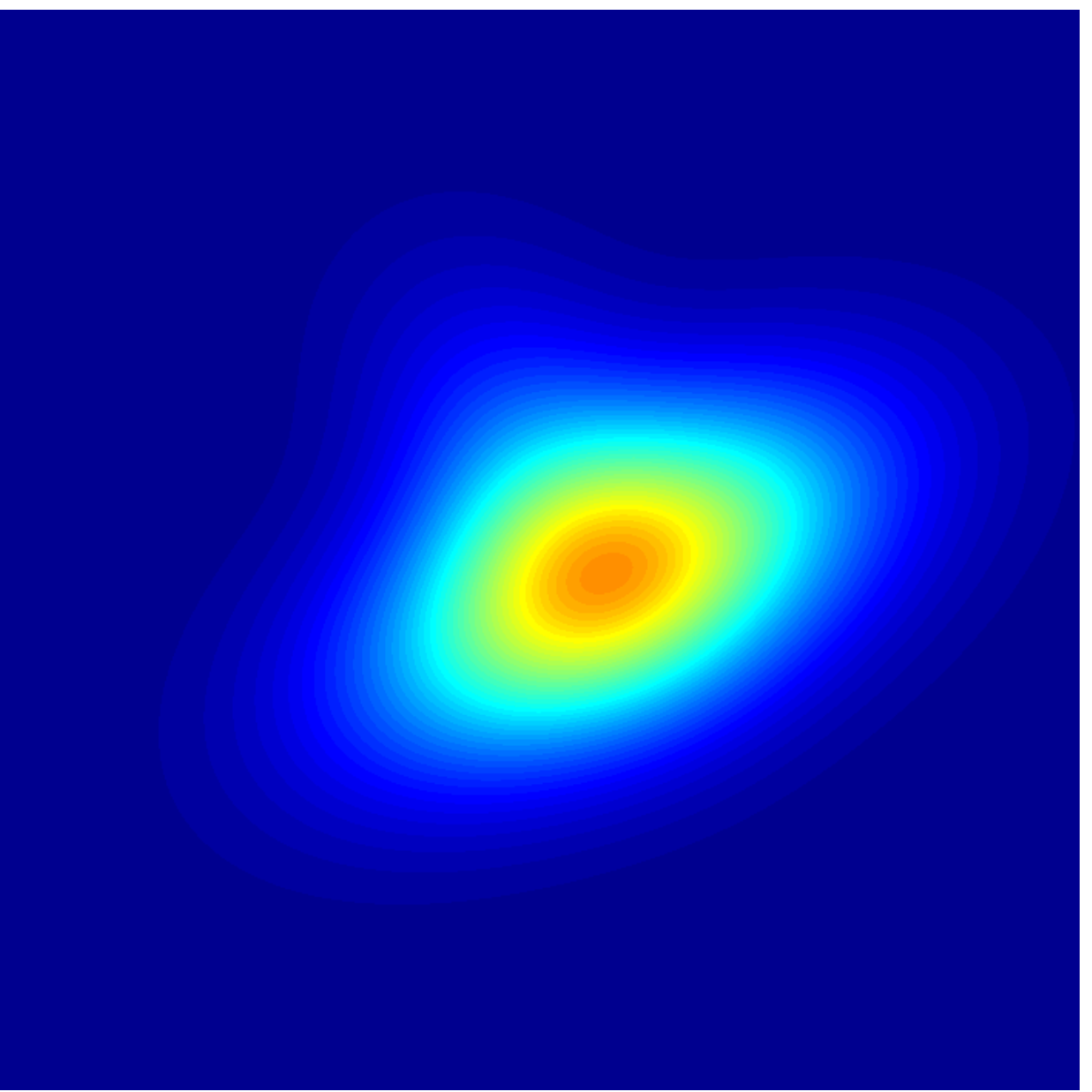}&\includegraphics[width=.5in]{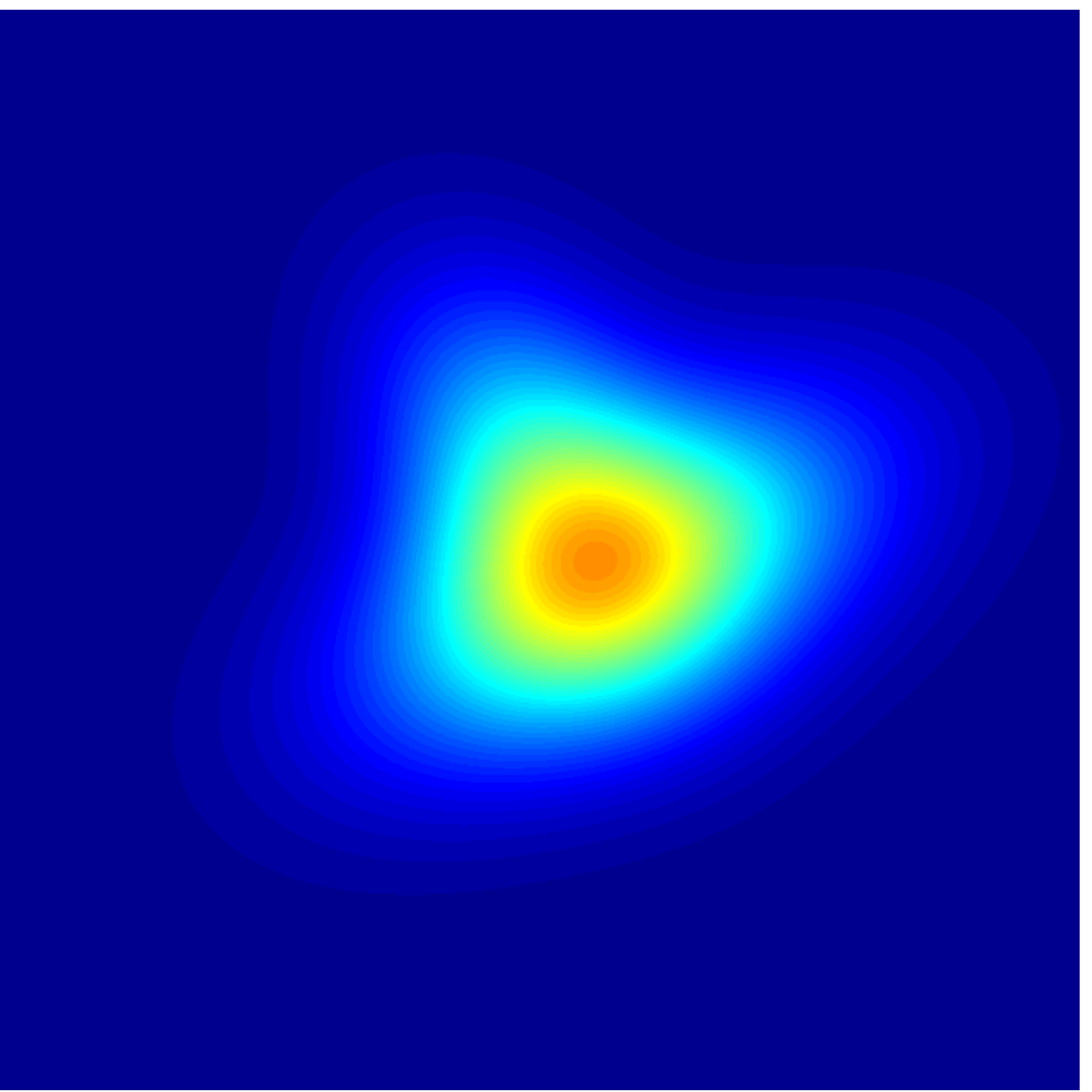}&\includegraphics[width=.5in]{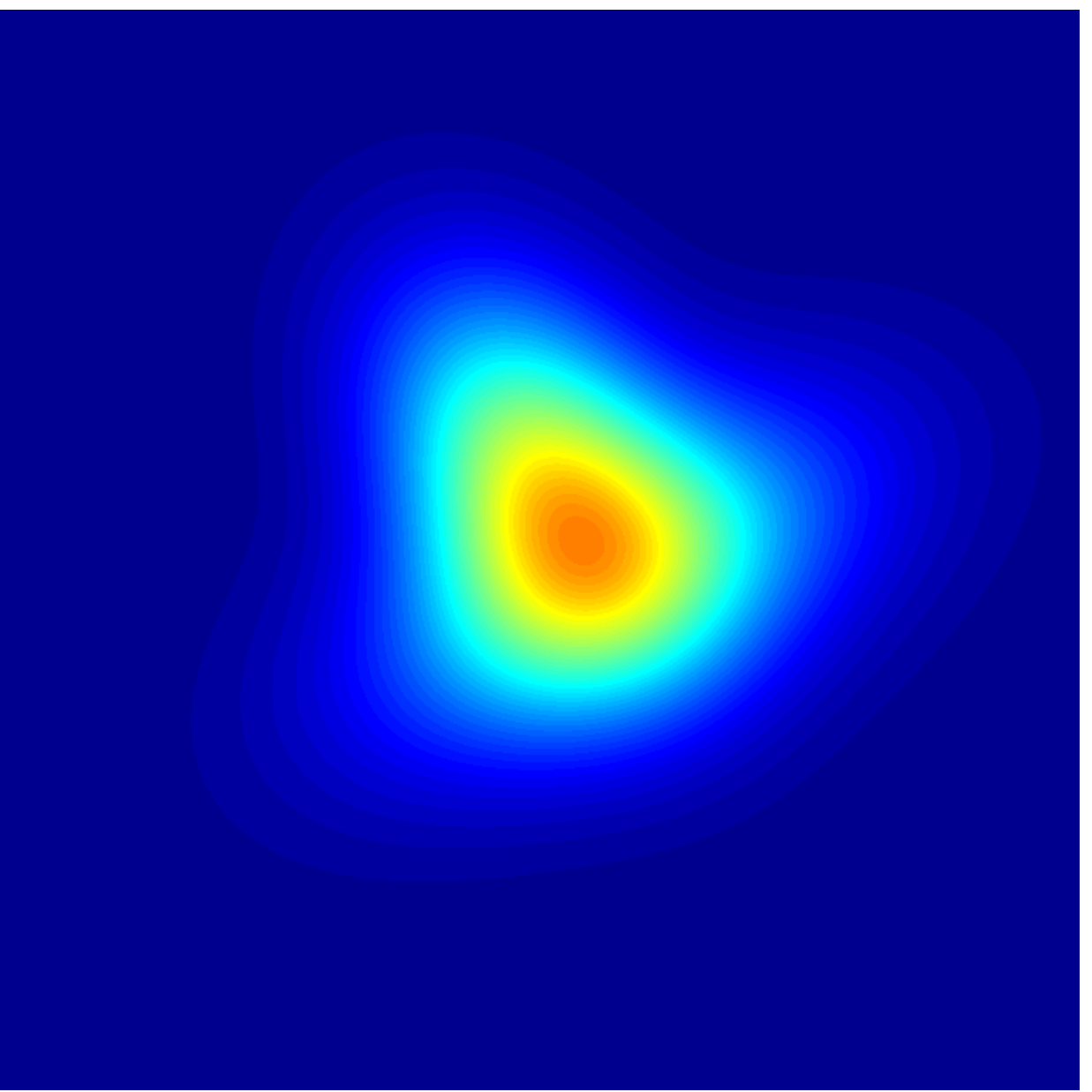}&\includegraphics[width=.5in]{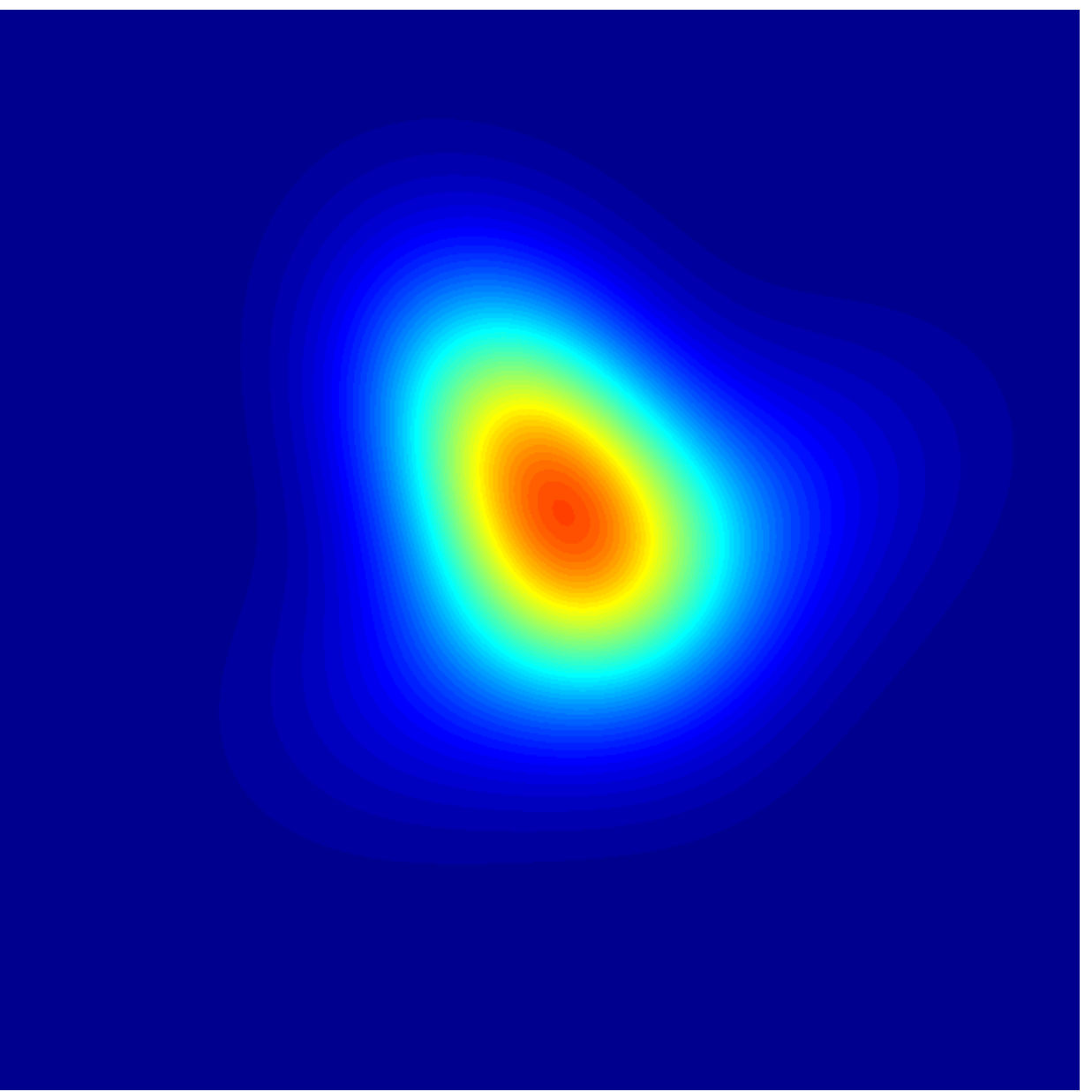}&\includegraphics[width=.5in]{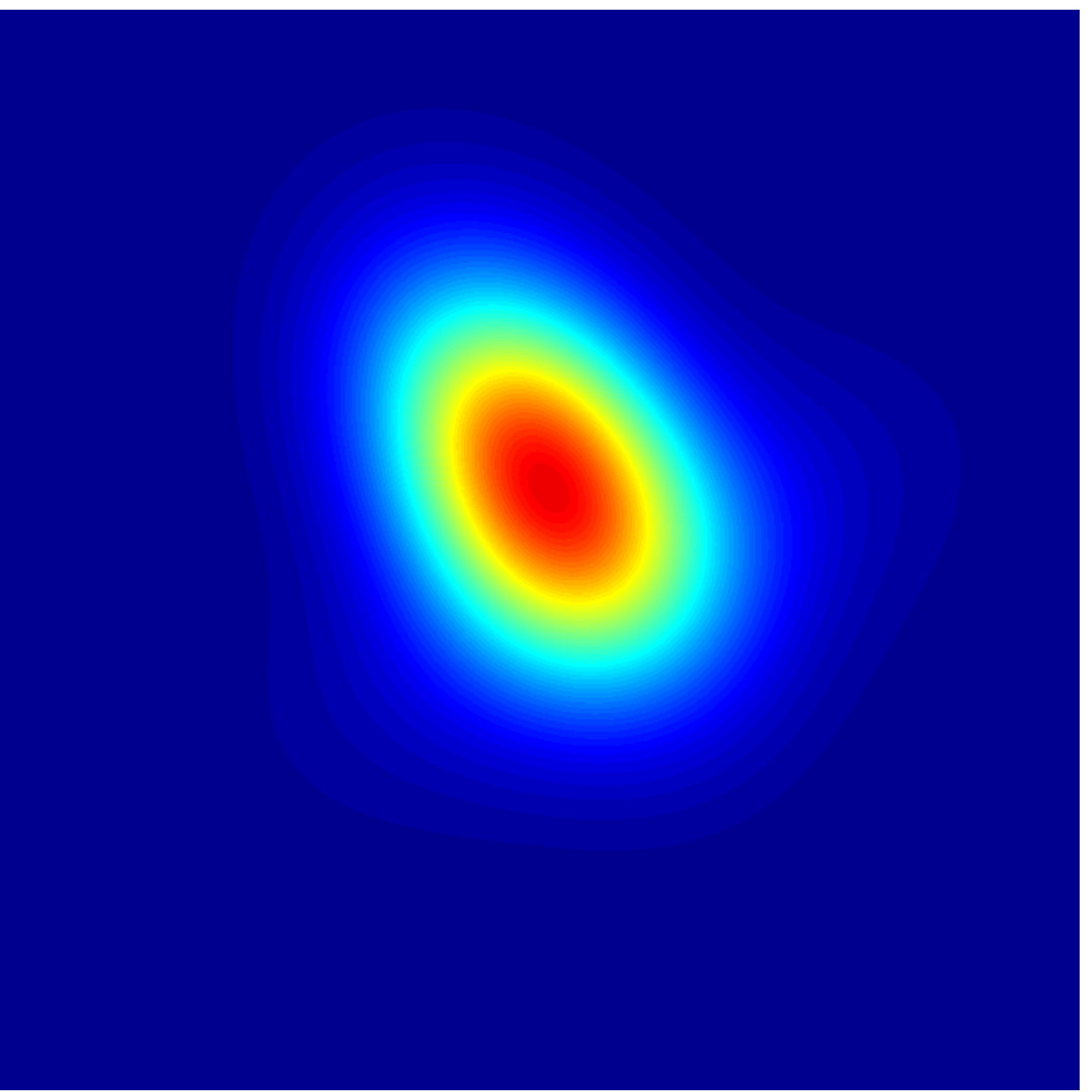}\\
\multicolumn{7}{c}{$d_{FR}(p_1,p_2)=0.7157$; $KL(p_1,p_2)=1.2522$; $KL(p_2,p_1)=1.3653$.}\\
\end{tabular}
\caption{\label{fig:geodex} Top: Comparison of a normal and a skew normal distribution using the Fisher-Rao metric. Bottom: Comparison of two bivariate normals with different means and covariances using the Fisher-Rao metric. The top path is the geodesic and the bottom path is the straight line interpolation (the start and end points are the same in both cases).}
\end{figure}

For illustrative purposes, we consider two simple examples where we compare probability density functions using geodesic paths and distances. We compare our approach with a straight line interpolation between the densities which contains no geometric information of the underlying space. We also plot the midpoints of the two paths (in the first example only) to highlight the difference. In addition, we contrast the values of the geodesic Fisher-Rao distance $d_{FR}$ with the Kullback-Leibler divergence (KL). The first example, presented in the top row of Figure \ref{fig:geodex}, considers comparing the standard normal distribution, $p_1 \sim N(0,1)$, to a skew normal distribution with skewness parameter 5, $p_2 \sim SN(5)$. The Fisher-Rao distance between $p_1$ and $p_2$ is 0.67, while the KL divergence between them is 6.6692. When we switch the arguments the Fisher-Rao distance remains the same while the KL divergence changes drastically to 0.5520. Thus, it is difficult to reconcile the two dissimilarity values provided by the KL divergence. We also note the difference between the linear interpolation path and the geodesic path between these two densities. The geodesic path accounts for the nonlinearity of the underlying space. The second example, presented in the bottom panel of Figure \ref{fig:geodex}, considers comparing two bivariate Gaussian densities emphasizing the generality of this approach to finite dimensional densities. In particular, we consider two bivariate normal distributions $p_1 \sim N(\mu_1,\Sigma_1)$ and $p_2 \sim N(\mu_2,\Sigma_2)$ where $\mu_1=\begin{bmatrix}
.5\\
.2
\end{bmatrix}
,
\Sigma_1=\begin{bmatrix}
1.2 & .4\\
.4 & .6
\end{bmatrix} \text{ and}
\quad
\mu_2=\begin{bmatrix}
0\\
.5
\end{bmatrix},
\Sigma_2=\begin{bmatrix}
.5 & -.2\\
-.2 & .7
\end{bmatrix}$.
The Kullback Leibler divergence is close to being symmetric in this case. The Fisher Rao distance is 0.7151. As in the previous example, the geodesic path differs significantly from the linear interpolation.

\section{Geometric Perturbation Class}\label{prior}

We are now in a position to develop a geometric framework for assessing prior and likelihood robustness based on the notion of $\epsilon$-contamination. We use the following notation in the rest of this paper: $X$ denotes the observable random variable which will be assumed to have a density $f(x|\theta)$ with respect to Lebesgue measure where $\theta$ is a vector (finite or infinite) of unknown parameters lying in a parameter space $\Theta$. A prior density on $\Theta$ is denoted by $\pi$ and the resulting posterior distribution of $\theta$ obtained by the Bayes rule, assuming it exists, is denoted by $p_{\pi}(\cdot|x)$ and is defined by $p_{\pi}(\theta|x)=\frac{f(x|\theta)\pi(\theta)}{m(x|\pi)}$; here, $m(x|\pi)$ is the marginal density of $X$ obtained by averaging over the prior ($m(x|\pi)=\int_{\Theta} f(x|\theta)\pi(\theta)d\theta$). In this section, we define the geometric perturbation class for the baseline prior and note that the likelihood perturbation class can be formed in the same manner.

\begin{figure}[!ht]
\begin{center}
\includegraphics[scale=0.12]{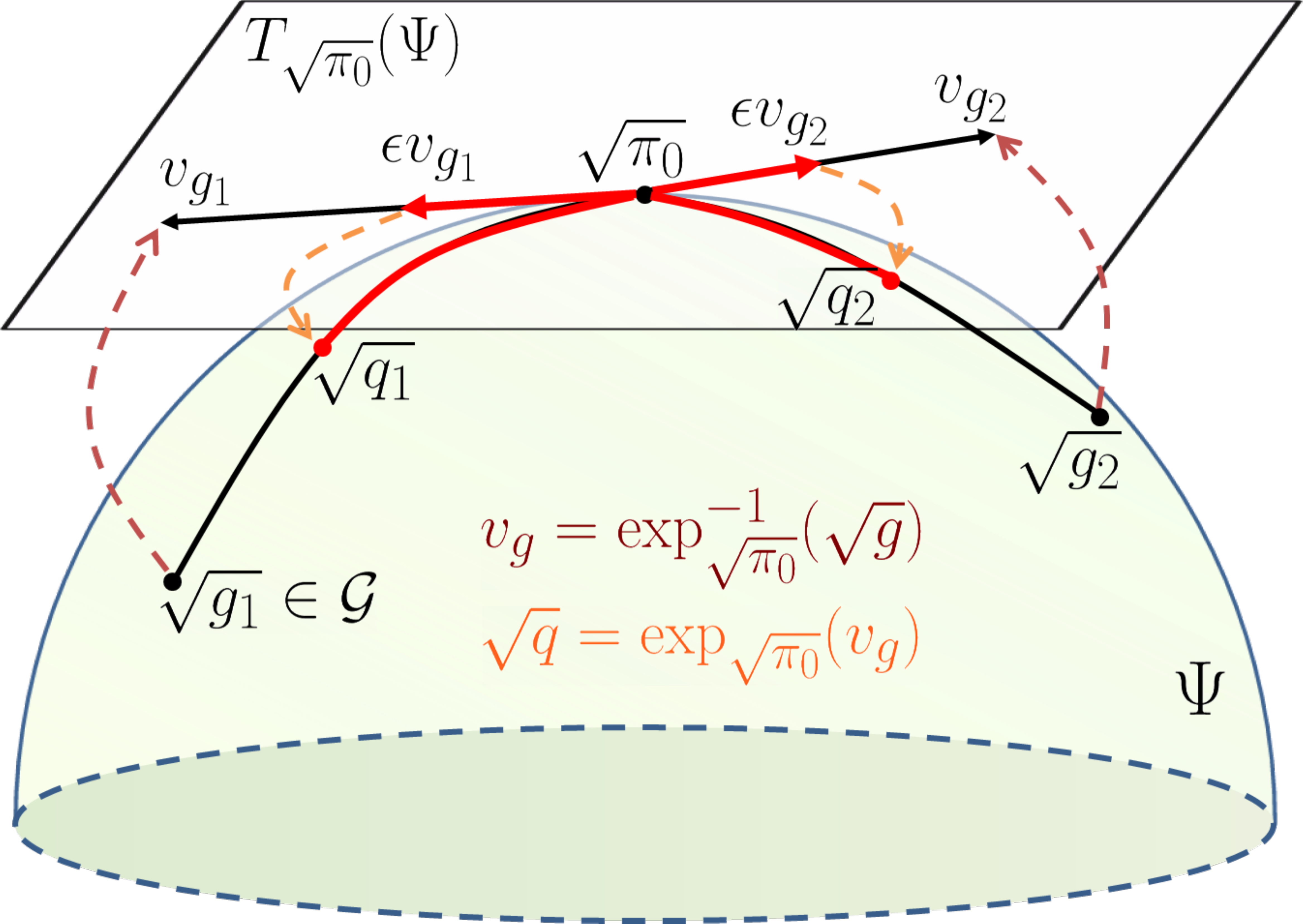}
\caption{\label{geo_prior} Using the SRT representation of the baseline prior $\sqrt{\pi_0}$ and the SRT representation of a contamination density $\sqrt{g}$, one can form a perturbation vector, $v_g$, by mapping $\sqrt{g}$ to the tangent space of $\sqrt{\pi_0}$ using the inverse exponential map. $\sqrt{\pi_0}$ is contaminated by mapping the point $\epsilon v_g$ onto $\Psi$ using the exponential map. The resulting point $\sqrt{q}$ is then mapped from $\Psi$ to ${\cal P}$ using the inverse of the mapping $\phi$.}
\end{center}
\end{figure}

Let $\pi_0$ represent a baseline prior probability density on the parameter $\theta$. Also, let $\mathcal{G}=\{g_1,\ldots,g_n\}$ denote a finite class of contaminants. We construct a set of tangent vectors $v_{g_1},\dots,v_{g_n}\in T_{\sqrt{\pi_0}}(\Psi)$ using the inverse exponential map as $v_{g_i}=\exp^{-1}_{\sqrt{\pi_0}}(\sqrt{g_i}),\ i=1,\dots,n$. This provides a finite class of perturbations of the baseline prior, leading to the following definition.
\begin{definition}\label{egdef}
For a class of densities $\mathcal{G}=\{g_1,\ldots,g_n\}$, the geometric $\epsilon$-contamination class corresponding to the baseline prior $\pi_0$ is defined as
\begin{equation}
\Gamma=\{(\exp_{\sqrt{\pi_0}}(\epsilon v_{g_i}))^2; 0\leq\epsilon\leq 1,\ g_i\in \mathcal{G}, 1 \leq i \leq n\}.\label{epsilongeom}
\end{equation}
\end{definition}
The interpretation of this set is as follows: for an element $g_i\in{\cal G}$, by varying $\epsilon$ from 0 to 1, one traces the geodesic path from $\pi_0$ to $g_i$. Thus, if we fix a value for $\epsilon$, we will obtain a finite set of priors that were contaminated in the directions of $g_1,\dots,g_n$. This is further described in Figure \ref{geo_prior}. The class $\mathcal{G}$ is appropriately constructed based on the problem of interest and the baseline prior. In this paper, we consider finite perturbation classes, although in principle, these methods can be extended to infinite classes as well. An advantage of using such a perturbation class is the natural incorporation of the geometry of the space of densities. As will be seen in later sections, this results in geometrically calibrated local sensitivity measures computed using directional derivatives on $\Psi$. This avoids having to artificially scale the commonly used measures using the geodesic distance as done in \cite{HZ}.

A note on the linear contamination class $\Gamma_{lin}=\{(1-\epsilon)\pi+\epsilon g; 0\leq\epsilon\leq 1,\ g\in{\cal G}\}$: the class has a nice interpretation in terms of mixtures of densities, but it disregards the underlying geometry of the space. This class can be rewritten as $\Gamma_{lin}=\{\pi+\epsilon (g-\pi); 0\leq\epsilon\leq 1,\ g\in{\cal G}\}$. While it is tempting to interpret each element of $\Gamma$ as a small perturbation of magnitude $\epsilon\|g-\pi\|$ {\it along} the direction $(g-\pi)$, one needs to tread with caution since the space of densities is not a linear space in general. In this regard, $\Gamma_{lin}$ technically does not represent a true perturbation class in the geometric sense. On the other hand, in the proposed framework, the $\epsilon$ is tied to actual distances, and it is sensible, loosely speaking, to {\it perturb $\pi_0$  by moving away from it in the direction $v_g$ by a small distance $\epsilon \|v_g\|$}. Since $\epsilon \in [0,1]$, $\epsilon \|v_g\|$ represents the fraction of $\|v_g\|$ along the direction $v_g$, which enables us to retain the mixture interpretation, albeit under the Fisher-Rao metric.

Owing to its linear structure, it is easy to see that the class $\Gamma_{lin}$ induces the same kind of contamination on the marginal and the posterior. One interpretation of $\epsilon$ is as a measure of uncertainty regarding the choice of the original prior $\pi$ (\cite{EM}). If one were to adopt this interpretation, then under the linear $\epsilon$-contamination class, the amount of uncertainty regarding $\pi$ carries over {\it exactly} to the amount of uncertainty regarding the marginal which is averaged over the prior. If one is uncertain regarding the choice of the likelihood as well, then such a phenomenon is quite undesirable. On the other hand, under geometric perturbation of the prior, the interpretation of $\epsilon$ does not carry over to the marginals and the posteriors in the same way due to the nonlinearity in the perturbation.

\subsection{Properties of Fisher-Rao Metric for Bayes Robustness}

In this section we verify that the Fisher-Rao metric satisfies two fundamental properties crucial in Bayes robustness analysis: First, any perturbation of the baseline prior should not have an effect on the sampling distribution; secondly, when considering simultaneous perturbations of the prior and likelihood, one should be able to separate their effects on the joint distribution. We show that these properties are satisfied under our framework. As before, let $f$ be the likelihood function, $\pi_0$ be the baseline prior, and $g\in{\cal G}$ represent a contamination density.

Under the geometric perturbation class $\Gamma$, we write a perturbation of the baseline prior (using the SRT representation) as $\delta_g\sqrt{\pi_0}=\exp^{-1}_{\sqrt{\pi_0}}(\sqrt{g})$. Then the SRT representation of the contaminated prior is given by $\sqrt{\pi_{g}}=\exp_{\sqrt{\pi_0}}(\epsilon\delta_g\sqrt{\pi_0})$. The exponential and inverse exponential maps that are used here were defined in Section \ref{sec:geometry-background}. The SRT representation of the contaminated joint density is then given by $\sqrt{p_{g}(x,\theta)}=\sqrt{f(x|\theta)\pi_{g}(\theta)}$. Thus, we can compute the perturbation vector on the space of SRT representations of joint densities as follows:
\begin{align*}
&v_{g}(x,\theta)=\frac{d}{d\epsilon}\sqrt{f(x|\theta)\pi_{g}(\theta)}|_{\epsilon=0}=\frac{d}{d\epsilon}\sqrt{f(x|\theta)}\exp_{\sqrt{\pi_0}}(\epsilon\delta_{g}\sqrt{\pi_0})(\theta)\Big|_{\epsilon=0}\\
&=\frac{d}{d\epsilon}\sqrt{f(x|\theta)}(\cos(\epsilon\|\delta_{g}\sqrt{\pi_0}(\theta)\|)\sqrt{\pi_0(\theta)}+
    \sin(\epsilon\|\delta_{g}\sqrt{\pi_0}(\theta)\|)\frac{\delta_{g}\sqrt{\pi_0}(\theta)}{\|\delta_{g}\sqrt{\pi_0}(\theta)\|})\Big|_{\epsilon=0}\\
&=\sqrt{f(x|\theta)}(-\sin(\epsilon\|\delta_{g}\sqrt{\pi_0}(\theta)\|)\sqrt{\pi_0(\theta)}\|\delta_{g}\sqrt{\pi_0}(\theta)\|
    +\cos(\epsilon\|\delta_{g}\sqrt{\pi_0}(\theta)\|)\delta_{g}\sqrt{\pi_0}(\theta))\Big|_{\epsilon=0}\\
&=\sqrt{f(x|\theta)}\delta_{g}\sqrt{\pi_0}(\theta)
\end{align*}
Given two geometric perturbations of the baseline prior $\delta_{g_1}\sqrt{\pi_0},\ \delta_{g_2}\sqrt{\pi_0}$ ($g_1,\ g_2\in{\cal G}$), we compute the corresponding perturbations of the joint density under the SRT representation and derive the corresponding Riemannian metric on that space:
\begin{equation}
\langle v_{g_1},v_{g_2} \rangle=\int_\Theta\int_\real \delta_{g_1}\sqrt{\pi_0}(\theta)\delta_{g_2}\sqrt{\pi_0}(\theta)f(x|\theta)dxd\theta=\int_\Theta \delta_{g_1}\sqrt{\pi_0}(\theta)\delta_{g_2}\sqrt{\pi_0}(\theta)d\theta.\label{eqn:metper}
\end{equation}
The quantity in Equation \ref{eqn:metper} is independent of the sampling distribution, verifying our claim. Furthermore, we notice that if the sampling distribution is fixed and the geometric perturbation model is used, the Riemannian metric on the space of joint densities is the same as that on the space of priors (isometry). Intuitively, one would expect this to be the case, and thus this is an attractive property of our framework. We note that isometry does not hold when the linear perturbation class is used, which has been shown in \cite{HZ}.

We now turn our attention to the second property. Let $f_0$ and $\pi_0$ be the baseline likelihood and prior, respectively. Also, let $q$ represent a likelihood contaminant density and $g$ represent a prior contaminant density. Then, as previously, the SRT representations of the contaminated likelihood and contaminated prior are given by $\sqrt{f_{q}}=\exp_{\sqrt{f}}(\epsilon\delta_q\sqrt{f})$ and $\sqrt{\pi_{g}}=\exp_{\sqrt{\pi_0}}(\epsilon\delta_g\sqrt{\pi_0})$, where $\delta_q\sqrt{f}=\exp^{-1}_{\sqrt{f}}(\sqrt{q})$ and $\delta_g\sqrt{\pi_0}=\exp^{-1}_{\sqrt{\pi_0}}(\sqrt{g})$. The resulting perturbations on the space of SRT representations of joint densitites are $v_{q}=\frac{d}{d\epsilon}\sqrt{f_q\pi_{0}}|_{\epsilon=0}=\delta_{q}\sqrt{f_0}\sqrt{\pi_0}$ and $v_{g}=\frac{d}{d\epsilon}\sqrt{f_0\pi_{g}}|_{\epsilon=0}=\sqrt{f_0}\delta_{g}\sqrt{\pi_0}$. Plugging these two quantities into the expression of the Fisher-Rao Riemannian metric we obtain:
\begin{align}
\langle v_g,v_q\rangle &= \int_\Theta\int_\real \sqrt{f_0(x|\theta)}\delta_{g}\sqrt{\pi_0}(\theta)\delta_{q}\sqrt{f_0}(x|\theta)\sqrt{\pi_0(\theta)} dxd\theta \nonumber\\
&= \int_\Theta \delta_{g}\sqrt{\pi_0}(\theta)\sqrt{\pi_0(\theta)} \int_\real \sqrt{f_0(x|\theta)}\delta_{q}\sqrt{f_0}(x|\theta) dxd\theta=0,\label{eqn:secprop}
\end{align}
because $\langle f_0,\delta_{q}\sqrt{f_0} \rangle=0$ (i.e. perturbations are orthogonal to the representation space). This result is important in that it leads to a natural decomposition of the metric on the space of joint densities. To show this, consider simultaneous perturbations of the likelihood and prior. It is easy to show that the resulting perturbation on the space of SRT representations of joint densities is given by $v=\delta_{q}\sqrt{f_0}\sqrt{\pi_0}+\sqrt{f_0}\delta_{g}\sqrt{\pi_0}$. If we are given two such simultaneous perturbations, the resulting Riemannian metric is:
\begin{align}
\langle v_1,v_2\rangle &= \langle \sqrt{f_0}\delta_{g_1}\sqrt{\pi_0},\sqrt{f_0}\delta_{g_2}\sqrt{\pi_0} \rangle + \langle \delta_{q_1}\sqrt{f_0}\sqrt{\pi_0},\sqrt{f_0}\delta_{g_2}\sqrt{\pi_0} \rangle \nonumber\\
&\quad + \langle \sqrt{f_0}\delta_{g_1}\sqrt{\pi_0},\delta_{q_2}\sqrt{f_0}\sqrt{\pi_0} \rangle + \langle \delta_{q_1}\sqrt{f_0}\sqrt{\pi_0},\delta_{q_2}\sqrt{f_0}\sqrt{\pi_0} \rangle \nonumber\\
&= \langle \sqrt{f_0}\delta_{g_1}\sqrt{\pi_0},\sqrt{f_0}\delta_{g_2}\sqrt{\pi_0} \rangle + \langle \delta_{q_1}\sqrt{f_0}\sqrt{\pi_0},\delta_{q_2}\sqrt{f_0}\sqrt{\pi_0} \rangle.
\end{align}
The last equality holds due to Equation \ref{eqn:secprop}. Thus, the Riemannian metric on the space of joint densities can be written down as a sum of likelihood and prior perturbation terms.

\section{Bayesian Sensitivity Analysis}

Armed with the tools from the preceding section, we now focus on the three tasks set out in the introduction. Throughout this section, we utilize the geometric perturbation class for local and global sensitivity. In the interests of brevity, we describe the framework for prior perturbations only and note that it is easily extended to handle likelihood perturbations, in which case the perturbations are defined on the space of sampling distributions with minimal change in computations.

\subsection{Global Sensitivity Analysis}
\label{sec:globsen}

Given a likelihood function $f(x|\theta)$ and a baseline prior $\pi_0$, one can define the baseline posterior density, when it exists, as $p_0(\theta|x)=\frac{f(x|\theta)\pi_0(\theta)}{m(x|\pi_0)}$. If it is the case that the posterior is given within the constant represented by the integral in the denominator of this expression, we can evaluate it using a numerical integral or by Monte Carlo methods. Note that under the SRT representation, this operation is the same as a straight-line projection from $\ltwo$ to $\Psi$. In order to compute distances between posterior probability density functions we will again utilize the space $\Psi$. We are now given $p_0(\cdot|x)$, the baseline posterior, and $p_{g_1}(\cdot|x),\ldots,p_{g_n}(\cdot|x)$, the set of posteriors generated from the $\epsilon$-contaminated priors.
\begin{definition}\label{pmeasure}
For a class of contamination densities $\mathcal{G}$ consider the geometric $\epsilon$-contamination class given by Definition \ref{egdef}. Then, a measure of sensitivity with respect to the geometric perturbation of $\pi_0$ is defined as
\begin{equation}
S(\epsilon,\pi_0,\mathcal{G})=\max\big\{d_{FR}(p_0(\cdot|x),p_{g_i}(\cdot|x)); g_i \in \mathcal{G}, 1 \leq i \leq n\big\}.
\end{equation}
\end{definition}
Guided by the measure $S(\epsilon,\pi_0,\mathcal{G})$, we can additionally compute posterior functionals with respect to the `nearest' and `farthest' posteriors. Note that using the geodesic distance as a measure of robustness in our framework is meaningful because all of the distances are bounded above by $\pi/2$. Furthermore, this is the intrinsic metric on the space of densities and thus takes into account the geometry of that space. We demonstrate the utility of the proposed global sensitivity measure through three examples. The first example is a simple one where one is able to assess the properties of the defined method, and contrast our results to those obtained using linear contamination and KL divergence. The second example considers a Bayesian model for directional data; the third example is more general pertaining to a mixed effects model, where we study robustness under perturbation of the prior for the variance of the random effects.

\begin{example}

Consider the following baseline model:
\begin{align*}
x_i|\theta&\overset{\text{i.i.d.}}\sim f=N(\theta,1), \quad i=1,\ldots,50;\\
\theta&\sim \pi_0=N(0,1)
\end{align*}
In this example, we consider a skew normal contamination class, parameterized by a shape parameter $\alpha\in[-5,5]$. Figure \ref{fig:skewn} displays the considered $\epsilon$-contaminated prior set under the linear and geometric frameworks by fixing $\epsilon=0.5$ and $\alpha$ in the set $\{\pm1,\pm2,\pm3,\pm4,\pm5\}$. We begin by simulating data $x_1,\dots,x_{50}$ from the baseline model and generating a set of contaminated priors for 31 equally spaced values $\epsilon\in [0,1]$ and 101 equally spaced values $\alpha\in [-5,5]$ using the two different types of contamination methods. First, the baseline posterior $p_0$ is computed, where the normalizing constant is calculated numerically. In similar fashion, we compute the posterior density resulting from any of the contaminated priors. Denote any of the contaminated posteriors by $p(\theta|x)$. In this example, we utilize two approaches: (1) geometric contamination with $d_{FR}$ between posteriors as a measure of global robustness (Figure \ref{fig:nsnrobus}(a)); and (2), linear contamination with KL divergence as a global robustness measure (Figure \ref{fig:nsnrobus}(b),(c)). Since the Kullback Leibler divergence is not symmetric in its argument we compute it in both ways. We also compute the posterior mean for $\epsilon=0.5$ based on the same set of contaminated models. Note that when $\alpha=0$, the contaminated posterior is the same as the baseline posterior. This procedure is performed on three simulated datasets corresponding to the three rows in Figure \ref{fig:nsnrobus}.

\begin{figure}[!ht]
\begin{center}
\scalebox{0.8}{
\begin{tabular}{c c }

\includegraphics[width=2in]{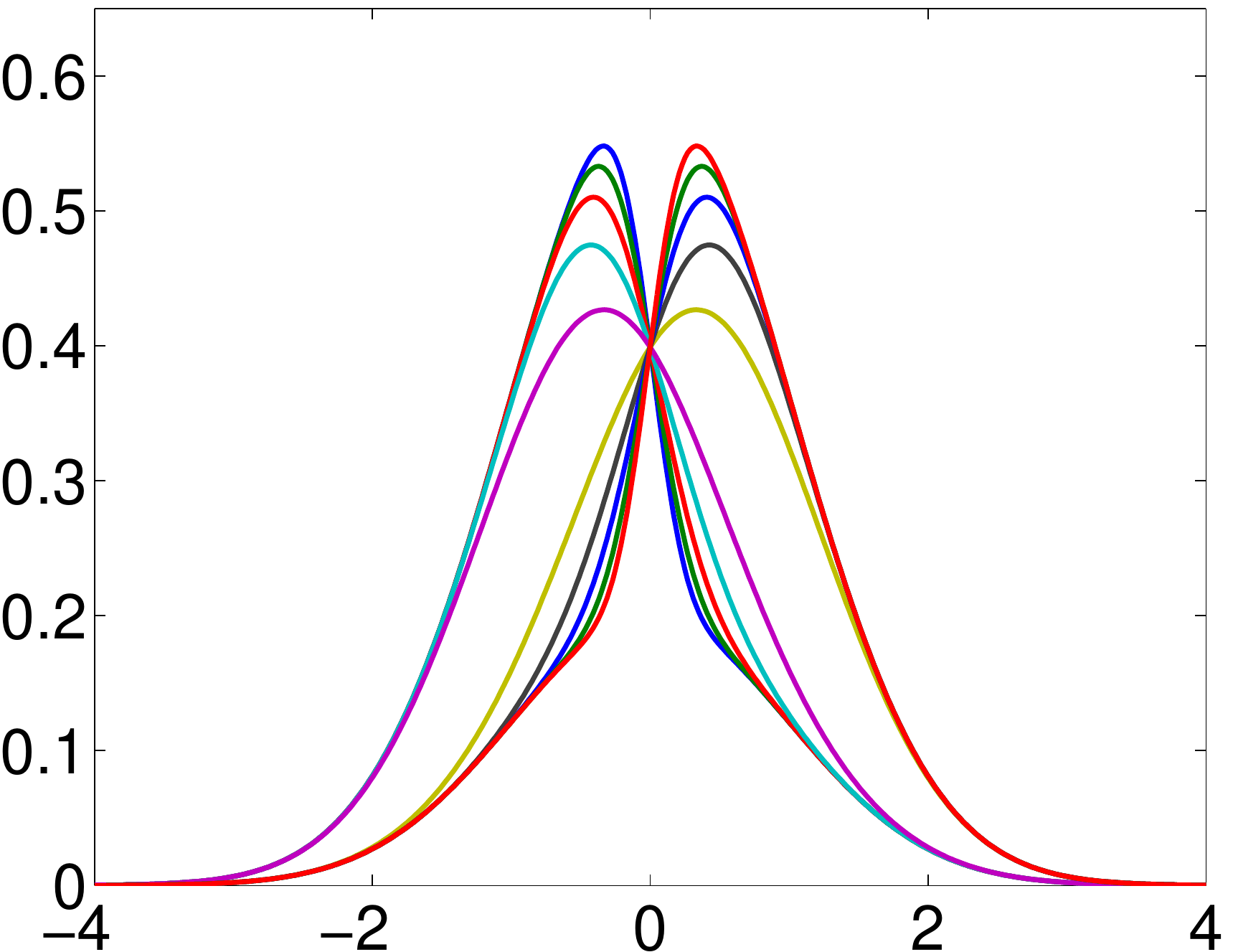}&\includegraphics[width=2in]{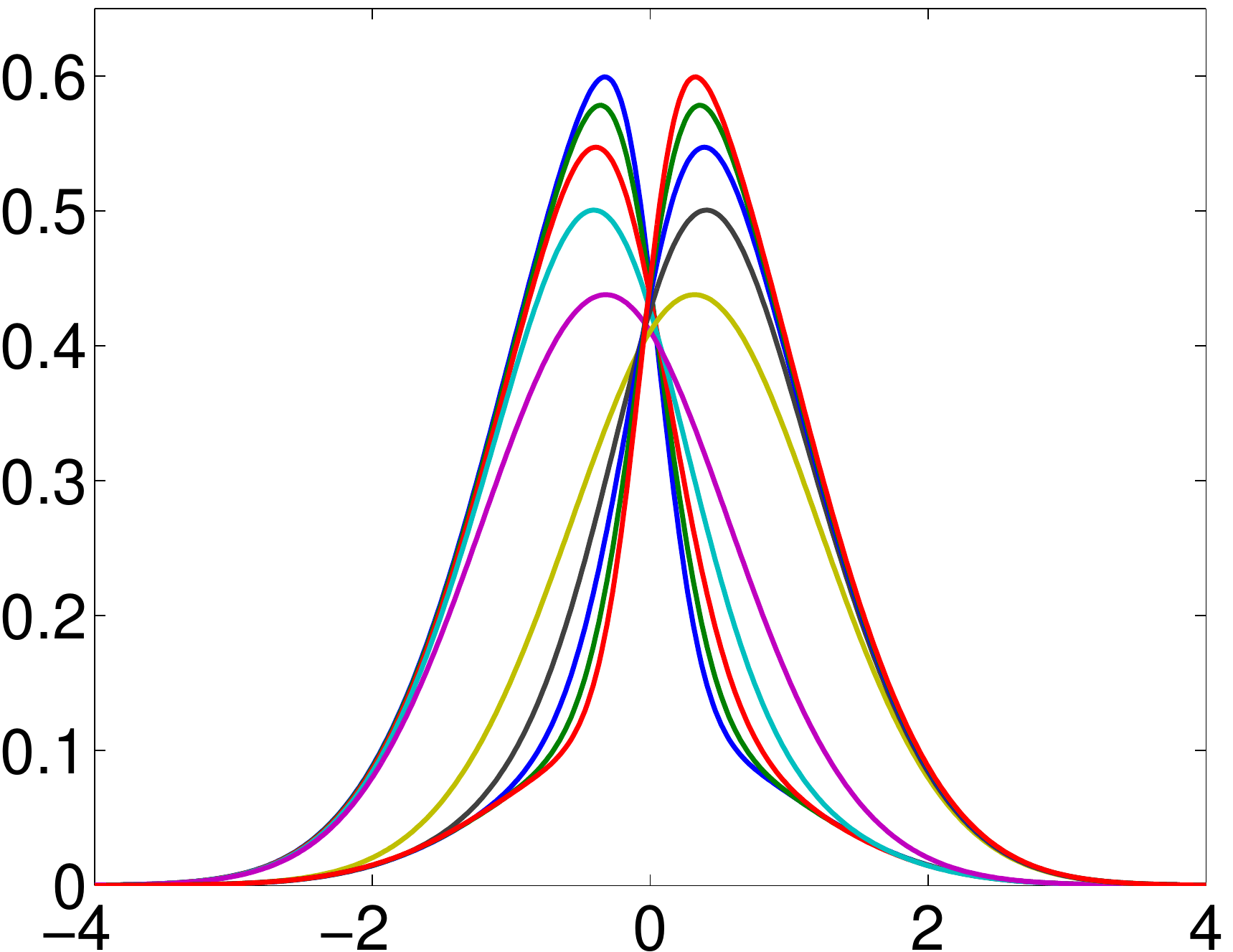}\\
(a)&(b)\\
\end{tabular}
}
\caption{\label{fig:skewn} Normal prior contaminated using the skew normal contamination class under the (a) linear and (b) geometric frameworks. Observe how the tails are more separated under the geometric framework than in the linear one illustrating the difference between the two methods.}
\end{center}
\end{figure}

\begin{figure}[!ht]
\begin{center}
\begin{tabular}{c c c c }
(a)&(b)&(c)&(d)\\
\includegraphics[width=1.2in]{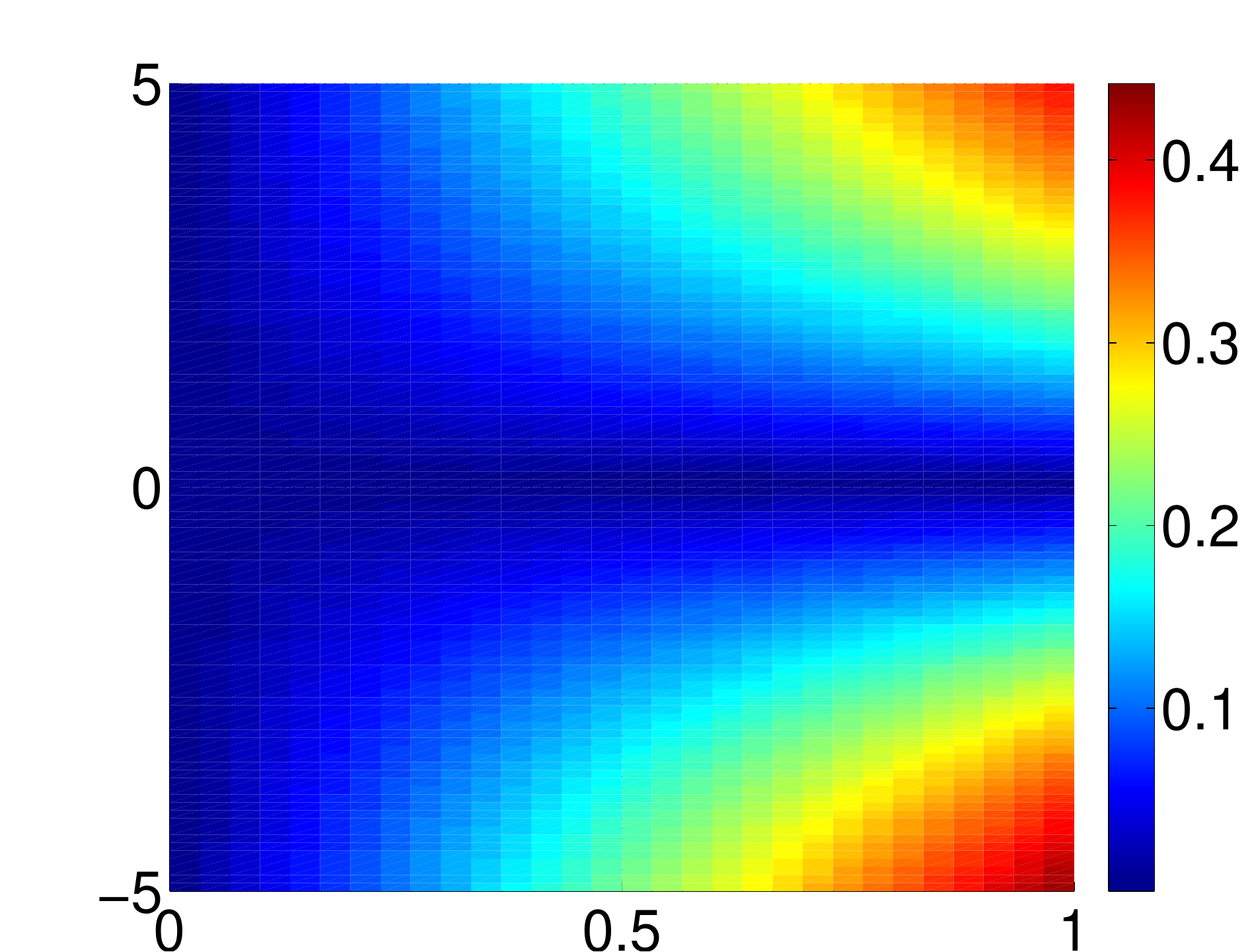}&\includegraphics[width=1.2in]{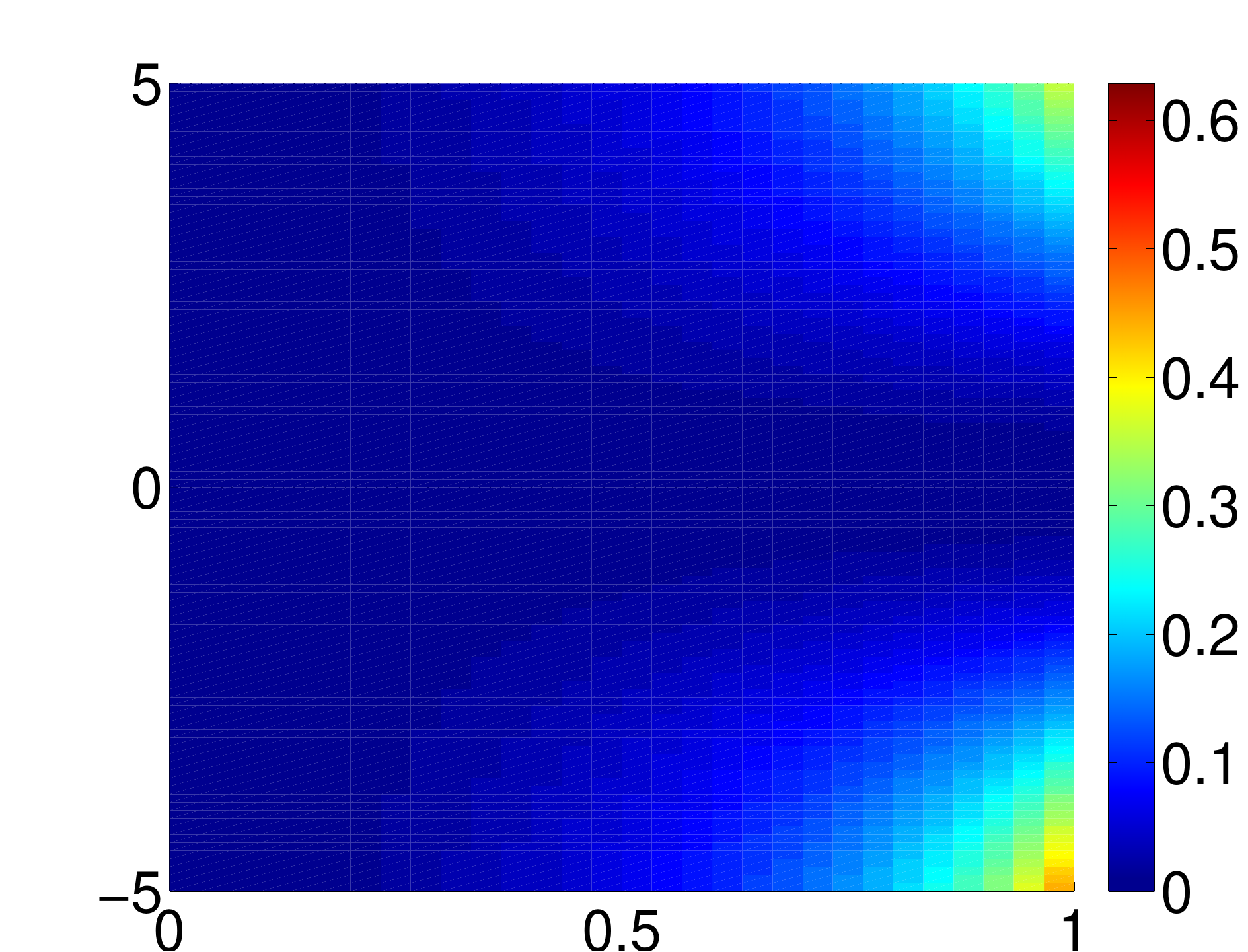}&\includegraphics[width=1.2in]{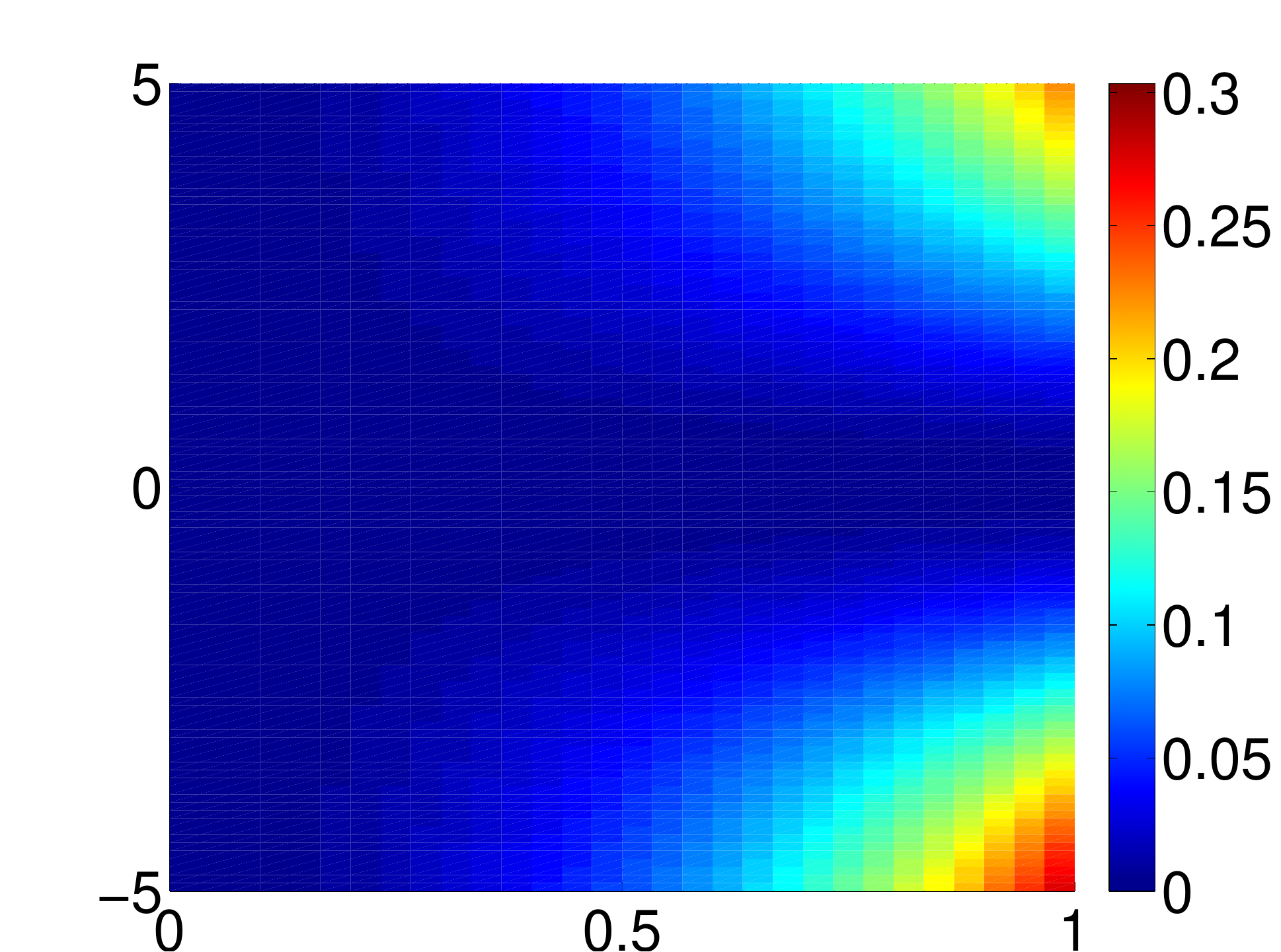}&\includegraphics[width=1.2in]{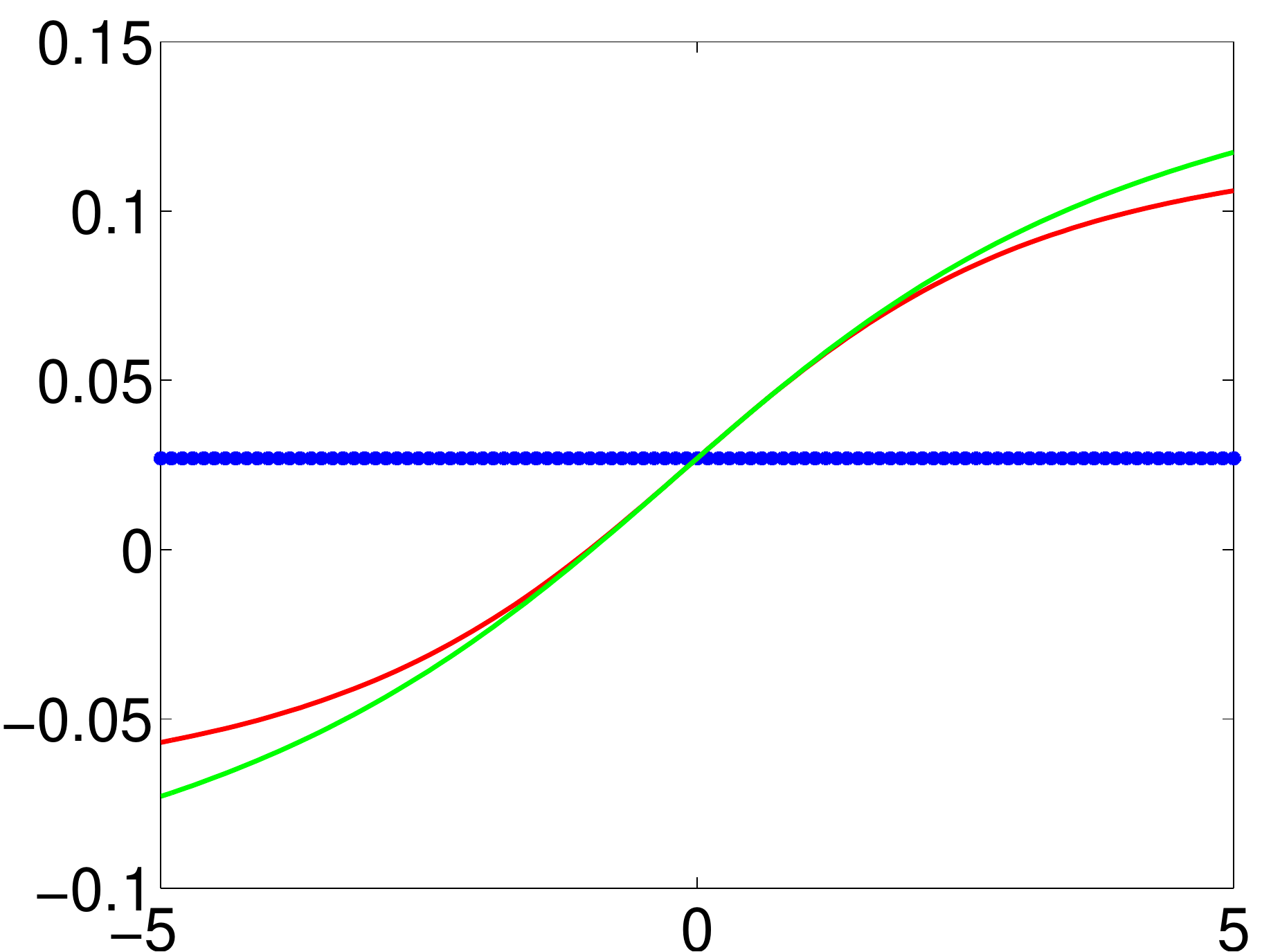}\\

\includegraphics[width=1.2in]{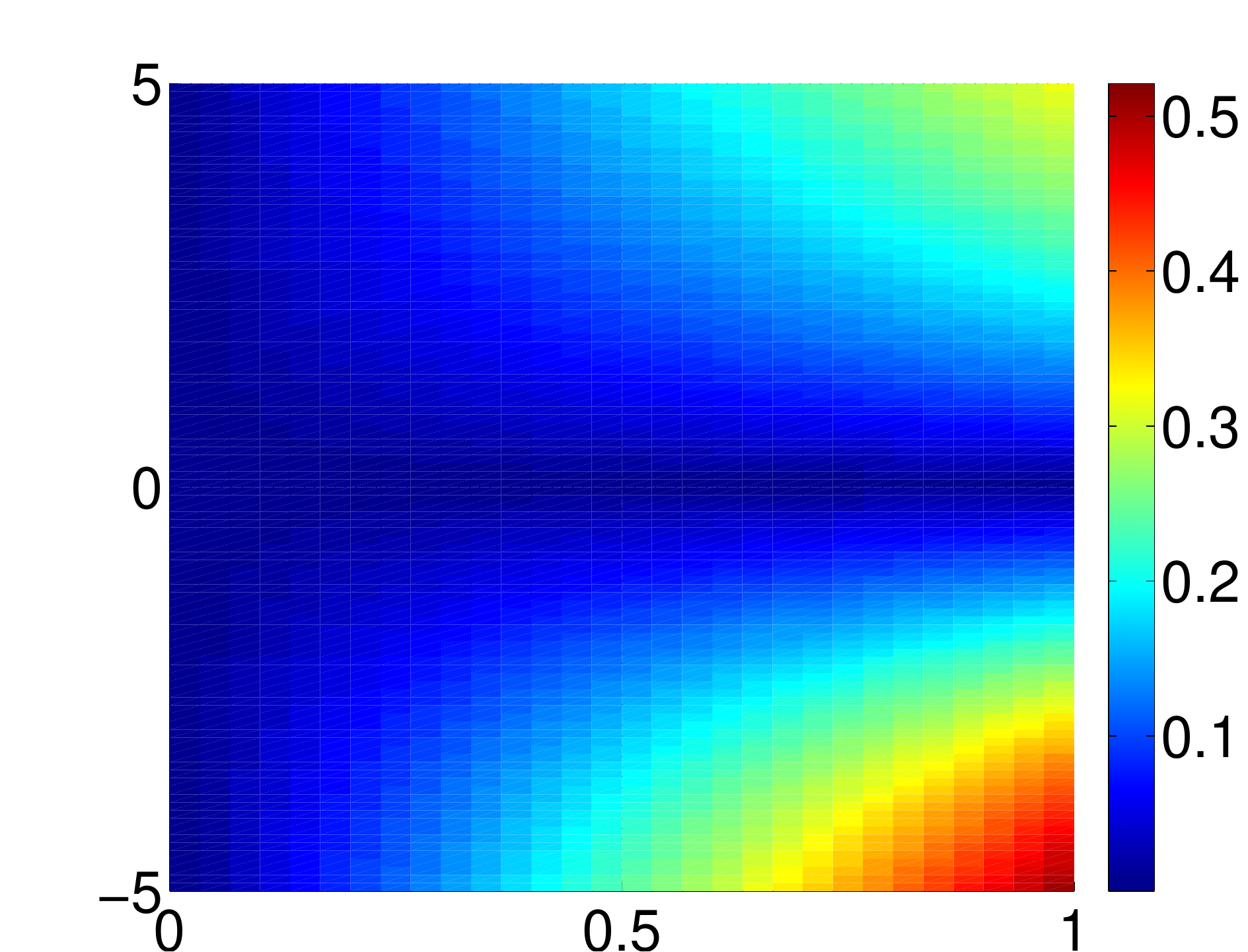}&\includegraphics[width=1.2in]{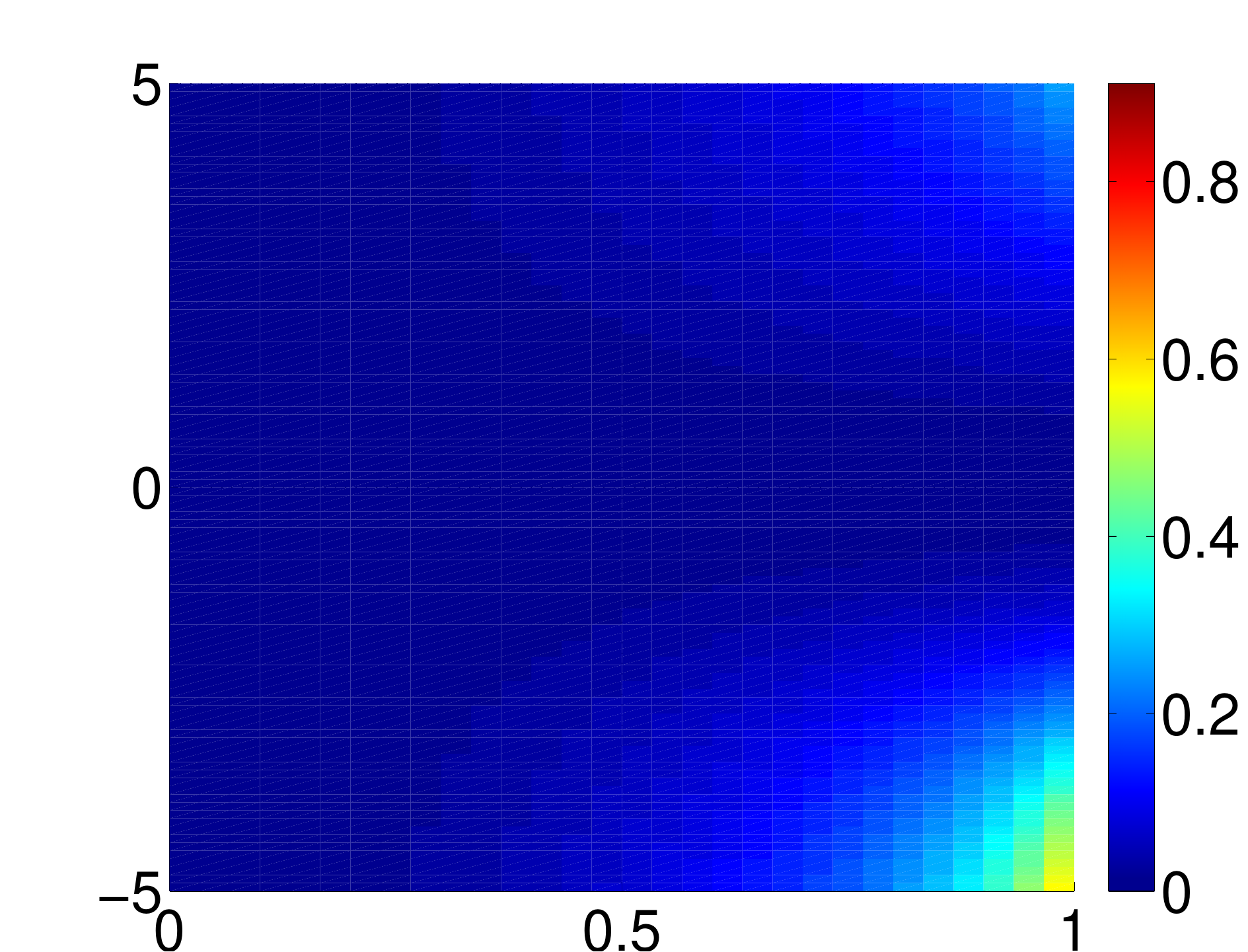}&\includegraphics[width=1.2in]{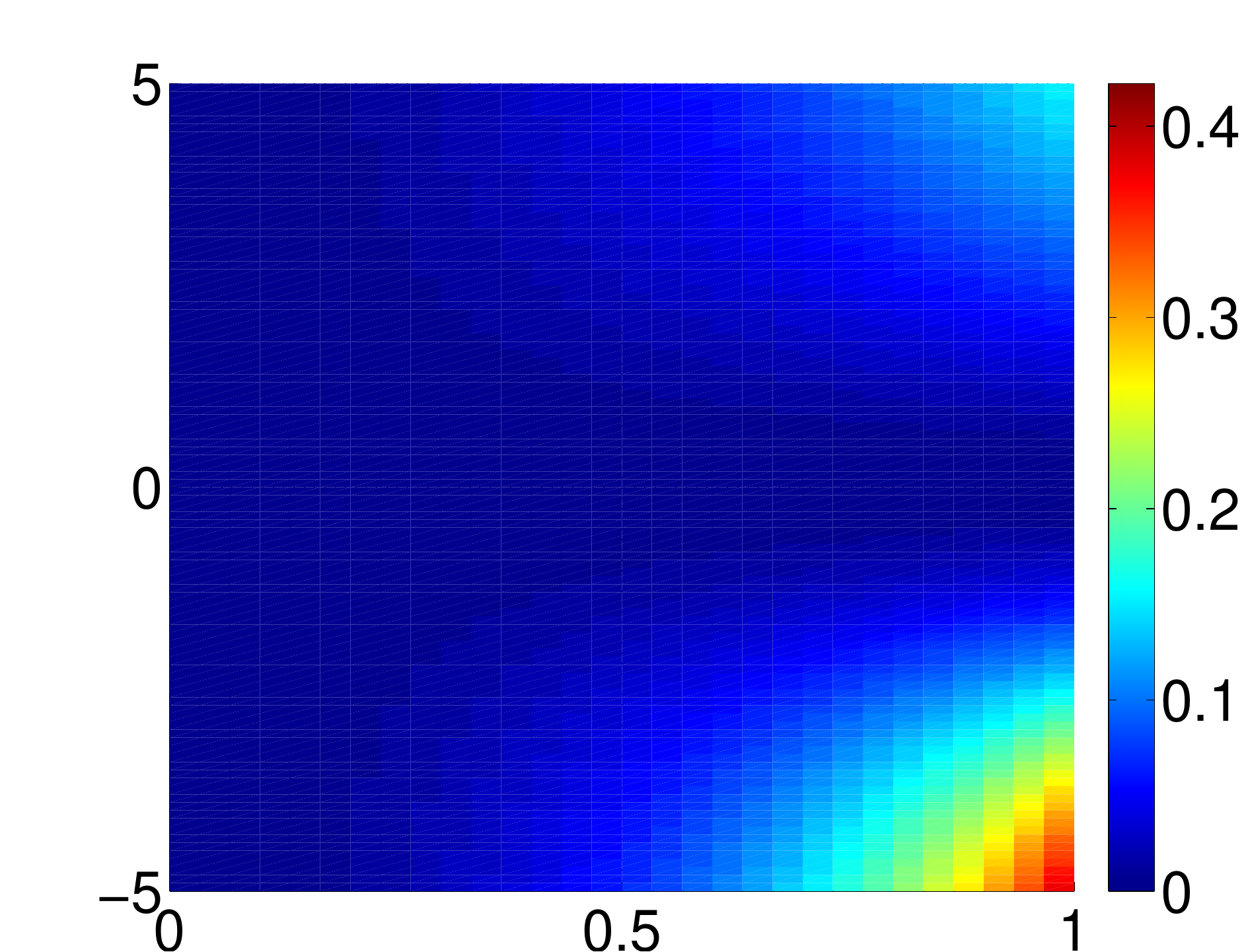}&\includegraphics[width=1.2in]{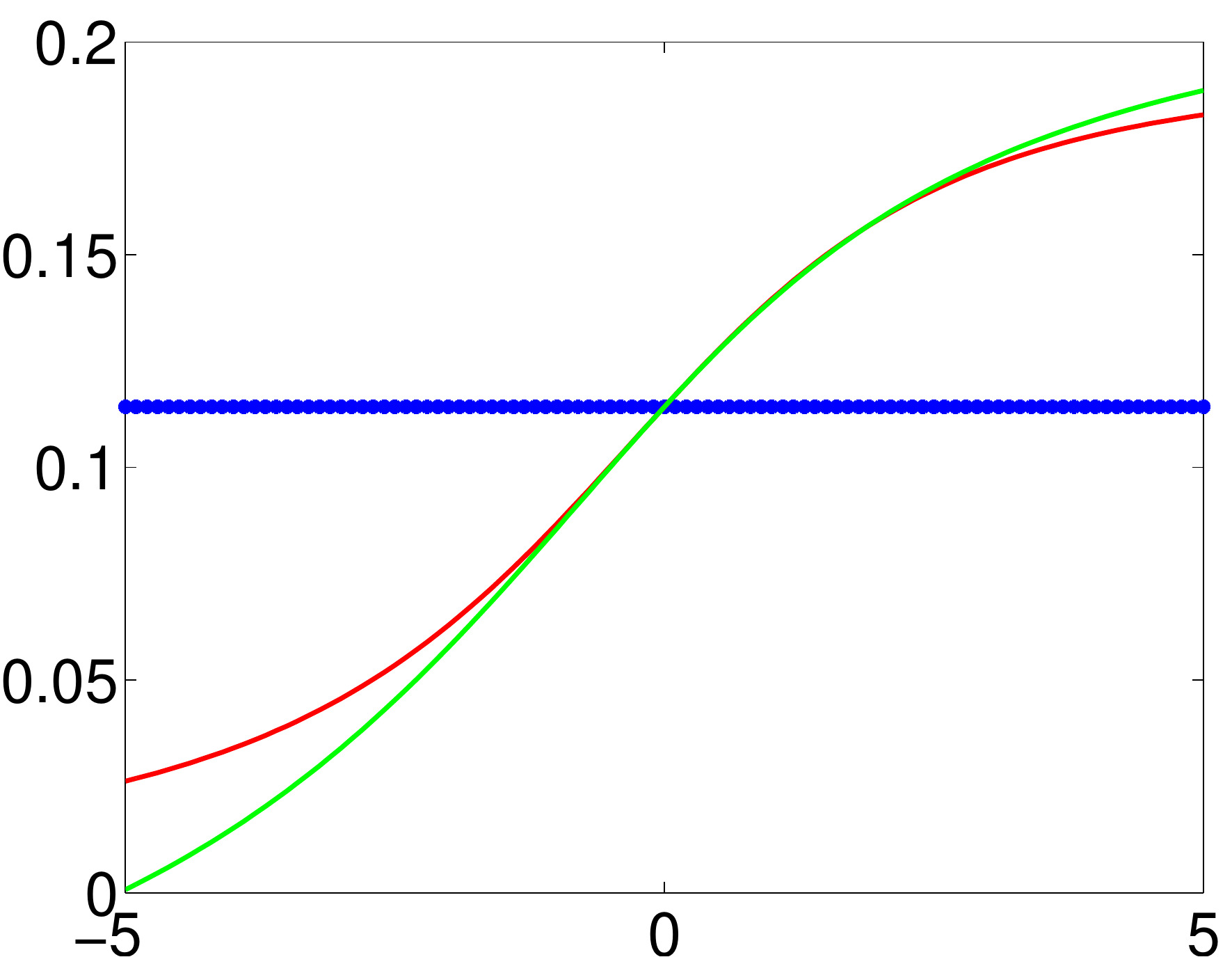}\\

\includegraphics[width=1.2in]{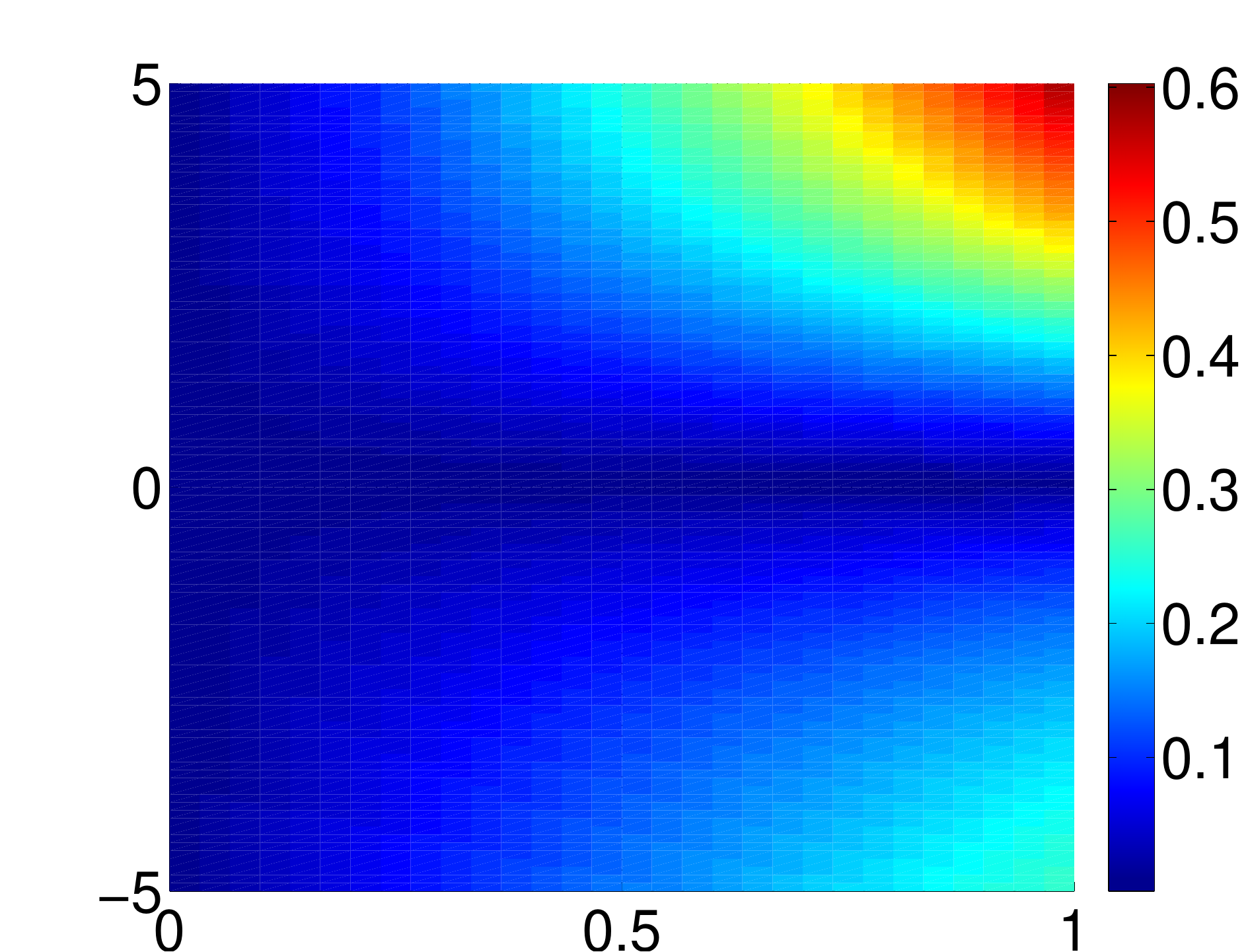}&\includegraphics[width=1.2in]{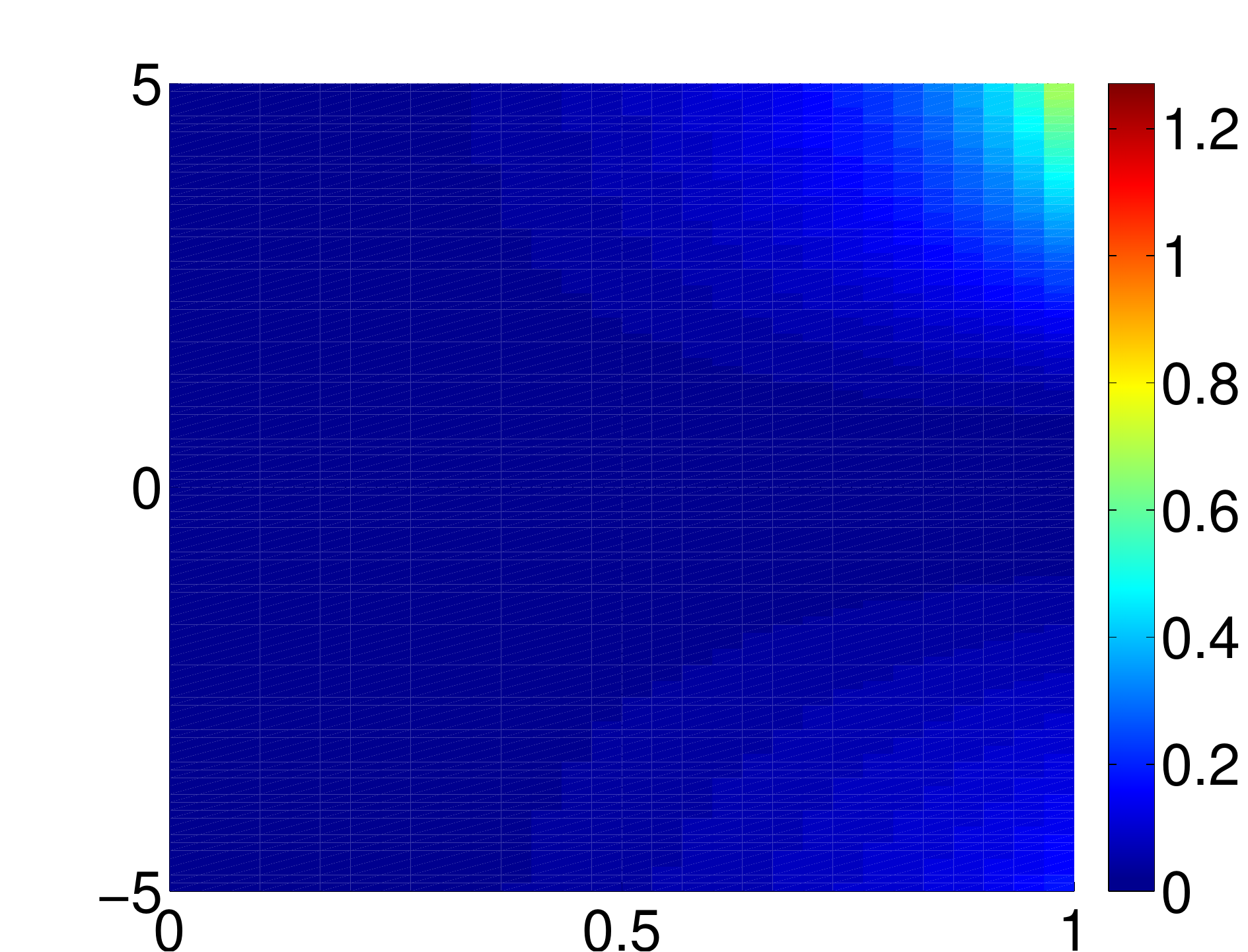}&\includegraphics[width=1.2in]{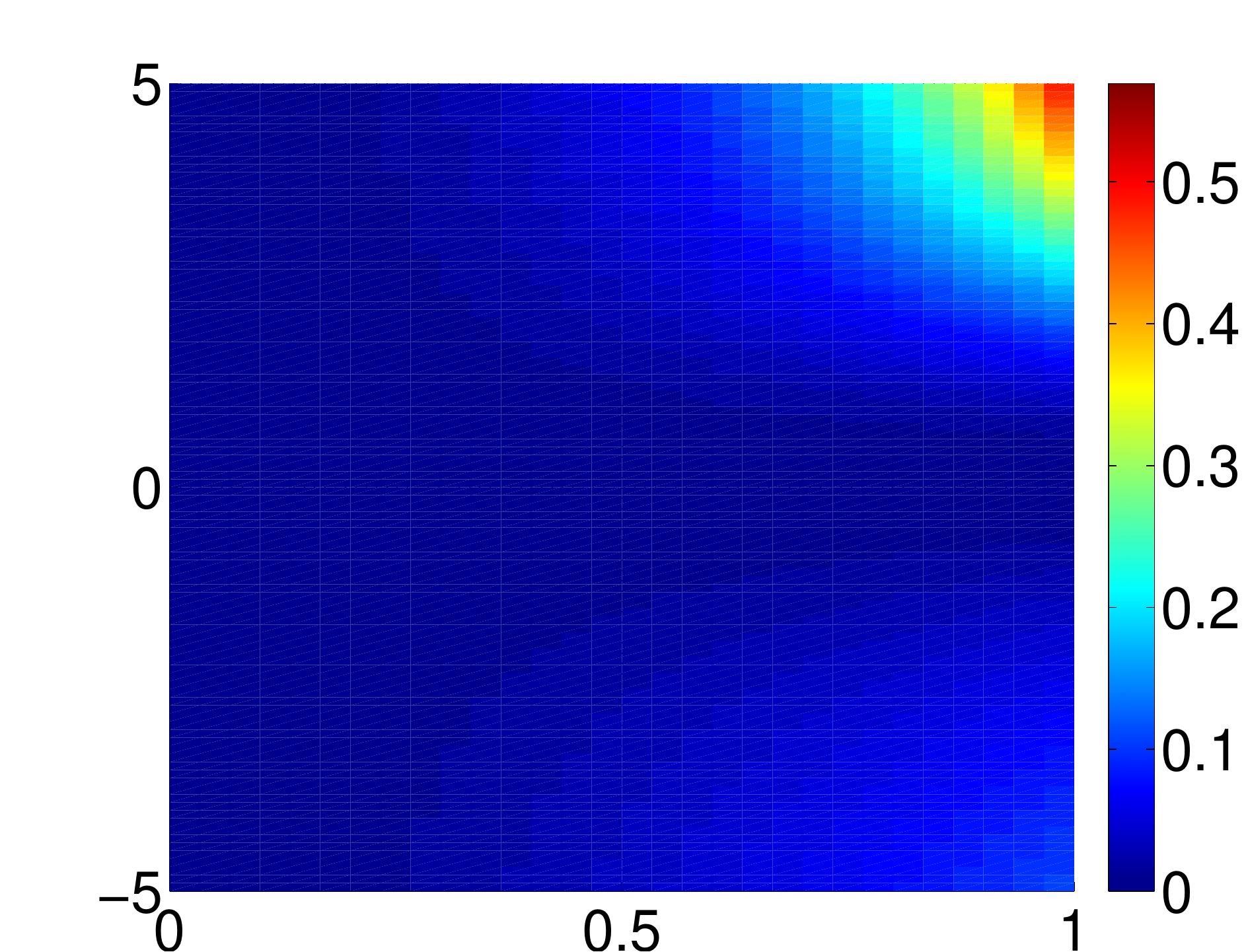}&\includegraphics[width=1.2in]{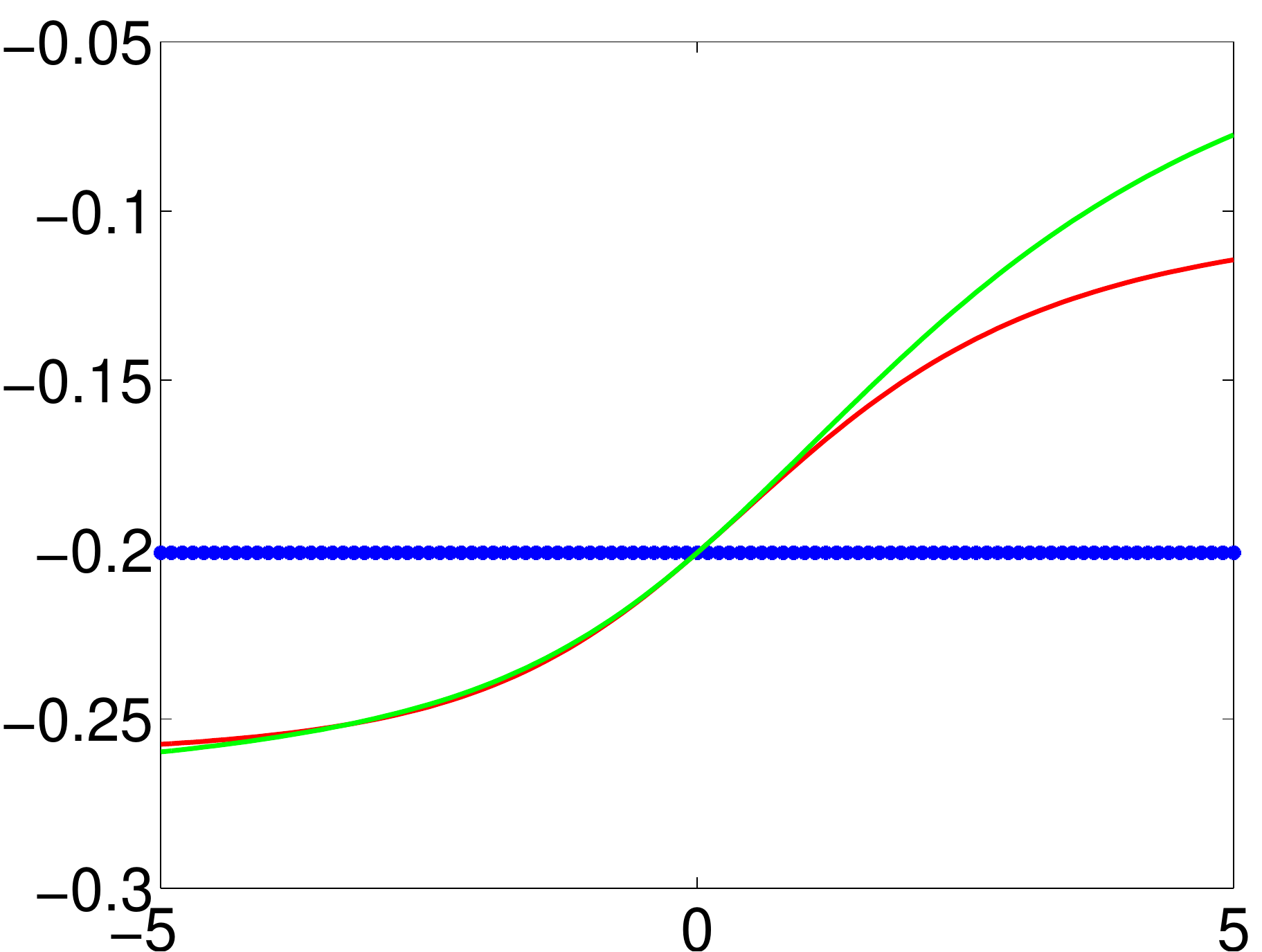}\\

\end{tabular}
\caption{Assessment of Bayesian prior robustness for $\epsilon$-contamination of a Gaussian prior with a skew normal distribution. (a) Image of Fisher-Rao distances between baseline and geometrically contaminated posteriors for different values of $\epsilon$ ($x$-axis) and $\alpha$ ($y$-axis) for 3 simulated datasets. (b) Image of KL divergences (expectation computed with respect to $p_0$) between baseline and linearly contaminated posteriors for different values of $\epsilon$ and $\alpha$. (c) Same as (b) but the expectation was computed with respect to $p$. (d) Posterior means for varying values of $\alpha$ and $\epsilon=0.5$ (baseline=blue, geometric contamination=green, linear contamination=red).\label{fig:nsnrobus}}
\end{center}
\end{figure}

We make a few key observations about the results presented in Figure \ref{fig:nsnrobus}. As expected, the KL divergence is asymmetric discouraging its use as a robustness measure. In all cases, the Fisher-Rao distance suggests that the posterior is fairly robust to geometric contamination of the Gaussian prior using skew normal distributions, especially if one takes $\epsilon$ to be small ($<0.5$). The distance becomes relatively high only when $\epsilon$ approaches one and for large $\alpha$. Note that when $\epsilon$ equals one, the baseline Gaussian prior is entirely replaced with the skew normal. It can also be seen that the Fisher-Rao distance is more sensitive to departures from $N(0,1)$ than the KL divergence; the KL divergence appears to pick up departures only for $\epsilon$ exceeding 0.75.

We also notice an interesting result from panel (d). When the baseline posterior mean is close to zero, the geometric and linear contamination methods result in similar values of the contaminated posterior mean (difference $<0.02$). When the posterior mean is greater than zero, the linear and geometric contamination classes yield very similar contaminated posterior means in the positive direction ($\alpha>0$). On the other hand, the geometrically contaminated posterior means portray a more severe departure from the baseline model than the linear contamination class in the negative direction ($\alpha<0$). The opposite result is also observed in the third row of panel (d). We posit that this phenomenon is due to the nonlinear structure of the geometric contamination class and is consistent with intuition.

\end{example}

\begin{example}

In this example, we utilize a Bayesian model to analyze directional data on $\mathbb{S}^1$. This dataset consists of 76 directions of turtle movement after a certain treatment is applied; the raw data is displayed in Figure \ref{fig:kappert}(a) in red with the origin (0 radians) in green. Recently, Goh and Dey (\cite{GD}) considered the following baseline model for this data:
\begin{align*}
x_i|\theta&\overset{\text{i.i.d.}}\sim f=vM(\theta,\hat{\kappa}), \quad i=1,\ldots,76;\\
\theta&\sim \pi_0=vM(0,0.01),
\end{align*}
where $vM(\mu,\kappa)$ is the von Mises distribution with mean $\mu$ and concentration $\kappa$, and $\hat{\kappa}=1.1423$ is the MLE of the concentration parameter based on the given data. In \cite{GD}, the authors considered the problem of identifying influential observations based on the functional Bregman divergence. For simplicity, they set the unknown likelihood concentration parameter to its MLE. We propose to assess the global sensitivity of the posterior distribution of $\mu$ to this choice via perturbations of this concentration parameter. For this purpose, we consider 100 different values of $\kappa$ for the likelihood ranging from 0.01 to 10. As before, the global sensitivity measure is the Fisher-Rao distance between the baseline posterior and the posterior under the perturbed likelihood. Note that all of the posteriors in this example are also von Mises and we utilize numerical integration to compute the distance. The results of our analysis are shown in Figure \ref{fig:kappert}(b). The minimum distance, as expected, is achieved when $\kappa=1.12$, which is very close to the baseline (MLE). Also, it appears that the posterior distribution of $\mu$ is highly sensitive to the choice of the likelihood concentration parameter, especially when perturbed toward 0. This suggests that one should also utilize a prior for this parameter rather than choosing it based on the data, which can lead to misleading inference.
\begin{figure}[!ht]
\begin{center}
\begin{tabular}{ c c }
\small{(a) Turtle Data}&\small{(b) Global Sensitivity}\\

\includegraphics[scale=0.35]{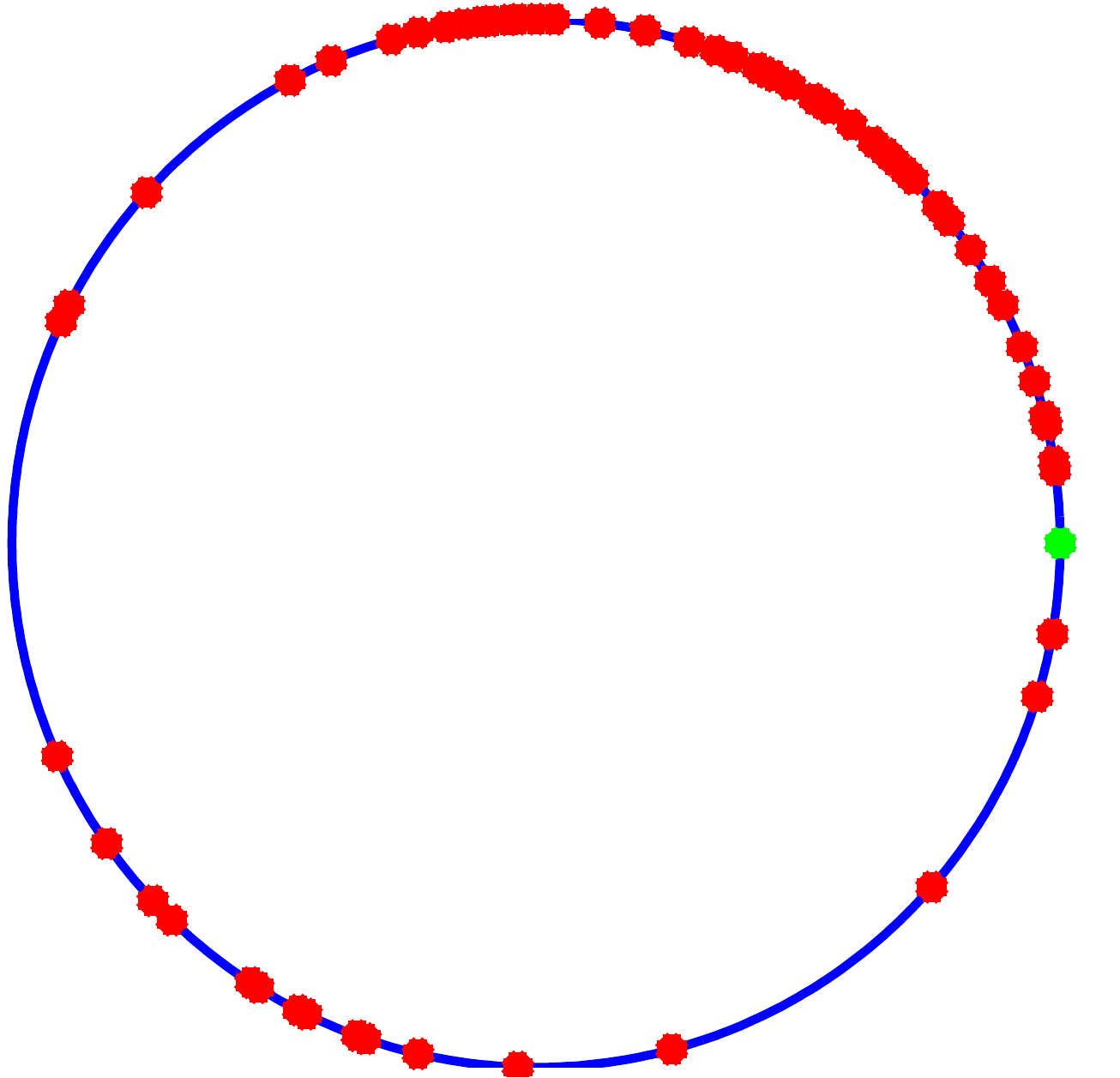}&\includegraphics[scale=0.3]{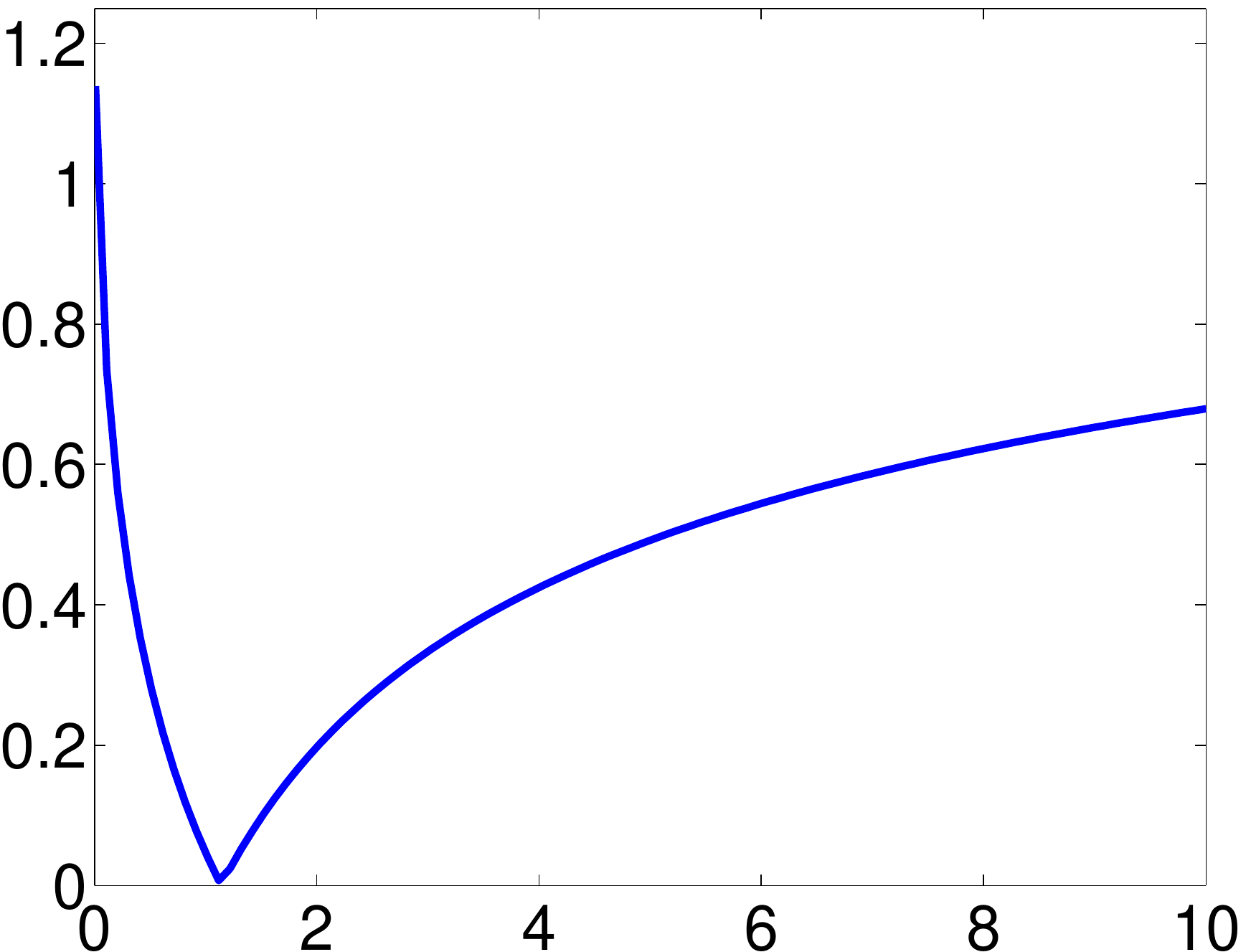}\\

\end{tabular}
\caption{(a) Turtle directional data in red with the origin in green. (b) Global sensitivity based on the Fisher-Rao metric to perturbations of the likelihood concentration parameter ($x$-axis) away from the MLE. \label{fig:kappert}}
\end{center}
\end{figure}

\end{example}

\begin{example}

The final example in this section considers modeling data related to presence or absence of bacteria in persons monitored through a fixed time window using a Bayesian generalized linear mixed effects model; this data, available in the \texttt{MASS} package in \texttt{R}, was previously used for illustrative purposes in \cite{BrownZhou}. The predictors are treatment (placebo, drug, drug$+$) and week of test. We use the following baseline logistic mixed effects model for this data:
\begin{align*}
Y_{ij}&\sim Bernoulli(p_{ij});\\
logit(p_{ij})&=\mu+\sum_{k=1}^3 x_{ij}^k\beta^k+V_i;\\
\mu\sim N(0,100); &\qquad \beta^k\overset{\text{i.i.d.}}\sim N(0,100);\\
V_i\overset{\text{i.i.d.}}\sim N(0,\sigma^2);& \qquad
\tau=\frac{1}{\sigma^2}\sim \Gamma(0.01,0.01),
\end{align*}
where the response $Y_{ij}$ indicates the presence or absence of bacteria in person $i$ at week $j$, $x_{ij}$ are the week of test and indicator variables for the treatment, $p_{ij}$ is the probability of bacteria presence and $V_i$ are subject random effects; here $\Gamma(a,b)$ denotes the Gamma distribution with $a$ and $1/b$ as shape and scale parameters respectively. In recent work, \cite{Roos11} and \cite{lunn2009bugs} argue that a $\Gamma(\epsilon,\epsilon)$, for a small $\epsilon$, prior on the precision of the random effects may be inappropriate and one should instead use a half normal or a half Cauchy prior on the standard deviation of the random effects. Furthermore, it was noted that of particular interest is the effect of the choice of prior for the precision parameter of the random effects on the posterior distribution of the fixed effects. Consequently, in this example, our interest is in utilizing the proposed framework for assessing global robustness to such choices of prior. Aside from the baseline, we consider five other choices: (1) half normal with variance 100; (2) half Cauchy with scale parameter 100; (3) uniform on $(0,100)$ (all for the standard deviation parameter of the mixed effects); (4) $\Gamma(1,2)$; and (5) $\Gamma(9,0.5)$ (all for the precision parameter of the mixed effects). We note that (5) is not a good prior and is included here for comparison purposes only. For all models, we use MCMC to generate 9500 samples from the posterior (after a burn-in of 1000) and use these samples to generate individual kernel density estimates for the marginal posteriors for all of the coefficients of the fixed effects (displayed in Figure \ref{fig:postplot}). We then compute the Fisher-Rao distance between each baseline marginal posterior and the corresponding posterior resulting from the perturbation of the prior. These results are reported in Table \ref{tab:dists}. In this case, the baseline model as well as models with priors (1)-(4) all yield very similar marginal posteriors, which is confirmed by the very small Fisher-Rao distances. For comparison, the unreasonable prior choice in (5) results in much larger distances between the marginal posteriors. It might be reasonable to conclude that, for the available dataset, Bayesian analysis is insensitive to any reasonable choice of the prior for the variance of the random effects.

\begin{figure}[!ht]
\begin{center}
\begin{tabular}{c c c c }

\small{intercept}&\small{drug}&\small{drug$+$}&\small{week}\\

\includegraphics[width=1.3in]{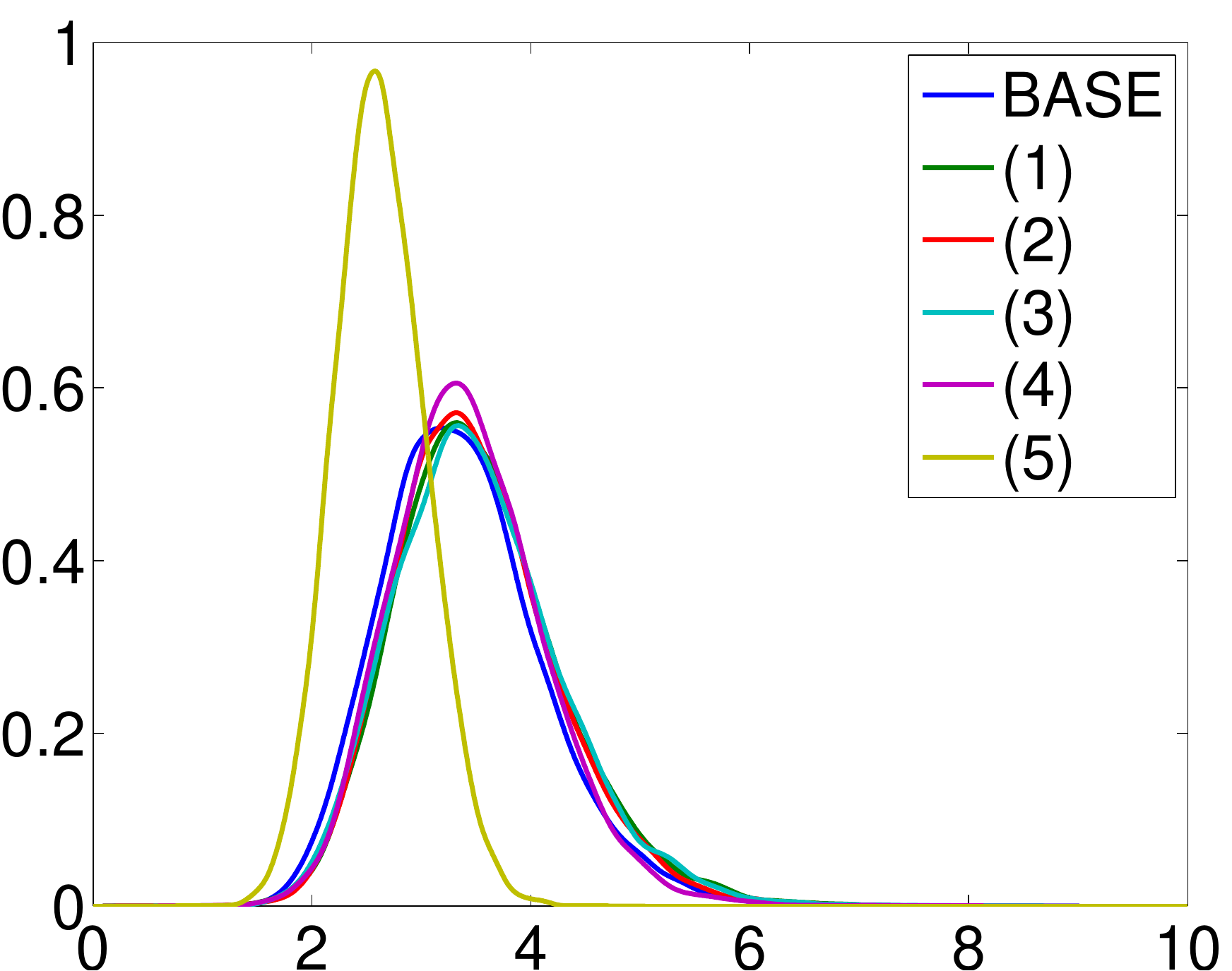}&\includegraphics[width=1.3in]{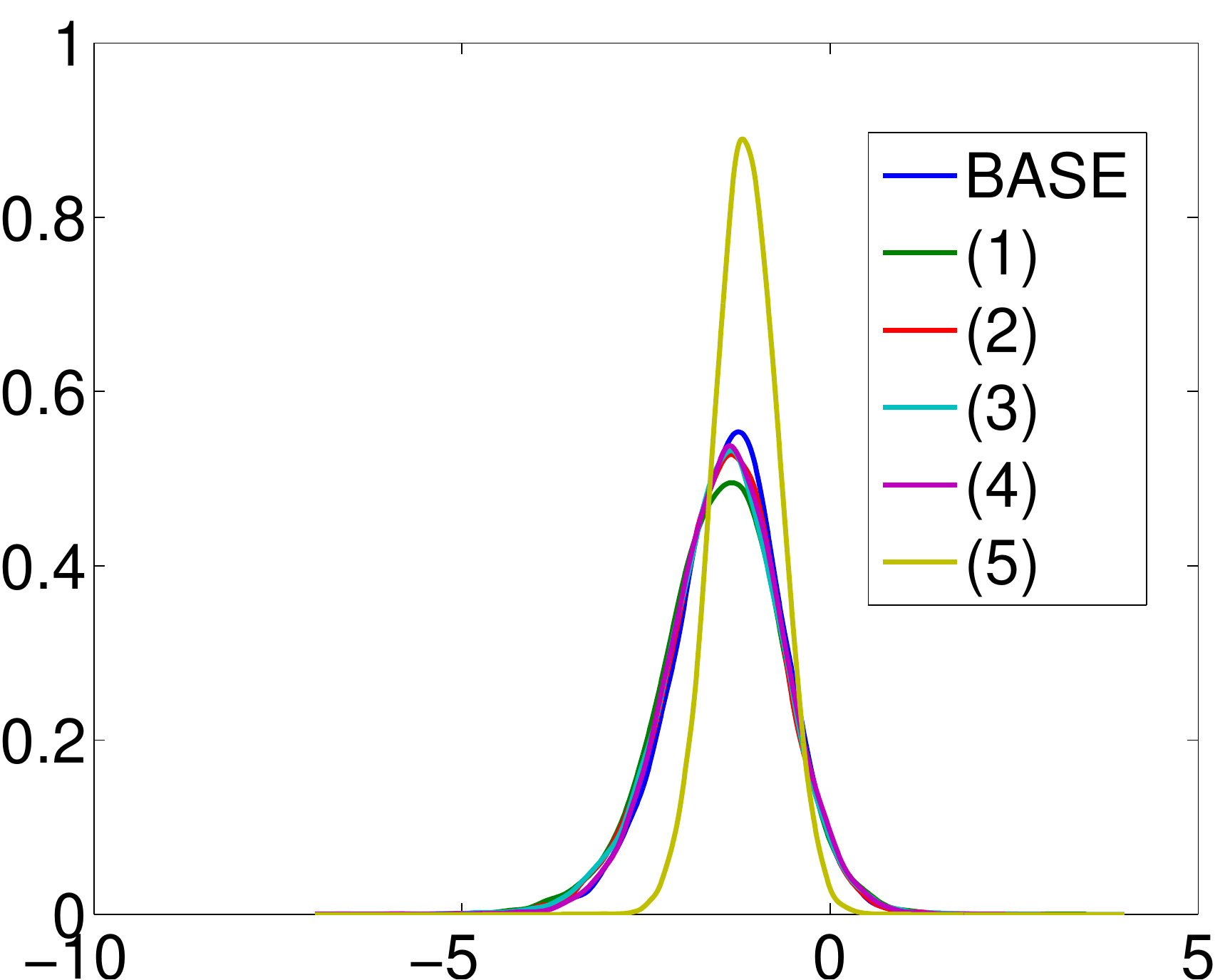}&\includegraphics[width=1.3in]{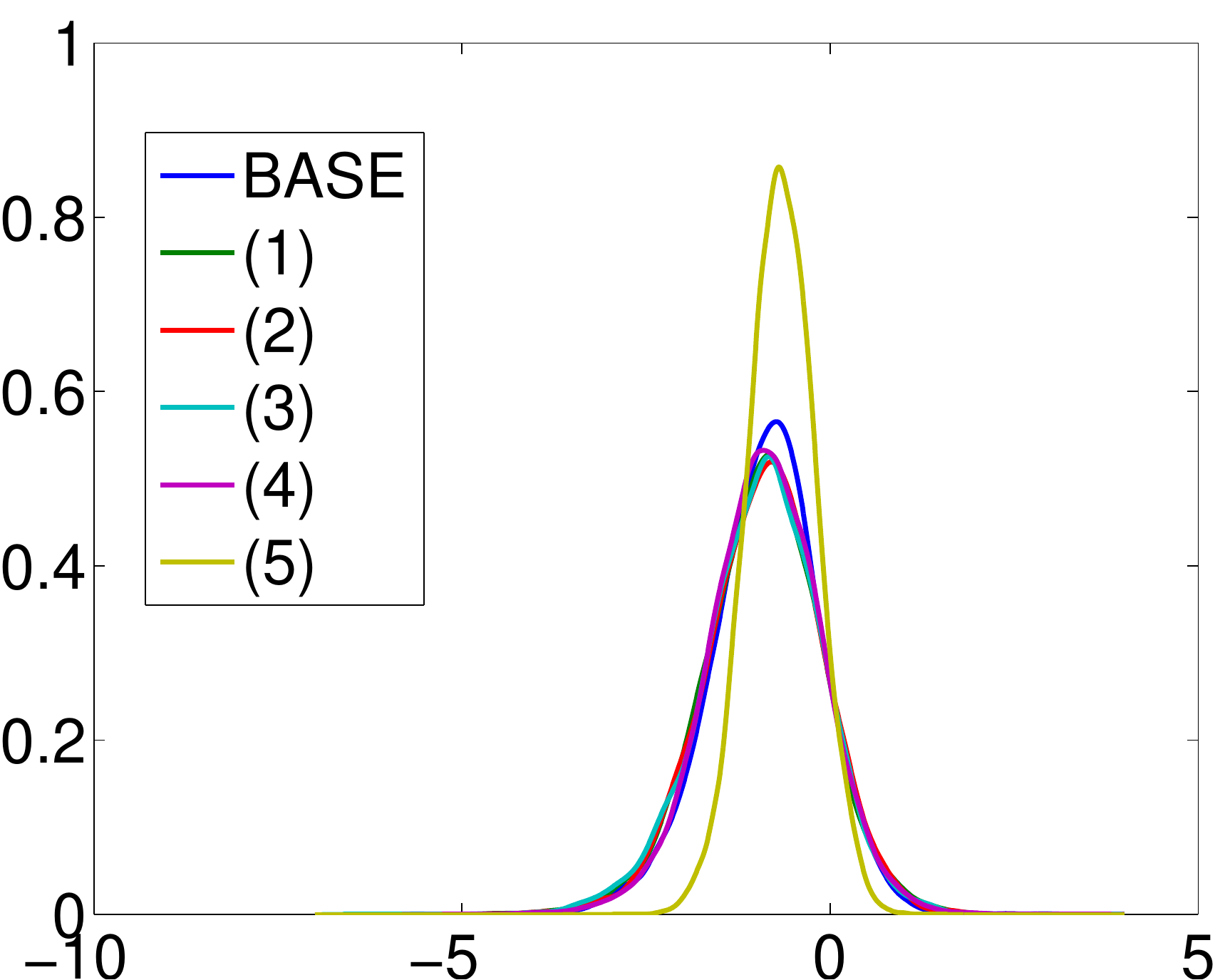}&\includegraphics[width=1.3in]{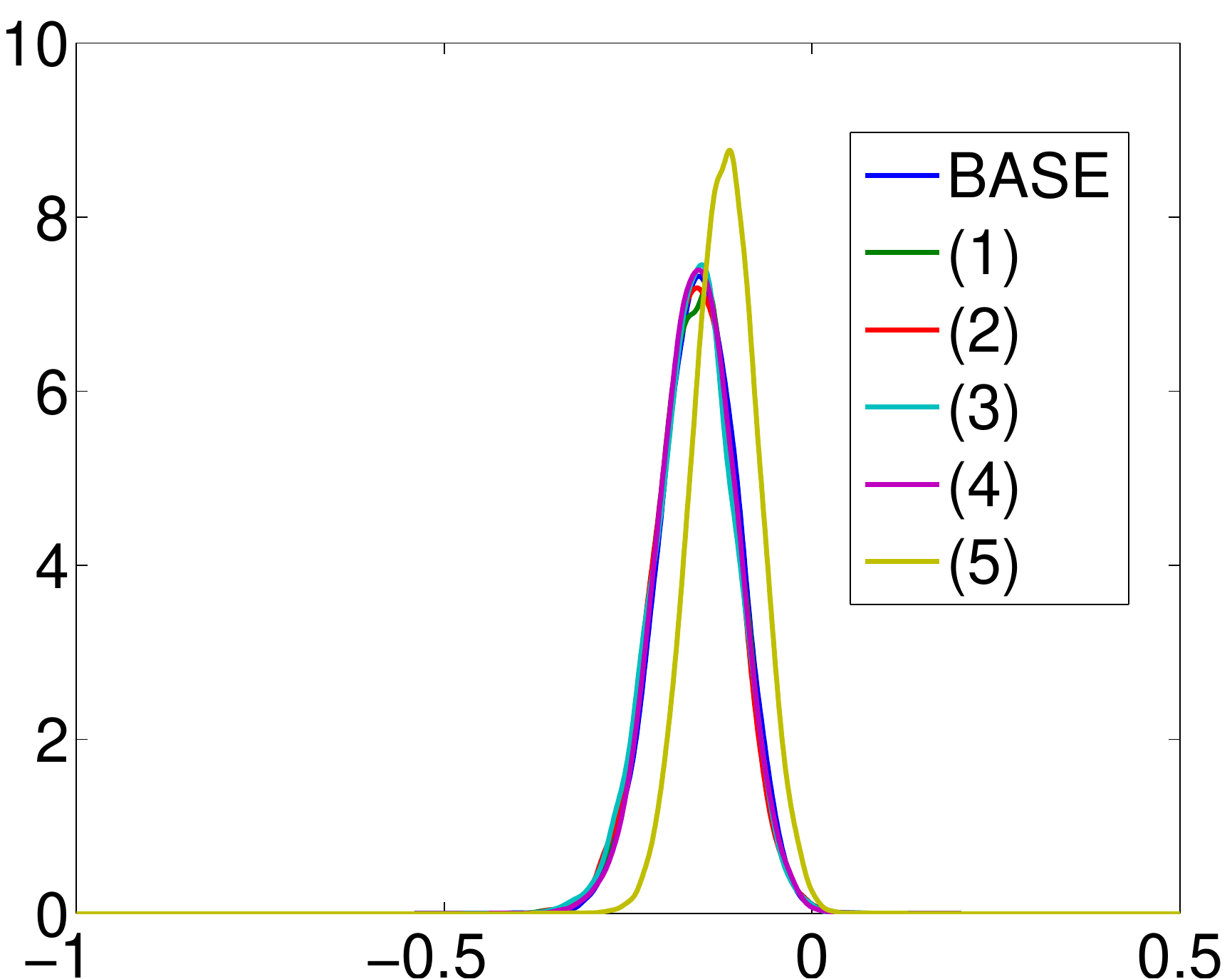}\\

\end{tabular}
\caption{Kernel density estimates of marginal posteriors under baseline (BASE) and perturbed models ((1)-(5)). \label{fig:postplot}}
\end{center}
\end{figure}

\begin{table}[!ht]
\begin{center}
\begin{tabular}{|c|c|c|c|c|c|}
\hline
\multirow{2}{*}{\textit{Fixed Effect}}&\multicolumn{5}{|c|}{\textit{Model}}\\
\cline{2-6}
&(1)&(2)&(3)&(4)&(5)\\
\hline
intercept&0.1054&0.0864&0.0982&0.0740&0.6716\\
drug&0.0716&0.0499&0.0590&0.0435&0.3835\\
drug$+$&0.0666&0.0580&0.0683&0.0445&0.3432\\
week&0.0524&0.0572&0.0630&0.0311&0.3670\\
\hline
\end{tabular}
\caption{Fisher-Rao distances between the baseline marginal posteriors and marginal posteriors under perturbation of the prior of the precision parameter for the random effect. \label{tab:dists}}
\end{center}
\end{table}

\end{example}

\subsection{Local Sensitivity Analysis}\label{sec:locsen}

In this section we define first order local sensitivity measures based on the commonly used Bayes factor and a general posterior functional represented via an integral. We then propose a second order local sensitivity measure based on the Fisher-Rao geodesic distance between posterior densities. All of the local sensitivity measures are derived under the geometric $\epsilon$-contamination class. First, we introduce some notation. Let $p_0$ be the baseline posterior, $\pi_0$ be the baseline prior, $\pi_1$ be another candidate prior (model selection setup), $f$ be the likelihood, and $v_g=\exp^{-1}_{\sqrt{\pi_0}}(\sqrt{g})\in T_{\sqrt{\pi_0}}(\Psi)$ be a perturbation of the baseline prior in the direction of a prior contaminant $g$. Let $m(x|\pi_0)$ denote the marginal with respect to the baseline prior, and define $m(x|\epsilon g)=\int_{\Theta}f(x|\theta)(\exp_{\sqrt{\pi_0}}(\epsilon v_g)(\theta))^2 d\theta$ and $\tilde{m}(x|v_g)=\int_{\Theta}f(x|\theta)\sqrt{\pi_0(\theta)}v_g(\theta) d\theta$. We use $F$ to denote a general functional of interest and $p_{\epsilon g}$ to denote the posterior obtained from a member of the geometric $\epsilon$-contamination class.
\begin{proposition}\label{local}
Under the notation described above, the local sensitivity measures based on the Bayes factor, posterior functional and geodesic distance, respectively, are:
\begin{enumerate}
 \setlength{\itemsep}{8pt}
\item If $F_{\pi_0,\pi_1}(v_g)=\frac{m(x|\epsilon g)}{m(x|\pi_1)}$ denotes the Bayes factor for comparing the marginals of the contaminated baseline prior and another candidate prior, then the corresponding local sensitivity measure is $dF_{\pi_0,\pi_1}(v_g)|_{\epsilon=0}=2\frac{\tilde{m}(x|v_g)}{m(x|\pi_1)}$.
\item
Suppose $F_{\pi_0,h}(v_g)$ is the expectation of $h(\theta)$ with respect to $p_{\epsilon g}$. Then, $dF_{\pi_0,h}(v_g)|_{\epsilon=0}=\frac{2}{m(x|\pi_0)}\int_{\Theta}h(\theta)f(x|\theta)\sqrt{\pi_0(\theta)}v_g(\theta)d\theta-\frac{2\tilde{m}(x|v_g)}{m(x|\pi_0)}\int_{\Theta}h(\theta)p_0(\theta|x)d\theta.$
\item 
Let $F_{\pi_0}(v_g)$ represent the squared geodesic distance between the posteriors $p_0$ and $p_{\epsilon g}$. Then,
$d^2F_{\pi_0}(v_g)|_{\epsilon=0}=\frac{4\tilde{m}(x|v_g)}{m(x|\pi_0)}\int_\Theta\frac{v_g(\theta)}{\sqrt{\pi_0(\theta)}}p_0(\theta|x)d\theta-2\int_\Theta\frac{v_g(\theta)^2}{\pi_0(\theta)}p_0(\theta|x)d\theta-\frac{2\tilde{m}(x|v_g)^2}{m(x|\pi_0)^2}$.
\end{enumerate}
\end{proposition}

The key point here is that these local sensitivity measures are defined using directional derivatives, where the directions are the perturbations defined using the proposed geometric $\epsilon$-contamination method. We are hence able to incorporate the geometry of the space of densities in the definition of the measure. In other words, this approach unifies the local diagnostic measures with the geometry of the space under consideration, and therefore possesses a natural geometric calibration. In the following, we present two examples showcasing these sensitivity measures.

\begin{example}

First, consider the following baseline (and data generating) model:
\begin{align*}
x_i|\theta&\overset{\text{i.i.d.}}\sim f=N(\theta,1), \quad i=1,\ldots,50;\\
\theta&\sim \pi_0=N(0,1)
\end{align*}
We simulate 50 observations from this model to use as the given data. We consider a family of $t$ prior contaminations, parameterized by the degrees of freedom, $df=3,\dots,100$. We compute the three different local sensitivity measures given in Proposition \ref{local}. The plots of these sensitivity measures are provided in Figure \ref{fig:localsen}. In the case of the Bayes factor, we use $\pi_1=N(0,5)$. This simulation example allows us to easily interpret the effectiveness of the proposed method. Suppose the true model is the baseline model; as we perturb away from the baseline model, the Bayes factor should decrease, indicated by a negative sign in its local sensitivity measure for all degrees of freedom of the contaminating $t$. Furthermore, as the degrees of freedom increase, this local measure should tend to zero, because the contaminating densities look more and more like the baseline prior. The same trend should hold for the second order local sensitivity measure based on the geodesic distance. This is easy to see because as one increases the degrees of freedom of the $t$ contaminant, the perturbed posterior approaches the baseline posterior, collapsing the geodesic to a single point. The local sensitivity of the posterior mean is harder to interpret in this case because the trend is dependent on the simulated data. In general, the sign of this measure should differ from that of the sample mean of the simulated data and the measure should tend to zero for increasing degrees of freedom of the $t$. Overall, we expect the local sensitivity of the posterior mean to be small since both the standard normal and the $t$ have mean zero.
\begin{figure}[!ht]
\begin{center}
\begin{tabular}{c c c }
\small{Bayes Factor}&\small{Posterior Mean}&\small{Geodesic}\\

\includegraphics[width=1.7in]{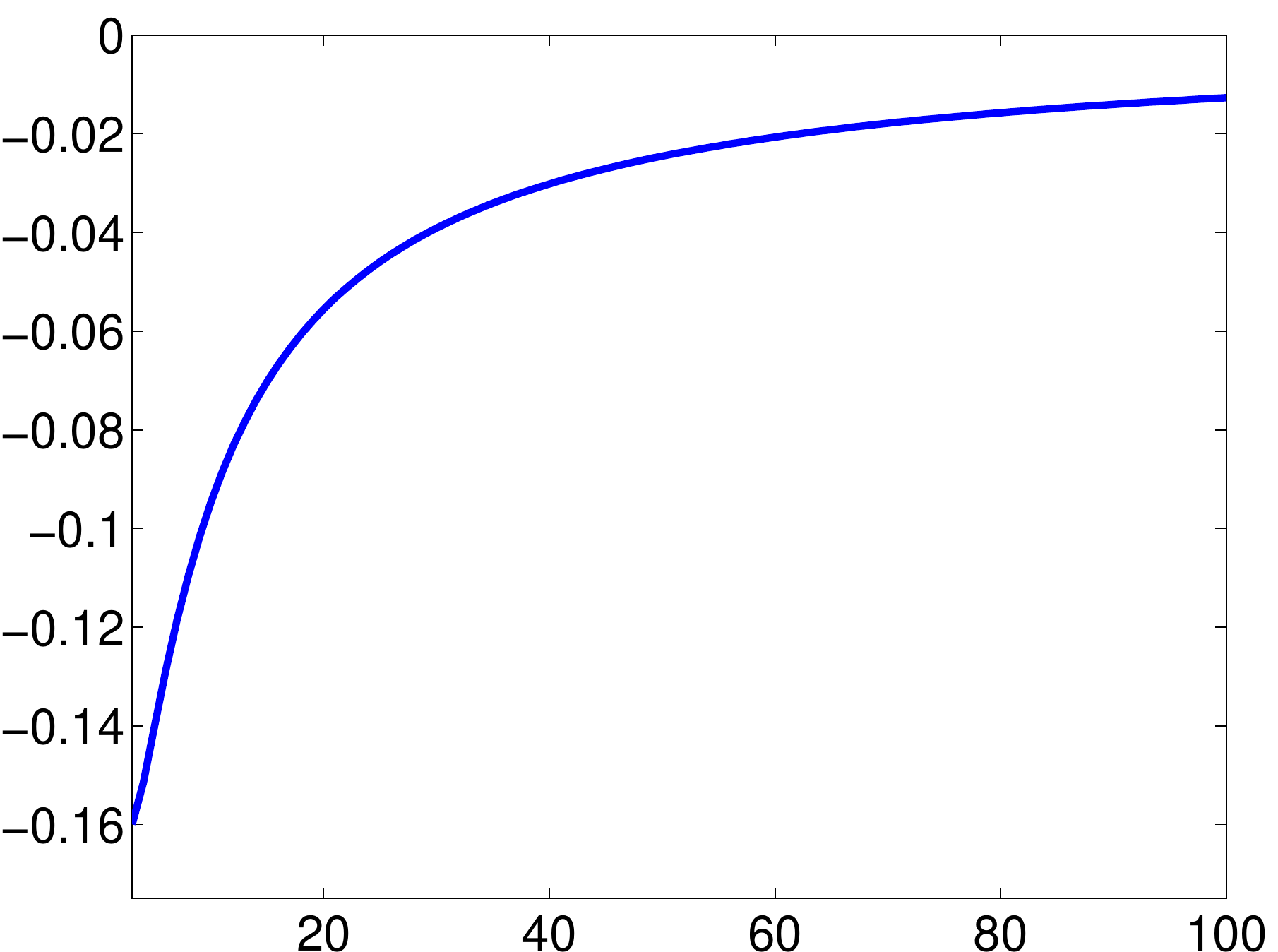}&\includegraphics[width=1.7in]{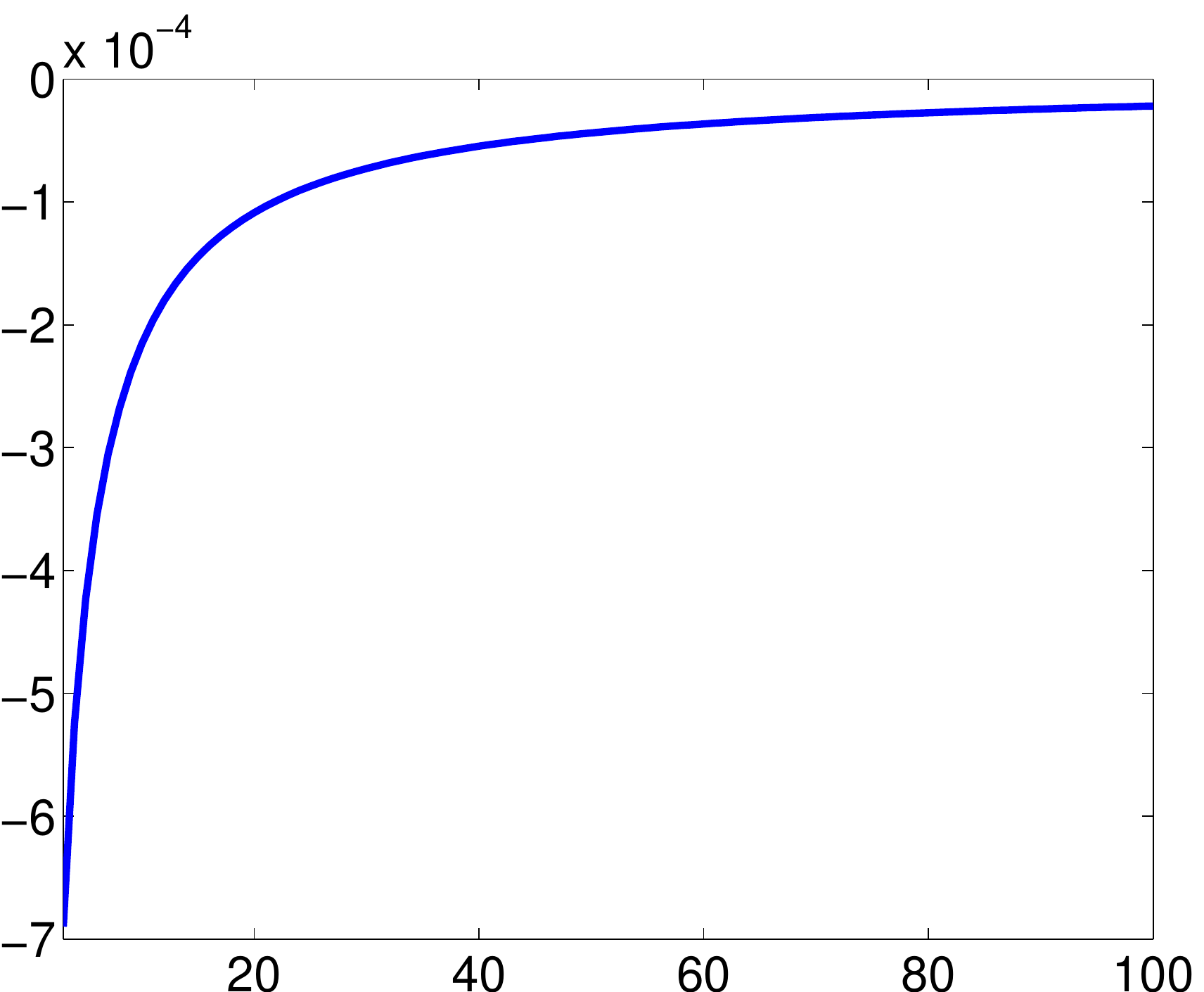}&\includegraphics[width=1.7in]{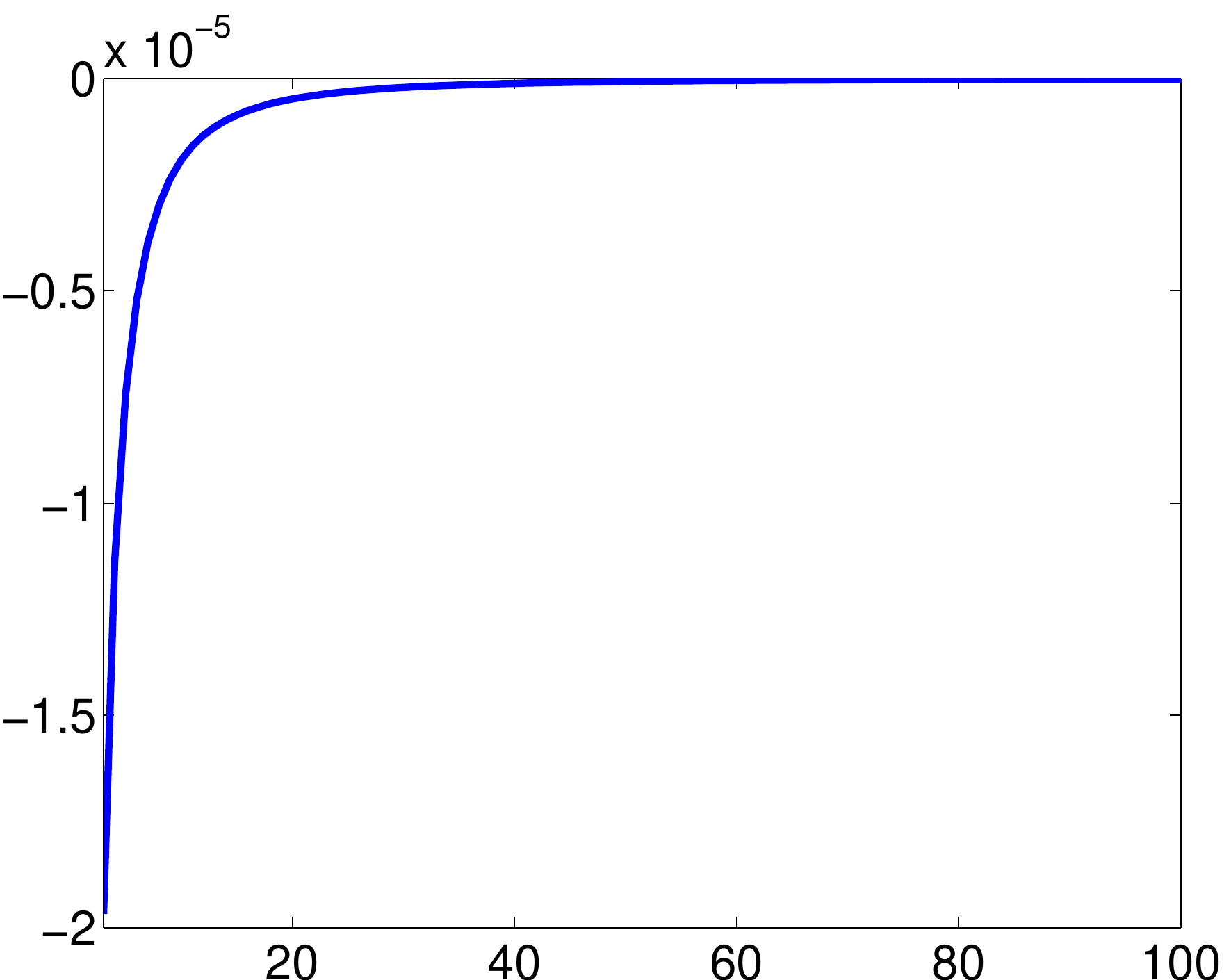}\\
\end{tabular}
\caption{Local influence analysis based on the Bayes factor, posterior mean and geodesic distance. Here, we consider perturbing the baseline standard normal prior with a $t$ with increasing degrees of freedom ($x$-axis, $df=3,\dots,100$). \label{fig:localsen}}
\end{center}
\end{figure}
\end{example}

\begin{example}

To compare and analyze the performance of our methodology for local and global perturbations, we consider the same turtle directional data previously used in Example 2 under the same baseline model:
\begin{align*}
x_i|\theta&\overset{\text{i.i.d.}}\sim f=vM(\theta,\hat{\kappa}), \quad i=1,\ldots,76;\\
\theta&\sim \pi_0=vM(0,0.01),
\end{align*}
where $\hat{\kappa}=1.1423$. In this example, we consider local sensitivity to $\epsilon$-contamination of the prior. The contamination class we consider is a family of wrapped Laplace distributions with zero mode developed by \cite{JKozub}. This family can be parameterized by a concentration parameter $\lambda$ and a skewness parameter $\eta$. For $\eta<1$ $(>1)$, the wrapped Laplace distribution is skewed in the counterclockwise (clockwise) direction, and when $\eta=1$ we obtain the symmetric wrapped Laplace distribution. Thus, our contamination class is formed by jointly varying the parameters $\lambda$ from 0.2 to 10 and $\eta$ from 0.2 to 5. For the local Bayes factor measure, we set $\pi_1=vM(\pi/2,0.01)$; Figure \ref{fig:localtur} displays the results. The local Bayes factor measure is insensitive to perturbations of the prior only when the contaminant is approximately symmetric $\nu\thickapprox 1$. When it is highly skewed in the counterclockwise direction, the measure is positive and vice versa when $\nu>1$. In terms of the concentration parameter, the local Bayes factor measure tends to zero as the concentration goes to zero. This is sensible since both the von Mises and the wrapped Laplace converge to the uniform in that case. As a result, the perturbations have little effect on the baseline model. The posterior mean is fairly sensitive for moderately concentrated ($\lambda>1$) and highly-counterclockwise skewed wrapped Laplace distributions. In these cases, it decreases from the baseline posterior mean of 1.1198 radians as indicated by the negative sign of the local sensitivity measure. On the other hand, when we perturb the prior using highly-clockwise skewed wrapped Laplace distributions, the posterior mean is fairly insensitive (local sensitivity measure is close to zero). Finally, the second order measure for the geodesic distance shows a similar trend to the first order measure for the posterior mean.

\begin{figure}[!ht]
\begin{center}
\begin{tabular}{c c c }

\small{Bayes Factor}&\small{Posterior Mean}&\small{Geodesic distance}\\

\includegraphics[width=1.7in]{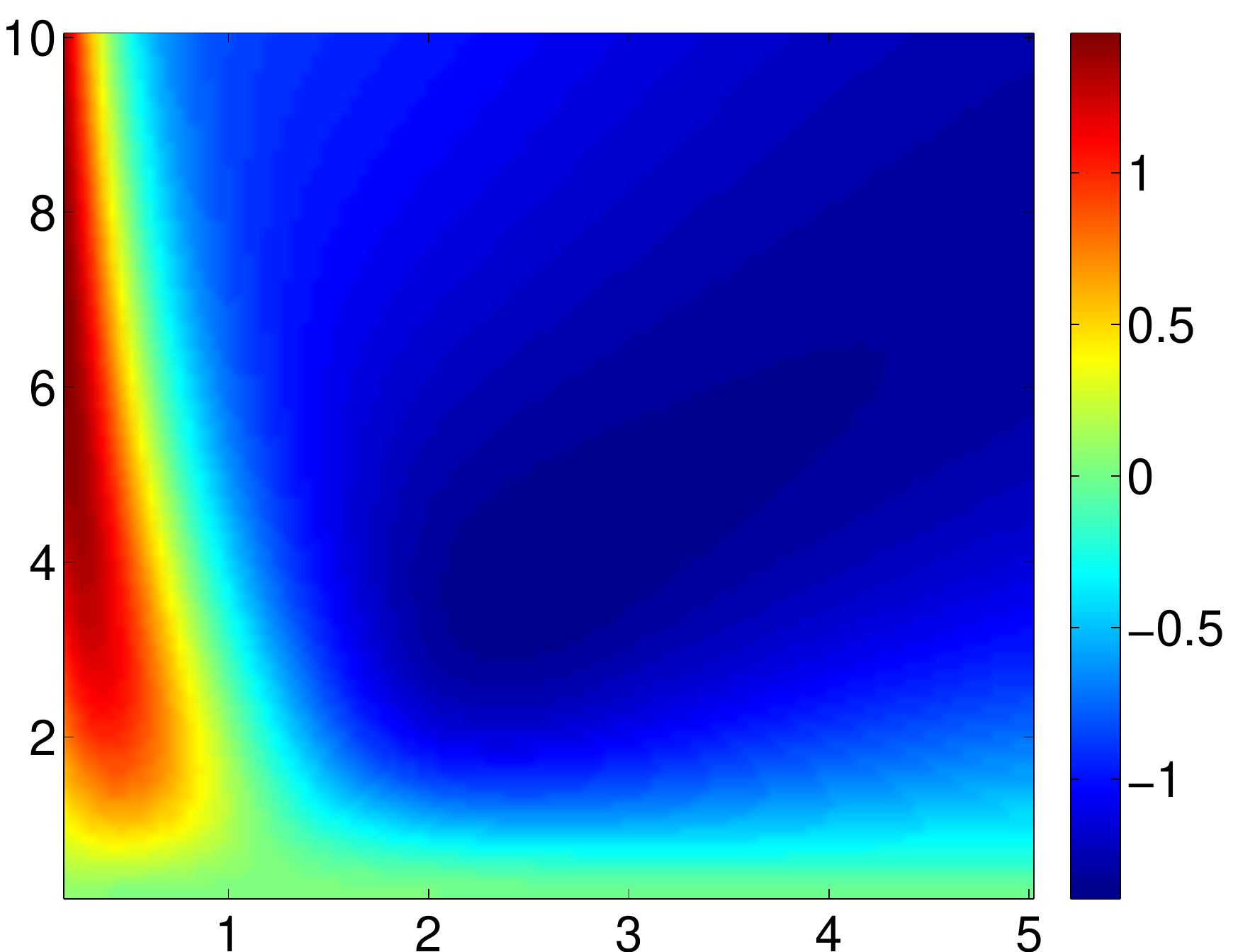}&\includegraphics[width=1.7in]{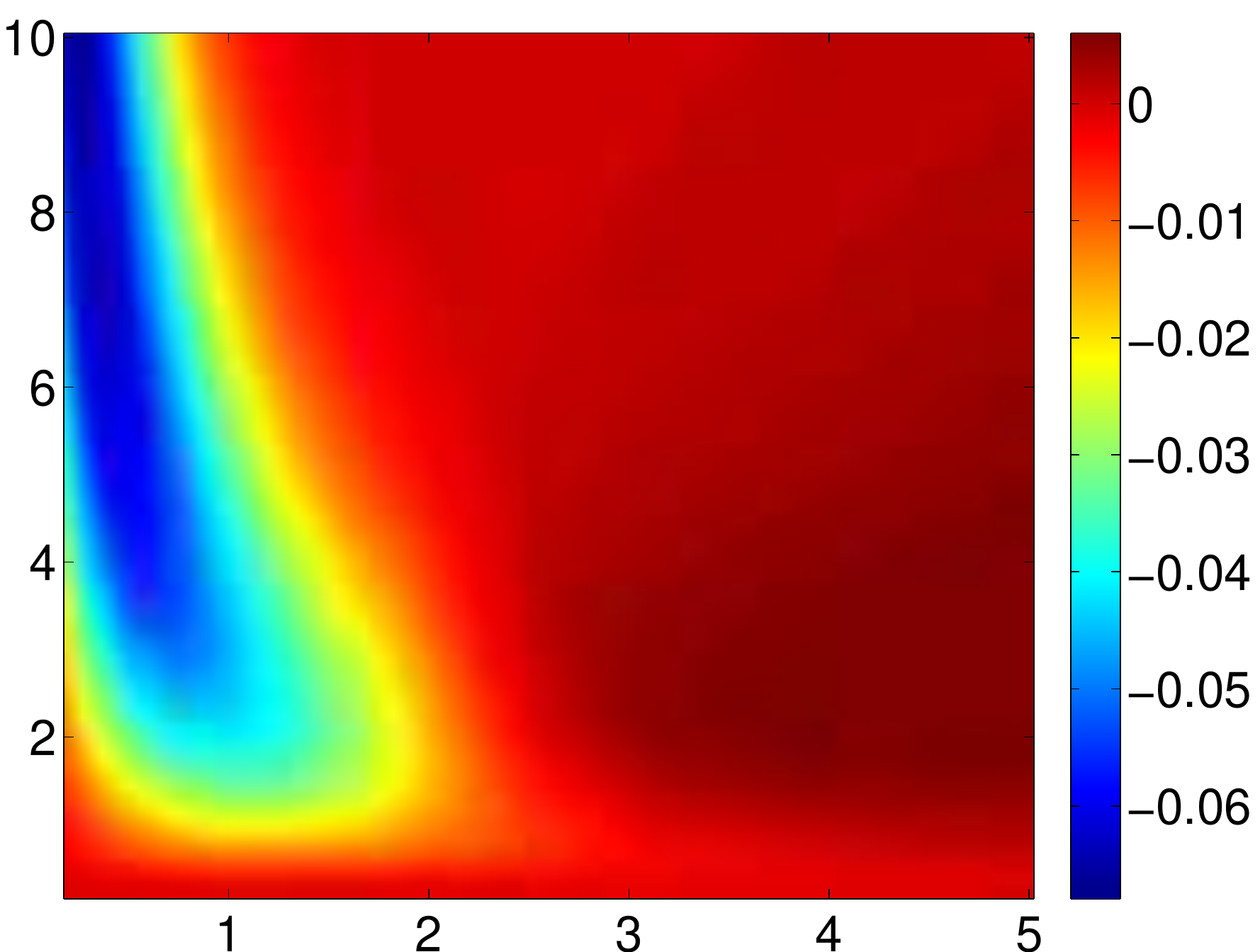}&\includegraphics[width=1.7in]{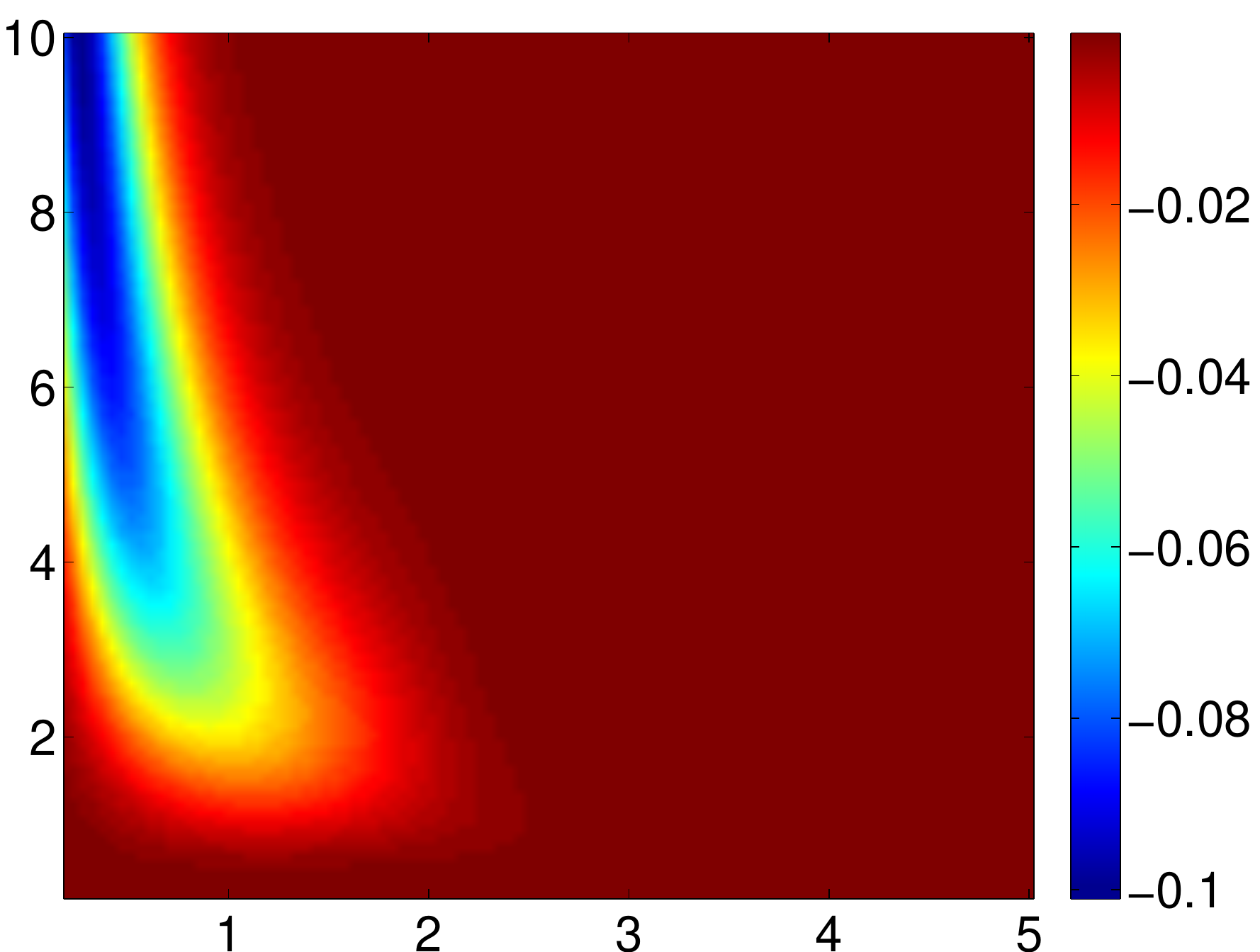}\\

\end{tabular}
\caption{Local influence analysis based on the Bayes factor, posterior mean and geodesic distance for the turtle directional data. We perturb the baseline von Mises prior by a class of wrapped Laplace priors with different concentration ($y$-axis) and skewness ($x$-axis) parameters. \label{fig:localtur}}
\end{center}
\end{figure}

\end{example}

\subsection{Influential Observations}\label{influence}

The general methodology in identifying influential observations is similar to that introduced in Section \ref{sec:globsen} in the sense that we define the influence measure based on distances between posteriors. We again denote the baseline posterior as $p_0$. One can evaluate the influence of the $k$th observation on the posterior distribution by removing it from the observation set and estimating the posterior distribution using the remaining observations. This results in a new posterior distribution $p_k$ leading to the following definition.
\begin{definition}
Given the baseline posterior $p_0$ and the posterior under case deletion $p_k$, the influence of observation $k$ is defined as $I(k)=d_{FR}(p_0,p_k)$. \label{imeasure}
\end{definition}
Note that this distance is symmetric and has an upper bound of $\pi/2$, which avoids the ambiguity present in many divergence measures and provides a natural scale for evaluating influence. When the posterior density is unavailable in closed-form, and computing the marginal likelihood numerically is infeasible, it becomes necessary to estimate the quantity in Equation \ref{imeasure} using Monte Carlo. We propose as estimator based on samples from the baseline posterior (generated using either direct sampling or MCMC) to evaluate the Fisher-Rao distance between the baseline posterior and the posterior under case-deletion.
\begin{proposition}\label{MCMC}
Suppose $p_k$ is the posterior density under case-deletion and $p_0$ is the baseline posterior density. Correspondingly, let $f_k$ and $f_0$ be the case-deletion and baseline likelihoods with $\pi$ representing the prior on the parameters. Let $x_k$ denote the $k$th observation and $x_{(k)}$ denote the set of observations not containing the $k$th one. Then, $$I(k)=d_{FR}(p_0,p_k)=\left[\displaystyle \int_{\Theta}\frac{1}{f(x_k|x_{(k)},\theta)}p_0(\theta|x)d\theta \right]^{-1/2}\displaystyle\int_{\Theta}\left[\frac{f_k(x|\theta)}{f(x|\theta)}\right]^{1/2}p_0(\theta|x)d\theta.$$ Given a sample from the baseline posterior density, $\{\theta_1,\dots,\theta_N\}$, the Monte Carlo estimate of $I(k)$ is given by
\begin{equation}
\hat{I}(k)=\cos^{-1}\Big[\frac{b}{N}\displaystyle\sum_{i=1}^N a_i\Big],\quad
\text{where  } a_i=\left[\frac{f_k(x|\theta_i)}{f(x|\theta_i)}\right]^{1/2}, b=\left[\frac{1}{N}\sum_{i=1}^N\frac{1}{f(x_k|x_{(k)},\theta_i)}\right]^{-1/2}.
\end{equation}
\end{proposition}
It is routine to show that the estimate $\hat{I}$ is a consistent estimator of $I$ using the Ergodic theorem. Also, note that only one posterior sample needs to be generated to evaluate the influence measure for all observations making this approach computationally tractable. While beyond the scope of this paper, we plan to study the properties of this estimator as was done for other case-deletion importance sampling estimators in \cite{epifani2008}. Next, we consider three different examples, linear regression, logistic regression, and mean shape estimation, to illustrate the performance of the proposed influence measure.

\begin{example}

We first consider influence analysis in a Bayesian multiple linear regression setting. The data analyzed here comes from the book by \cite{Kutner} containing 54 test cases. The response $y$ is the natural logarithm of survival time. There are eight predictors: blood-clotting score, prognostic index, enzyme test, liver test, age, gender (binary), moderate alcohol use (binary), and heavy alcohol use (binary). Due to large differences in predictor scales, we standardize the response and predictor variables. We use $X$ to denote the standard design matrix and $\theta$ to denote the nine-dimensional vector of unknown regression coefficients. We utilize the following baseline Bayesian model:
\begin{align*}
y|\theta,X&\sim f=N(X^T\theta,\sigma^2I_{54})\\
\theta&\sim \pi=N(0,1000I_{9}).
\end{align*}
For simplicity, instead of placing a prior on $\sigma$, we estimate it from the given data. Because we have chosen a conjugate prior for $\theta$, the posterior density is also a Gaussian distribution. Note that if one deletes a case from this data, the resulting posterior distribution is again Gaussian. In this setup, we are faced with computing the Fisher-Rao distance between two multivariate Gaussian posteriors. This requires the computation of a high-dimensional integral and we will utilize Monte-Carlo and importance sampling to approximate it. We note that it is easy to sample from the baseline posterior density; thus, we can use it as a natural importance sampling density to estimate the integral in the expression of the Fisher-Rao distance. We rewrite the inner product between the baseline posterior and the posterior under case-deletion as $\langle \sqrt{p_0},\sqrt{p_k} \rangle = \int_{\Theta} \sqrt{p_0(\theta|y,X)}\sqrt{p_k(\theta|y,X)}d\theta = \int_{\Theta}\sqrt{\frac{p_k(\theta|y,X)}{p_0(\theta|y,X)}}p_0(\theta|y,X)d\theta$. Thus, our approach is to generate a large sample, $\{\theta_1,\dots,\theta_N\}$, from the baseline posterior and then estimate the distance using the following Monte Carlo estimate: $\hat{I}(k)=\hat{d}_{FR}(p_0,p_k)=\cos^{-1}\left[\frac{1}{N}\sum_{i=1}^N\sqrt{\frac{p_k(\theta_i|y,X)}{p_0(\theta_i|y,X)}}\right]$. In this example, we set $N=100000$. Convergence of the estimate $\hat{I}$ follows via the Ergodic theorem and a continuous mapping argument for $\cos^{-1}$. Note that this approximation is possible because we can easily evaluate the posterior density of each $\theta$ under $p_k$ and $p_0$. In other cases, we would be forced to resort to the estimate given in Proposition \ref{MCMC}.

\begin{figure}[!ht]
\begin{center}
\begin{tabular}{c c c}
$\small{\hat{I}}$&\small{Cook's Distance}&\small{\cite{PengDey}}\\
\includegraphics[width=1.7in]{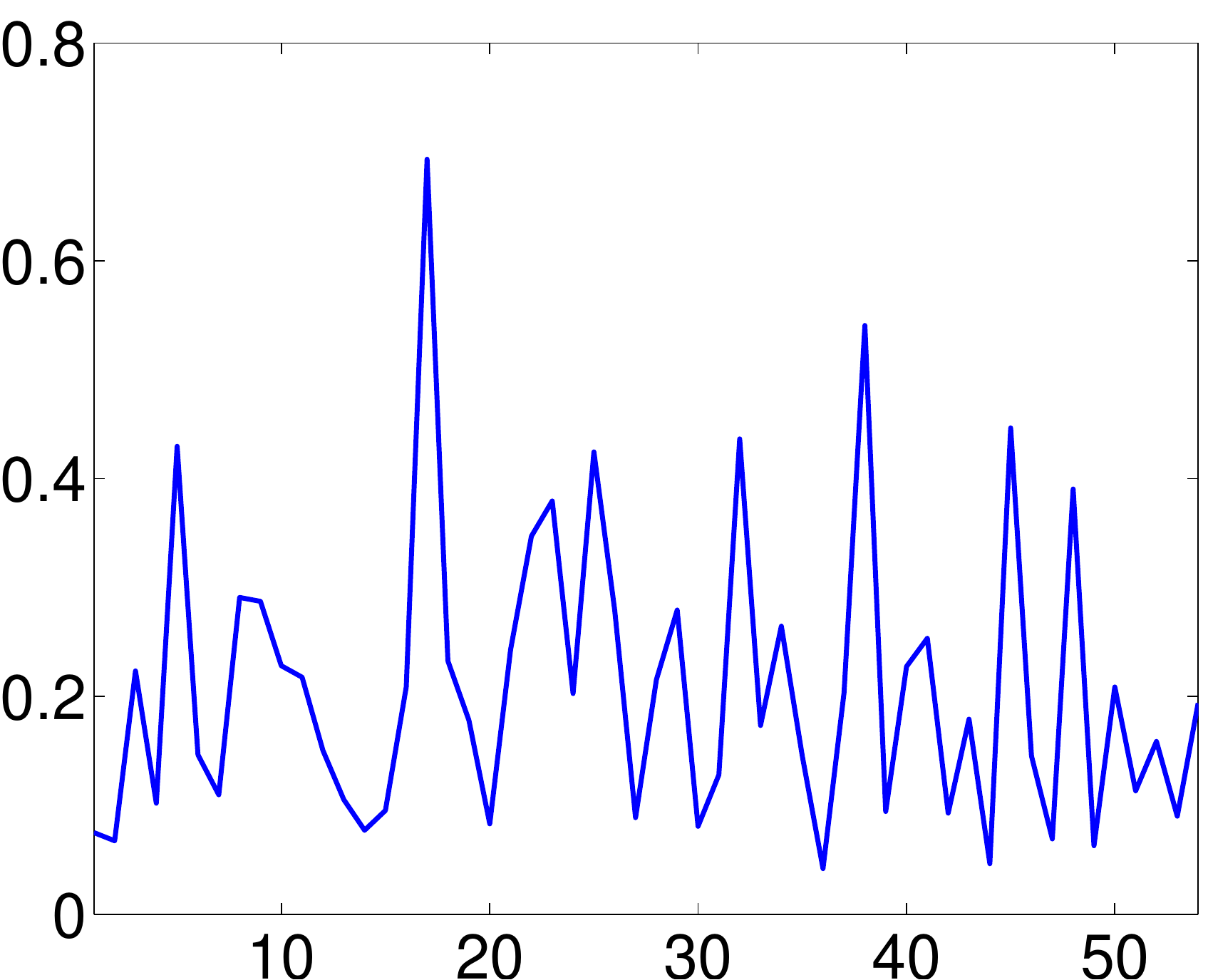}&\includegraphics[width=1.7in]{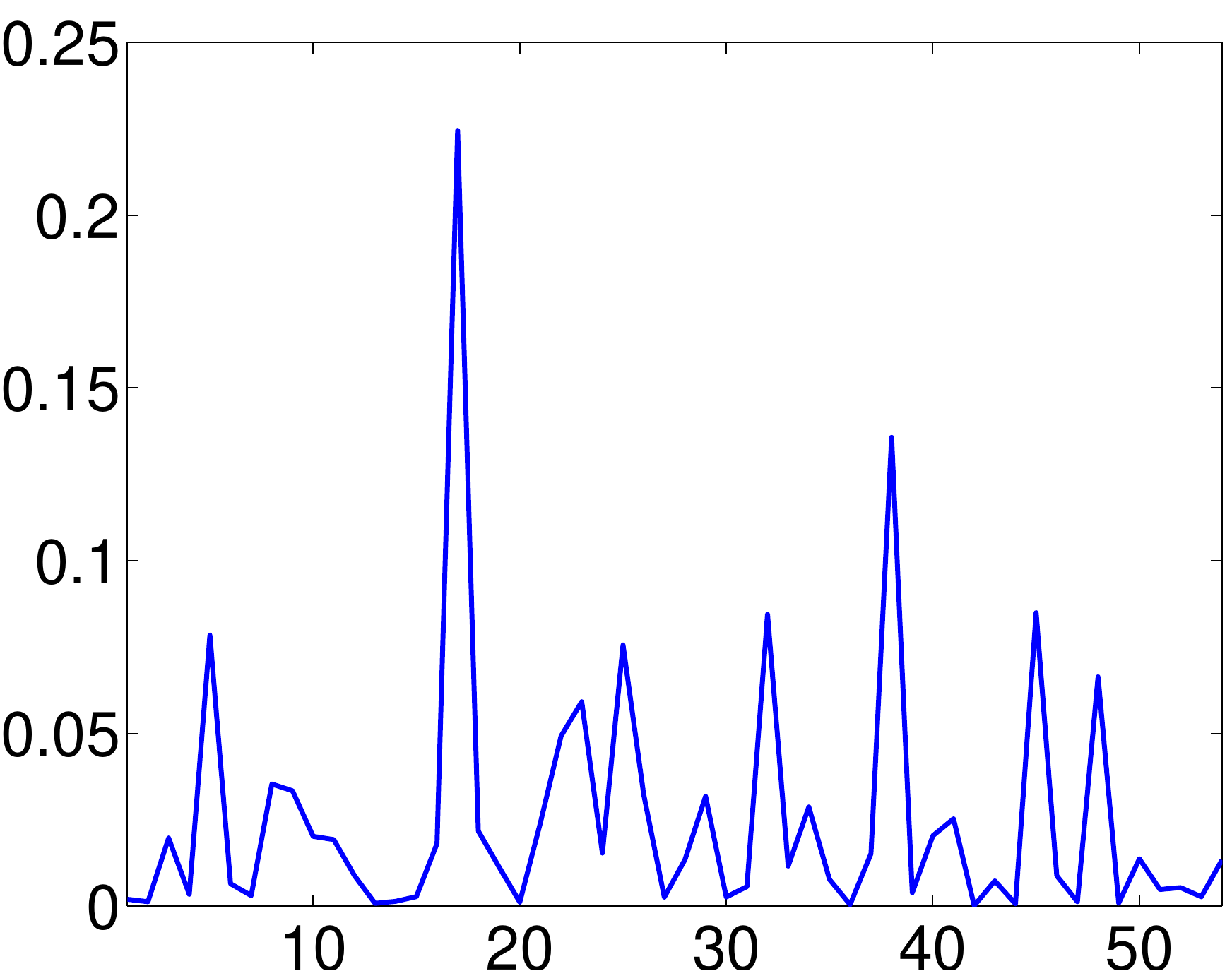}&\includegraphics[width=1.7in]{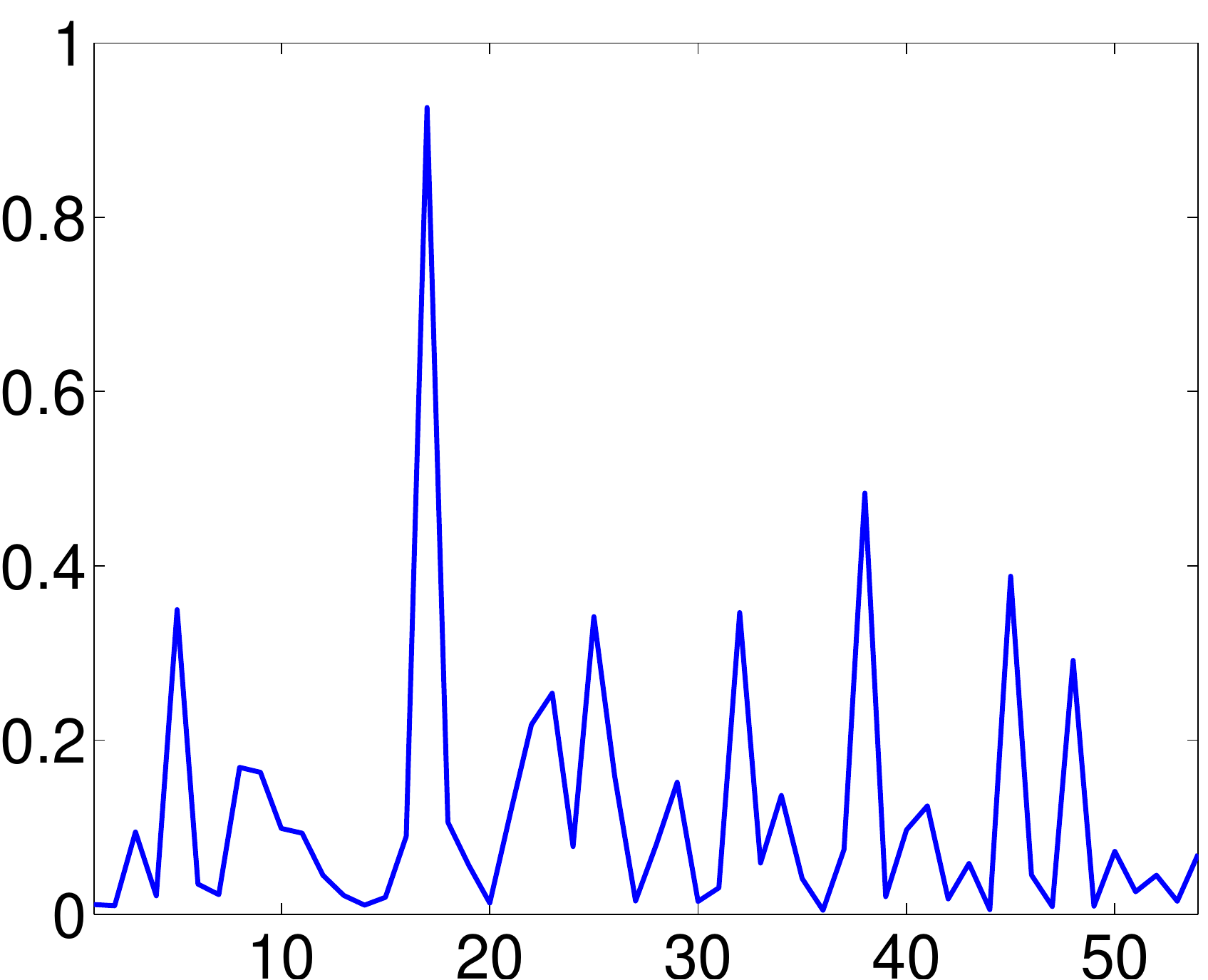}\\
\end{tabular}
\caption{Influence analysis in Bayesian multiple linear regression. The 54 observations are listed on the $x$-axis with the corresponding measure of influence on the $y$-axis. \label{fig:cooklin}}
\end{center}
\end{figure}

 In the left panel of Figure \ref{fig:cooklin}, we display the Fisher-Rao distances between the baseline posterior and the posterior under deletion of each case; the middle panel displays the standard Cook's distance in a frequentist setting; finally, in the right panel we have computed the influence measure proposed in \cite{PengDey} based on the KL divergence. Based on the $F$ statistic, Cook's distance does not flag any of the observations as influential, even though visually, observation 17 appears influential. Peng and Dey suggest flagging all observations, which yield a ``distance" greater than 0.25 as influential under their measure. Under their framework, one would consider seven observations as influential, with 17 being highly influential. A similar result can be seen when using our influence measure: observation 17 is again highly influential (distance greater than 0.7), and there are eight other observations which can be considered as possibly influential (distance is higher than 0.3).

In this example, we used Monte Carlo to estimate the Fisher-Rao distance. In order to numerically assess the convergence of this estimator we plot the estimate as a function of the number of samples from the baseline posterior. We do this for observations 2, 17, 38 and 52 as shown in Figure \ref{fig:impsampest}. From these plots it is evident that the estimator used in this example has good convergence properties and can be reliably used for detecting influential observations. In fact, for all of the presented cases, the estimate of the Fisher-Rao distance has converged with approximately 30000 samples from the baseline posterior. We also assess the quality of our estimator by generating 50 different samples of size 100000 from the baseline posterior and reporting the variance of the estimated Fisher-Rao distances. For all observations, the variance of these estimates was less than $1\times 10^{-5}$, supporting the claim that the estimates are fairly good and have very low variability.

\begin{figure}[!h]
\begin{center}
\begin{tabular}{c c c c }
Case 2&Case 17&Case 38&Case 52\\

\includegraphics[width=1.2in]{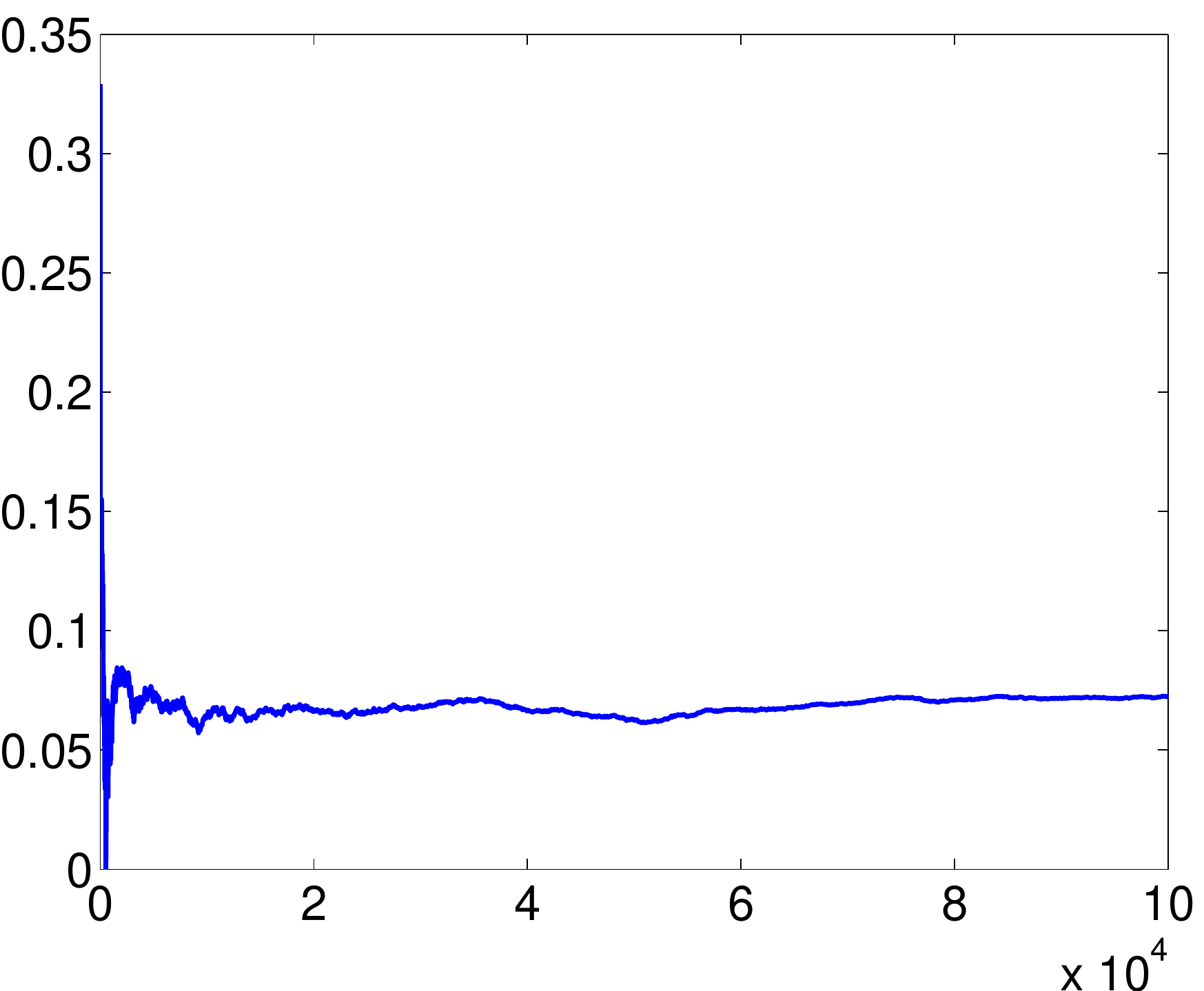}&\includegraphics[width=1.2in]{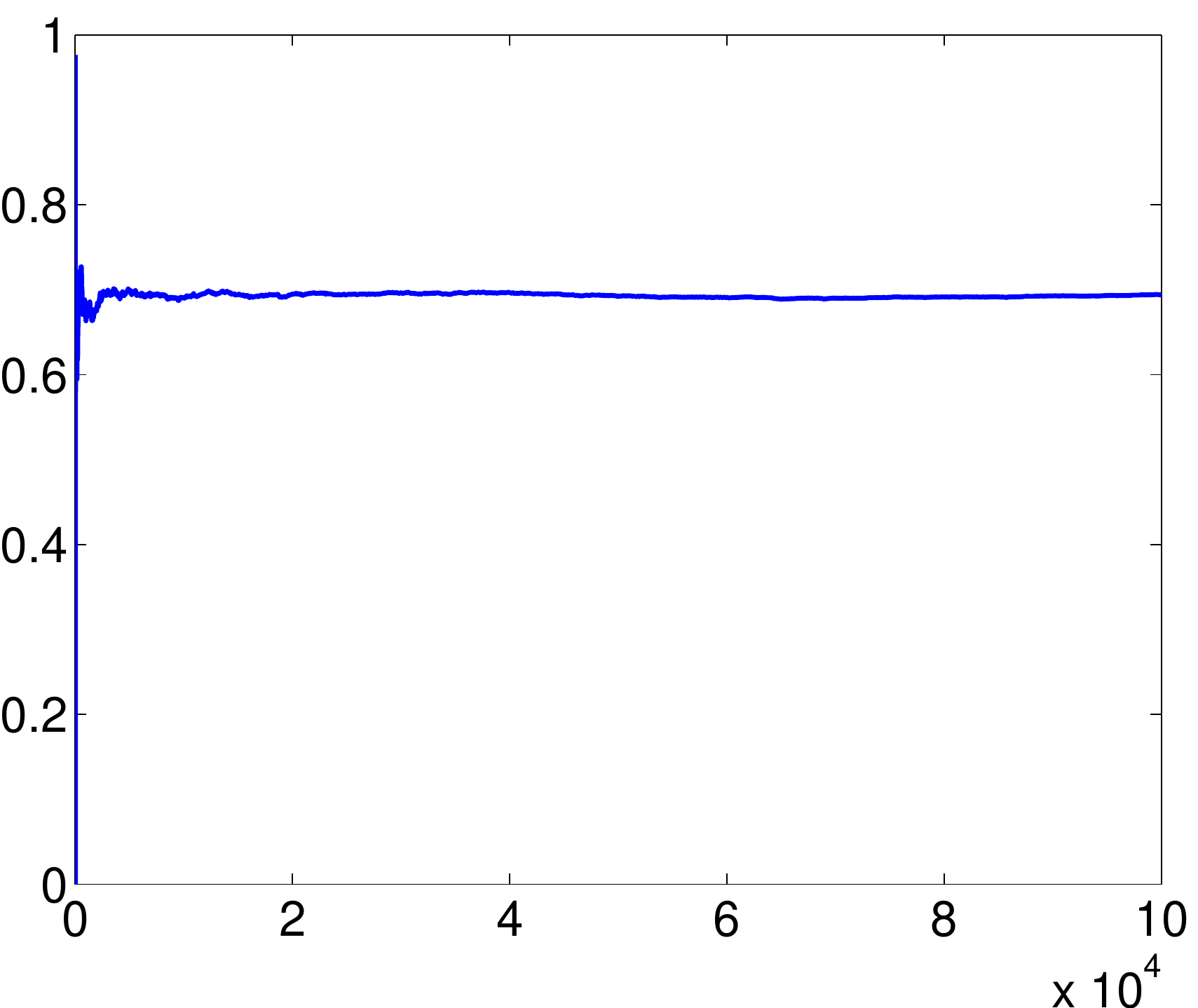}&\includegraphics[width=1.2in]{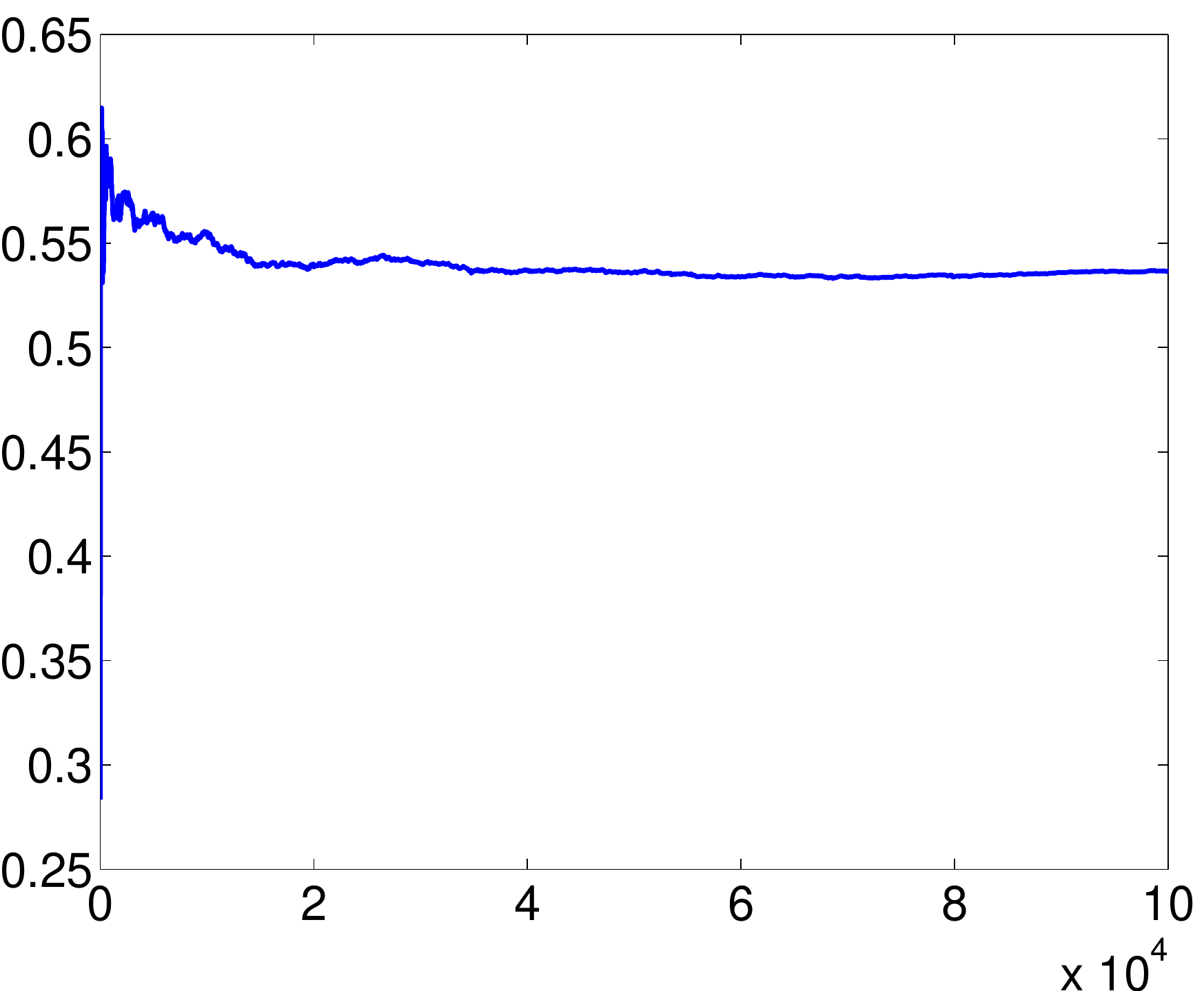}&\includegraphics[width=1.2in]{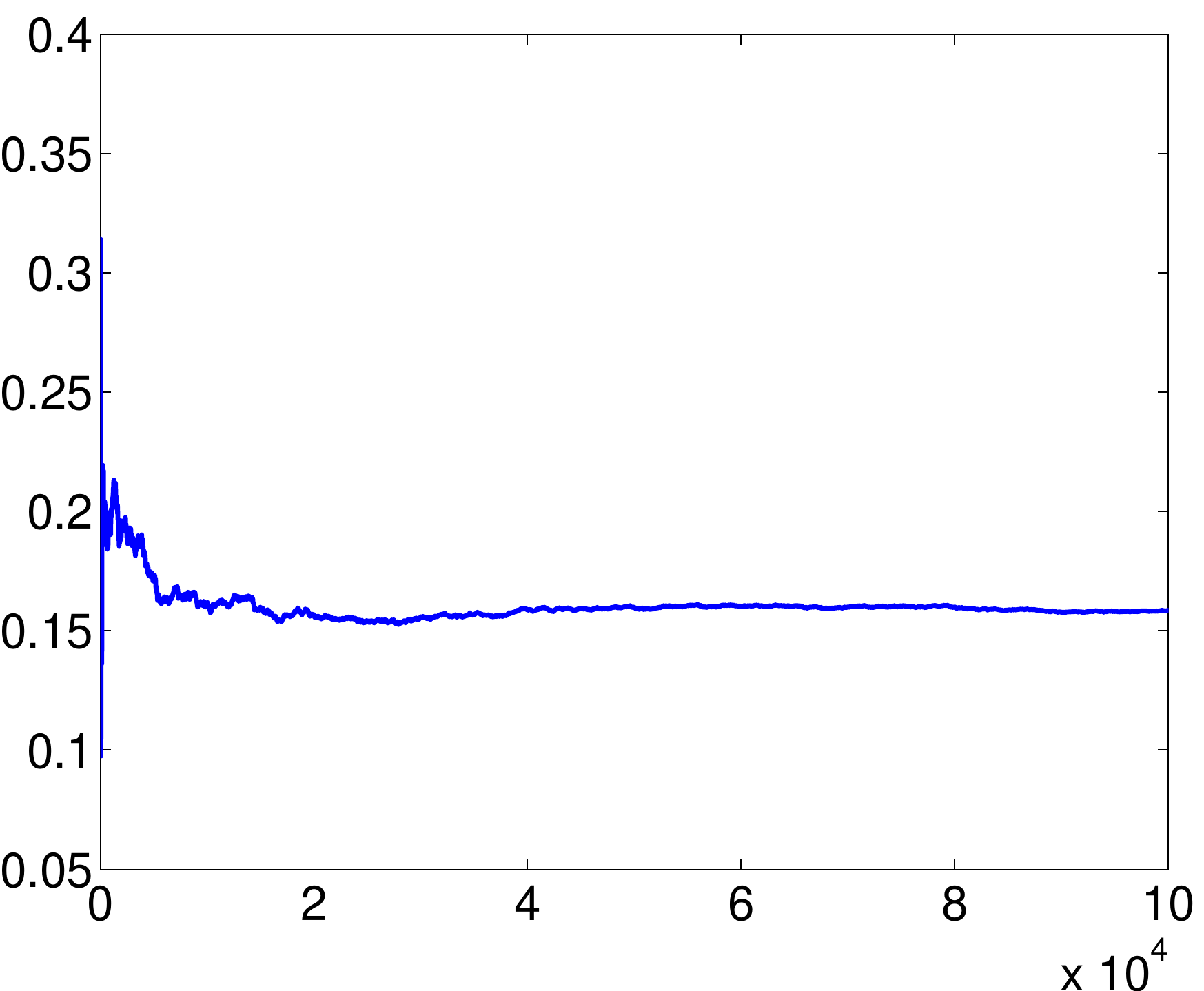}\\


\end{tabular}
\caption{Convergence plots for the Monte Carlo estimate of the Fisher-Rao distance for cases 2, 17, 38 and 52. \label{fig:impsampest}}
\end{center}
\end{figure}

\end{example}

\begin{example}

In this illustration, we identify influential observations in the Bayesian logistic regression setup. The dataset used here was previously analyzed by \cite{Fin} and was studied by \cite{PengDey} under the influence analysis setting. There are 39 cases in this data, where the response $y$ is a vector of binary outcomes indicating whether or not vasoconstriction occurred. The two predictor variables are the volume of air inspired and the rate of air inspiration. We use $X$ to denote the standard design matrix and consider the logistic model for this data: $P(Y_i=1)=p_i=\frac{exp(X_i^T\theta)}{1+exp(X_i^T\theta)},$ where $\theta$ is the unknown vector of regression coefficients. We assume a multivariate normal prior on $\theta$, $\pi=N(1,1000I_{3})$. Then, the baseline posterior distribution is given by $p_0(\theta|y,X)\propto f(y|X,\theta)\pi(\theta)\propto exp(-\frac{0.5}{1000^2}(\theta-1)^{T}(\theta-1)) + \sum_{i=1}^{39}(y_i X_i^{T}\theta-\log(1+exp(X_i^T\theta)))$. In this problem $\theta$ is only three-dimensional; thus, we use numerical integration to obtain the normalizing constant to specify all posterior distributions, and to compute the Fisher-Rao distance.

Figure \ref{fig:cooklog} presents the results of our analysis. The influence measures indicate four clear influential observations (4, 18, 13 and 32 in order of decreasing influence). Observations 4 and 18 appear to have the most severe effect on the posterior distribution of $\theta$ with resulting distances close to 0.6 or nearly half of the maximal distance on the space of probability densities. The remaining 35 observations yield influence measures lower than 0.2, which we consider as having low influence. We compare our result to that provided in \cite{PengDey}. We refer the reader to their paper for a similar figure as Figure \ref{fig:cooklog} generated under their framework. We note that their influence measure, based on divergence measures, is not symmetric and possesses no natural scale. The authors suggest a strategy to calibrate the proposed divergence measures but a choice of this calibration is rather arbitrary. Their method flags observations 4 and 18 (in decreasing order of influence) as infleuntial and many other observations as weak outliers. It appears that our approach provides a clearer separation of the influential versus the non-influential observations in this example.

\begin{figure}[!h]
\begin{center}
\includegraphics[width=1.7in]{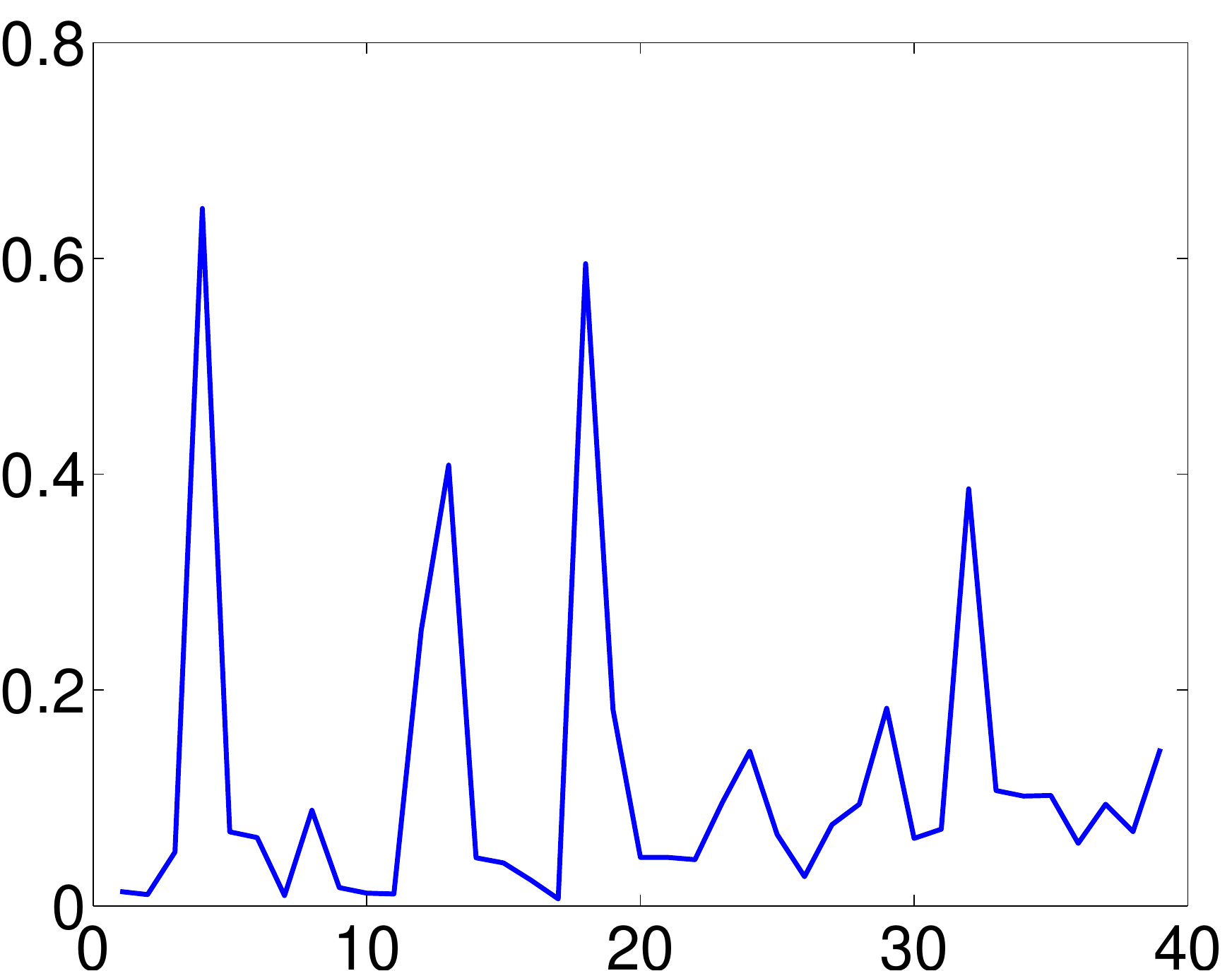}
\caption{Influence analysis in Bayesian logistic regression. Each of the 39 observation is listed on the $x$-axis with the corresponding measure of influence on the $y$-axis. \label{fig:cooklog}}
\end{center}
\end{figure}

\end{example}

\begin{example}

In the final example, we consider a common problem in statistical shape analysis of examining the effects of deleting an observation from a sample of shapes on the estimated posterior distribution of the mean shape. For this purpose, we utilize a handwritten digit dataset from \cite{alimoglu}. The full database was created by collecting 250 writing samples of numerical digits from 44 writers. The raw data is provided as $(x,y)$ coordinates of 8 landmark points on each digit. An arbitrary example of each digit is shown in Figure \ref{fig:digits}, with the landmarks in the plot connected by straight lines for improved visualization.

\begin{figure}[!h]
\begin{center}
\begin{tabular}{c c c c c}
\includegraphics[width=0.5in]{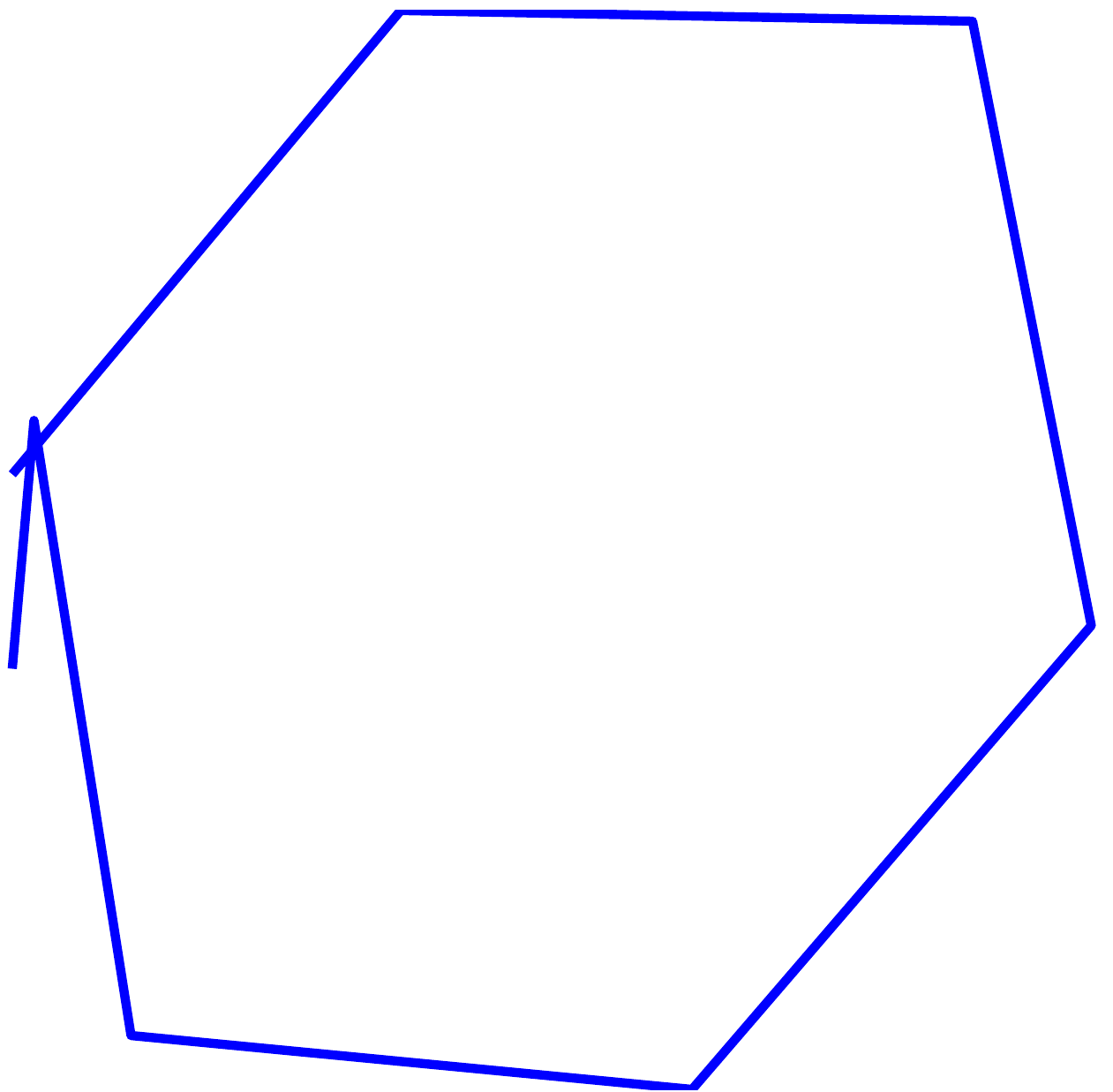}&\includegraphics[width=0.5in]{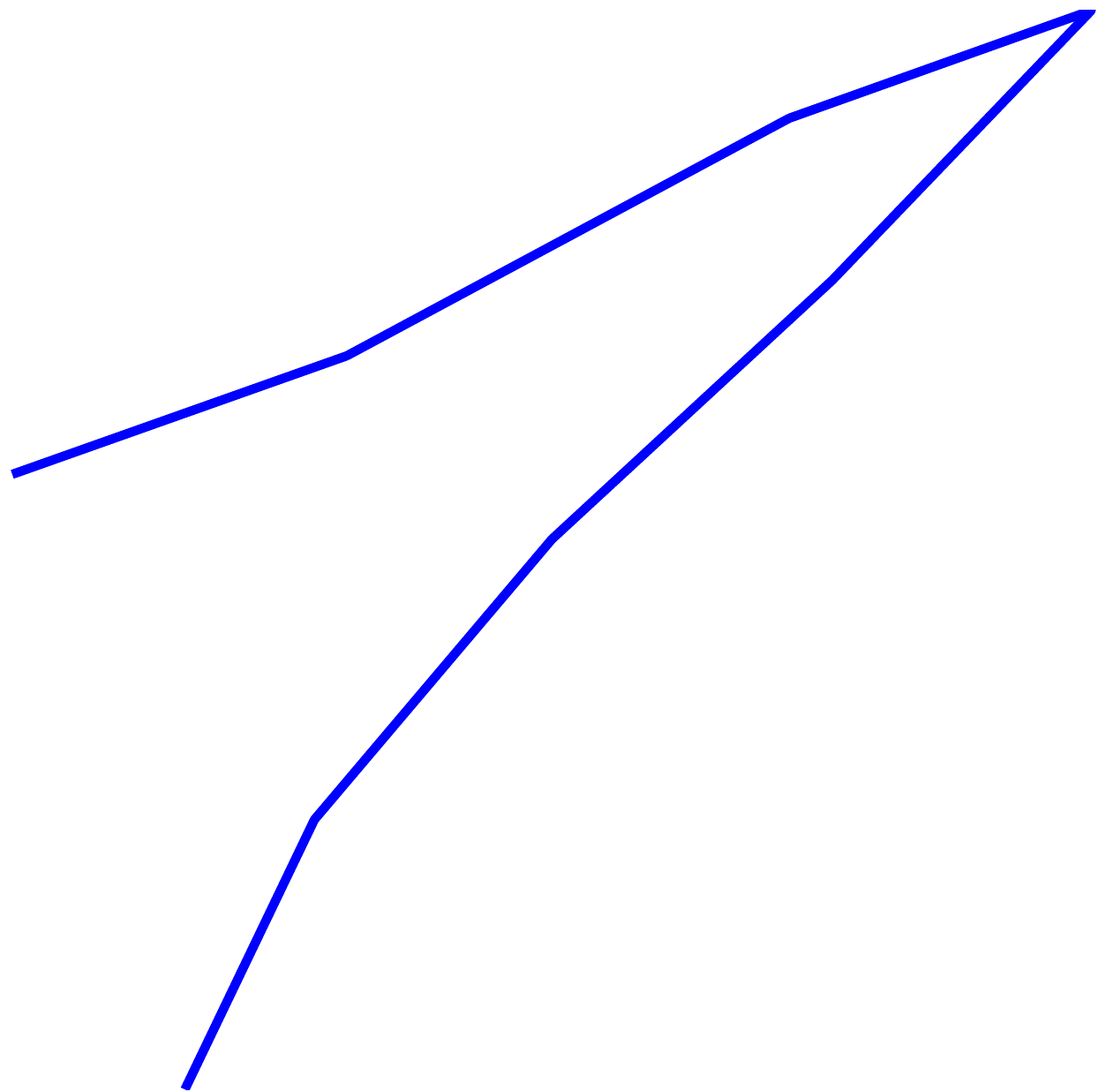}&\includegraphics[width=0.5in]{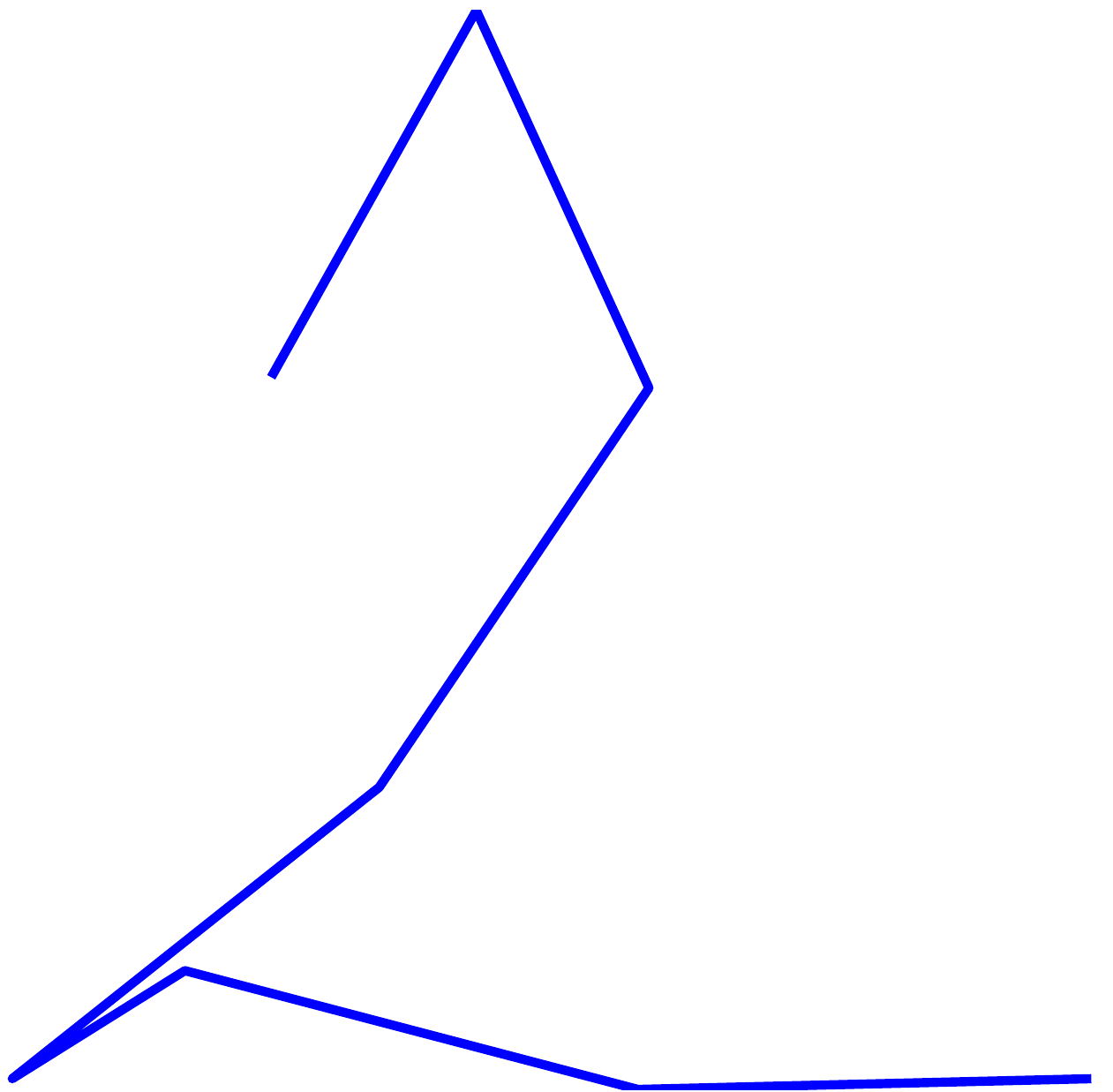}&\includegraphics[width=0.5in]{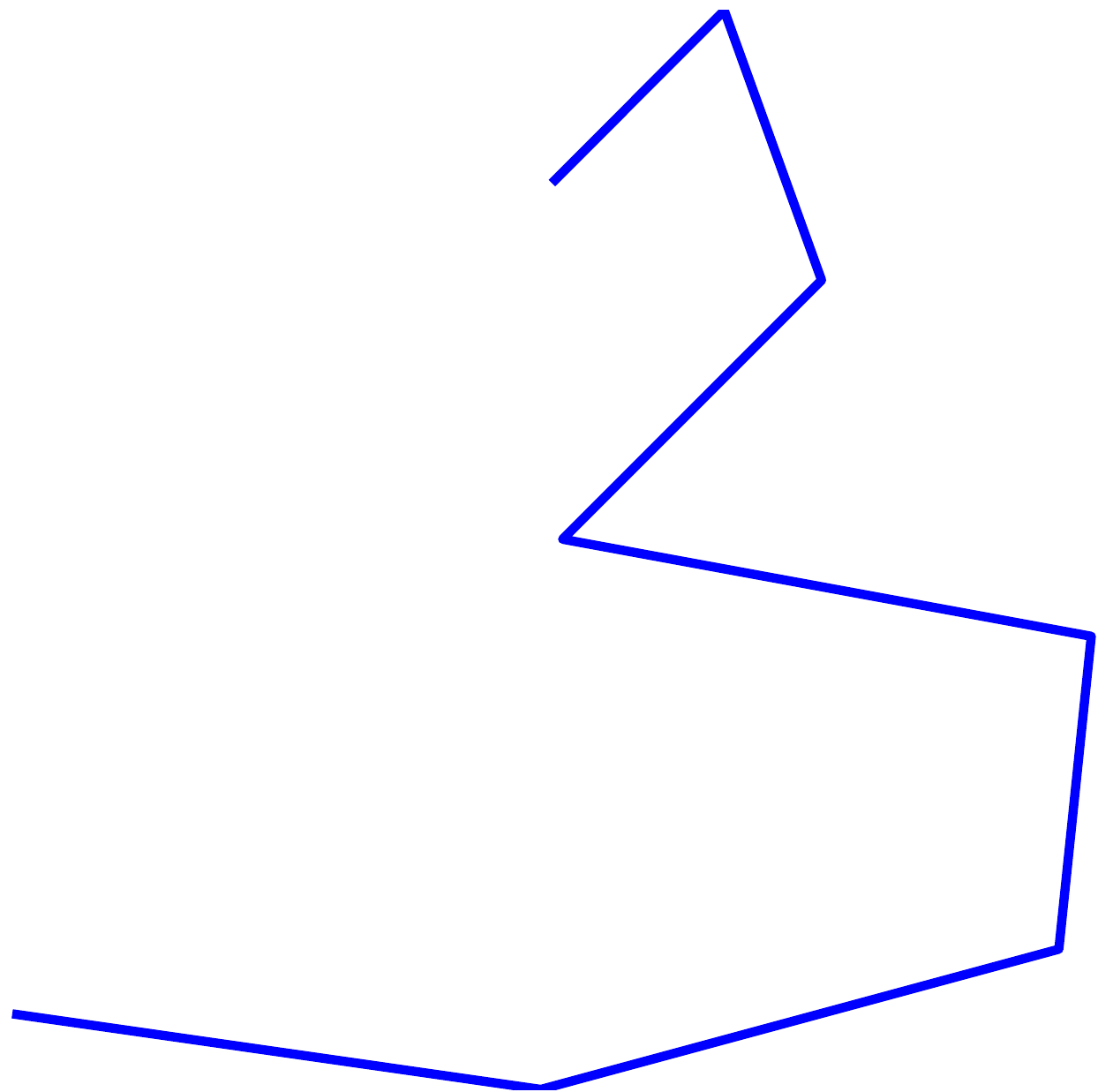}&\includegraphics[width=0.5in]{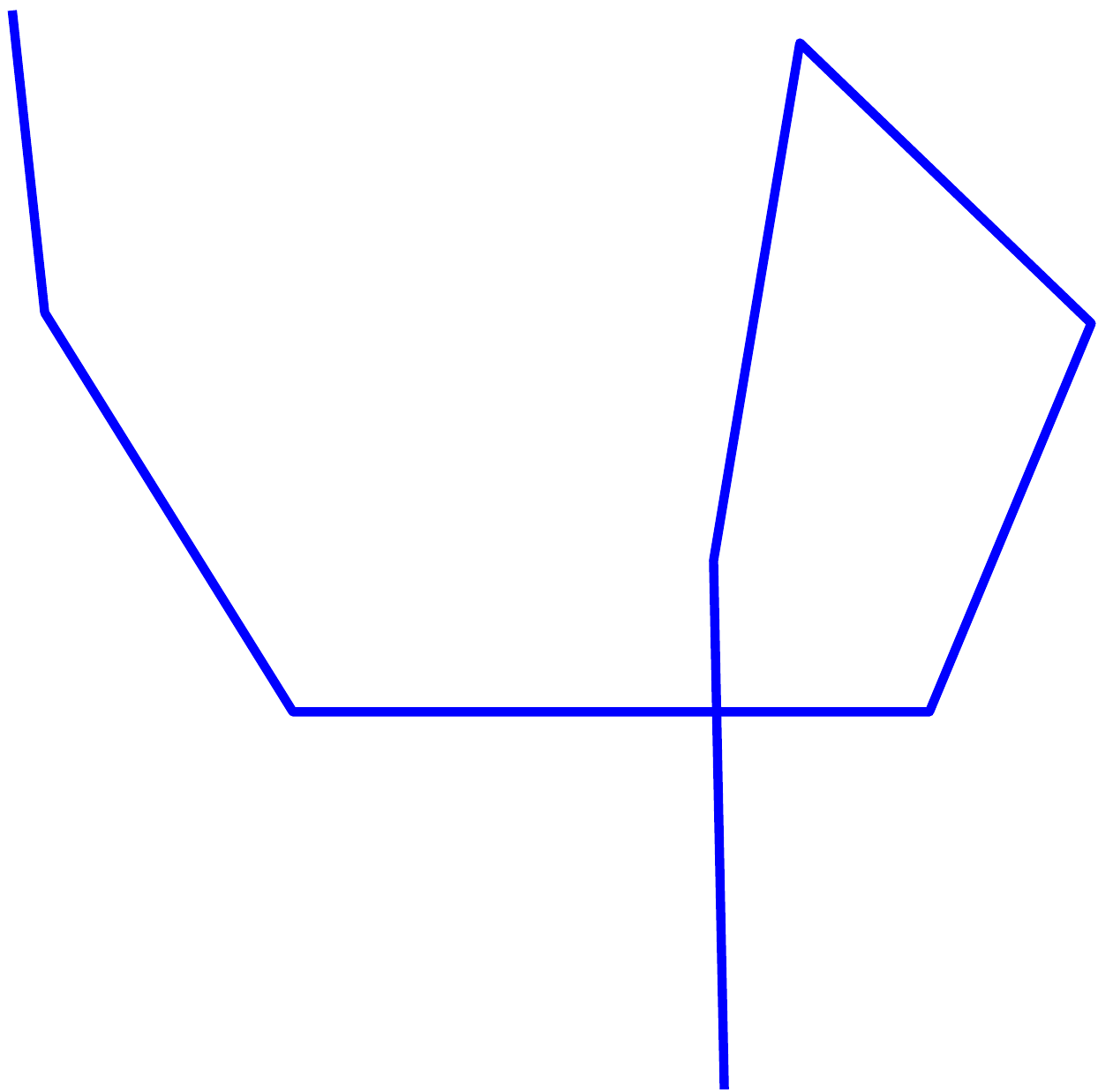}\\

\includegraphics[width=0.5in]{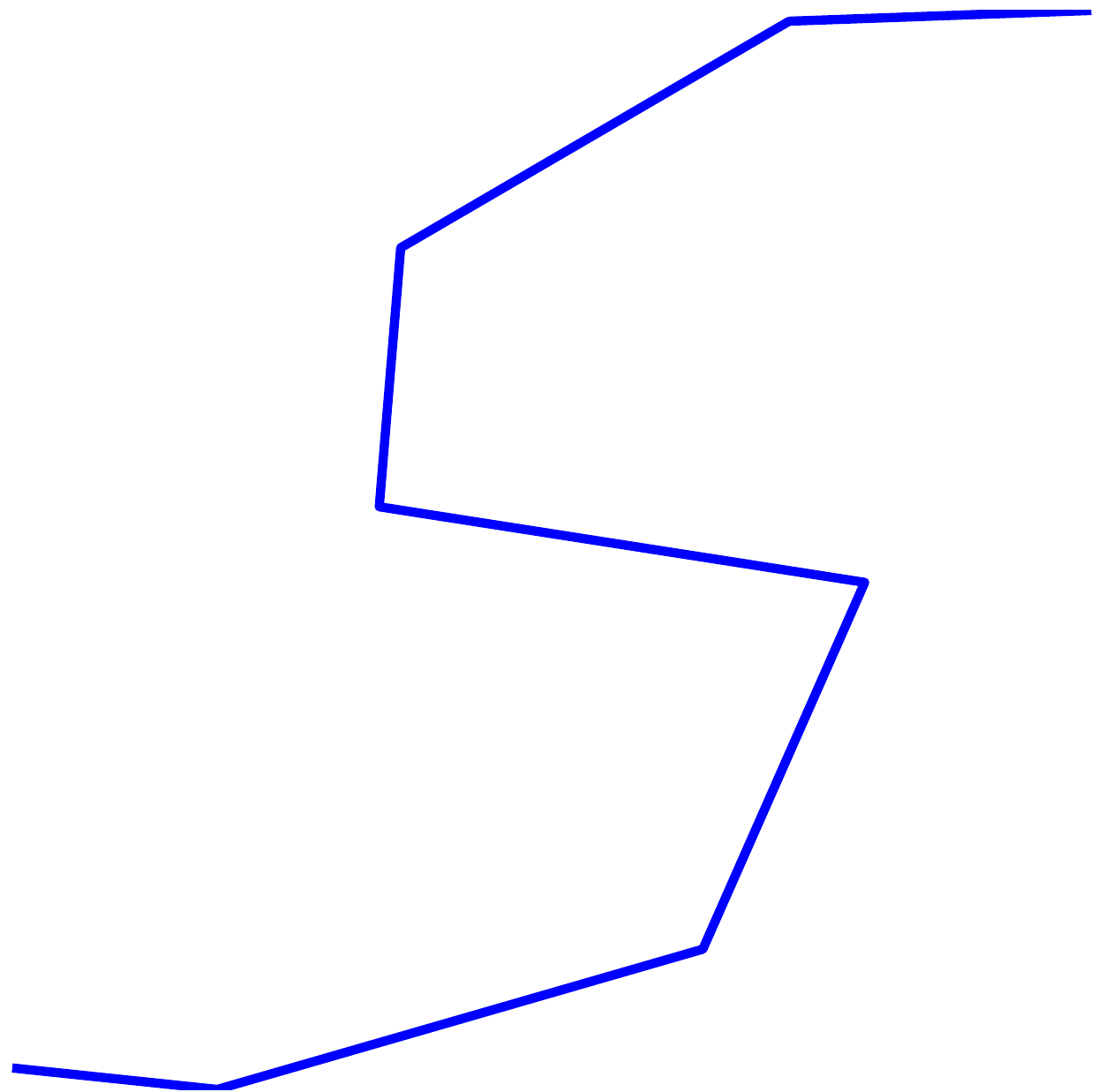}&\includegraphics[width=0.5in]{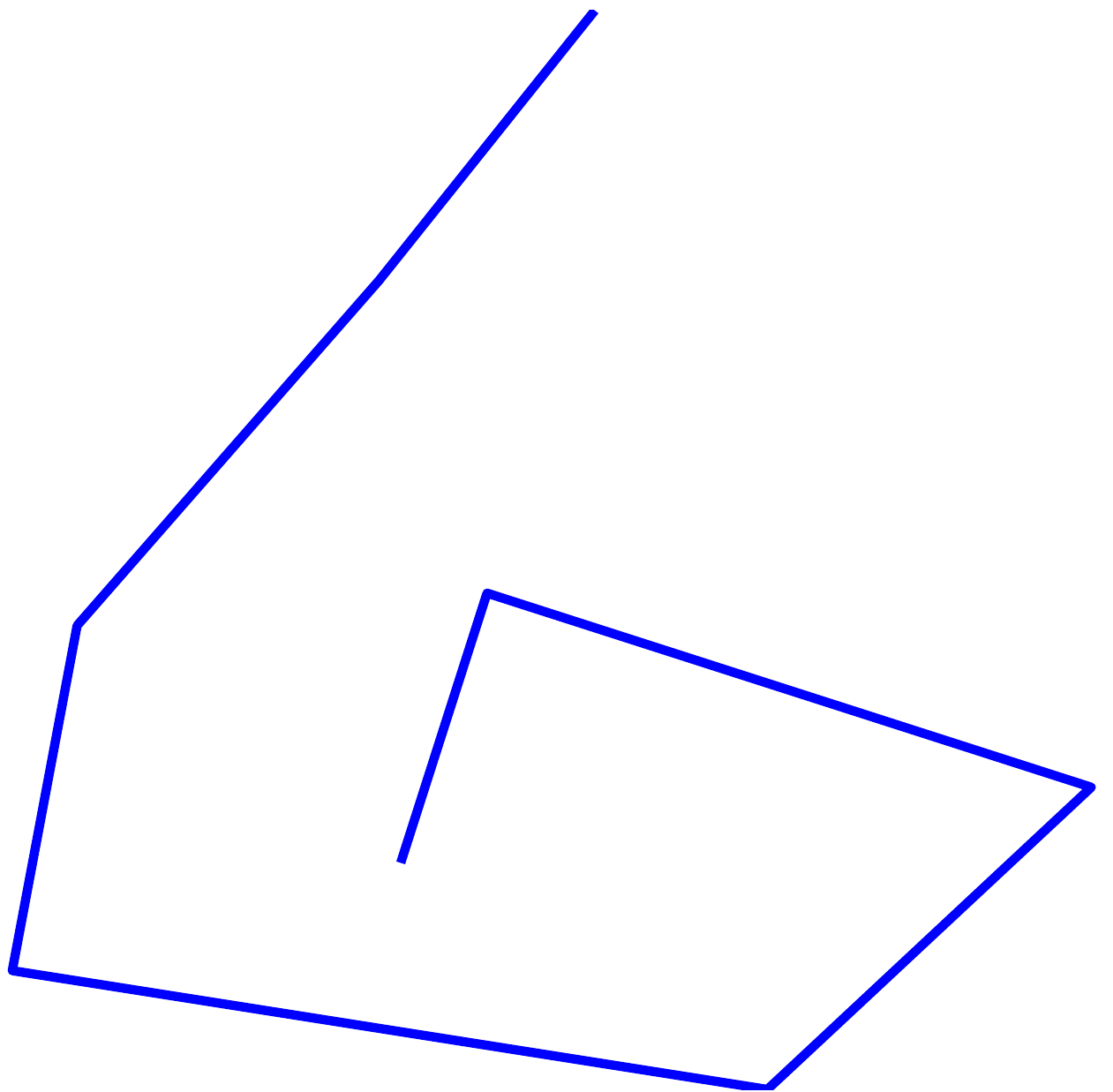}&\includegraphics[width=0.5in]{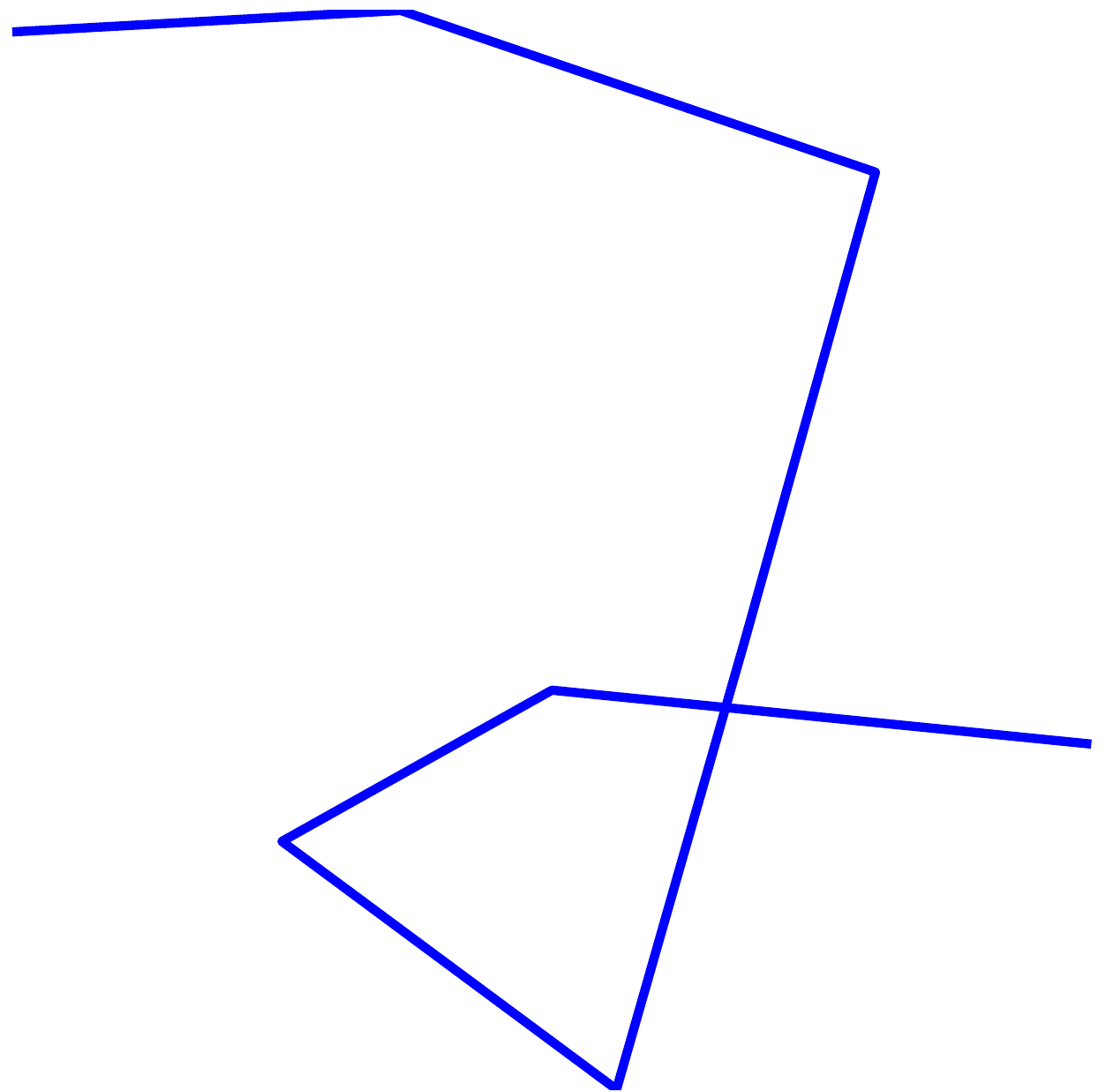}&\includegraphics[width=0.5in]{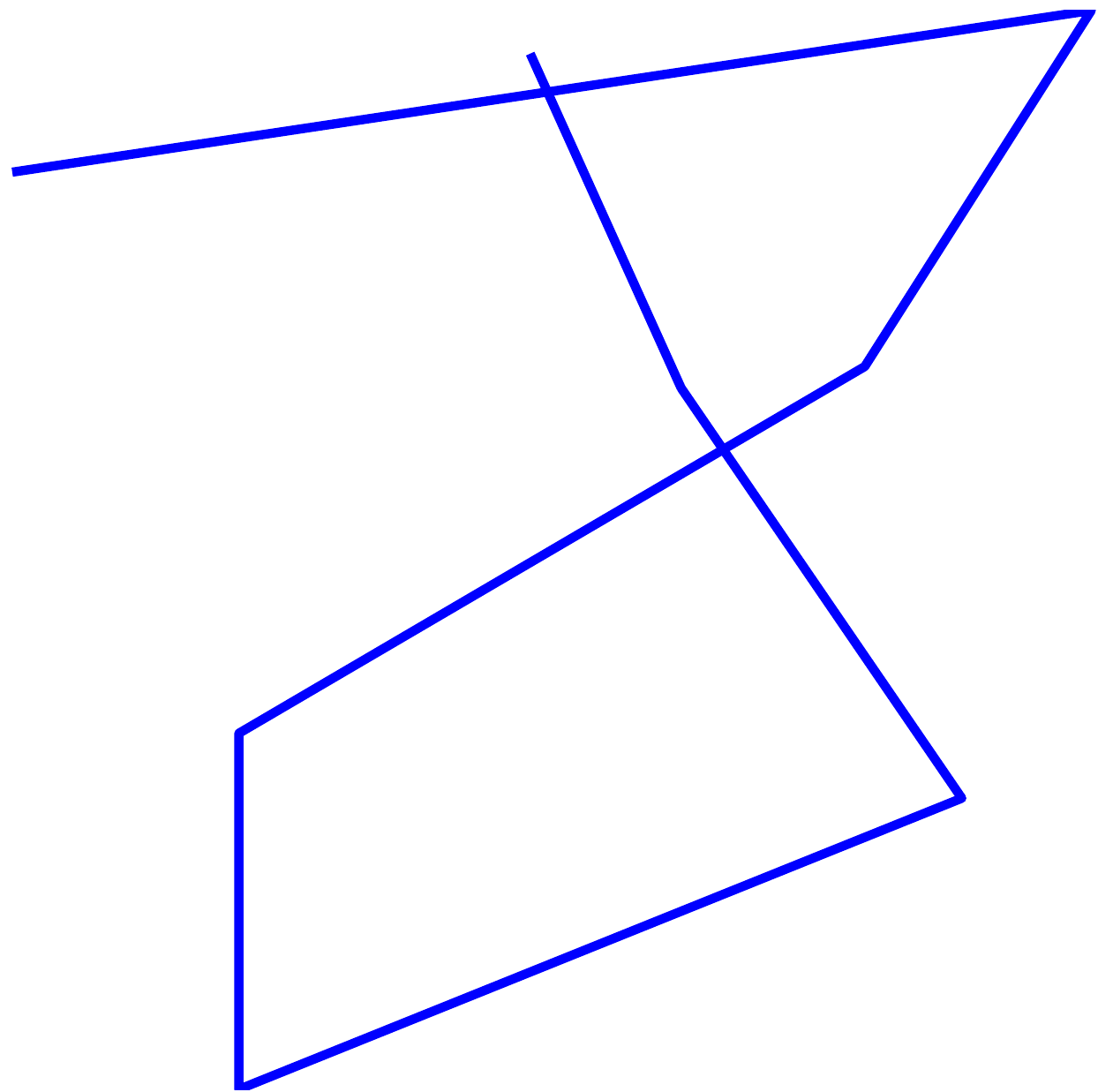}&\includegraphics[width=0.5in]{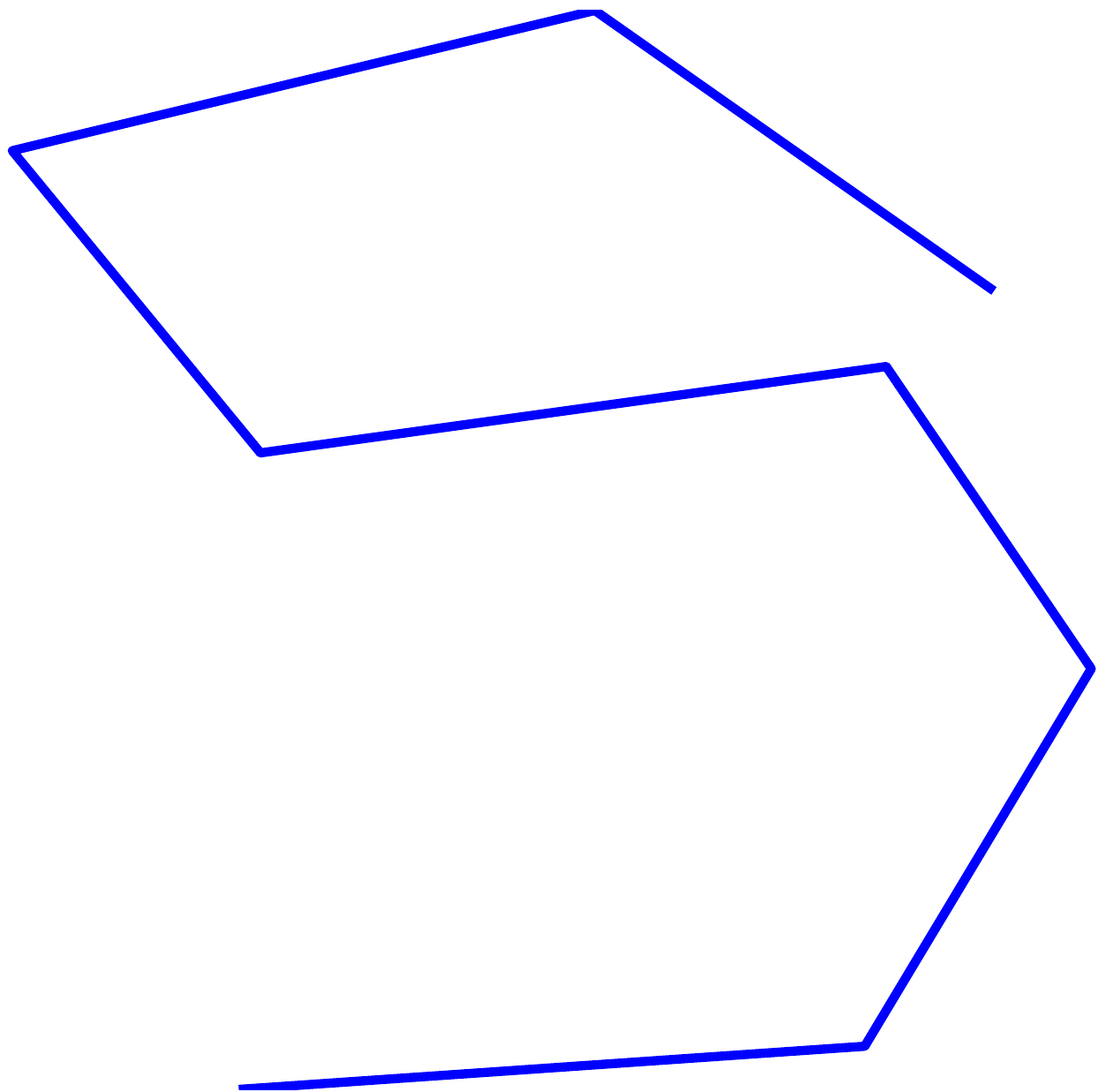}\\

\end{tabular}
\caption{One example of each digit in this dataset. \label{fig:digits}}
\end{center}
\end{figure}

Kendall defined shape as a mathematical property that remains unchanged under rotation, translation, and global scaling. In this example, we use his definition and use a Bayesian approach to estimate the mean shape of a digit class. Throughout this description, we use material from the book by \cite{dryden-mardia_book:98}. Let a configuration of landmarks denoting a digit be represented using a complex vector $x\in\mathbb{C}^8$. In order to remove the translation variability from the representation space, we pre-multiply each of the landmark configurations with a Helmert submatrix resulting in $x_H=Hx\in\mathbb{C}^7$. Then, the pre-shape of a landmark configuration is defined as $z=x_H/\|x_H\|\in\mathbb{C}S^{6}$. The pre-shape is invariant to translation and scaling of the original landmark configurations. The set of all pre-shapes is the complex unit sphere and it is called the pre-shape space. In order to remove rotational variability from the data, we align all of the pre-shapes to a randomly chosen observation. Then, given a sample of digit shapes $z_1,\dots,z_n\in\mathbb{C}S^{6}$, we define the likelihood as the complex Watson distribution with mode (mean) $\mu$ and a known concentration parameter $\kappa$. In this example, we estimate $\kappa$ from the data using Equation 6.14 in \cite{dryden-mardia_book:98}. As the prior distribution for $\mu$ we choose the complex Bingham distribution with parameter matrix $A=I_7$. The advantage of using the complex Bingham distribution as a prior in shape analysis is that it is invariant to rotation and it is a conjugate prior for the complex Watson. The resulting posterior distribution for the mode $\mu$ is a complex Bingham with the parameter matrix $\kappa\sum_{i=1}^nz_iz_i^*+I_7$, where $z^*$ denotes the conjugate transpose of $z$.

As mentioned earlier, we are interested in identifying observations that have a high influence on the posterior distribution of $\mu$. We again utilize Definition \ref{imeasure} for this purpose. In order to compute this influence measure we use the estimator of the Fisher-Rao distance given in Proposition \ref{MCMC}. Because the baseline posterior distribution for $\mu$ is a complex Bingham, we can sample from it directly using the methods given in \cite{kent}. In other cases, we would have to resort to MCMC methods. Figure \ref{fig:alldig} provides plots of sorted influence measures for all observations in the case of digits $\mathit{0, 1 ,4, 6, 7}$ (each digit class was considered separately). First, we note that there is a lot of variability in each of the digit classes due to significantly different handwriting styles of the subjects. As a result, none of the observations in each of the classes are highly influential on their own; all influence measures were less than 0.3. We hypothesize that if we considered removing blocks of observations, this result would change. Nonetheless, we plot the three least and three most influential observations in each of the considered digit classes in Figure \ref{fig:inflobs}. Consistent with intuition, the shapes of the three least influential digits look very similar, while the three most influential digits look like potential outlying shapes.

\begin{figure}[!h]
\begin{center}

\begin{tabular}{c c c c c }

$\mathit{0}$&$\mathit{1}$&$\mathit{4}$&$\mathit{6}$&$\mathit{7}$\\

\includegraphics[width=1in]{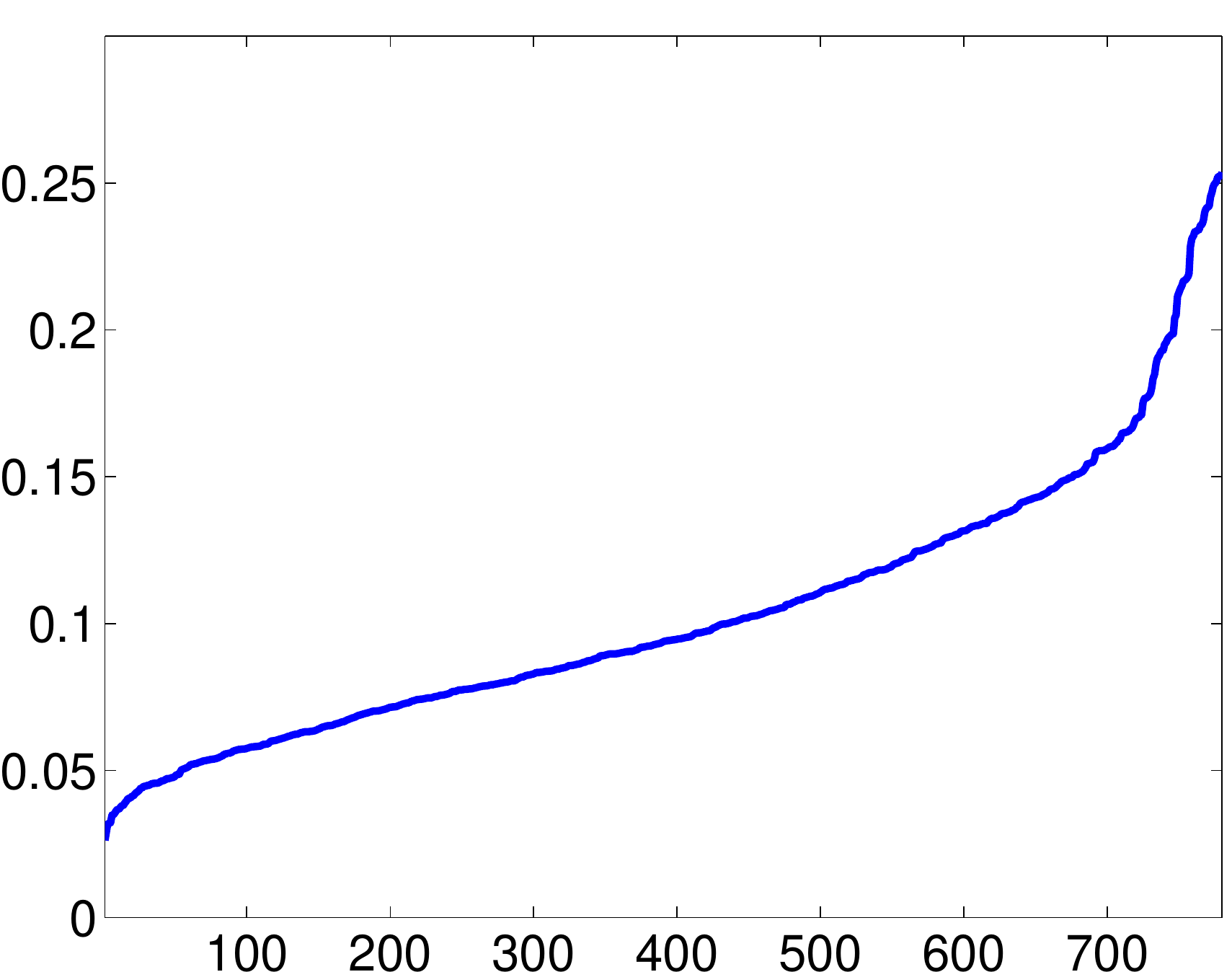}&\includegraphics[width=1in]{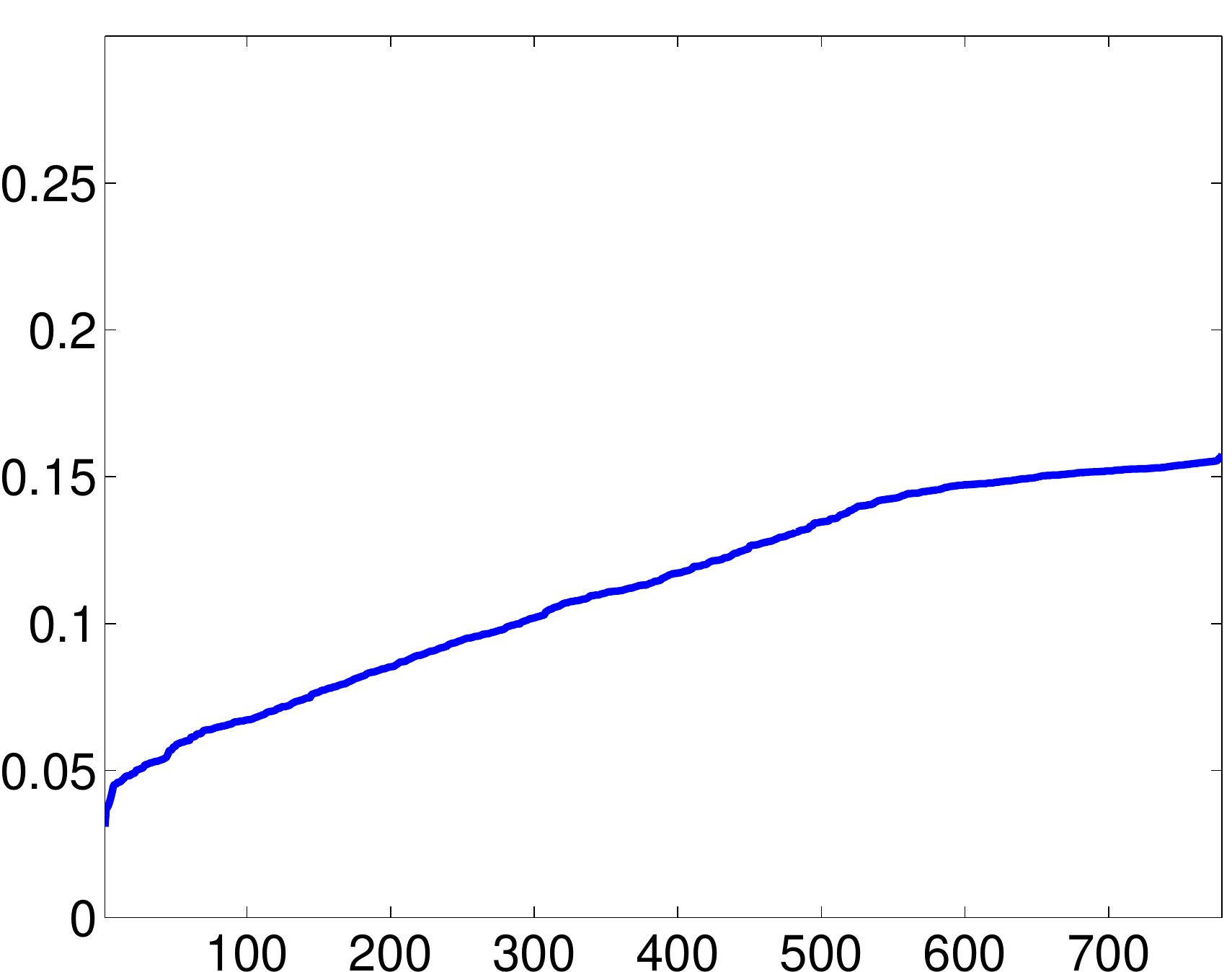}&\includegraphics[width=1in]{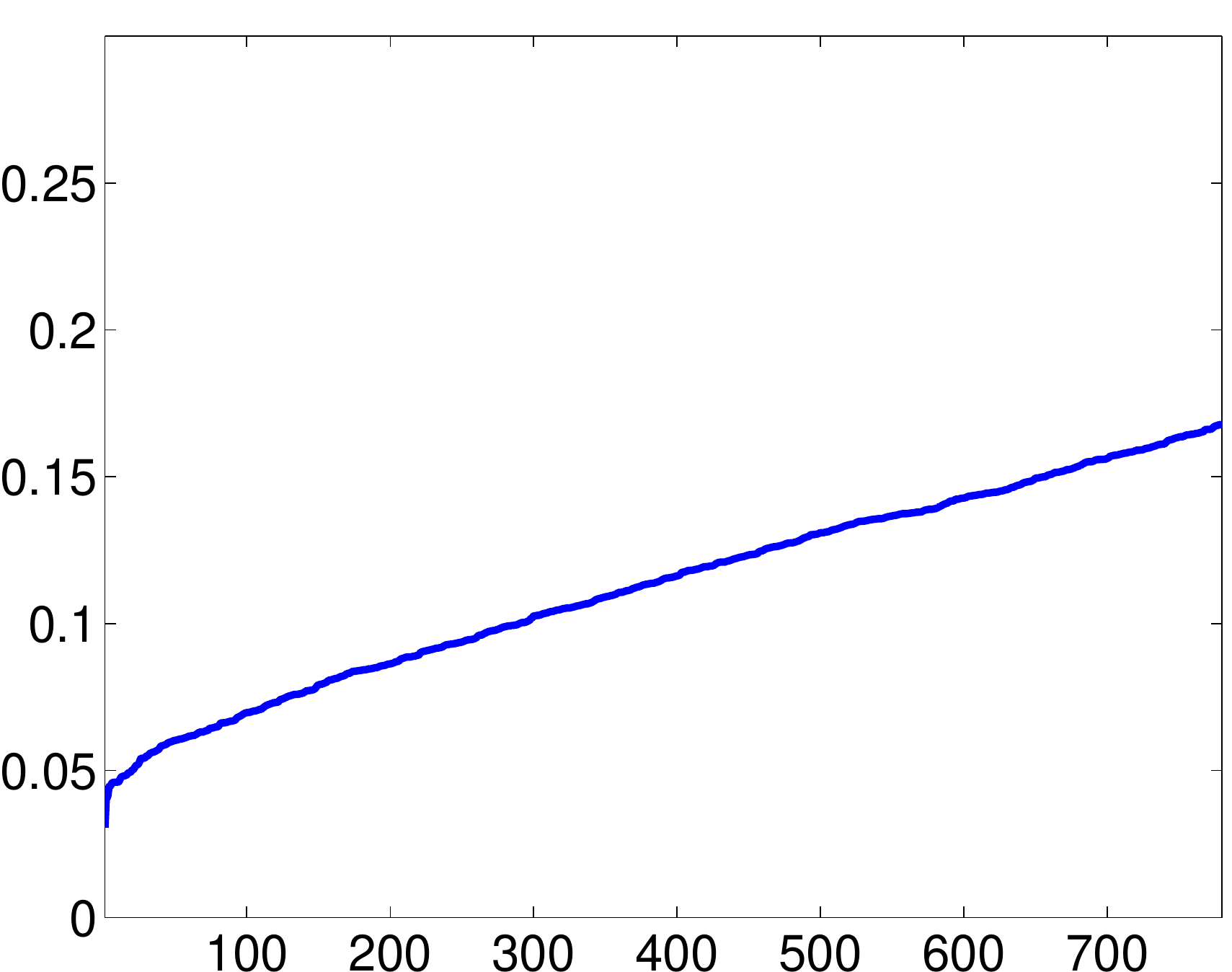}&\includegraphics[width=1in]{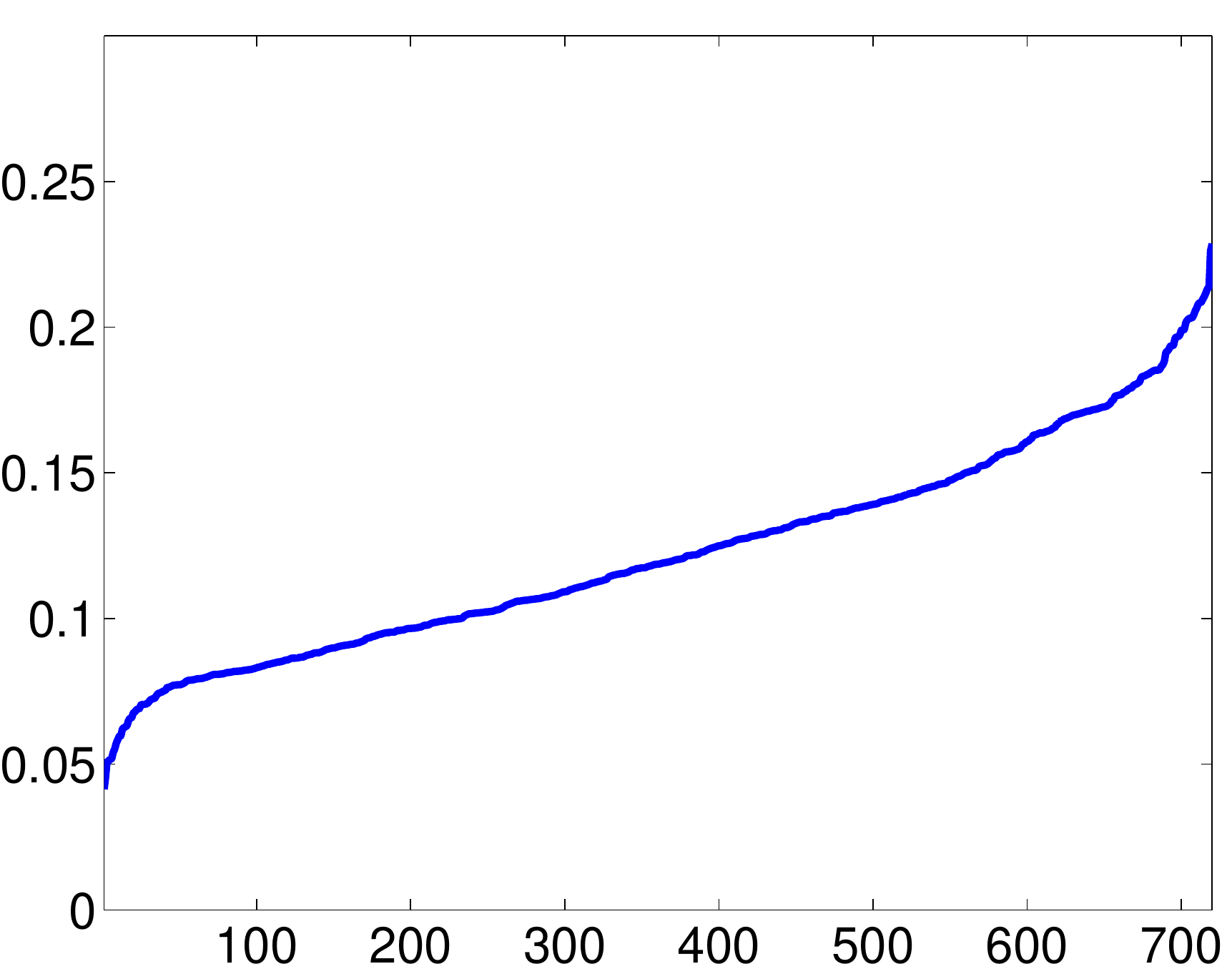}&\includegraphics[width=1in]{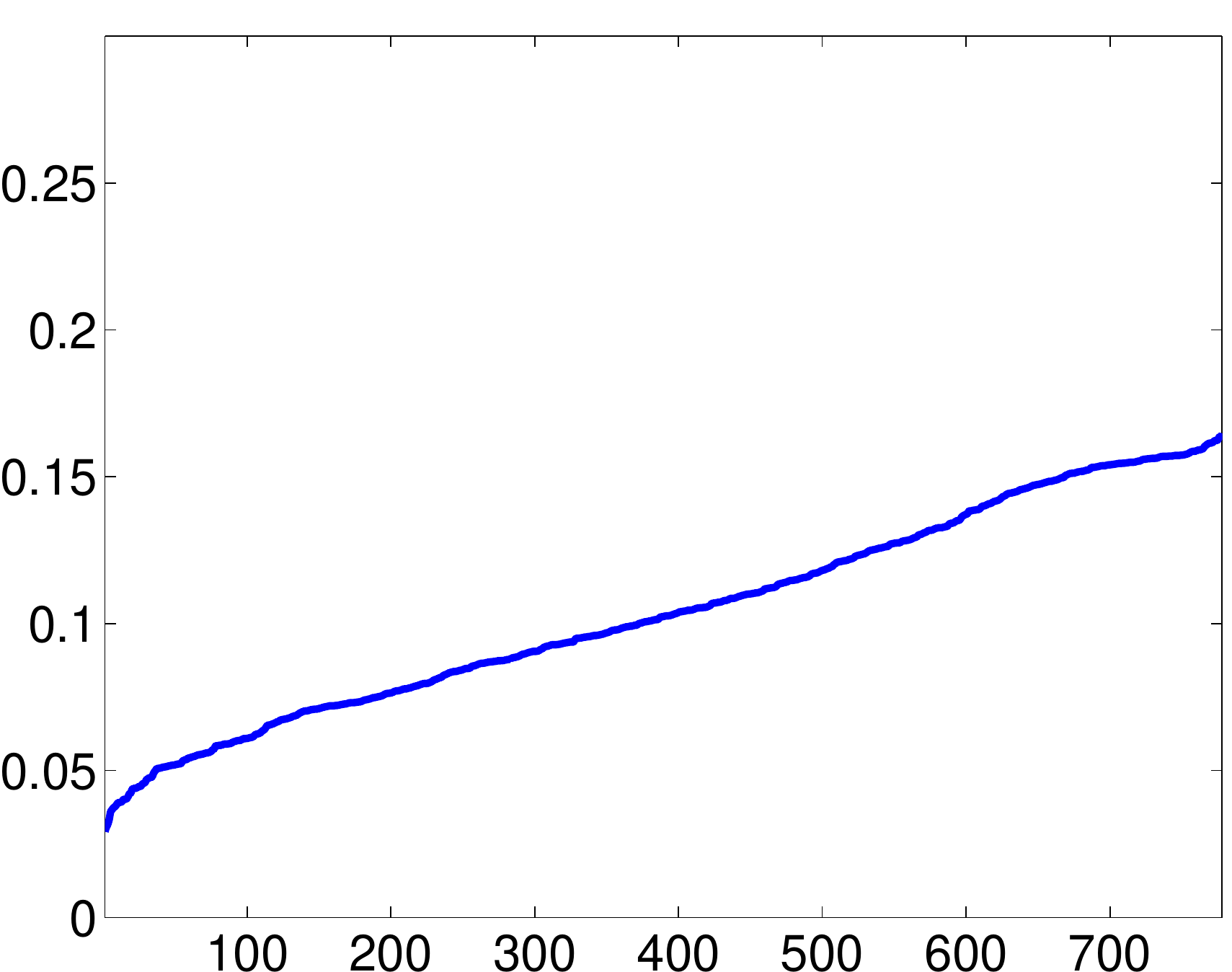}\\

\end{tabular}
\caption{Sorted influence measures for all instances of digits $\mathit{0, 1, 4, 6, 7}$ in the dataset. \label{fig:alldig}}
\end{center}
\end{figure}

\begin{figure}[!h]
\begin{center}
\scalebox{0.8}{
\begin{tabular}{c|ccc||ccc|}
\multicolumn{1}{r}{}&\multicolumn{3}{c}{\small{Less Influential}}&\multicolumn{3}{c}{\small{More Influential}}\\
\cline{2-7}
&&&&&&\\
\scalebox{1.3}{$\mathit{0}$}&\includegraphics[width=0.65in]{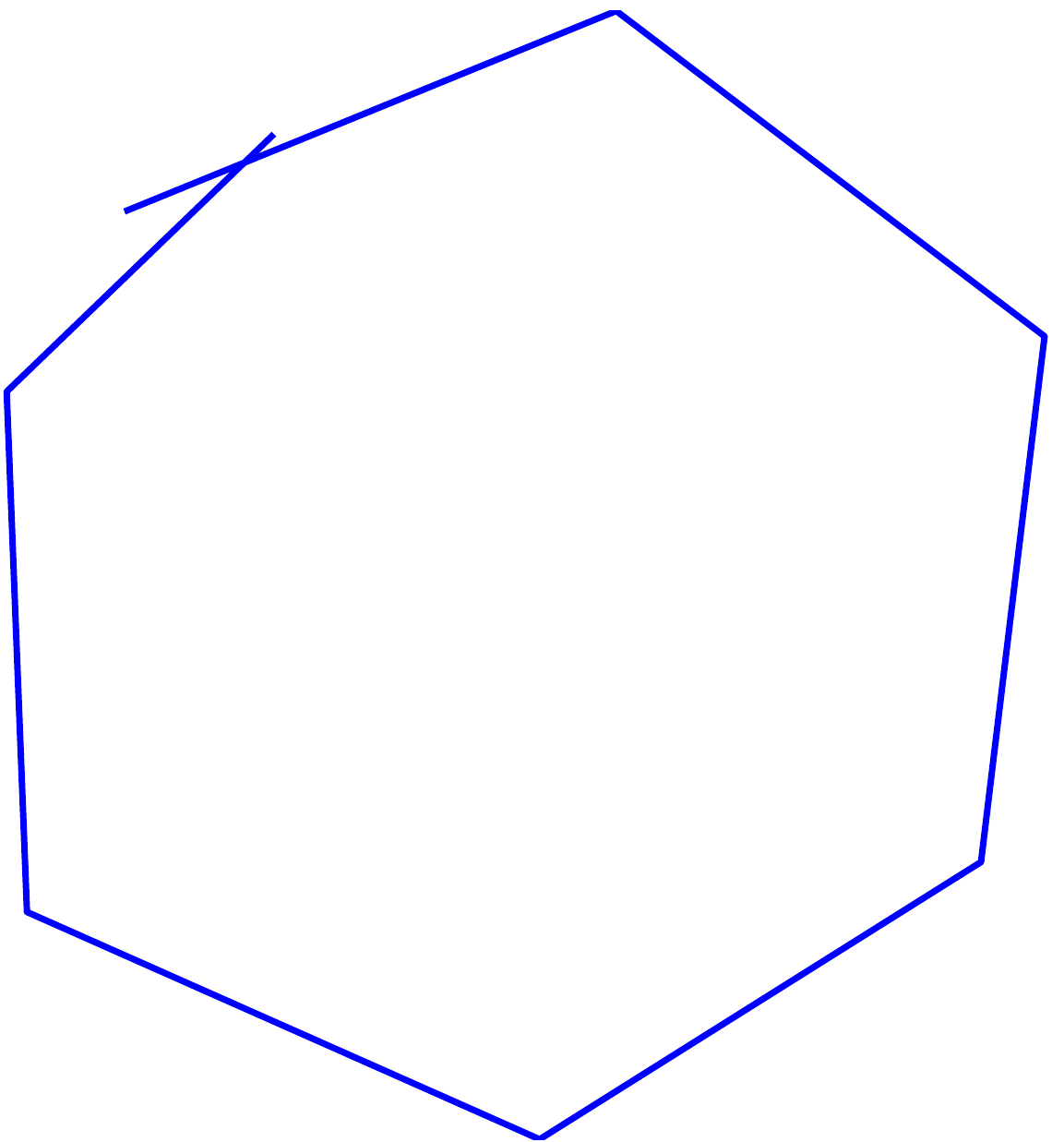}&\includegraphics[width=0.65in]{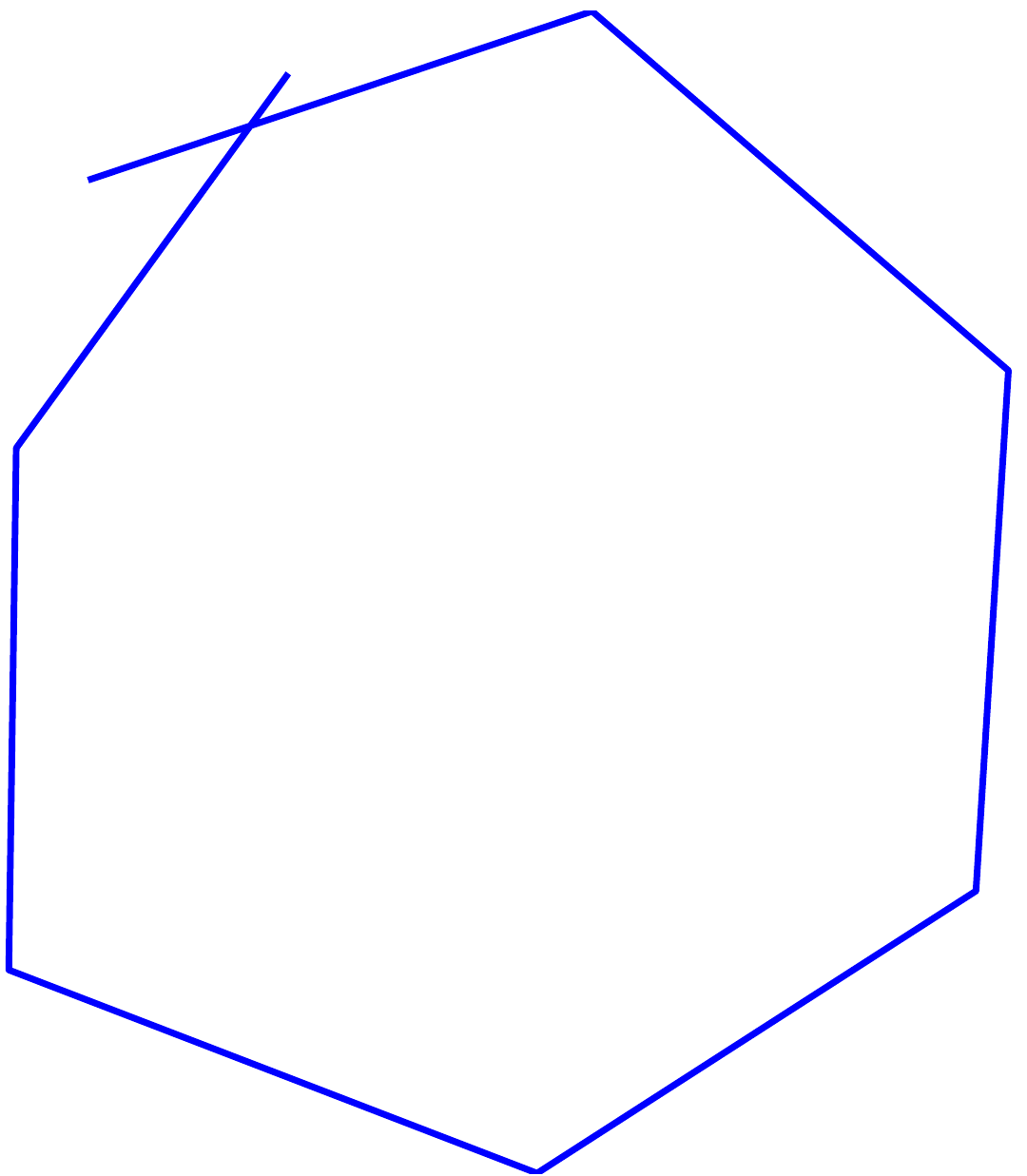}&
\includegraphics[width=0.65in]{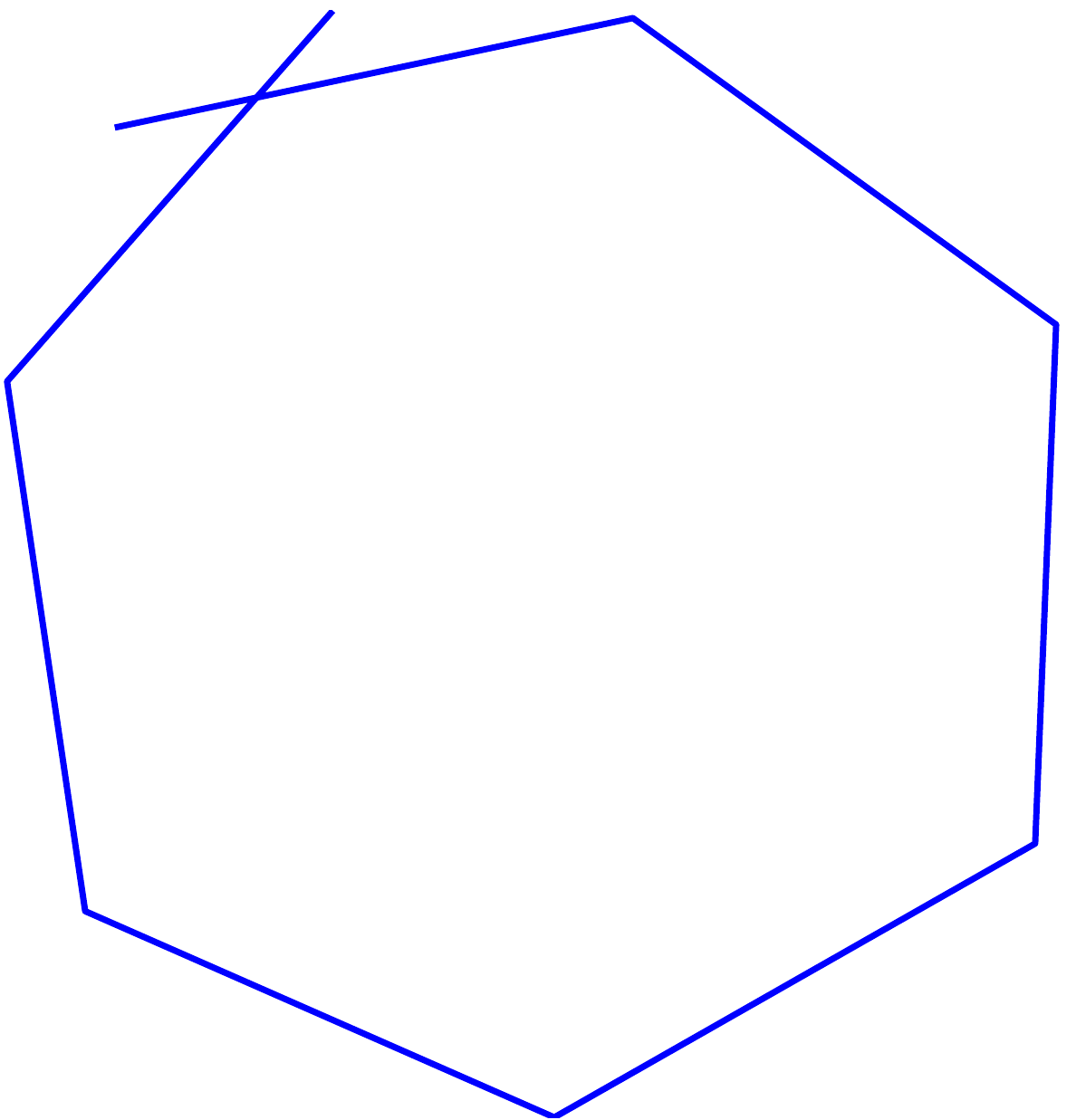}&\includegraphics[width=0.65in]{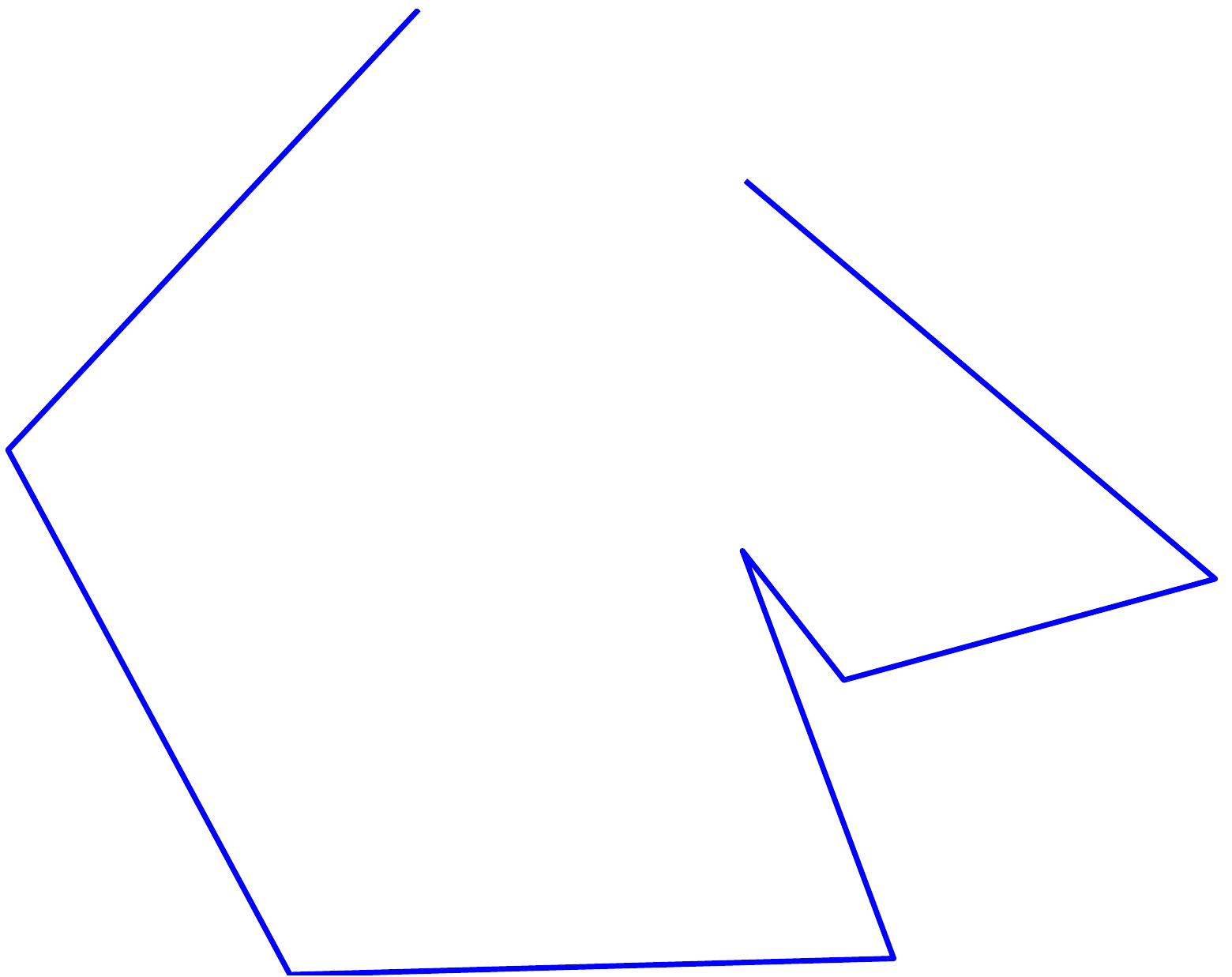}&
\includegraphics[width=0.65in]{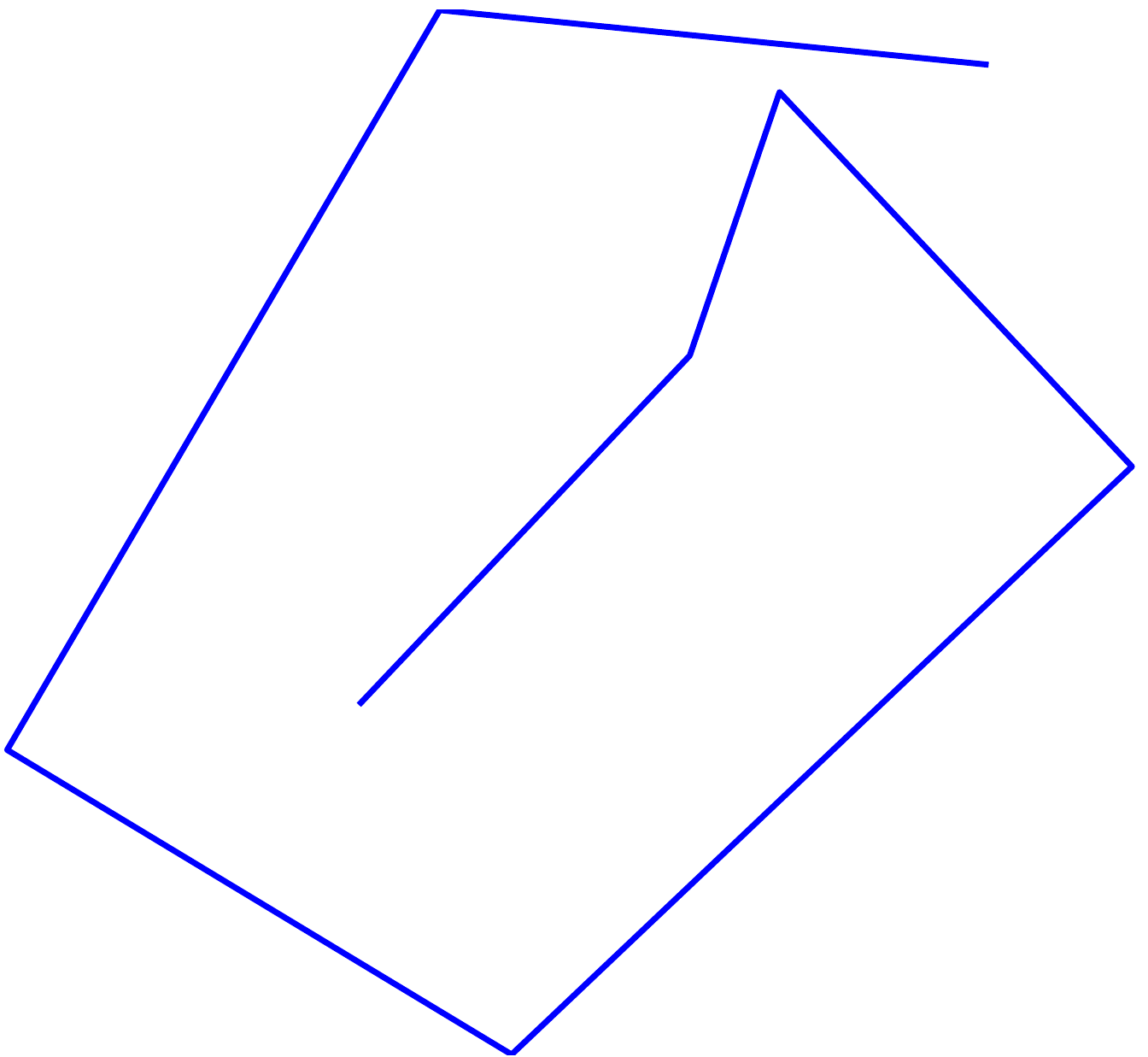}&\includegraphics[width=0.65in]{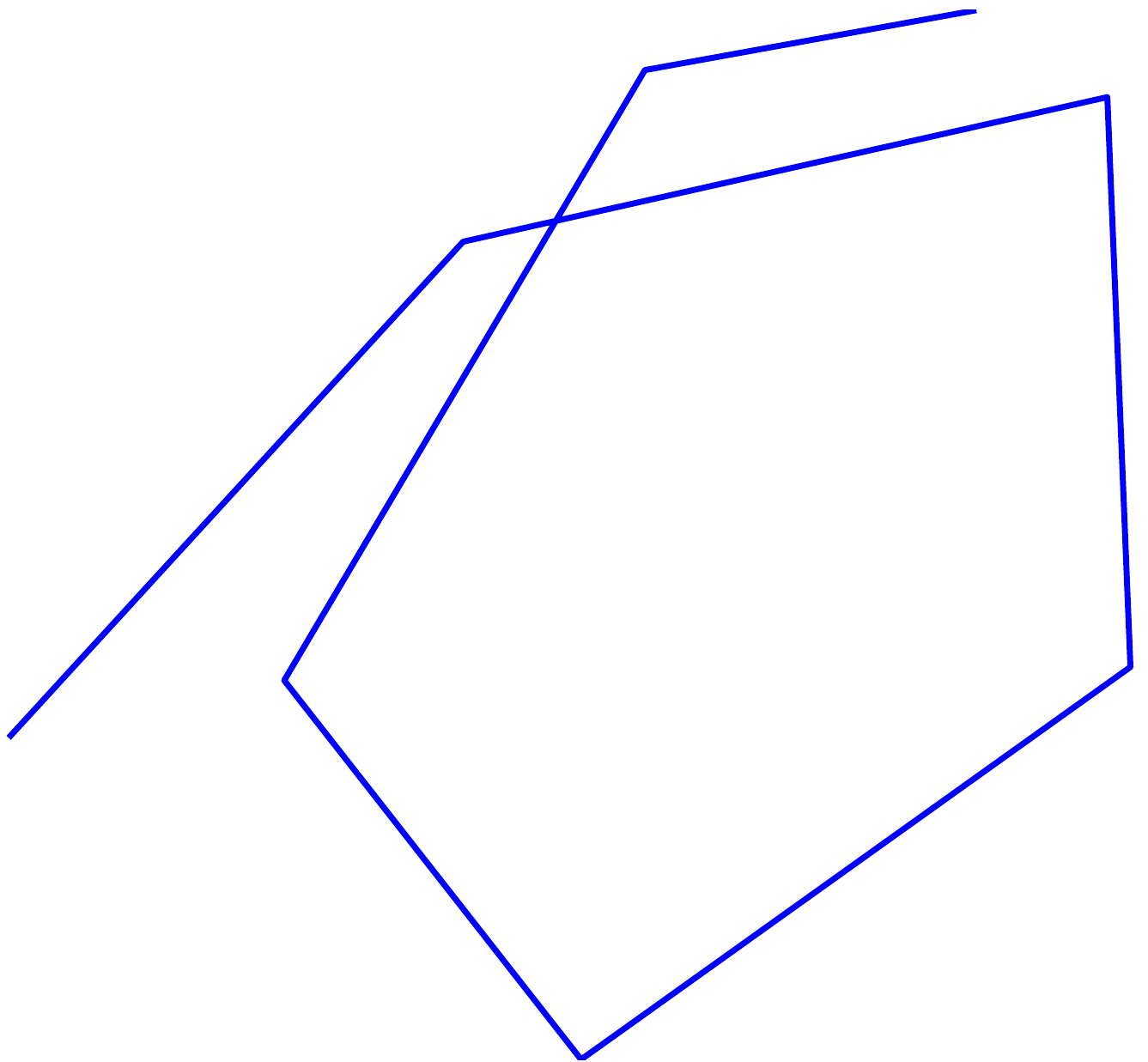}\\
\cline{2-7}
&&&&&&\\
\scalebox{1.3}{$\mathit{1}$}&\includegraphics[width=0.65in]{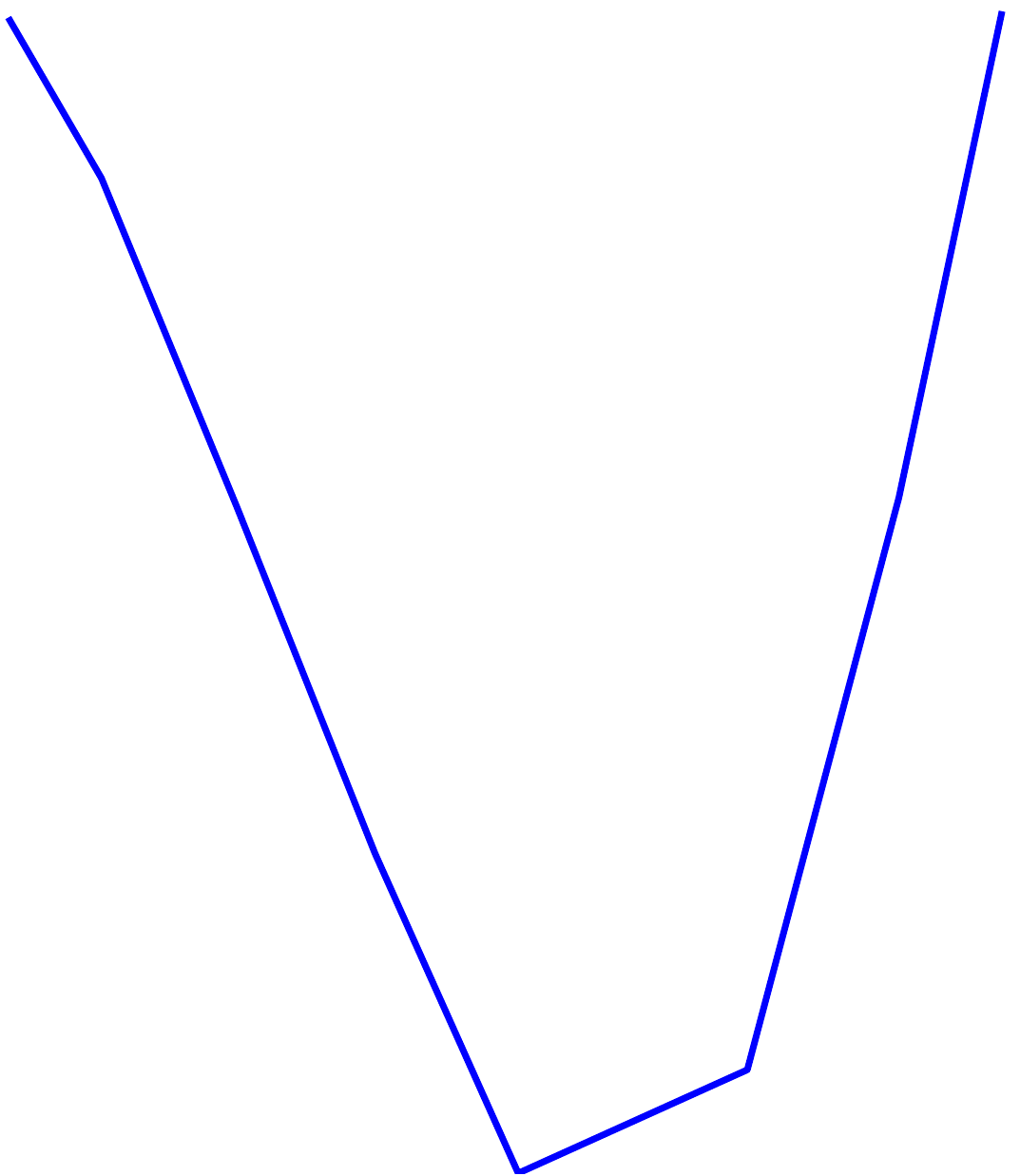}&\includegraphics[width=0.65in]{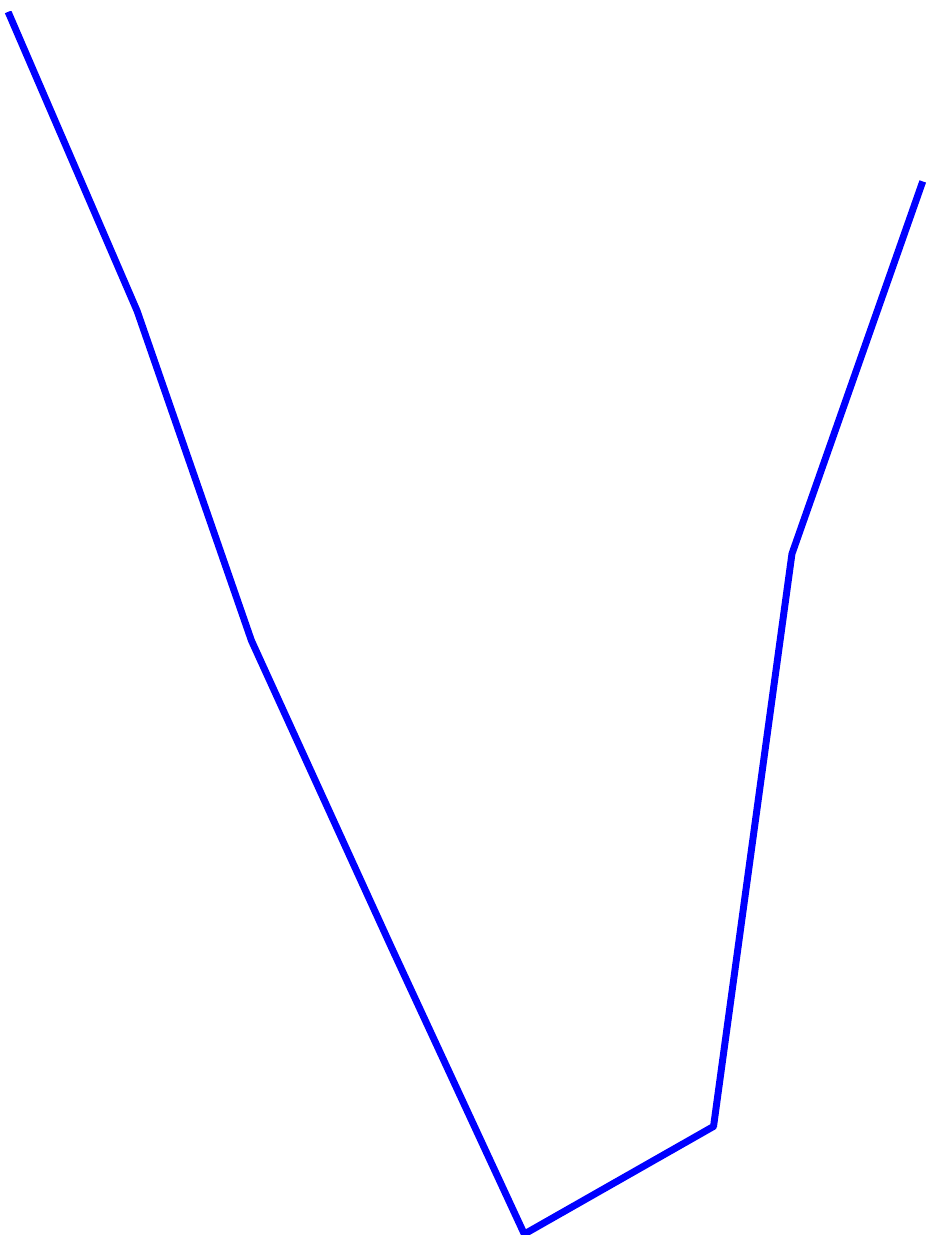}&\includegraphics[width=0.65in]{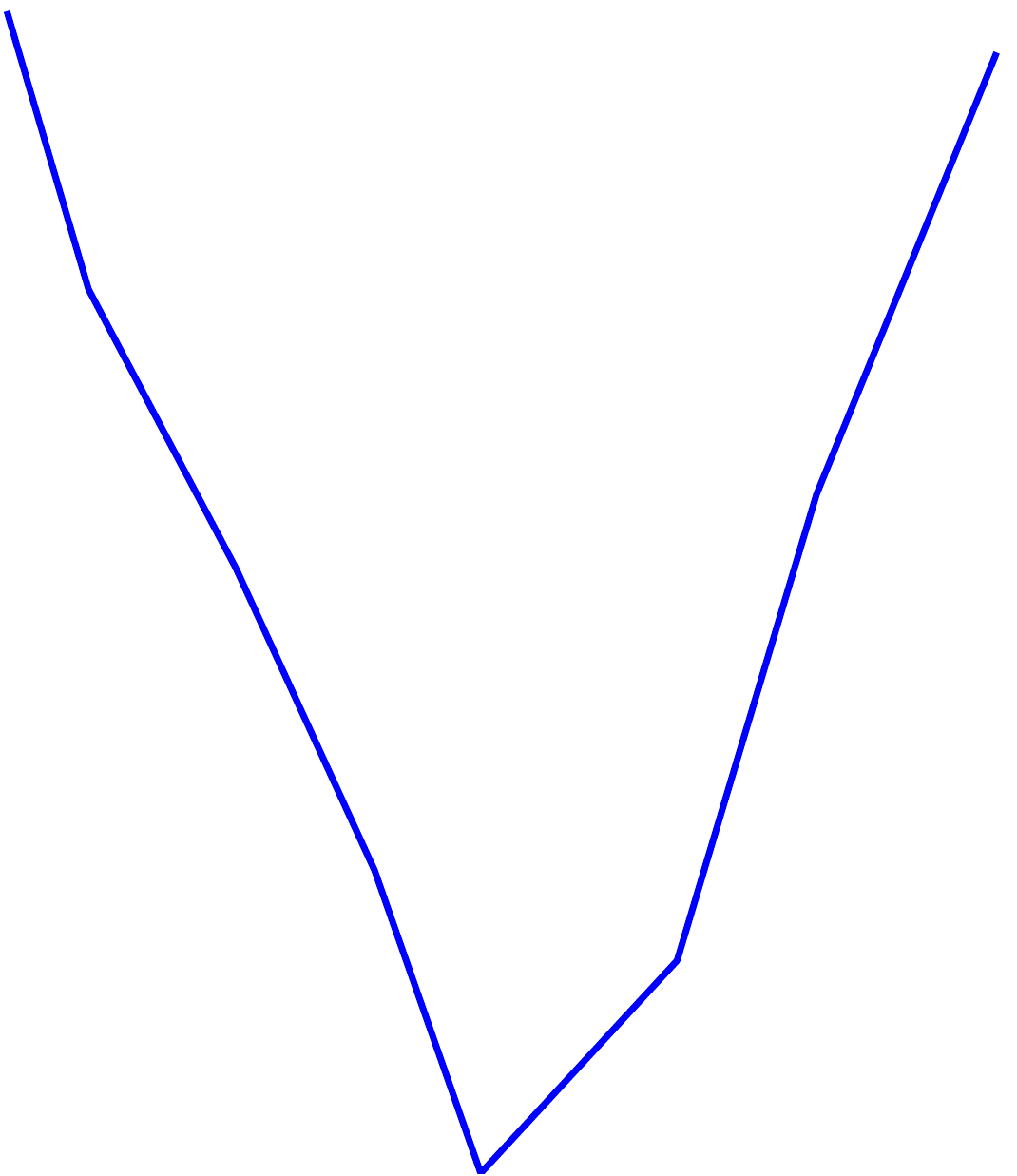}&\includegraphics[width=0.65in]{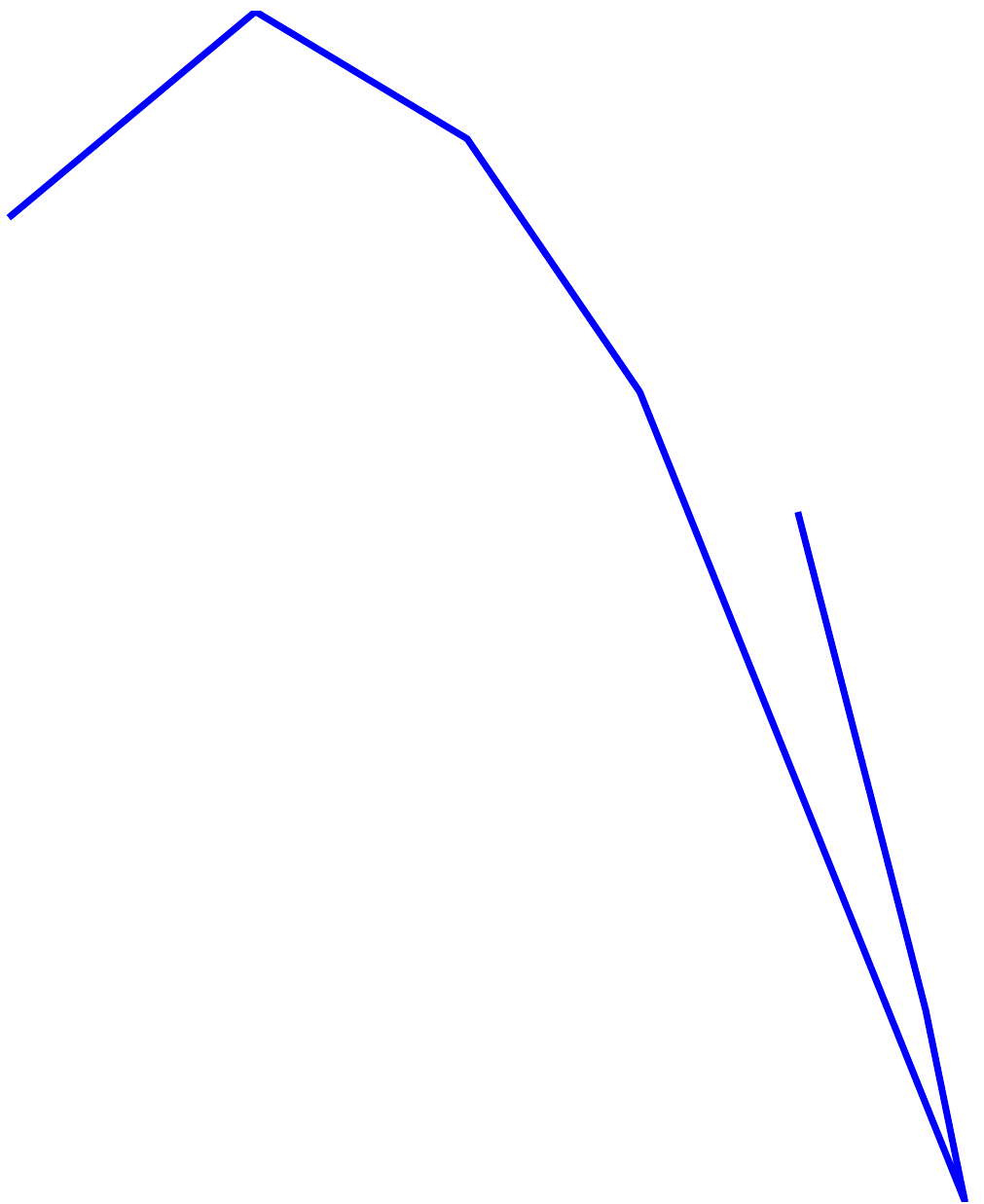}&\includegraphics[width=0.65in]{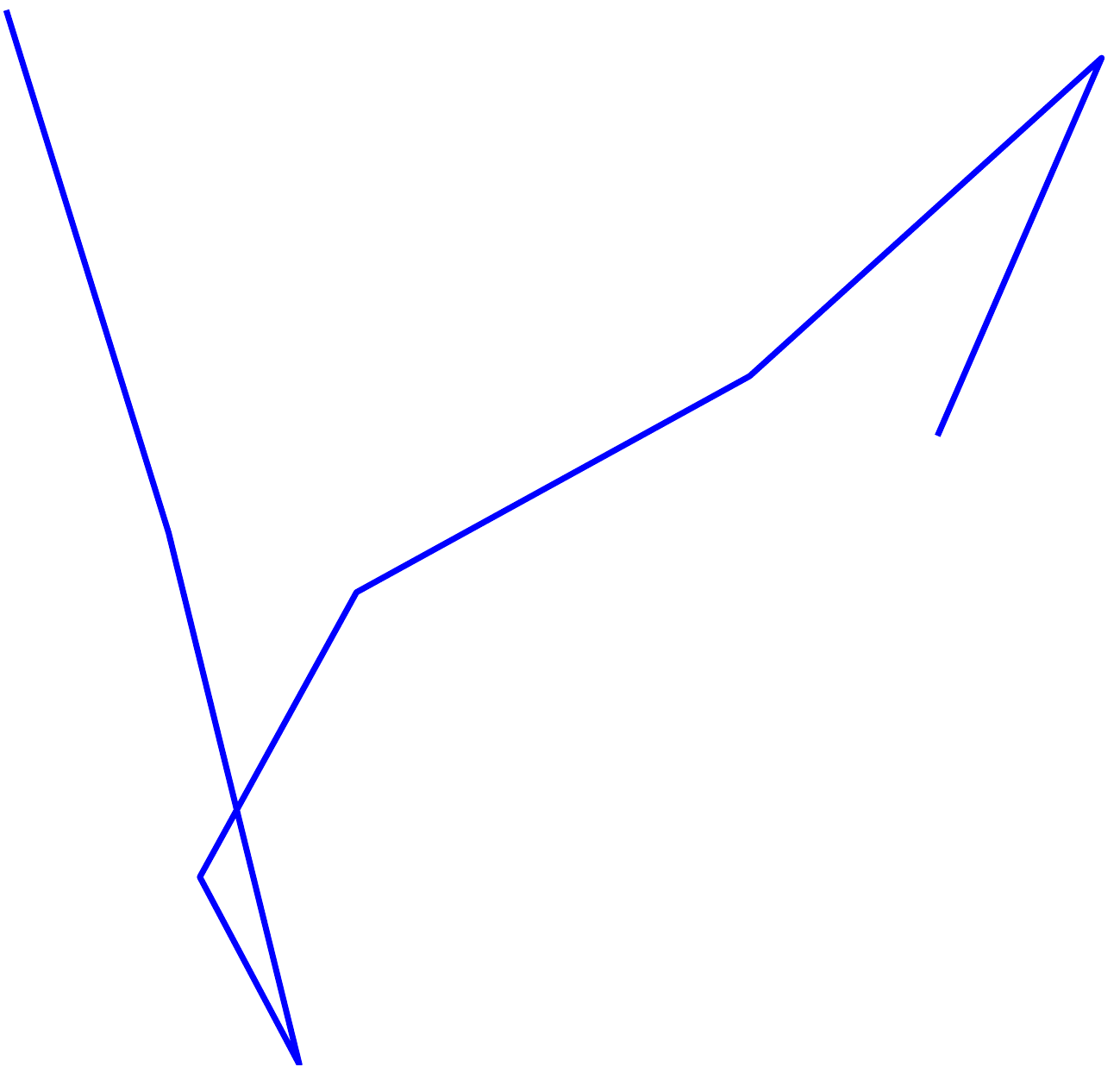}&\includegraphics[width=0.65in]{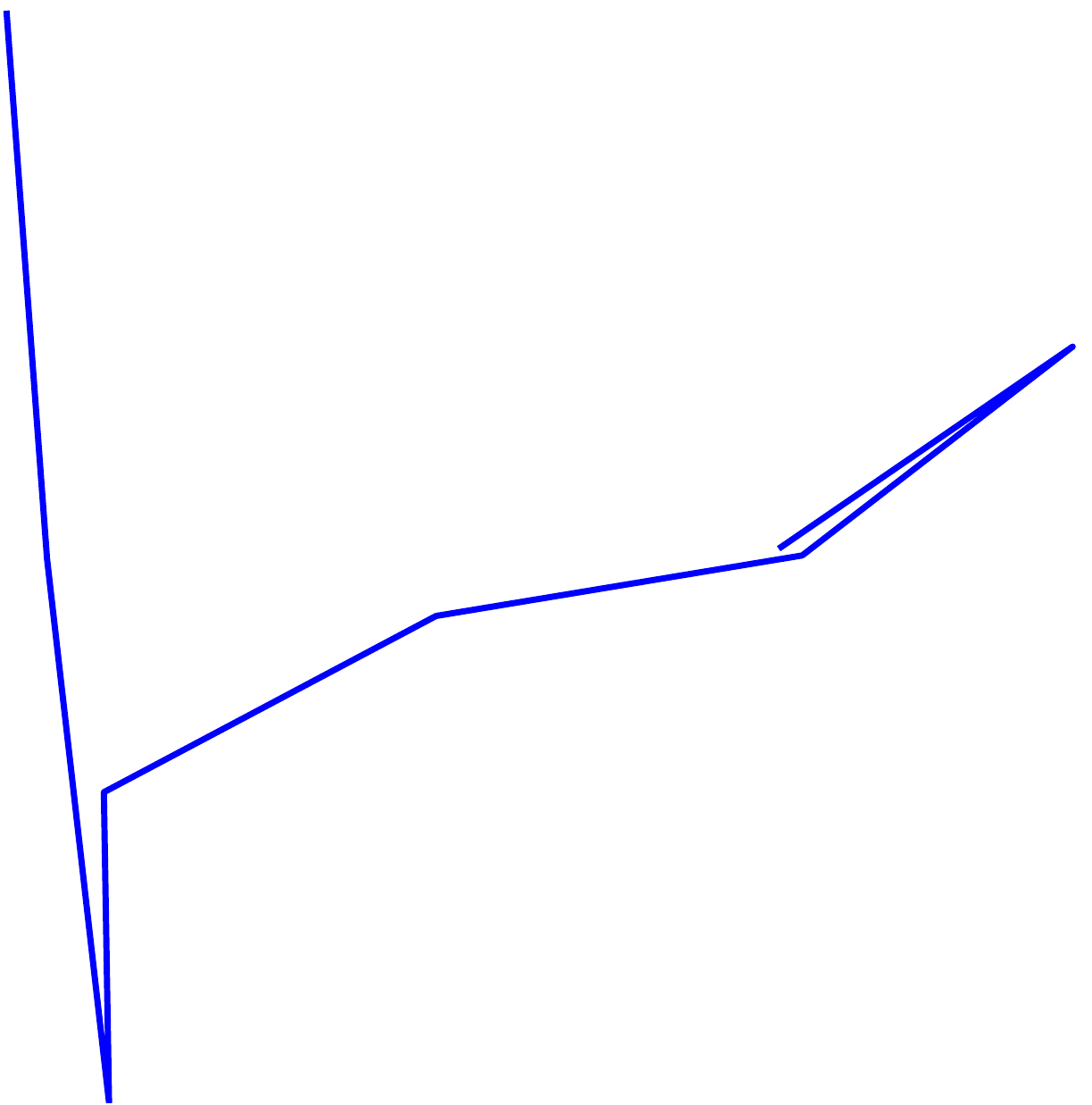}\\
\cline{2-7}
&&&&&&\\
\scalebox{1.3}{$\mathit{4}$}&\includegraphics[width=0.5in]{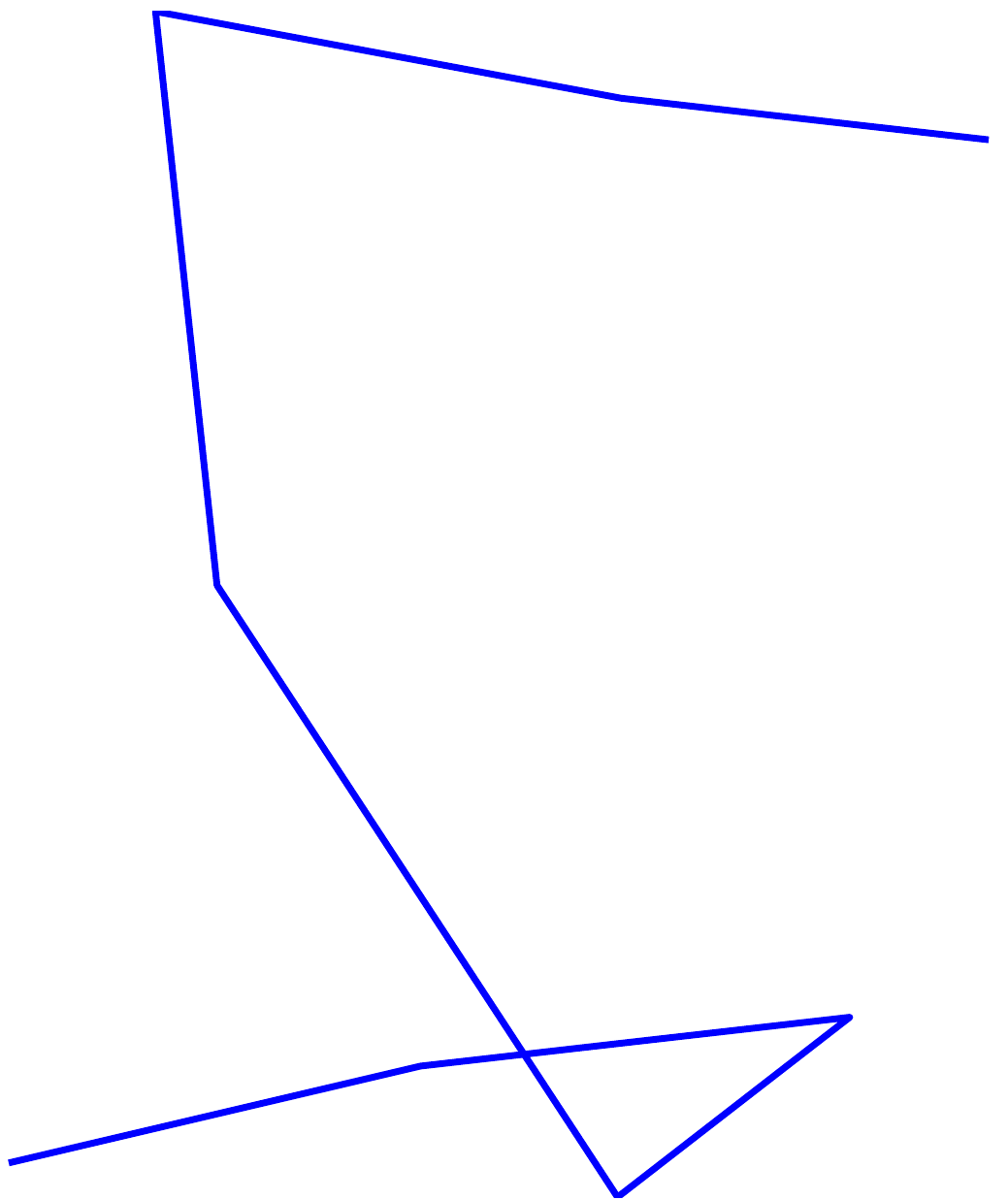}&\includegraphics[width=0.5in]{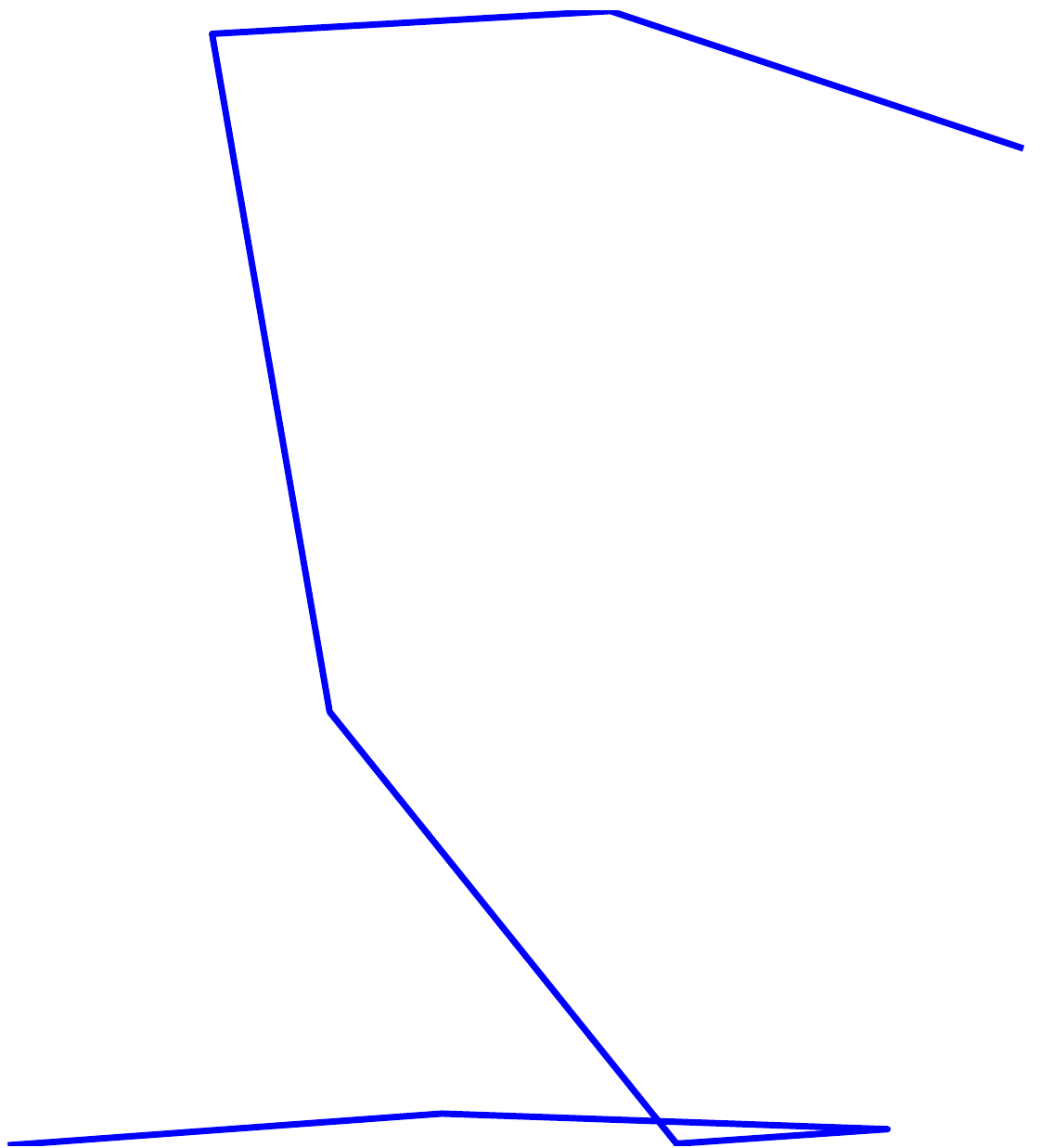}&\includegraphics[width=0.5in]{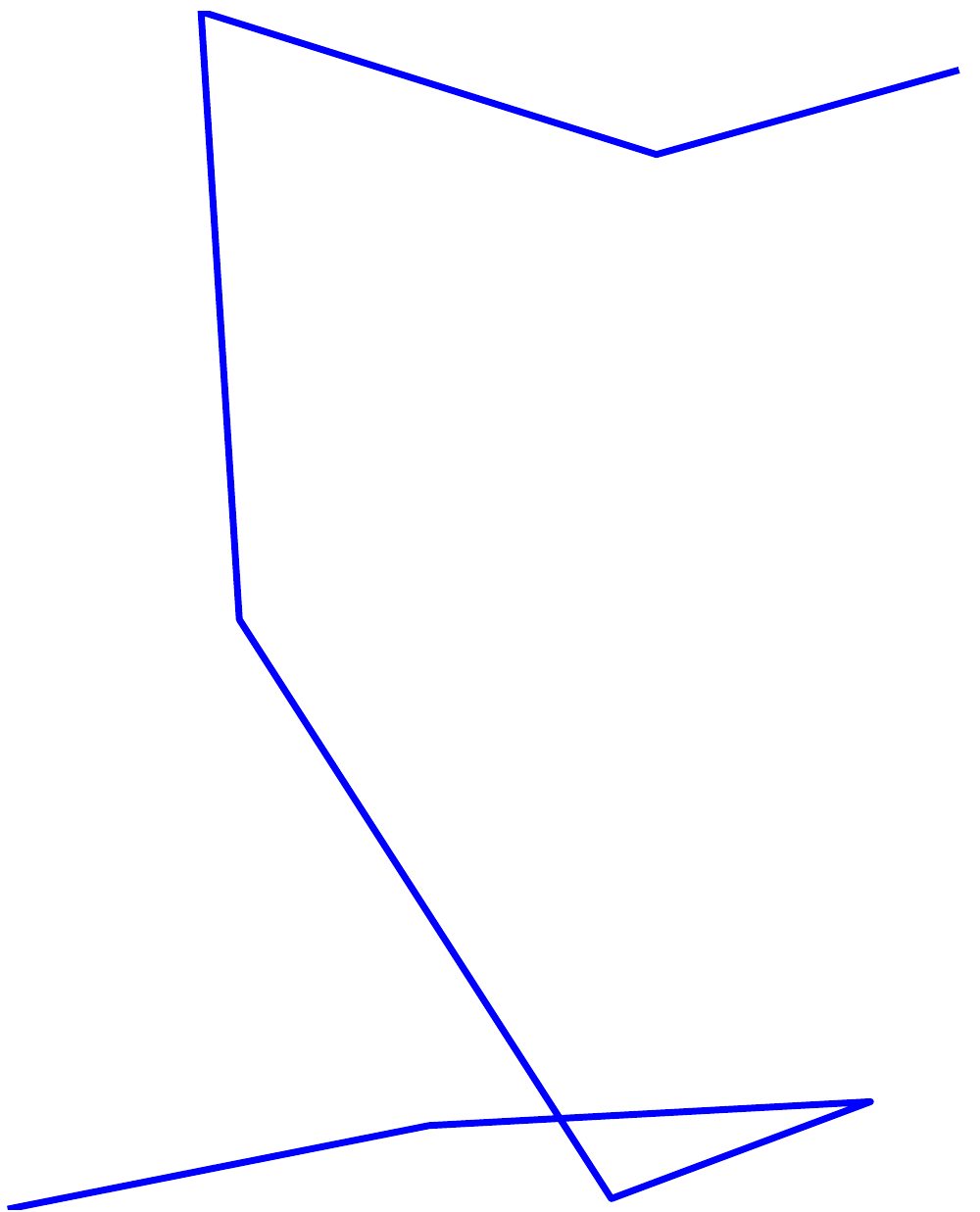}&\includegraphics[width=0.65in]{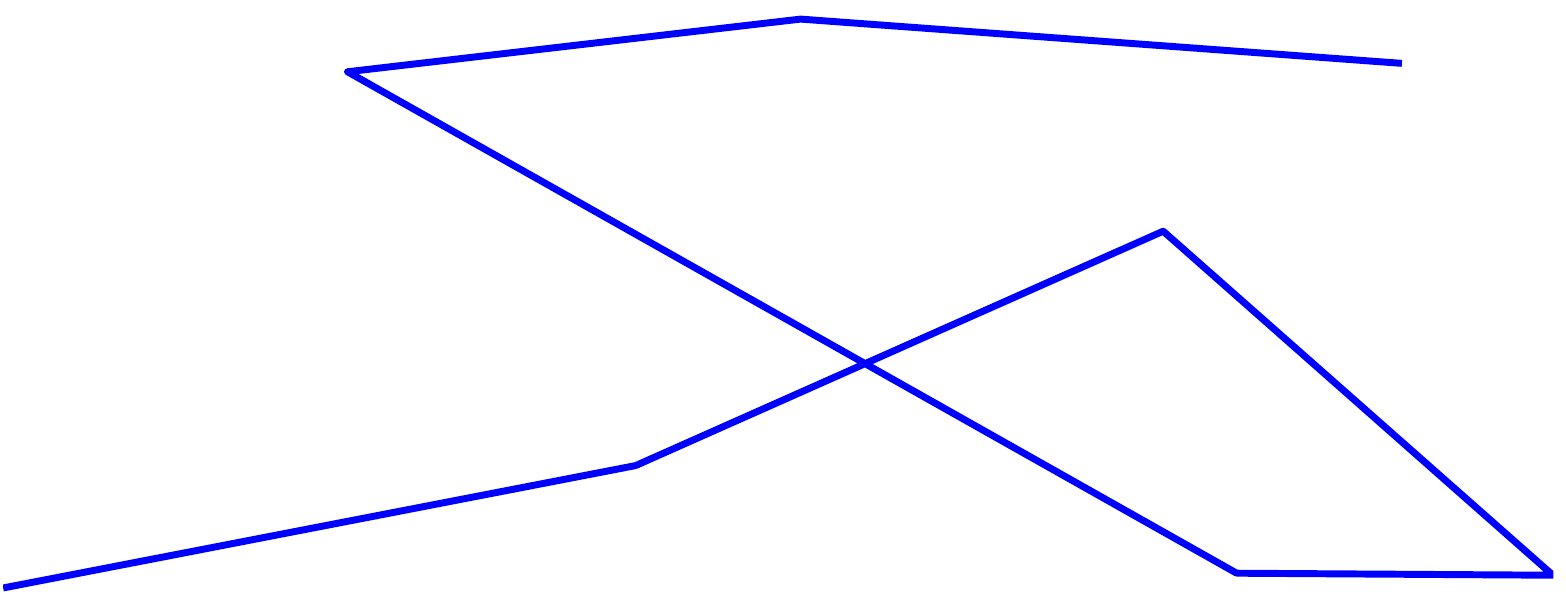}&\includegraphics[width=0.65in]{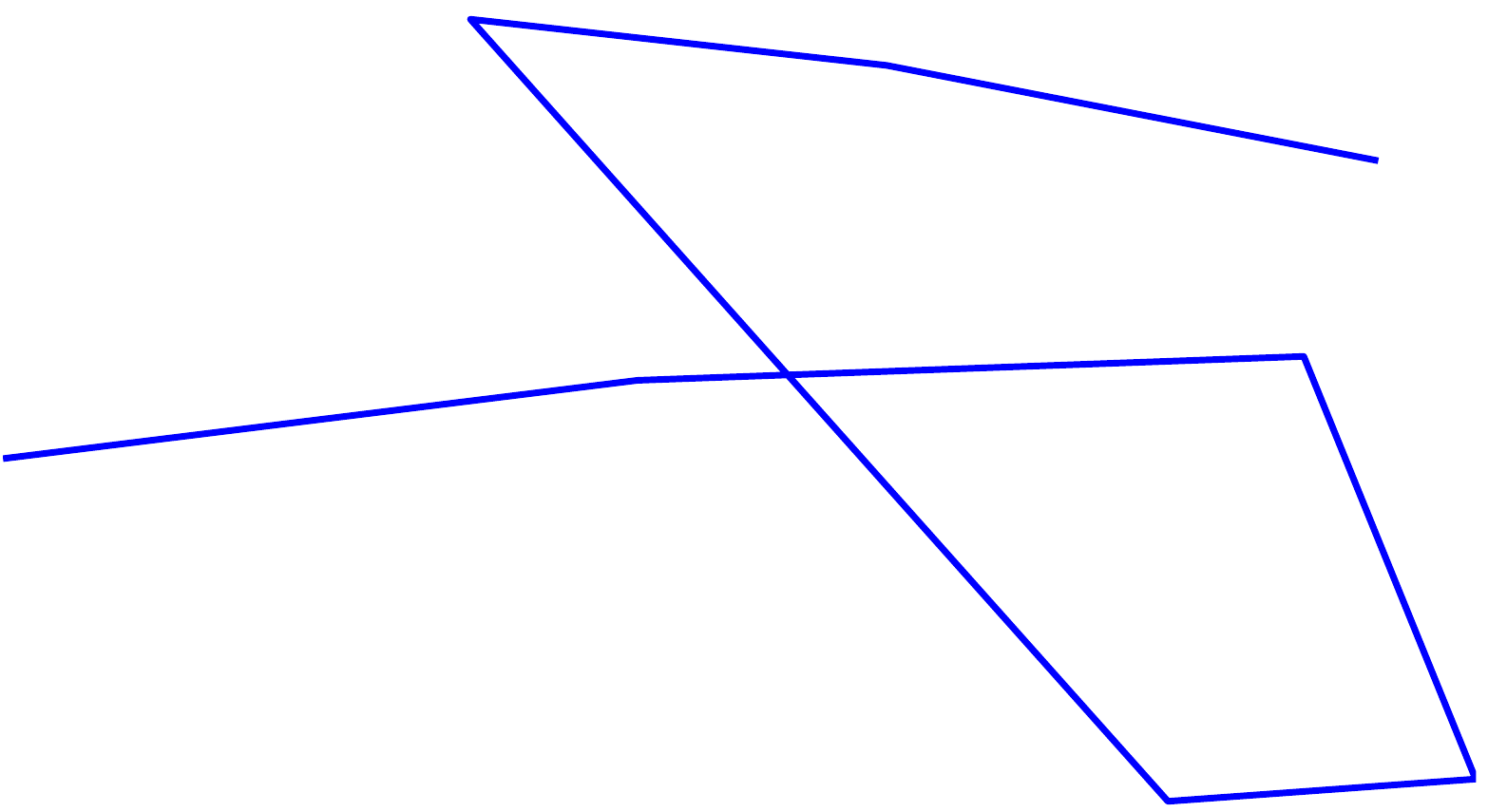}&\includegraphics[width=0.65in]{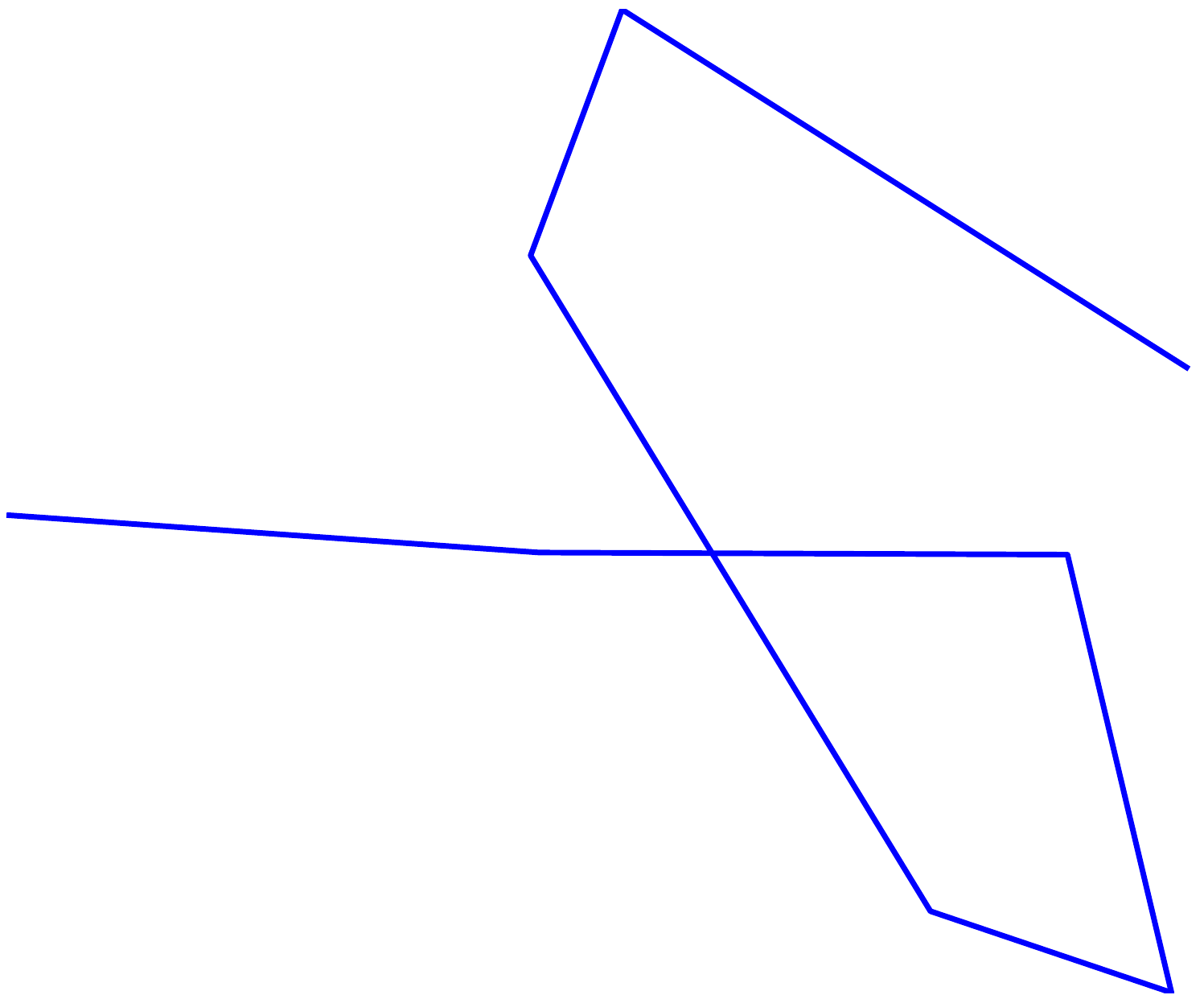}\\
\cline{2-7}
&&&&&&\\
\scalebox{1.3}{$\mathit{6}$}&\includegraphics[width=0.65in]{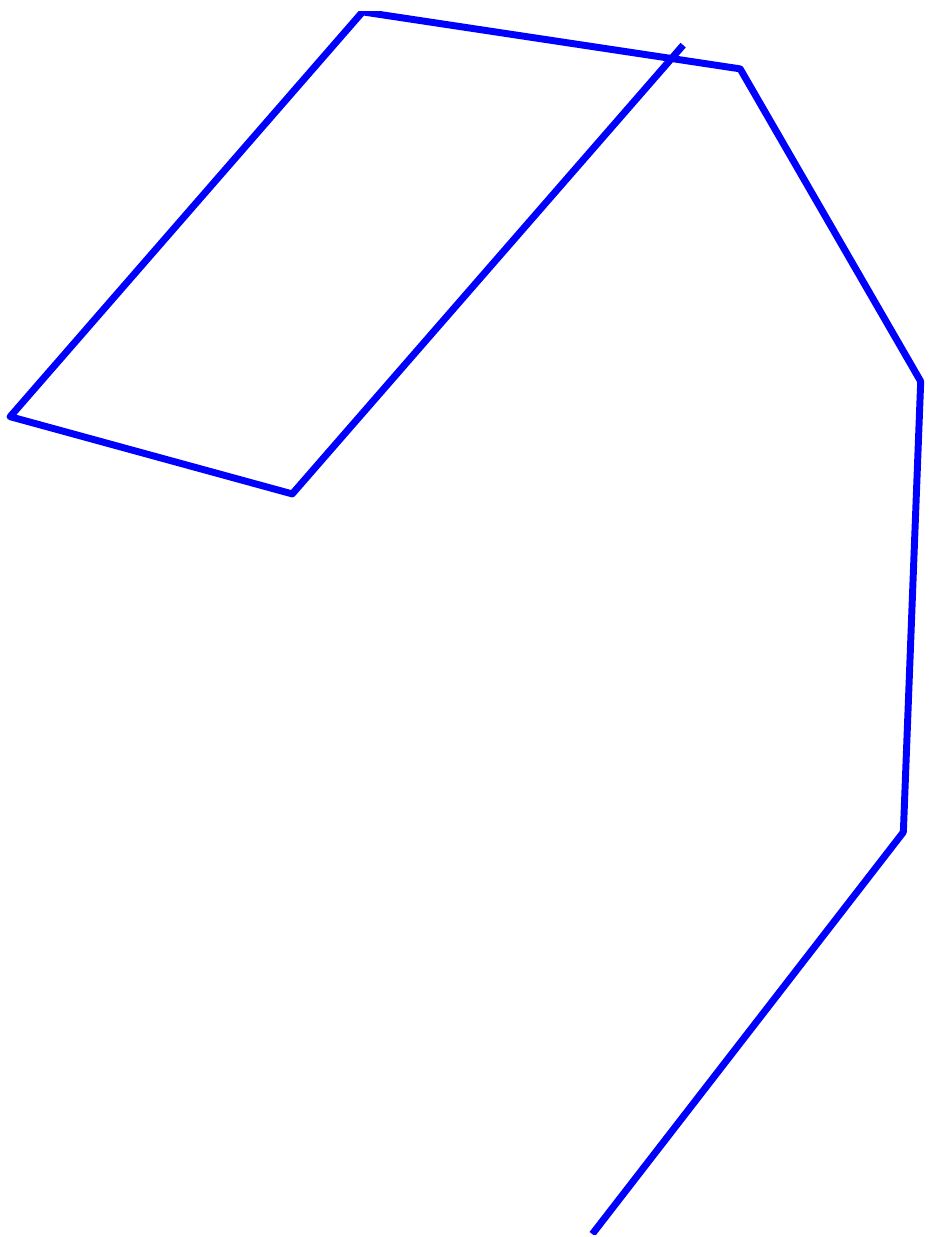}&\includegraphics[width=0.65in]{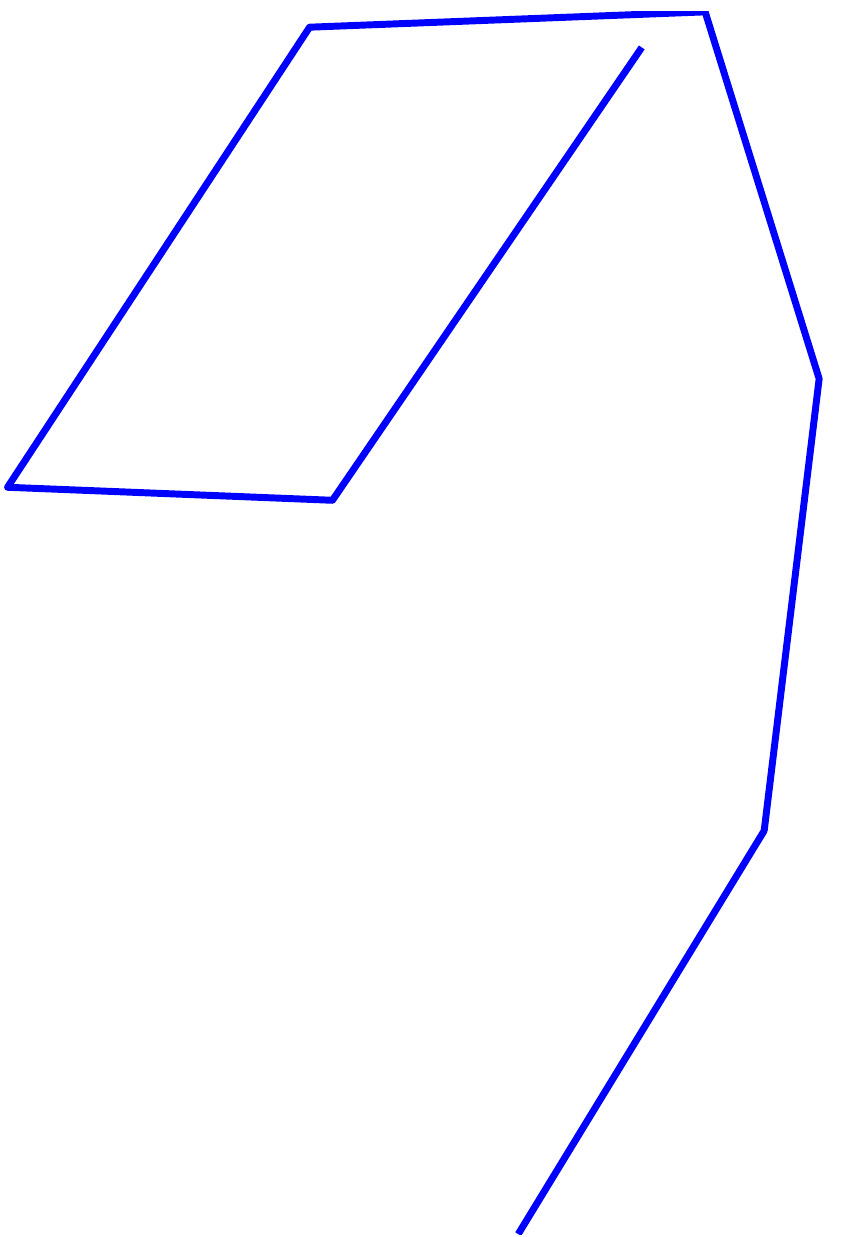}&\includegraphics[width=0.65in]{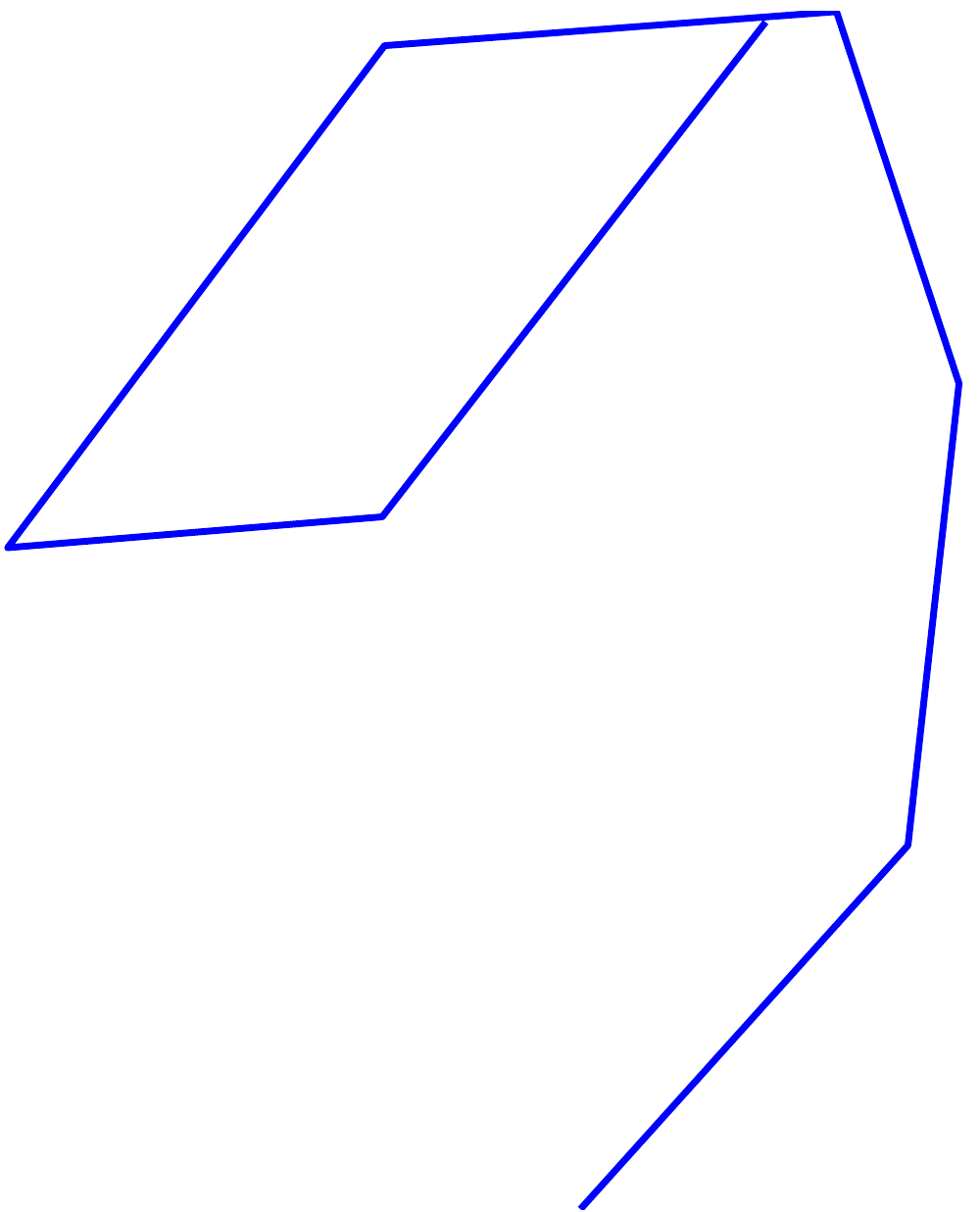}&\includegraphics[width=0.65in]{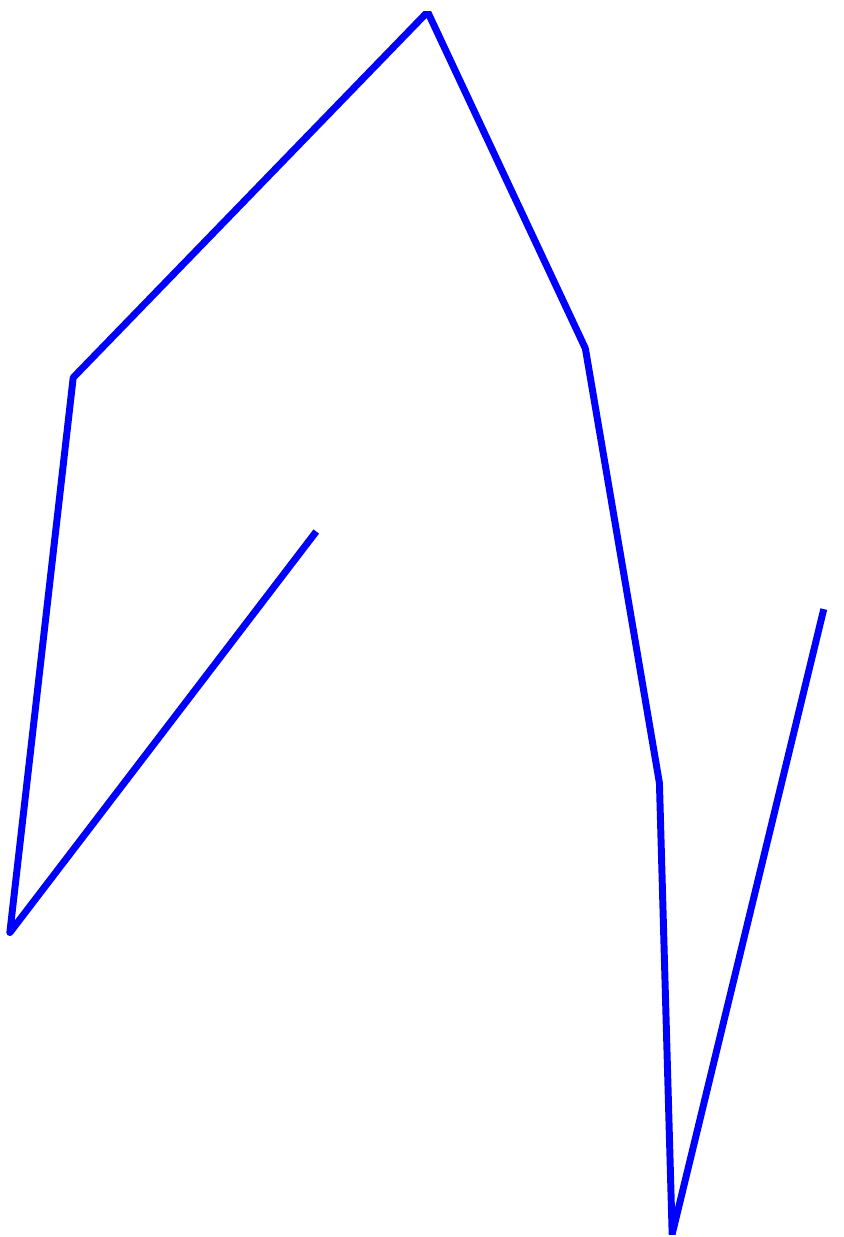}&\includegraphics[width=0.65in]{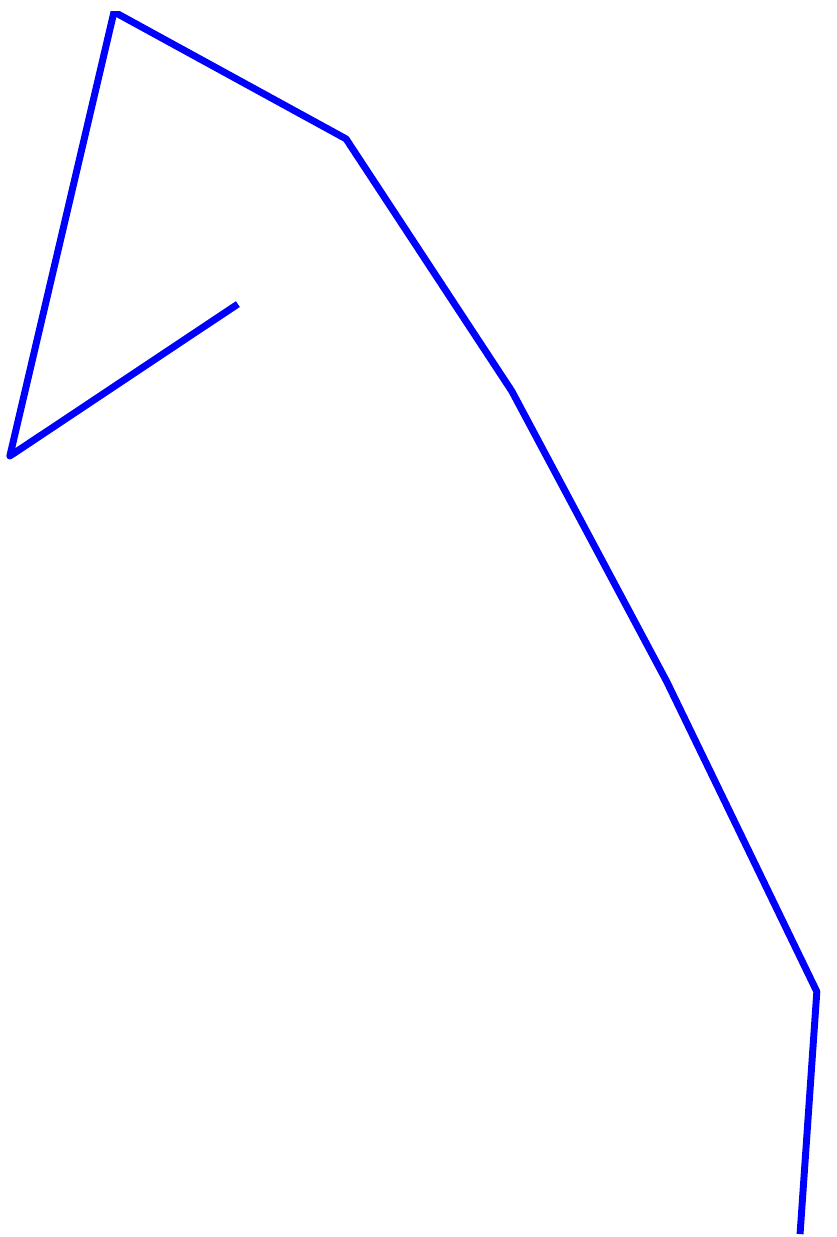}&\includegraphics[width=0.65in]{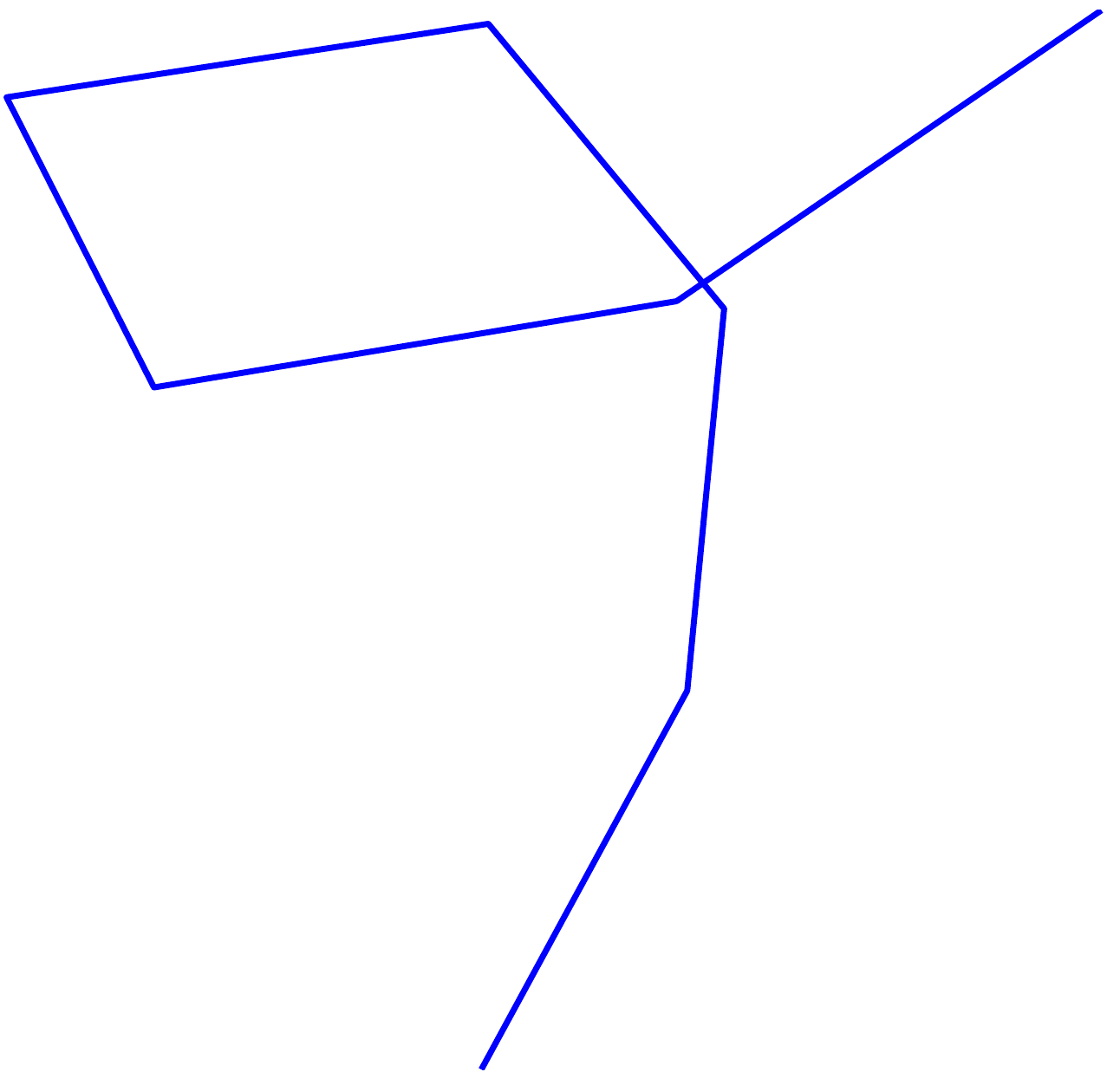}\\
\cline{2-7}
&&&&&&\\
\scalebox{1.3}{$\mathit{7}$}&\includegraphics[width=0.65in]{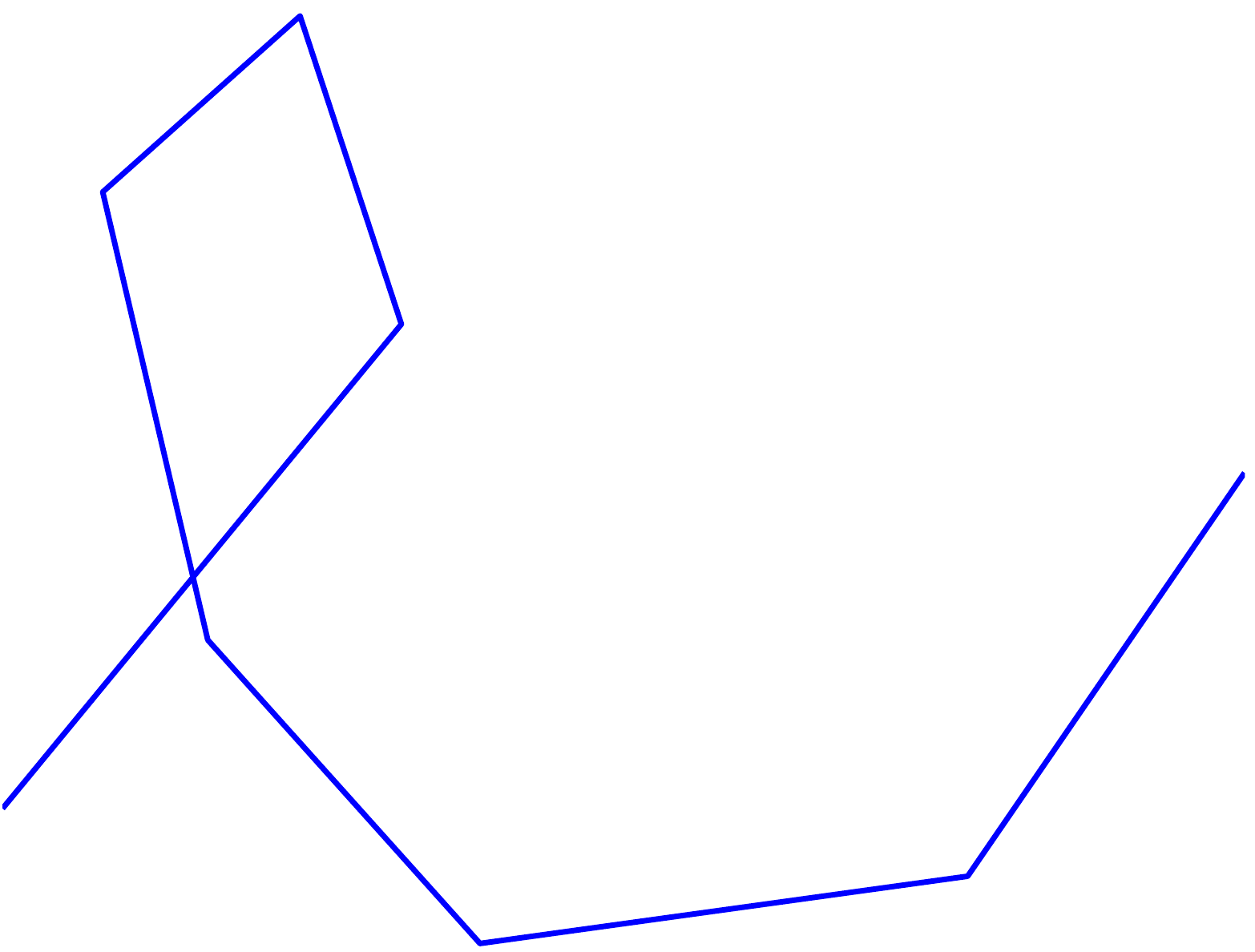}&\includegraphics[width=0.65in]{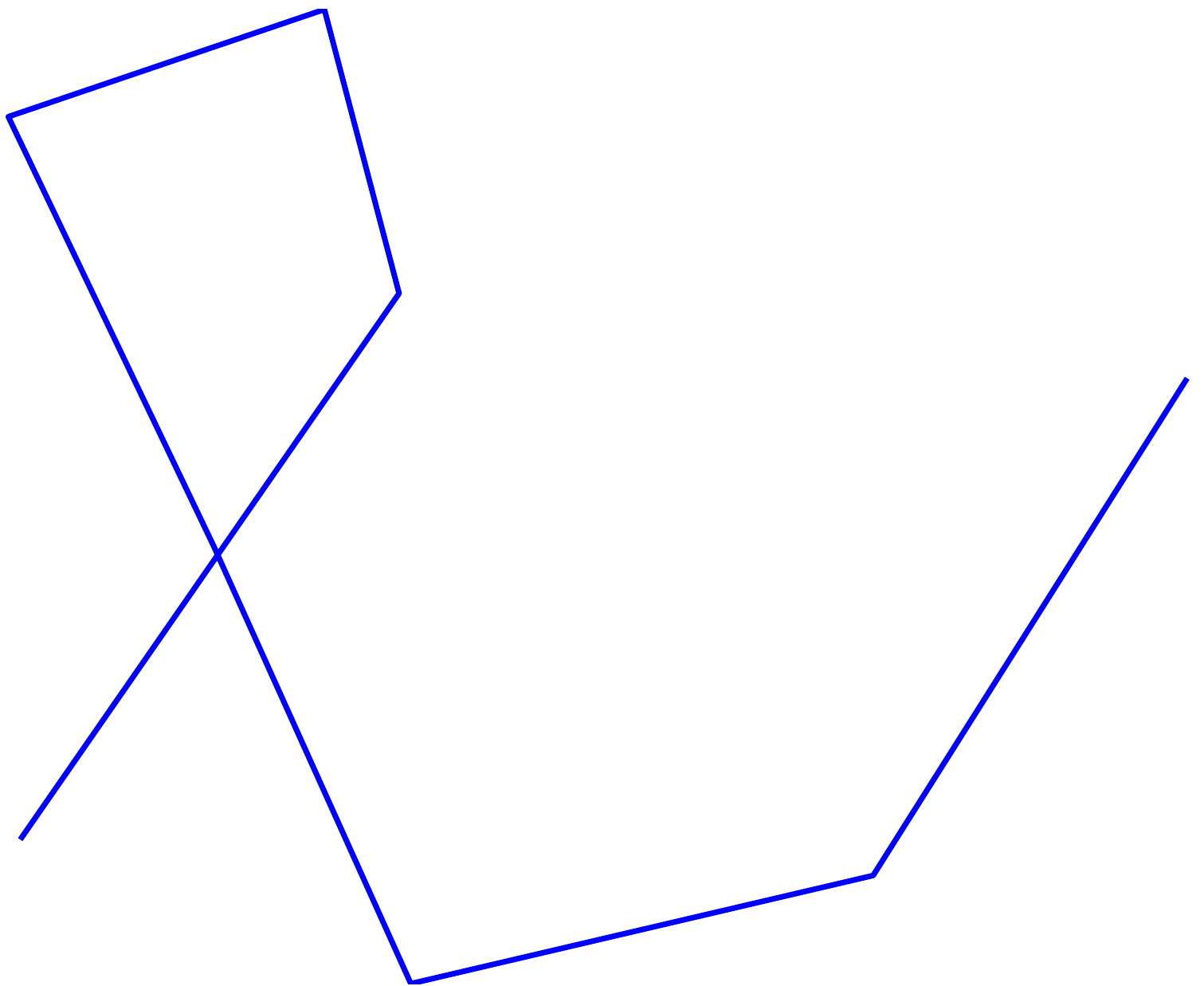}&\includegraphics[width=0.65in]{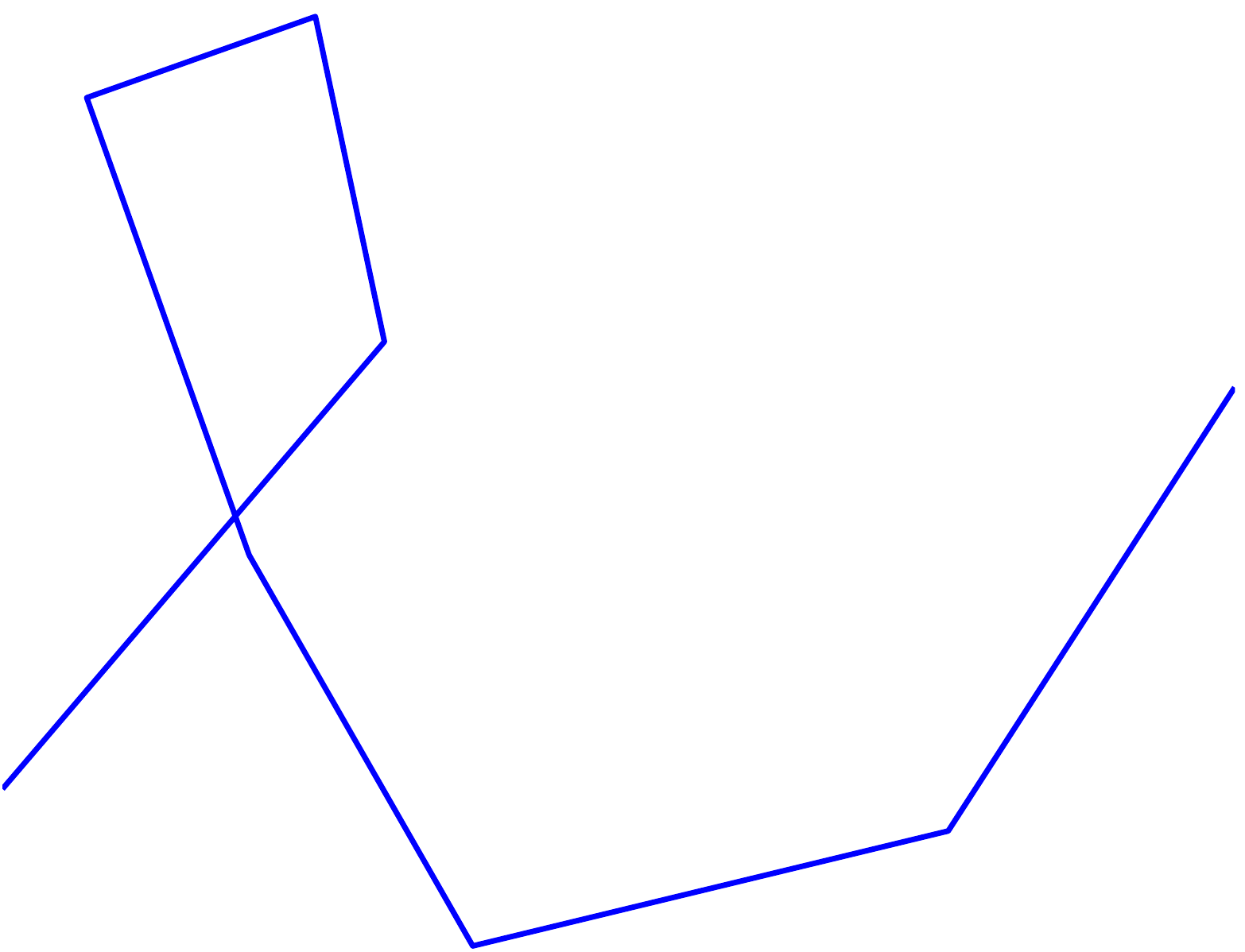}&\includegraphics[width=0.5in]{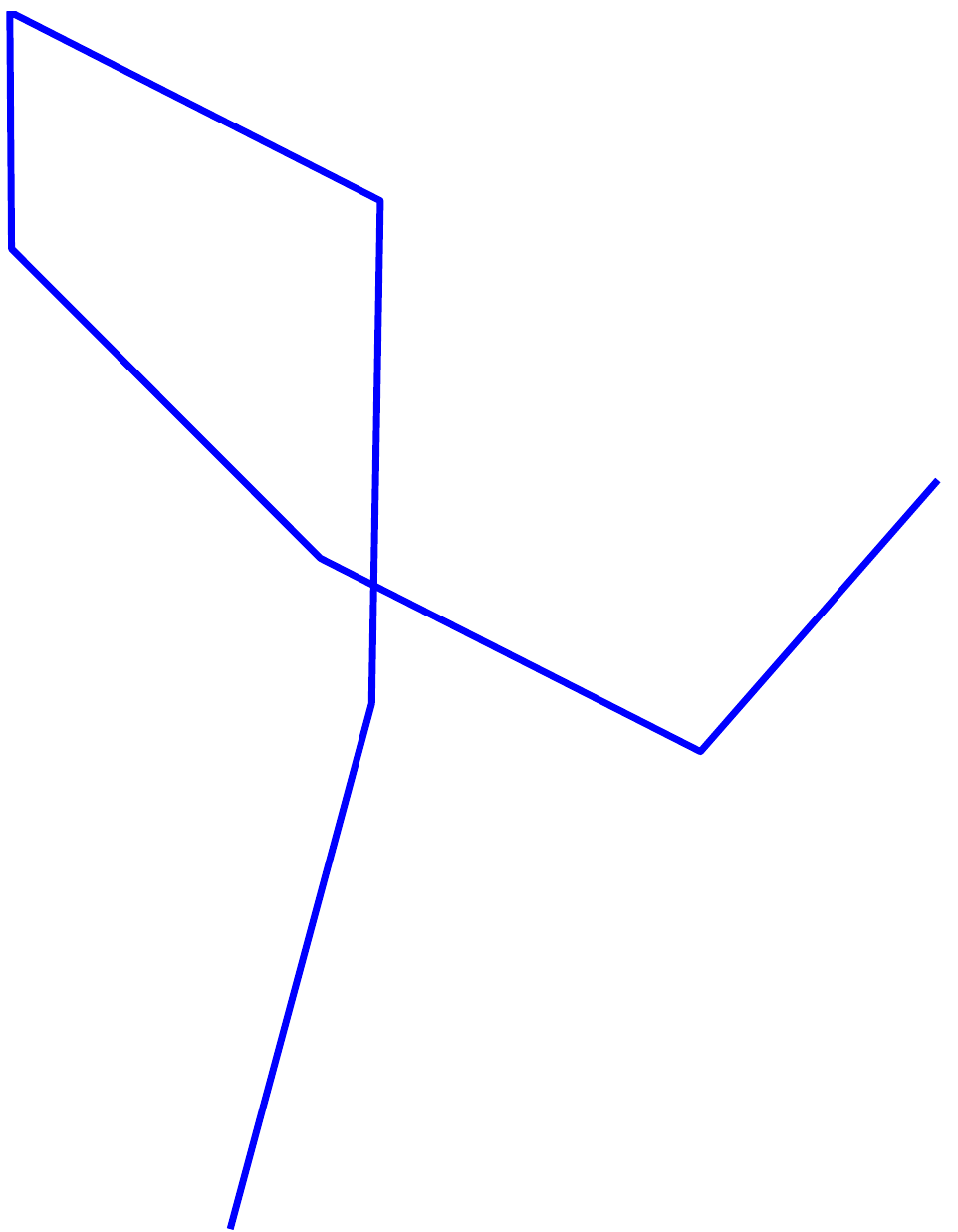}&\includegraphics[width=0.65in]{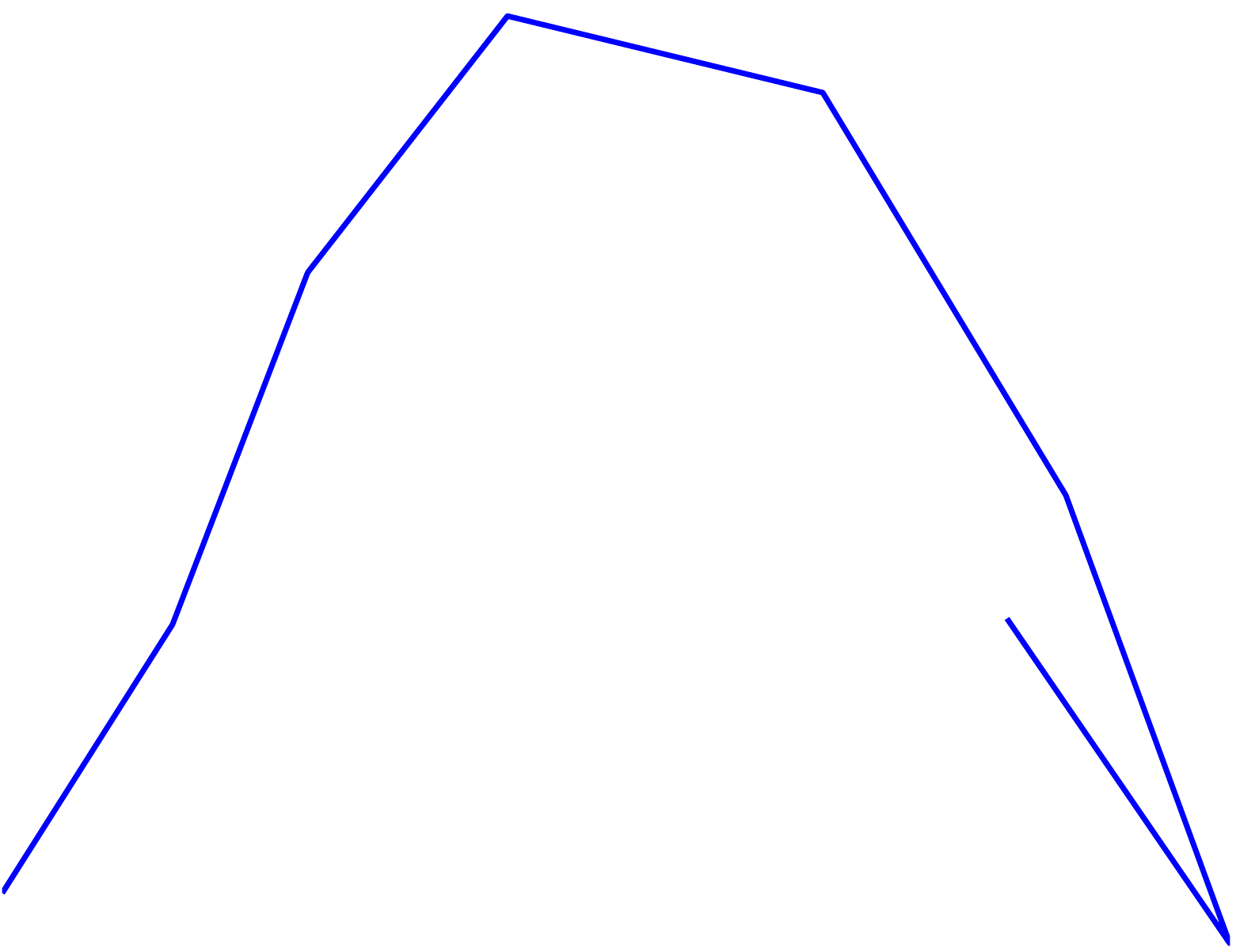}&\includegraphics[width=0.65in]{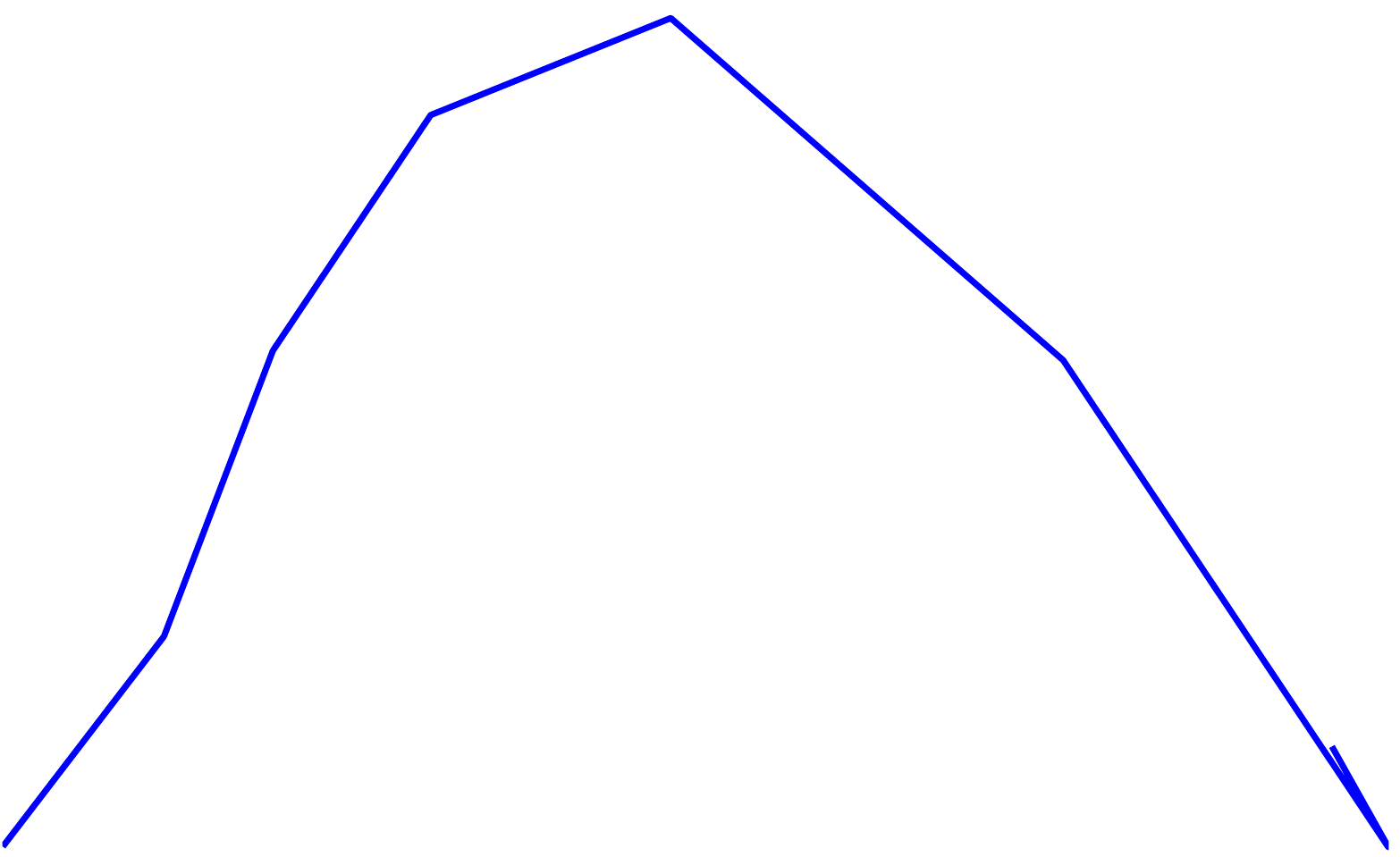}\\
\cline{2-7}
\end{tabular}
}
\caption{Least and most influential digit shapes for digits $\mathit{0, 1, 4, 6, 7}$. Note that the shapes are invariant to rotations. \label{fig:inflobs}}
\end{center}
\end{figure}

\begin{figure}
\begin{center}
\includegraphics[width=1.7in]{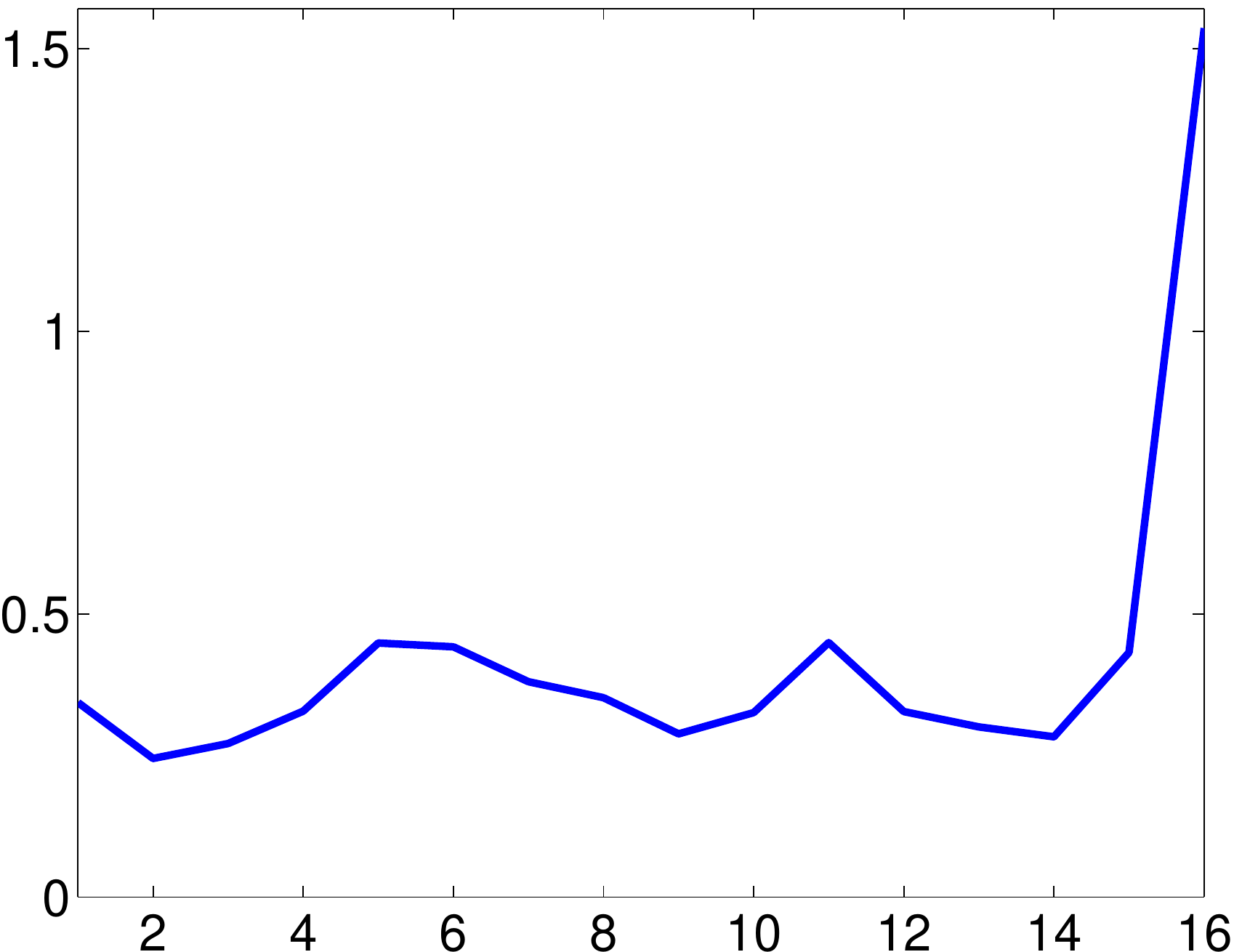}
\caption{Influence measures for 15 shapes of digit $\mathit{0}$ (first 15 cases), and one shape of digit $\mathit{1}$ (16th case).  \label{fig:artexsh}}
\end{center}
\end{figure}

In order to further assess the effectiveness of the proposed influence measure, we generated a dataset consisting of the fifteen least influential shapes for digit $\mathit{0}$ (based on the previous example) and a random shape of digit $\mathit{1}$, and computed the influence measures for this data. We expect the shape of digit $\mathit{1}$ (16th shape) to be highly influential in this case. Figure \ref{fig:artexsh} shows the influence measures of all 16 shapes used in this simulation. While some of the $\mathit{0}$ digit shapes have influence scores close to 0.5, there is one clear highly influential shape in this set with a score close to $\pi/2$, which is the maximum of the scale. To assess the quality of the proposed estimator in this setting, we estimated the influence measure for 50 randomly chosen shapes of digits $\mathit{0, 1, 4, 6, 7}$ based on 50 samples of size 100000 from the baseline posterior. The mean variances (across the 50 randomly chosen shapes) of our estimator were $6.9\times 10^{-8}$ ($\mathit{0}$), $2.3\times 10^{-4}$ ($\mathit{1}$), $1.5\times 10^{-4}$ ($\mathit{4}$), $1.5\times 10^{-5}$ ($\mathit{6}$) and $4.9\times 10^{-4}$ ($\mathit{7}$). These numbers are all very small, indicating that the proposed estimator is appropriate in this setting.

\end{example}

\section{Summary and Future Work}\label{discussion}
In summary, we have proposed a novel approach to Bayesian sensitivity analysis based on the nonparametric Fisher-Rao metric and the square-root representation of densities, using tools from differential geometry. The SRT representation is quite advantageous in practice: there is no requirement to consider only positive densities, and geodesic distances between densities or their estimators (kernel density estimators, for instance) can be calculated despite one or both of them being zero for certain values on their domains. This not the case with the log representation used in \cite{HZ}. Specifically, we defined a new geometric $\epsilon$-contamination class for the likelihood and prior, define local and global sensitivity measures, and considered the task of identifying influential observations under the case-deletion setup. The main advantage of our framework is that all quantities of interest are intrinsic to the space of probability density functions providing a natural scale (upper and lower bound on distances between posteriors) and geometric calibration. We have applied this framework in a number of different settings including simple simulation studies, a generalized linear mixed effects model, directional data analysis, and statistical shape analysis. Through these examples, we have shown the benefits of the proposed methodology.

The natural next step would be to test the effectiveness of our framework under the nonparametric Bayesian setting. The SRT representation, in principle, would make a seamless transition to that setting from the parametric setup since the manifold of parametric densities is a submanifold of $\mathcal{P}$ considered here; expressions for geodesic paths and distances remain unaltered. Also, neighborhoods based on the KL divergence or the Hellinger distance are commonly used while assessing posterior consistency. It would be interesting to examine consistency in a geometric neighborhood such as the one considered in this paper; much work remains to be done in this direction.

When posterior densities are unavailable in closed-form, good estimators of the geodesic distance are imperative. Excepting the setting of influence analysis under case-deletion, this has not been explored in this article and is of importance. Methods of incorporating the calculation of the geodesics into existing MCMC procedures would be greatly beneficial. However, under the parametric setting when the unknown parameter vector is of small dimension, similar to the settings considered in this article, the geodesic distances can be calculated with a fair degree of accuracy.

\section{Acknowledgments}
We are indebted to Steve MacEachern for the encouragement and discussions. We also thank Dipak Dey and Anuj Srivastava for some useful suggestions.

\section*{Appendix}\label{appendix}
\noindent \textbf{SRT representation}:
Let $r$ be a small positive scalar and $\delta p\in T_p({\cal P})$. We begin by computing the differential of the mapping $\phi$, $\phi_*: T_p({\cal P})\to T_{\phi(p)}(\Psi)$:
\begin{equation*}
\phi_*(\delta p)=\frac{d}{dr}\phi(p+r\delta p)\Big|_{r=0}=\frac{d}{dr}\sqrt{p+r\delta p}\Big|_{r=0}=\frac{\delta p}{2\sqrt{p+r\delta p}}\Big|_{r=0}=\frac{\delta p}{2\sqrt{p}}.
\end{equation*}
Plugging this expression into the standard $\ltwo$ metric, for two tangent vectors $\delta p_1,\ \delta p_2\in T_p({\cal P})$, we obtain the following:
\begin{align*}
\langle \phi_*(\delta p_1) , \phi_*(\delta p_2) \rangle = \left\langle\frac{\delta p_1}{2\sqrt{p}} , \frac{\delta p_2}{2\sqrt{p}}\right \rangle = \frac{1}{4}\int_\real \delta p_1(x) \delta p_2(x) (1/p(x)) dx = \frac{1}{4}\langle\langle \delta p_1 , \delta p_2 \rangle\rangle_p.
\end{align*}
\\
\noindent \textbf{Proof of Proposition \ref{local}}:\\
\noindent\emph{Proof of 1}: For notational convenience we use the same notation as given in the main text and set $e(\theta)$ to denote $\exp_{\sqrt{\pi_0}}(\epsilon v_g)(\theta)$, $de(\theta)$ to denote $\frac{d}{d\epsilon}\exp_{\sqrt{\pi_0}}(\epsilon v_g)(\theta)$, and $d^2e(\theta)$ to denote $\frac{d^2}{d\epsilon^2}\exp_{\sqrt{\pi_0}}(\epsilon v_g)(\theta)$. We use the following results:
\begin{align*}
\exp_{\sqrt{\pi_0}}(\epsilon v_g)\Big|_{\epsilon=0}&=\cos(\epsilon\|v_g\|)\sqrt{\pi_0}+\sin(\epsilon\|v_g\|)\frac{v_g}{\|v_g\|}\Big|_{\epsilon=0}=\sqrt{\pi_0};\\
\frac{d}{d\epsilon}\exp_{\sqrt{\pi_0}}(\epsilon v_g)\Big|_{\epsilon=0}&=-\sin(\epsilon\|v_g\|)\sqrt{\pi_0}\|v_g\|+\cos(\epsilon\|v_g\|)v_g\Big|_{\epsilon=0}=v_g;\\
\frac{d^2}{d\epsilon^2}\exp_{\sqrt{\pi_0}}(\epsilon v_g)\Big|_{\epsilon=0}&=-\cos(\epsilon\|v_g\|)\sqrt{\pi_0}\|v_g\|^2-\sin(\epsilon\|v_g\|)v_g\|v_g\|\Big|_{\epsilon=0}=-\sqrt{\pi_0}\|v_g\|^2.
\end{align*}
Then, it is straightforward to show the following result:
$$\frac{d}{d\epsilon}F_{\pi_0,\pi_1}(v_g)\Big|_{\epsilon=0}=\frac{d}{d\epsilon}\frac{m(x|\epsilon g)}{m(x|\pi_1)}|_{\epsilon=0}=2\frac{\int_\Theta f(x|\theta)de(\theta)e(\theta)d\theta}{m(x|\pi_1)}\Big|_{\epsilon=0}=2\frac{\tilde{m}(x|v_g)}{m(x|\pi_1)}.
$$

\noindent\emph{Proof of 2}: Here we use the same notation as in the proof of $\mathit{1}$.
\begin{align*}
&\frac{d}{d\epsilon}F_{\pi_0}(v_g)\Big|_{\epsilon=0}=\frac{d}{d\epsilon}\int_{\Theta}h(\theta)\frac{f(x|\theta)e(\theta)^2}{\int_\Theta f(x|\theta)e(\theta)^2d\theta}d\theta\Big|_{\epsilon=0}\\
&=2\int_\Theta h(\theta)\frac{f(x|\theta)e(\theta)de(\theta)\int_\Theta f(x|\theta)e(\theta)^2d\theta-f(x|\theta)e(\theta)^2\int_\Theta f(x|\theta)e(\theta)de(\theta)d\theta}{(\int_\Theta f(x|\theta)e(\theta)^2d\theta)^2}d\theta\Big|_{\epsilon=0}\\
&=\frac{2}{m(x|\pi_0)}\int_{\Theta}h(\theta)f(x|\theta)\sqrt{\pi_0(\theta)}v_g(\theta)d\theta-\frac{2\tilde{m}(x|v_g)}{m(x|\pi_0)}\int_{\Theta}h(\theta)p_0(\theta|x)d\theta.
\end{align*}

\noindent\emph{Proof of 3}: We note that since we are dealing with infinitesimal quantities, we make a simplification using the local Euclidean structure of $\Psi$, by approximating the arc-length distance using a chord-length distance, which locally are essentially the same (see equation 2.9 in \cite{RK}). An important property of a manifold is its locally Euclidean structure. We we exploit this property in the proof in the sense that the geodesic distance based on the Fisher-Rao metric is locally well approximated by the $\ltwo$ distance (Hellinger distance under our representation) $\Big\|\sqrt{p_0}-\sqrt{\frac{f\exp_{\sqrt{\pi_0}}(\epsilon v_g)^2}{m(x|\epsilon g)}}\Big\|^2$. Therefore,
\begin{align*}
&\frac{d}{d\epsilon}F_{\pi_0}(v_g)\Big|_{\epsilon=0}=\frac{d}{d\epsilon}\int_\Theta\left[\sqrt{p_0(\theta|x)}-\sqrt{\frac{f(x|\theta)e(\theta)^2}{m(x|\epsilon g)}}\right]^2d\theta\Big|_{\epsilon=0}.
\end{align*}
Taking the derivative inside the integral, the RHS is now
\begin{align*}
&=\int_\Theta \left[\sqrt{p_0(\theta|x)}-\sqrt{\frac{f(x|\theta)e(\theta)^2}{m(x|\epsilon g)}}\right]\sqrt{\frac{m(x|\epsilon g)}{f(x|\theta)e(\theta)^2}}\\
&\quad \left[\frac{f(x|\theta)e(\theta)de(\theta)\int_\Theta f(x|\theta)e(\theta)^2d\theta-f(x|\theta)e(\theta)^2\int_\Theta f(x|\theta)e(\theta)de(\theta)d\theta}{m(x|\epsilon g)^2}\right]d\theta\Big|_{\epsilon=0}\\
&=\int_\Theta \left[\sqrt{p_0(\theta|x)}\sqrt{\frac{m(x|\epsilon g)}{f(x|\theta)e(\theta)^2}}-1\right]T(\theta)d\theta\Big|_{\epsilon=0}\\
&=\int_\Theta \left[\sqrt{\frac{p_0(\theta|x)}{p_0(\theta|x)}}-1\right]\left[\frac{f(x|\theta)\sqrt{\pi_0(\theta)}v_g(\theta)-p_0(\theta|x)\tilde{m}(x|v_g)}{m(x|\pi_0)}\right]d \theta=0,\\
\end{align*}
where $$T(\theta)=\left[\frac{f(x|\theta)e(\theta)de(\theta)\int_\Theta f(x|\theta)e(\theta)^2d\theta-f(x|\theta)e(\theta)^2\int_\Theta f(x|\theta)e(\theta)de(\theta)d\theta}{m(x|\epsilon g)^2}\right].$$
This result is expected since the distance is minimized at 0. This compels us to consider the second derivative to obtain a finer measure. The second derivative with respect to $\epsilon$ is
\begin{align*}
\frac{d^2}{d\epsilon^2}F_{\pi_0}(v_g)\Big|_{\epsilon=0}&=\frac{d^2}{d\epsilon^2}\int_\Theta\left[\sqrt{p_0(\theta|x)}-\sqrt{\frac{f(x|\theta)e(\theta)^2}{m(x|\epsilon g)}}\right]^2d\theta\Big|_{\epsilon=0}\\
&= 2\int_\Theta \sqrt{\frac{m(x|\pi_0)e(\theta)^2}{m(x|\epsilon g)\pi_0(\theta)}}\Big[\frac{m(x|\pi_0)\pi_0(\theta)e(\theta)^2\int_\Theta f(x|\theta)e(\theta)de(\theta)d\theta}{m(x|\pi_0)^2e(\theta)^4}\\
&\quad -\frac{m(x|\pi_0)\pi_0(\theta)e(\theta)de(\theta)m(x|\epsilon g)}{m(x|\pi_0)^2e(\theta)^4}\Big]T(\theta)d\theta\\
&\quad +2\int_\Theta\left[\sqrt{\frac{m(x|\epsilon g)\pi_0(\theta)}{m_0(x|\pi_0)e(\theta)^2}}-1\right]\frac{dT(\theta)}{d\epsilon}d\theta\Big|_{\epsilon=0}.\\
\end{align*}
Further calculations yield the RHS to be
\begin{align*}
&= 2\int_\Theta \left[\sqrt{\frac{m(x|\pi_0)\pi_0(\theta)}{m(x|\pi_0)\pi_0(\theta)}}\frac{\pi_0(\theta)\tilde{m}(x|v_g)-\sqrt{\pi_0(\theta)}v(\theta)m(x|\pi_0)}{m(x|\pi_0)\pi_0(\theta)}\right]\\
&\qquad \left[\frac{f(x|\theta)\sqrt{\pi_0(\theta)}v_g(\theta)m(x|\pi_0)-f(x|\theta)\pi_0(\theta)\tilde{m}(x|v_g)}{m(x|\pi_0)^2}\right]d\theta\\
&\qquad +2\int_\Theta\left[\sqrt{\frac{m(x|\epsilon g)\pi_0(\theta)}{m_0(x|\pi_0)e(\theta)^2}}-1\right]\frac{dT(\theta)}{d\epsilon}d\theta\Big|_{\epsilon=0}\\
&=4\frac{\tilde{m}(x|v_g)}{m(x|\pi_0)}\int_\Theta\frac{v_g(\theta)}{\sqrt{\pi_0(\theta)}}p_0(\theta|x)d\theta-2\int_\Theta\frac{v_g(\theta)^2}{\pi_0(\theta)}p_0(\theta|x)d\theta-\frac{2\tilde{m}(x|v_g)^2}{m(x|\pi_0)^2}.
\end{align*}

\noindent \textbf{Proof of Proposition \ref{MCMC}}:\\
A key observation here is that under the case deletion setup, the prior on the parameters does not change. Thus, we have the following:
\begin{align*}
&d_{FR}(p_0,p_k)=\int_{\Theta}\sqrt{p_k(\theta|x)}\sqrt{p_0(\theta|x)}d\theta=\int_{\Theta}\sqrt{p_k(\theta|x)}\frac{1}{\sqrt{p_0(\theta|x)}}p_0(\theta|x)d\theta\\
&=\int_{\Theta}\sqrt{\frac{f_k(x|\theta)\pi(\theta)}{\int_{\Theta}f_k(x|\theta)\pi(\theta)d\theta}}\sqrt{\frac{\int_{\Theta}f_0(x|\theta)\pi(\theta)d\theta}{f_0(x|\theta)\pi(\theta)}}p_0(\theta|x)d\theta\\
&=\int_{\Theta}\sqrt{\frac{f_k(x|\theta)}{f_0(x|\theta)}}\sqrt{\frac{\int_{\Theta}f(y|\theta)\pi(\theta)d\theta}{\int_{\Theta}f_k(y|\theta)\pi(\theta)d\theta}}p(\theta|y)d\theta\\
&=\int_{\Theta}\sqrt{\frac{f_k(x|\theta)}{f_0(x|\theta)}}\Big[\int_{\Theta}\frac{1}{f(x_k|x_{(k)},\theta)}p_0(\theta|x)d\theta\Big]^{-1/2}p_0(\theta|x)d\theta.
\end{align*}

\bibliography{biblio}
\bibliographystyle{plainnat}
\end{document}